\documentclass[a4paper,english,english,12pt,twoside]{report}
\usepackage[T1]{fontenc}
\usepackage[latin1]{inputenc}
\usepackage{babel}
\usepackage{geometry}
\usepackage{setspace}
\usepackage{epsfig}
\usepackage{graphics}
\usepackage{amsmath}
\usepackage{amssymb}

\newcommand{\Niezurawski}{Nie\.zurawski}
\newcommand{\Zarnecki}{\.Zarnecki}

\def\figheight{0.7\textwidth} 
\def\figheightsmall{0.35\textwidth} 
\def\figheightsmaller{0.2\textwidth} 
\def\twofigheight{0.5\textwidth} 

\def\eg{\textit{e.g.\ }}
\def\ie{\textit{i.e.\ }}
\def\etal{\textit{et al.}}

\def\bc{\begin{center}}
\def\ec{\end{center}}

\def\bmp{\begin{minipage}}
\def\emp{\end{minipage}}

\def\ar{\rightarrow}
\def\ga{\gamma}
\def\gaga{\ga\ga}

\def\Qbar{\bar{Q}}
\def\QQbar{Q\Qbar}
\def\QQbarg{\QQbar (g)}

\def\qbar{\bar{q}}
\def\qqbar{q\qbar}
\def\ubar{\bar{u}}
\def\uubar{u\ubar}

\def\bbar{\bar{b}}
\def\cbar{\bar{c}}

\def\xs{\sigma}

\def\bbbar{ b\bbar }
\def\ccbar{ c\cbar }

\def\bbg{ \bbbar (g)}
\def\ccg{ \ccbar (g)}

\def\hSM{h}

\def\hbb{ \hSM \ar \bbbar }
\def\gagah{ \gaga \ar \hSM }
\def\gagahbb{ \gaga \ar \hbb }

\def\higgs{\textit{higgs}}
\def\higgses{\textit{higgses}}

\def\higgsm{\mathit{higgs}}

\def\higgsbb{ \higgsm \ar \bbbar }
\def\higgsgg{ \higgsm \ar g g }
\def\gagahiggsbb{ \gaga \ar \higgsbb }

\def\Dss{\Delta \xs / \xs }

\def\Brhiggsbb{{\rm BR}(\higgsbb)}

\def\gagaQQ{ \gaga \ar \QQbar }
\def\gagaQQg{ \gagaQQ (g) }
\def\Qcb{ Q \! = \! c,b }

\def\gagaqq{ \gaga \ar \qqbar }
\def\quds{ q \! = \! u,d,s }

\def\gagabb{ \gaga \ar \bbbar }

\def\gagacc{ \gaga \ar \ccbar }

\def\gagabbcc{ \gagabb , \, \ccbar }

\def\gagabbg{ \gagabb (g) }

\def\gagaccg{ \gagacc (g) }

\def\gagabbgccg{ \gagabbg , \, \ccbar(g) }

\def\sgagahbb{ \sigma( \gagahbb ) }

\def\hgaga{ \hSM \ar \gaga  }

\def\Gh{ \Gamma_{\hSM}}

\def\Brhgaga{{\rm BR}(\hgaga)}
\def\Brhbb{{\rm BR}(\hbb)}
\def\Brhbbinclusive{{\rm BR}(\hbb + \dots)}

\def\GhBrhgaga{\Gh \, \Brhgaga}

\def\Ghgaga{ \Gamma ( \hgaga ) }
\def\Ghgagahbb{ \Ghgaga \Brhbb }

\def\phihgaga{\phi_{h\gaga}}

\def\WW{W^{+} W^{-}}

\def\gagaWW{\gaga \ar \WW}

\def\Wgaga{W_{\gaga}}

\def\epem{ e^{+} e^{-} }
\def\emem{ e^{-} e^{-} }

\def\Mh{M_h}
\def\Mheq{$ \Mh = $ }

\def\Mhiggs{M_{\higgsm}}

\def\sqrtsee{ \sqrt{s_{ee}} }
\def\sqrtseeeq{$ \sqrtsee  = $ }

\def\sqrtspee{ \sqrt{s'_{ee}} }

\def\Lgaga{ L_{\gaga} }

\def\Emf{{E}^{\max{\!1}}_{\ga}}

\def\Wgagamf{\Wgaga^{\max{\!1}}}
\def\Wgagamin{\Wgaga^{\min}}

\newcommand{\gagahad}{\( \gaga \ar  \mathit{hadrons} \)}

\newcommand{\higgstagging}{\textit{higgs}-tagging}

\newcommand{\bbtagging}{\( \bbbar  \)-tagging}
\newcommand{\btagging}{\( b \)-tagging}

\newcommand{\ccmistagging}{\( \ccbar  \)-mistagging}

\def\AO{A}
\def\HO{H}

\def\MAO{ M_{\AO} }
\def\MHO{ M_{\HO} }

\def\MAOeq{$ \MAO = $ }

\def\AOHO{ \AO,\HO }
\def\MAOHO{ M_{\AOHO} }

\def\AHbb{ \AOHO \ar \bbbar }
\def\Abb{ \AO \ar \bbbar }
\def\Hbb{ \HO \ar \bbbar }

\def\gagaAH{ \gaga \ar \AOHO }

\def\gagaAbb{ \gaga \ar \Abb }
\def\gagaHbb{ \gaga \ar \Hbb }
\def\gagaAHbb{ \gaga \ar \AHbb }

\def\sgagaAHbb{ \sigma (\gagaAHbb)}

\def\DssgagaAHbb{ \Delta \sgagaAHbb / \sgagaAHbb }

\def\tanb{\tan \beta}
\def\tanbeq{$\tanb = $ }
\def\tbseven{\tanb = 7}

\def\HOgaga{\HO \ar \gaga}

\def\HObb{\HO \ar \bbbar}

\def\HOgg{\HO \ar gg}

\def\BrAOHObb{{\rm BR}(\AO / \HObb)}
\def\BrAOHOgaga{{\rm BR}(\AO / \HOgaga)}
\def\BrAOHOgg{{\rm BR}(\AO / \HOgg)}

\def\tautau{\tau^{+}\tau^{-}}
\def\gagatautau{\gaga \ar \tautau}

\def\thetamask{ \theta_{\mathit{mask}} }

\def\thetamindet{ \theta_{\mathnormal{\,TC}} }
\def\costhmindet{ \cos\thetamindet }
\def\thetamindeteq{$ \thetamindet = $ }
\def\costhmindeteq{$ \costhmindet = $ }
\def\ptjet{p_{T}^{jet}}
\def\ptjetET{\ptjet/E_{T}}

\def\Cct{\mathcal{C}_{\theta}}
\def\Cpz{\mathcal{C}_{P_z}}
\def\Cmj{\mathcal{C}_{M_{jet}}}
\def\Cec{\mathcal{C}_{E_{TC}}}
\def\Cnt{\mathcal{C}_{N_T}}

\newcommand{\pnfiggeneral}[5]{
\begin{figure}[#1]
{\centering \resizebox*{!}{#2}%
{#3} \par}
 
\caption{\label{#4}
#5
}
\end{figure}
}

\newcommand{\pnfig}[5]{
\pnfiggeneral{#1}{#2}{\includegraphics{#3}}{#4}{#5}
}

\def\Pythia{\textsc{Pythia}}

\def\Simdet{\textsc{Simdet}}
\def\Siver{4.01}

\def\Brahms{\textsc{Brahms}}

\def\Hdecay{\textsc{Hdecay}}

\def\CompAZ{\textsc{Comp\hspace*{-0.2ex}AZ}}

\def\Orlop{\textsc{Orlop}}

\def\ZBHT{\textsc{Zvtop-B-Hadron-Tagger}}

\title{Higgs-boson production \\ at the Photon Collider at TESLA}

\author{ Piotr \Niezurawski{} \\
         Institute of Experimental Physics, Warsaw University \\
         ul. Ho\.za 69, 00-681 Warsaw, Poland }

\begin{document}


\thispagestyle{empty}

\begin{center} 

\vspace*{1cm}
\textbf{  Warsaw University \\ Faculty of Physics \\ Institute of Experimental Physics }

\vspace*{3cm}
\textbf{\textsc{\Large Higgs-boson production \\ at the Photon Collider at TESLA } }

\vspace*{1cm}
Piotr \Niezurawski{}

\vspace*{1cm}
Thesis submitted to the Warsaw University \\
in partial fulfillment of the requirements \\
for the Ph.\ D.\ degree in Physics. \\
Prepared under supervision \\ 
of Dr.\ hab.\ Aleksander Filip \Zarnecki{}.

\vspace*{3cm}
Warsaw 2005

\end{center}




\newpage 
\thispagestyle{empty}
 \
\newpage 
\thispagestyle{empty}

\vspace*{0.80\textheight}
\noindent \begin{flushright}
 \emph{What exists is beyond reach and very deep.} \\ 
 \emph{Who can discover it?} \\
 \emph{\small Ecclesiastes 7,24}
\end{flushright}

\newpage 
\thispagestyle{empty}
 \


\begin{abstract}

In this thesis feasibility of the precise  measurement 
of the Higgs-boson production cross section
at the Photon Collider at TESLA is studied in detail. 
For the Standard-Model Higgs-boson production  the decay to $\bbbar$ pairs
is considered for the mass between 120 and 160~GeV.
The same decay channel is also studied for production of the heavy neutral Higgs bosons in MSSM,
for masses 200--350~GeV.
For the first time in this type of  analysis  all relevant experimental and theoretical effects,
which could affect the measurement,
are taken into account.
The study is based on the realistic $\gaga$-luminosity spectra simulation.
The heavy quark background $\gagaQQg$ is estimated using the dedicated code based on NLO QCD calculations.
Other background processes, which were neglected in the earlier analyses, are also studied:
$\gagaWW$, $\gagatautau$, and light-quark pair production $\gagaqq$.
Also the contribution from the so-called overlaying events, \gagahad{}, is taken into account; 
a dedicated package called \Orlop{} has been prepared for this task.
The non-zero beam crossing angle and the finite size of  colliding bunches
are included in the event generation.
The analysis is based on the full detector simulation with 
realistic \btagging{}, 
and 
the  criteria of event selection are optimized separately  for each considered Higgs-boson mass.
In spite of the significant background contribution
and deterioration of the invariant mass resolution
due to overlaying events,
precise measurement of the Higgs-boson production cross section is still possible.
For the Standard-Model Higgs boson with mass of 120 to 160~GeV
the partial width \( \Ghgagahbb \) can be measured 
with a statistical accuracy of 2.1--7.7\% after one year of the Photon Collider running.
The systematic uncertainties of the measurement are estimated to be of the order of 2\%.
%
For MSSM Higgs bosons $\AO$ and $\HO$, for \MAOeq 200--350~GeV and $\tbseven$, 
the statistical precision of the cross-section measurement is estimated  to be 8--34\%, 
for four considered MSSM parameters sets.
As heavy neutral Higgs bosons in this scenario may not be discovered at LHC 
or at the first stage of the $\epem$ collider, 
an opportunity of being a discovery machine is also studied for the Photon Collider.

%
%


\end{abstract}

\thispagestyle{empty}
\
\newpage 
\tableofcontents{}



\chapter{Introduction \label{ch_introduction}}

%
%
So far,  all experimental results concerning fundamental particles and their interactions
are well described by the Standard Model (SM),
 consisting of the electroweak theory (EWT) and the quantum chromodynamics (QCD).
These theories allow us to quantify electromagnetic, weak and strong processes;
among all known kinds of interactions only the gravity is not incorporated in the SM framework.
The very important ingredient of the SM is the so-called Higgs boson, $\hSM$, 
which is responsible for generating masses of all particles.
Although predicted by the model, the Higgs boson has not yet been experimentally detected. 
However, such a particle must exist if the SM is to remain a consistent theory. 
Consequently, a search for the Higgs boson is among the most important tasks 
of the present and future colliders. 
Once the Higgs boson is discovered, 
it will be crucial to determine its properties with high accuracy,
to understand the mechanism of the so-called electroweak symmetry breaking (EWSB).

%
The neutral Higgs boson couples to the photon pair only at the loop level, 
through loops of all massive charged particles. 
In the SM the dominant contribution is due to $W$ and $t$ loops.
This loop-induced $ \hSM \gaga $ coupling is sensitive to 
contributions of new particles which may appear in various extensions of the SM. 
Hence, the precise measurement of the Higgs-boson partial width $\Ghgaga$ 
can indicate existence of very heavy particles even if their direct production is not possible.
A photon-collider option%
\footnote{A photon collider option was foreseen for all projects of the $\epem$ linear collider:
TESLA \cite{TDR}, NLC \cite{NLCTDR} and GLC (earlier JLC) \cite{GLCTDR,JLCPC}.
In this work the photon collider at the TESLA  is considered.
The superconducting technology developed within the TESLA project 
has been recently selected as the best suited for the International Linear Collider.
Decision of the International Technology Recommendation Panel was presented
during the ICHEP2004 conference in Beijing \cite{ITRP}.
}
of the $\epem$ collider 
offers a unique possibility to produce the Higgs boson as an \( s \)-channel 
resonance in the process $\gagah$.  
As the SM Higgs boson with the mass%
\footnote{The energy unit [GeV] is used for masses and momenta, 
\ie the speed of light is set to 1. 
However, for lengths and times the corresponding units are [m] and [s].}
below \( \sim 140\)~GeV is expected
to decay predominantly into the \( \bbbar \) final state,
we consider the measurement of the cross section for the process \( \gagahbb \), 
shown in Fig.\ \ref{fig:diagram_gagahbbbar},
for the Higgs-boson mass in the range  \Mheq 120--160~GeV. 
The aim of this study is to estimate the precision with which this measurement 
and extraction of $\Ghgaga$
will be possible after one year of the TESLA Photon Collider running. 

\pnfig{t}{\figheightsmaller}{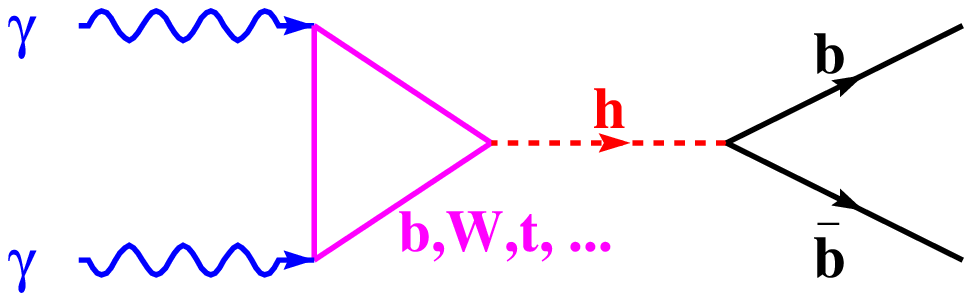}
{fig:diagram_gagahbbbar}
{A diagram of the process $\gagahbb$. 
The Higgs boson particle couples to photons through the loop of all massive and charged particles.}

Besides precision measurements, a photon collider can be also considered as a candidate for a discovery machine. 
In case of the Minimal Supersymmetric extension of the SM (MSSM) the photon collider will be able to
measure the production cross section of the heavy neutral Higgs bosons, $\AO$ and $\HO$,
covering the so-called ``LHC wedge'' in the MSSM parameter space, 
\ie region of intermediate values of $ \tan \beta$, $ \tan \beta \approx$ 4--10,
and masses $\MAOHO$ above 200~GeV.
For this part of parameter space, MSSM Higgs bosons $\AO$ and $\HO$  may not be discovered 
at the LHC \cite{ATLAS_CMS_TDR,SearchATLAS,CMSDiscovery} 
and at the first stage of the $\epem$ linear collider \cite{TESLATDR_part3}
because of small branching ratios into leptons or photons 
(which allow the efficient signal selection) 
and because of the kinematical limit $\MAOHO \lesssim \sqrt{s_{ee}}/2$ for pair production
process $\epem \ar \AO \, \HO$, respectively. 
Parameter range considered in this analysis corresponds to 
a SM-like scenario 
where the lightest MSSM Higgs boson $h$ has properties similar to the SM Higgs boson, 
while heavy neutral Higgs bosons are nearly degenerated in mass 
and have negligible couplings to the gauge bosons $W/Z$. 
We consider the process \( \gagaAHbb \) at the Photon Collider 
at TESLA for Higgs-boson masses \MAOeq 200--350~GeV.
The aim of the presented study is to evaluate the discovery potential of the considered experiment 
by estimating the statistical significance of the signal measurement
for the chosen region in the MSSM parameters space.
Also the precision of the $\gagaAHbb$ cross section measurement is estimated.
%

%

The measurements of $\Ghgagahbb$ and $\sgagaAHbb$ at a photon collider have already been studied before
\cite{Gunion:1990nx,Richard:1991ae,Borden:1992qd,Gunion:1992ce,Borden:1993cw,Borden:1994fv,Khoze:1995mn,
JikiaAndTkabladze,Jikia:1996bi,
Ohgaki2,JLCPC,Ohgaki1,
MellesStirlingKhoze,
MMuhlleitner_Thesis,JikiaAndSoldner,MMuhlleitner,
Asner,Velasco,
NZKhbbm120appb,NZKJeju,NZKMSSMeps2003,NZKSMeps2003,
Rosca}.
Although very promising estimates were obtained, many important aspects of the measurement
were not considered. 
This study is the first one to take all relevant experimental and theoretical effects into account.
Only results of such a realistic analysis can be used to support
%
the project of the Photon Collider
in the framework of the International Linear Collider.
The motivation for this study is outlined in Chapter \ref{ch_motivation}.
The proposed experimental setup and simulation tools are described in Chapter \ref{ch_colldet}.
In Chapter \ref{ch_signalbackground} details of the signal and background simulations 
are given. 
%
%
%
A discussion of the event selection and the final results for SM and MSSM scenarios 
are given in Chapters  \ref{ch_sm_analysis} and \ref{ch_mssm_analysis}, respectively.
%
%
All results presented in Chapters \ref{ch_signalbackground}, \ref{ch_sm_analysis} and \ref{ch_mssm_analysis},
and in Appendices were obtained by the author of this thesis.

\
\newpage 
\
\newpage 

\chapter{Motivation \label{ch_motivation}}

In this chapter our current understanding of  the Higgs mechanism
and prospects for a  discovery of the Higgs boson are outlined.
The Higgs sector in the SM  is discussed first,
then its extension to the  MSSM is shortly reviewed.
%
%
Current limits on the Higgs-boson mass from direct and indirect measurements are summarized.
Expected experimental results at future colliders, 
relevant to the presented study,  are also given.
%
%
%
An in-depth review of the Higgs-boson theory and phenomenology can be 
found, for example, in \cite{HiggsHuntersGuide,HiggsTheory_Carena_Haber}.
An extensive summary of experimental results concerning Higgs-boson searches
is presented in \cite{PDG2004}. 
\section{The Higgs sector in the SM}

Among all fundamental particles of the SM only the Higgs boson still remains hypothetical.
This neutral spinless particle is required in the model 
to \emph{break} the gauge symmetry of weak interactions.
%
%
Photon, which is a carrier of electromagnetic force, is massless.  
But three weak bosons $Z$, $W^+$ and $W^-$ are massive; 
this is a serious difficulty
as SM equations for interactions involving massive bosons lack a very basic property, 
the so-called \emph{gauge invariance}\footnote{
A principle of gauge invariance originates in the classical theory of electromagnetism
and reads: there is a transformation of electromagnetic four-potential $A^\mu$
after which physically relevant fields, $\vec{E}$ and $\vec{B}$ 
(or the tensor $F^{\mu\nu}=\partial^{\mu}A^{\nu}-\partial^{\nu}A^{\mu}$), remain unchanged.
In quantum field theories the invariance of equations
after simultaneous, special transformations of all fields is required.
Thus, the gauge invariance principle determines the allowed interaction terms.
%
%
%
}. 
Other problem emerges in cross section calculations for some weak processes, \eg $\epem \ar \WW$,
because \emph{unitarity} condition  is violated for this transition. 
Probability current is not conserved
unless we introduce new particles which couple to electrons and massive bosons.
One complex Higgs doublet (four real scalar fields) is introduced in the SM 
in order to describe experimental results and to preserve clear theoretical picture.
%
These new fields, filling the vacuum, couple to the massless vector bosons, 
giving them effective mass.
 This mechanism, introduced by P.\ Higgs \cite{PeterHiggs}, 
allows us to introduce    massive gauge bosons in the theoretical description   without
violating the gauge invariance (so-called spontaneous symmetry breaking).
%
%
One of the scalar fields is expected to exist
as a real particle, so-called Higgs boson, $\hSM$.
All couplings of the Higgs boson to other particles and its self-couplings
are predicted by the SM;
the couplings to bosons (fermions) are proportional 
to the mass squared of the boson (the mass of the fermion). 
The only unknown parameter of the theory is the Higgs-boson mass, $\Mh$.
An intelligible introduction to the Higgs mechanism can be found, for example, in \cite{AitchisonHey}.
The SM constitutes a complete effective theory of fundamental interactions
(excluding gravity).
Existence of the new particle, the Higgs boson,
explains how the electroweak symmetry (or gauge invariance) is broken
and solves the unitarity problem in weak reactions.
However, this great theoretical achievement is undermined by some unsolved problems.
On the way to the  Planck energy scale 
some new phenomena are expected to appear.
Otherwise, without unnatural tuning, higher order corrections to the Higgs-boson mass diverge 
as the energy scale increases (so called ``hierarchy problem'').
The second problem is due to our expectation that at some high energy scale
all interactions should unify (\ie their  couplings should be equal)
which is not exactly the case in the SM.%
\footnote{
 Only approximate unification is obtained in the SM.
 At the scale of $M_{\mathrm{GUT}} \sim 10^{15}$~GeV
 couplings are 'unified' to $\mathcal{O} ( 10 \% )$
 \cite{StephenMartin}.
} 
To fulfill this unification requirement new particles or interactions
have to be introduced. 
%
%


\section{Higgs sector in the MSSM }

The new symmetry between bosons and fermions, so-called \emph{supersymmetry} (SUSY),
could remove the two above-mentioned problems of the SM.
It guaranties cancellation of divergences in Higgs-mass calculation.%
\footnote{In fact, cancellation is not exact as the supersymmetry is broken, 
\ie particles have different masses than their superpartners.
This results in the prediction that masses of superpartners cannot be
heavier than a few TeV. 
Otherwise supersymmetry does not solve the hierarchy problem.}  
Also the unification of three fundamental couplings is realized.
However, in the general case of the Minimal Supersymmetric extension 
of the SM (MSSM) around 100 new parameters must be introduced
whose values are not predicted by the model.
All SM particles have their superpartners: 
fermions -- spin-zero bosons (\eg electron -- selectron),
bosons -- fermions (\eg higgs -- higgsino, $W$ -- $W$-ino, photon -- photino).
To generate masses for all particles and sparticles,
two Higgs doublets (\ie eight fields) have to be introduced.
As a result,  supersymmetric models contain five Higgs bosons (instead of one Higgs particle).
Two of them are neutral scalars and are denoted as $h$ and $\HO$.%
\footnote{
By definition, $h$ denotes the lighter scalar Higgs boson, 
and $\HO$ denotes the heavier one.
}
There is also one neutral pseudoscalar, $\AO$,
and two charged scalars: $H^{+}$ and $H^{-}$.
The Higgs sector of the MSSM is described by a subset of  parameters which includes:
\begin{enumerate}
\item $\tanb$ -- the ratio of vacuum expectation values 
                 of neutral Higgs fields coupling to \emph{up}- and \emph{down}-type fermions,
                 $\tanb = v_{u}/v_{d}$. 
\item $\MAO$ -- the mass of the neutral, pseudoscalar Higgs boson, $\AO$.
\item $\mu$ -- the supersymmetry-breaking higgs-higgsino mass parameter.
\item $M_2$ -- the supersymmetry-breaking universal gaugino mass parameter (mass of the $W$-ino; 
masses of other gauginos are related with $M_2$).
%
\item $M_{\widetilde{f}_L}$, $M_{\widetilde{f}_R}$ and $A_{\widetilde{f}}$ -- other supersymmetry-breaking parameters: 
masses of left- and right-handed supersymmetric partners of fermion $f$ and their coupling to Higgs bosons,
respectively.
\end{enumerate}
Only first two parameters, $\tanb$ and $\MAO$, influence the Higgs sector on the tree level.
Other parameters can affect properties of the Higgs bosons via radiative corrections.
In contrast to the SM, mass of the lightest Higgs boson, $h$, is constrained,
\ie $h$ cannot be heavier than around $150$~GeV.

%
%
\section{Status of the Higgs-boson searches}

In precise calculations of the SM predictions
the higher order corrections resulting from the Higgs boson contribution are sizable
and must be taken into account.
Expected results for many observables depend on the Higgs-boson mass, $\Mh$.
Thus, constraints on the value of $\Mh$ can be  obtained
from the analysis of  electroweak measurements.
The result of such analysis is shown in Fig.~\ref{fig:MhFitToEWPD} \cite{LEPEWG},
where the 
$\chi^{2}$ 
value from the SM fit to precise measurements at LEP, SLC, Tevatron and other experiments
is presented as a function of the Higgs-boson mass.
The best agreement is found for $\Mh \approx 126$~GeV, 
and with 95\% C.L.\ the upper limit on $\Mh$ is 280~GeV.
The best fit value is slightly above the lower mass limit from the direct searches at LEP;
excluded is the mass range $\Mh < 114.4$~GeV \cite{LEPHWG}.

%
\pnfig{!tb}{\figheight}{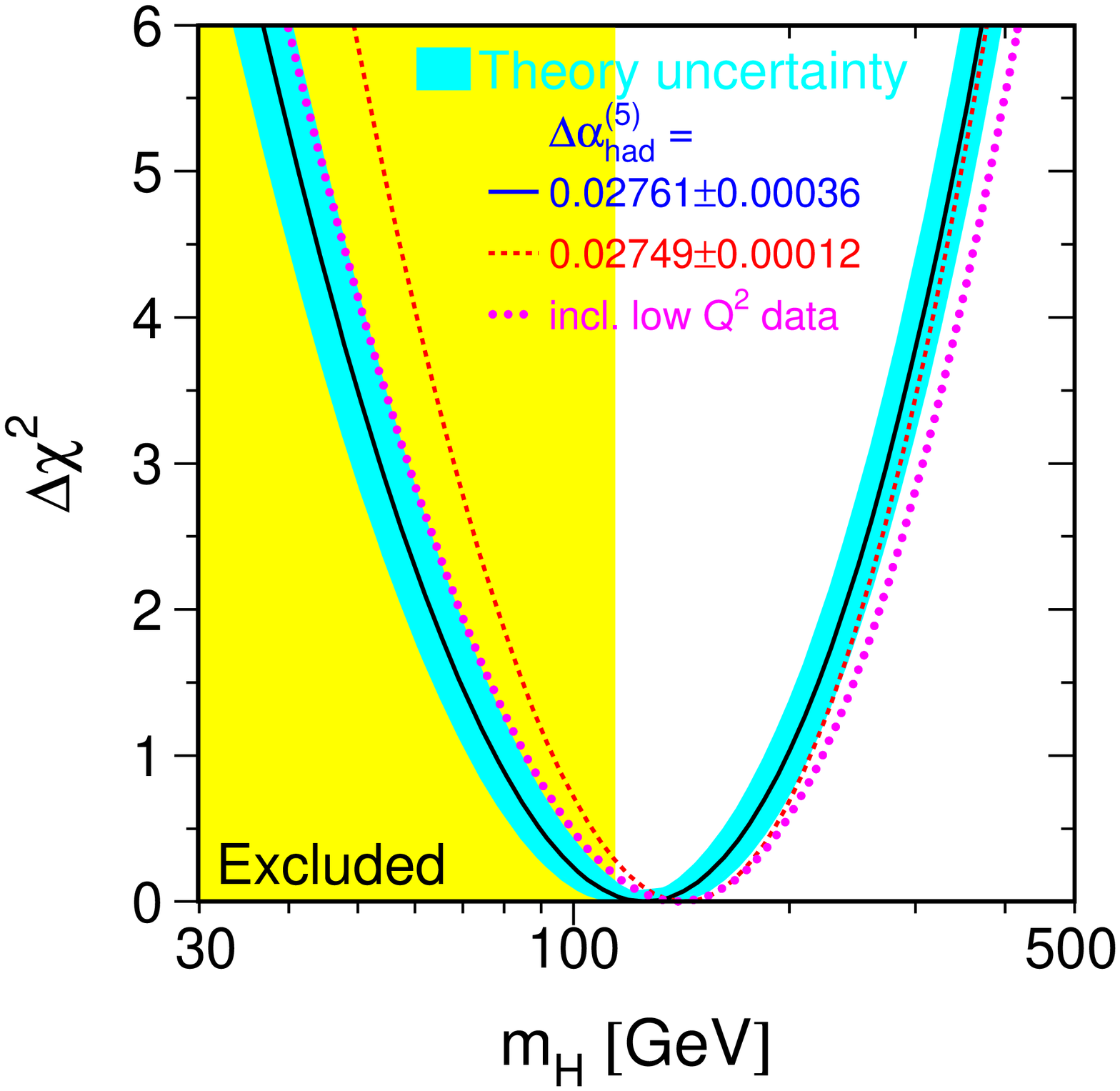}{fig:MhFitToEWPD}
{
The result of the SM fit to the precision electroweak measurements \cite{LEPEWG}.
The $\Delta\chi^{2}=\chi^{2} - \chi^{2}_{\min}$ value 
is presented as a function of the Higgs-boson mass (here denoted as $m_H$).
The best agreement is found for $\Mh \approx 126$~GeV, 
and with 95\% C.L.\ the upper limit on $\Mh$ is 280~GeV.
Excluded  mass range  from direct searches at LEP, $\Mh < 114.4$~GeV, is indicated in yellow.
The band represents the theoretical uncertainty due to missing higher order corrections.
}

In the MSSM case  limits for the Higgs-boson masses depend on other model parameters.
In the general approach, when other parameters are allowed to vary,
we can only conclude that all Higgs bosons must be heavier than 80--90~GeV
if model with no $CP$-violation in Higgs sector is assumed.
However, some MSSM parameter sets result in the lightest \higgs, $h$, 
 having couplings similar to those of the SM Higgs boson (SM-like scenarios).
In such cases the Higgs-boson mass constraints are  similar to those obtained in the SM.


%
%
\section{Prospects for Higgs-boson measurements }

If the mass of the SM Higgs boson is around 115~GeV
it is still possible that it will be discovered 
at the Tevatron.
However, only future machines will have sufficient \higgs{} production rates 
to measure precisely the mass and couplings of the Higgs boson(s).
All large accelerator projects aim at measurements of the Higgs-boson properties
from which the fundamental one is the mass, $\Mh$, being at the same time
the only unknown parameter in the SM.
Measurements of other parameters describing the Higgs boson 
(total  and partial widths, branching ratios, spin, parity)
are considered as the crucial tests of the SM and its extensions.
Below the estimated  precisions of future measurements are summarized,
in the expected order of appearance.
%

%
%


%
\subsection*{Large Hadron Collider} 

At the Large Hadron Collider (LHC), which should become operational in 2007,
Higgs boson(s) will be produced in processes of 
gluon fusion (about 80\% of the SM Higgs-boson production rate for $\Mh < 2 M_Z$) 
and $W$ boson fusion.
Both experiments, ATLAS and CMS, have presented detailed studies showing that
the SM Higgs boson will be discovered at LHC if it is lighter than about 1 TeV.
The Higgs-boson mass can be measured with precision about 0.1\% for $\Mh \lesssim $ 400~GeV \cite{ATLASZivkovic}.

Various ratios of Higgs partial widths can be determined with precisions about 10--20\%,
assuming integrated luminosity of 100 fb$^{-1}$  \cite{LHC1}.
For heavy SM \higgs{}, $\Mh \gtrsim 250 $~GeV,
also its total width, $\Gh$, can be measured as it becomes larger than the experimental mass resolution \cite{ATLASZivkovic}.
The heavy MSSM Higgs bosons, $\AO$ and $\HO$,
will be observed at the LHC for most of the allowed MSSM parameter space.
However, there is a region of  $\MAO-\tanb$ values where the LHC %
may not be able to discover heavy MSSM Higgs bosons.
This is the so called ``LHC wedge'' covering $\MAO > 200$~GeV and $\tanb \approx $ 4--10.

\subsection*{Linear $\epem$ Collider} 

According to the currently proposed schedule, 
the $\epem$  International Linear Collider (ILC) can become operational in 2015.
Two processes contribute to the Higgs-boson production at the ILC: 
Higgs-strahlung and vector boson fusion. 
For the SM Higgs boson with mass in the range $\Mh \approx $ 115--180~GeV
the expected precision of  the mass measurement  can be better than 0.05\%  \cite{TESLATDR_part3}.
As the background is much smaller than at the LHC,
Higgs-boson branching ratios can be determined with much better precision
and in the model-independent way. 
 Branching ratios $\Brhgaga$ and $\Brhbb$ 
may be determined with accuracy of about 10\% and 1.5\%, respectively,  
after one year of ILC running at nominal luminosity \cite{BRhgaga,BRhbb}. 

%

\subsection*{Experiments after LHC and LC}

After LHC and ILC  measurements there will still be some properties of the Higgs-boson(s)
which are poorly known and should be determined with greater precision at other experiments. 
One of the interesting quantities is the $\gaga$ partial width, $\Ghgaga=\GhBrhgaga$. 
As already mentioned in Chapter \ref{ch_introduction},
measurement of partial width $\Ghgaga$ is of special importance
as it is sensitive to all charged particles which
have mass generated by the Higgs mechanism.
Due to the so-called non-decoupling effect contributions of the heavy charged particles
to the $\hgaga$ loop are finite even in the limit of infinite mass of the particle.
Thus, the measurement of $\Ghgaga$ can indicate existence of particles
whose direct production will be impossible in available accelerators.

In the MSSM, 
the loop induced coupling is sensitive to contributions 
of supersymmetric particles \cite{LoopMSSM,LoopMSSMHaber,AsnerGaGaHGaGA}:
chargino and top squark loops can lead even to 60\% 
difference between the SM and the SUSY couplings. 
Scenarios, in which all new particles are very heavy, 
may be realised not only in the MSSM but also in other  models with extended Higgs sector, 
for example in the Two Higgs Doublet Model (2HDM). 
In this case the two-photon width of the Higgs boson 
will differ from the SM value due to the contributions 
of the heavy charged Higgs bosons,
even if all direct couplings 
to  gauge bosons and fermions 
are equal to the corresponding  SM couplings.
Different realizations of the 2HDM have been discussed in \cite{2HDM}. 
Assuming that the partial widths of the observed Higgs boson to quarks, 
$Z$ or $W$ bosons are close to their SM values (SM-like scenario), 
three different combinations of couplings are possible. 
Fig.\ \ref{fig:mk_2hdm_A_B} shows deviations of the two-photon Higgs width 
from the SM value for the three SM-like solutions considered in \cite{2HDM}.
For solution $B_u$ one expects significant deviation from the SM predictions
since, as compared to the SM, there is a change of the relative sign of the top-quark and the $W$
contributions.
Consequently, for solution $B_u$ the $\higgsm \ar \gaga$ width is significantly larger 
than in the SM, where these two contributions partly cancel each other.
However,  deviation due to the charged Higgs-boson contribution only (solution $A$)  
is much smaller, of the order of 5--10\%,
and 
the measurement with precision at the level of a few percent is required.

\pnfiggeneral{tb}{\twofigheight}{\includegraphics{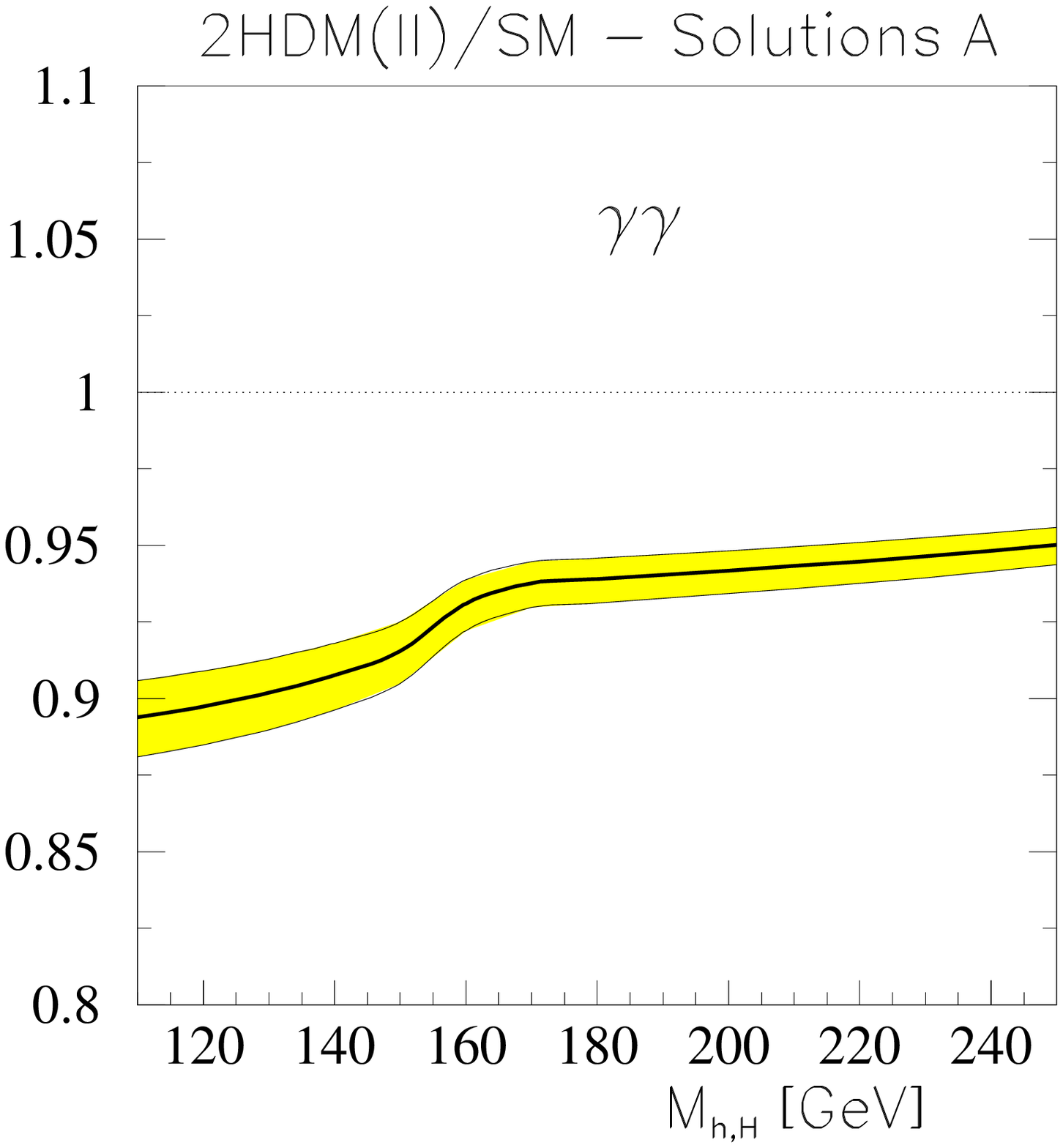} 
   \includegraphics{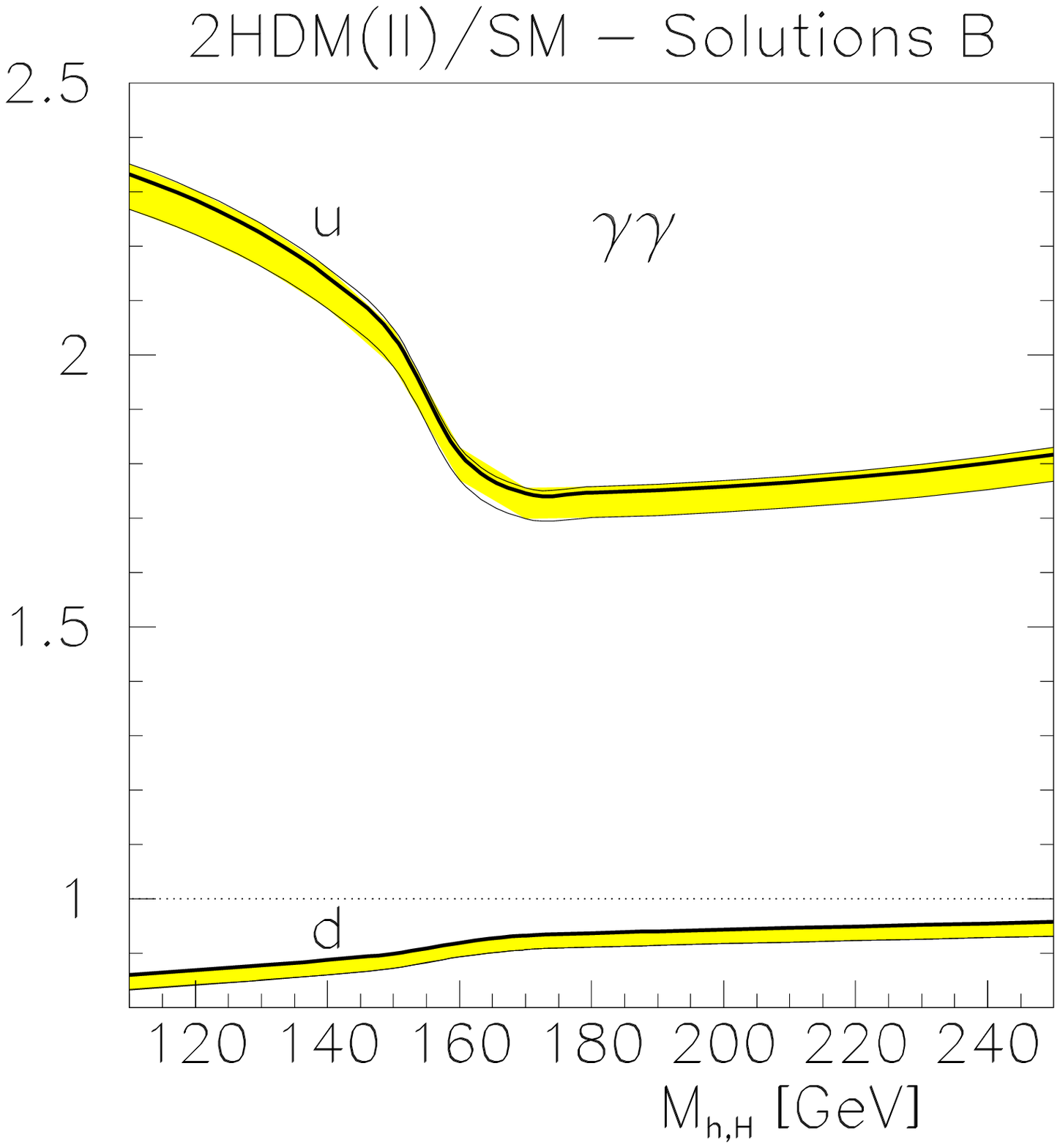}}{fig:mk_2hdm_A_B}{
Ratio of the $\higgsm \ar \gaga$
decay widths in the 2HDM and the SM as a function of the \higgs{} mass, $M_{h,H}$,
for  solutions~$A$  (left)  and $B$ (right) \cite{2HDM}.
The yellow bands correspond to the uncertainties expected at $M_{h,H}=$ 120~GeV.
}

The machine best suited for the measurement of the Higgs boson two-photon width is a photon collider.
All  photon collider  proposals  (within TESLA, NLC, GLC and CLIC projects) emphasize
the feasibility of a very  precise $\Ghgaga$ measurement.
For the light SM-like Higgs boson
the most promising process is $\gagahbb$
due to the very high branching ratio $\Brhbb$. 
The measurement of the cross section for this process 
has been studied in detail and is the main subject of this work.
One has to note that the final results on $\Ghgaga$ from a photon collider 
must rely on $\Brhbb$ measurement from other experiment
 -- this branching ratio should be determined at LC with precision of around 1.5\% \cite{BRhbb}.
Measurements of $\Ghgagahbb$ at the Photon Collider
and of the branching ratios $\Brhgaga$ and $\Brhbb$ at the $\epem$ collider 
can be used to determine the total Higgs width, $\Gh$,
in a model independent way.

The quantity $\Ghgaga\Brhgaga$ could also be unfolded
from the cross section measurement for the process $\gaga \ar \hSM \ar \gaga$. 
Unfortunately this process has a very low rate due to small $\gaga$ branching ratio.
Moreover, in addition to a one-loop background process $\gaga \ar \gaga$ also
'machine' background $e\ga \ar e\ga$ must be considered.
Even with optimistic assumptions about angular coverage
(down to 3$^\circ$) and high granularity of calorimeter the precision of the 
measurement has been estimated to be of the order of 30\% \cite{AsnerGaGaHGaGA}.
Therefore this measurement cannot be considered as an alternative to the analysis of the process $\gagahbb$.
The Photon Collider (PC) seems to be the only machine 
allowing precise measurement of $\Ghgaga$.
However, physics potential of the PC is much reacher and complementary to that of the LHC and the ILC.
For Higgs sector itself many interesting measurements can be considered:

\begin{itemize}
\item \underline{$CP$-parity of the Higgs boson.}
  The CP-parity of the Higgs boson can be determined 
  in a model independent way from
  analysis of angular correlations in 4-fermion decays 
  $\hSM \ar \WW \ar 4j$ and 
  $\hSM \ar ZZ \ar 2j \: 2l$ \cite{Niezurawski:2004ga}.
  Similar measurements have been also proposed for LHC and ILC.
\item \underline{Phase of the $\hgaga$ coupling.} 
  In addition to $\Ghgaga$, precise measurements at the PC are also sensitive 
  to the phase of the $\hgaga$ amplitude, $\phihgaga$. 
  The phase $\phihgaga$ can be extracted 
  from the measurement of the interference between
  the resonant \higgs{} production processes $\gaga \ar \hSM \ar \WW$ 
  and the background process $\gaga \ar \WW$ \cite{wwzz}.
  It turns out that for \higgs{} mass of the order of 300--400~GeV
  the phase $\phihgaga$ is more sensitive to possible contributions
  of heavy charged scalar particle than $\Ghgaga$. 
  Only by combining $\Ghgaga$ and $\phihgaga$ measurements  at the PC 
  with those at the LHC and at the LC
  unique determination of the Higgs-boson couplings and
  distinction between various models will be possible.


\item \underline{Charged Higgs-bosons production.} 
  For intermediate values of $\tanb$ the charged MSSM Higgs boson, $H^\pm$, 
  may not be
  discovered at the LHC, if its mass is greater than $\sim 150$~GeV \cite{SearchATLAS,CMSDiscovery}.
  In the photon collider $H^+ H^-$ pairs can be produced in QED process
  and the mass reach for discovery can be extended up to $\sim 300$~GeV 
  (if running at \sqrtseeeq 800~GeV).
  At the $\epem$ LC, running with \sqrtseeeq 800~GeV, almost an order of magnitude
  smaller number of \higgs{} pairs 
  could be produced during the same time due to smaller cross section \cite{TESLATDR_part3,TDR}.
\end{itemize}

The second generation  $\epem$ linear collider project 
for which the feasibility study is still in progress is CLIC \cite{CLIC1}. 
At this accelerator multi-TeV energies will be obtained, 
and the wider range of masses will be accessible for new-particle searches. 
The photon--photon collision mode has been proposed  for this machine as well,
but the detailed physics case studies are still missing.

Opportunity of studying elementary particle collisions 
at multi-TeV energies is the main reason for considering 
the next generation project of a muon collider.
As muons lose much less energy via bremsstrahlung than electrons
a circular accelerator option is preferable 
even for much higher beam energies than those accessible at LEP2.
Circulating beams would allow the operation with high luminosity.
However, progress must still be made 
in formation of high-intensity $\mu$-beams with low emittance.
At the muon collider the same channels could be used for Higgs-boson measurements
as at $\epem$ collider but with extended mass reach.
Moreover, the $s$-channel production in the process
$\mu\mu \ar \higgsm$  
has cross section sufficient for precision measurements
due to higher mass of the muon 
\cite{MuonCollider1}.
As the energy spread of muon beam is expected to be negligible,
energy scan at the muon collider would result in determination 
of the Higgs-boson mass and width with precision of the order of 2 MeV.
Construction of a photon collider based on the muon collider is also possible.
Unfortunately the high mass of the muon is a problem in this case.
The maximal energy of photons from Compton backscattering 
would be very small compared to the beam energy, 
\eg the maximal photon energy would be only around 75~GeV at 20 TeV $\mu$-beam 
as compared to about 400~GeV for 500~GeV electrons
with the same laser setup%
\footnote{Laser parameters of the TESLA Photon Collider design are assumed.}. 

\
\newpage 

\chapter{Collider and detector \label{ch_colldet}}

%
%
\section{The TESLA Linear $\epem$ Collider }


Results presented in this thesis are based on the design and machine parameters of
the TESLA (TeV-Energy Superconducting Linear Accelerator) $\epem$ collider  \cite{TESLATDR_part2}. 
Accelerator design is based on the superconducting technology,
recently accepted by International Technology Recommendation Panel
as the best suited for the ILC project \cite{ITRP}. %
%
Each of the two linear accelerators, 
which accelerate $e^+$ and $e^-$ towards the interaction region,
will consist of around 10000 one-meter-long superconducting cavities.
 Cavities made from niobium and cooled to temperature of 2~K
can provide accelerating field with gradient well above 35 MV/m.
As the average gradient equal to 23.4 MV/m is required for operation 
at the  nominal total collision energy of 500~GeV,
the opportunity emerges to increase the machine reach up to 800~GeV or even 1 TeV.
With  dumping rings and other additional accelerator components
the total length of the machine is 33 km.
Main machine parameters are listed in the table \ref{tab:LCParameters}.

\begin{table}[!b]
\label{tab:LCParameters}
\bc
\begin{tabular}{|l|c|}
\hline
Description of the parameter            & Value of the parameter \\
\hline
Accelerating gradient                   &     23.4 MV/m   \\
No.\ of accelerating structures         &  21024          \\
Train repetition rate                   &      5 Hz       \\
No.\ of bunches per train               &   2820          \\
Bunch spacing                           &    337 ns       \\
No.\ of $e^- (e^+)$ per bunch         &  $2 \cdot 10^{10}$      \\ 
Beam size at IP ($\sigma_x$;$\sigma_y$) &   553 nm; 5 nm      \\
Bunch length at IP ($\sigma_z$)         &    0.3 mm       \\
Luminosity                              &  $3.4 \cdot 10^{34}$ cm$^{-2}$s$^{-1}$ = 34 nb$^{-1}$s$^{-1}$ \\
Luminosity per year                     &  340 fb$^{-1}$y$^{-1}$ \\
\hline
\end{tabular}
\ec
\caption{Main parameters of TESLA Linear Collider for the 500~GeV design. 
 One accelerator year [y] is assumed to be equal to 10$^7$ s.}
\end{table}

%
%
\section{A photon collider as an extension of the LC \label{pc_as_ext_of_lc}}

Future linear $\epem$ colliders offer unique opportunities to study photon-photon interactions
if the idea of a photon collider is realized. 
In this option the energy of the primary electron-electron%
\footnote{
  For the photon collider positron beam is not needed. 
  Moreover, use of two electron beams has important advantages:
  higher photon polarization and reduced backgrounds 
  in the interaction region.}
beams  is ``transfered'' to photons in the process of Compton back-scattering
\cite{Ilya}.
Assuming the beam electron collides with one laser photon head-on, 
the highest energy of the scattered photon, $\Emf$, is equal to:   
\[
 \Emf =\frac{x}{x+1}\frac{E_{e}+p_{e}}{2}  \; ,
\]
\[
\textrm{where~~~~~} x=\frac{2E_{L }(E_{e}+p_{e})}{m_{e}^{2}} \; .
\]
 $E_{L }$, $E_{e}$, $p_{e}$ and $m_{e}$ 
are the energy of laser photon,
the energy of beam electron,
its momentum and mass, respectively.
%
%
As the energy of primary electrons will be of the order of 100~GeV,
one can use an approximation $p_{e} \approx E_{e}$.
In this case the simplified formulae are obtained:
\[
\Emf=\frac{x}{x+1} E_{e} \; ,
\]
\[
x=\frac{4 E_{L } E_{e}}{m_{e}^{2}} \;.
\]
Collision of  a high energy photon from Compton back-scattering with a laser photon  
in the conversion region 
can result in creation of $\epem$ pairs.
This process 
would significantly limit the $\gaga$-luminosity
as its cross section is comparable with that of the Compton scattering.  
Therefore, it is preferable to select laser parameters such that 
the threshold for  $\epem$ pair creation is not reached.
The condition which has to be imposed on the invariant mass of two photons is:
\[ 
 \sqrt{ s_{{\ga_{L}} \ga_{\max\!1}} } =  2 \sqrt{ E_{L } \Emf } < 2 m_{e}. 
\]
This is equivalent to the requirement: $ x < x_{thr} = 2 + 2\sqrt{2} \approx 4.83 $.
 Leading order Compton cross section formula
indicates that the approximate monochromaticity
of the $\ga$ beams can be achieved 
for high polarizations of colliding electrons and of laser photons.
Figures \ref{fig:compton_y} and \ref{fig:compton_pol_gamma} show 
the expected distributions of production probability  
and of the circular polarization of high-energy photons, $P_{\ga}$,
as a function of the photon energy  relative to the primary electron energy, $y = E_{\ga}/E_{e}$, 
for various 
combinations of laser photon polarization, $P_{c}$, and electron polarization, $P_{e}$. 
With the choice  $P_{e}=1$ and $P_{c}=-1$ most of the back-scattered photons
have energies close to $\Emf$ and are highly polarized.

\pnfig{!t}{\figheight}{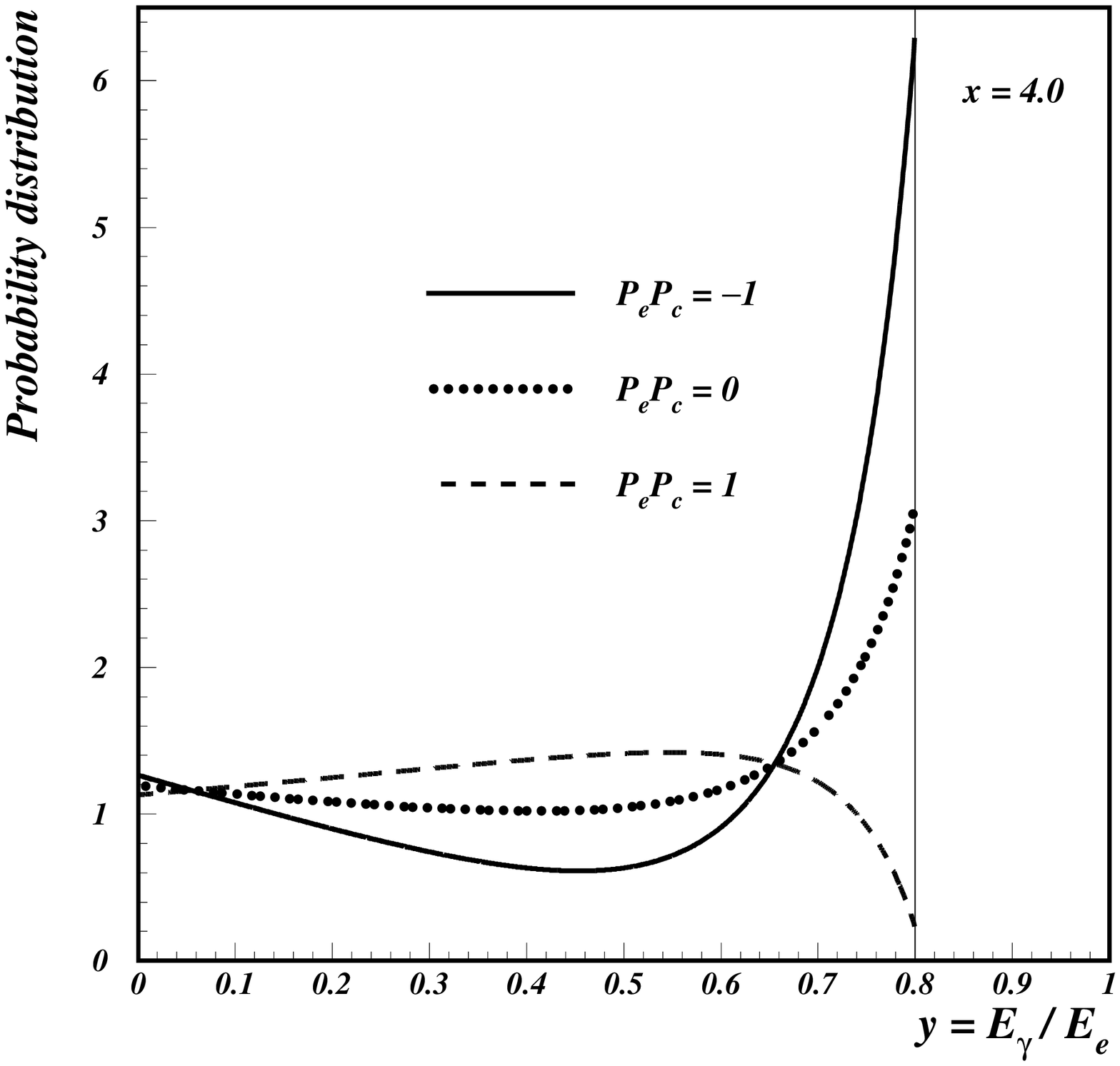}{fig:compton_y}{
Probability distributions for production of photon with relative energy  $y = E_{\ga}/E_{e}$. 
Results for various combinations of laser and electron polarizations, $P_{c}$ and $P_{e}$, are shown.}

\pnfig{!t}{\figheight}{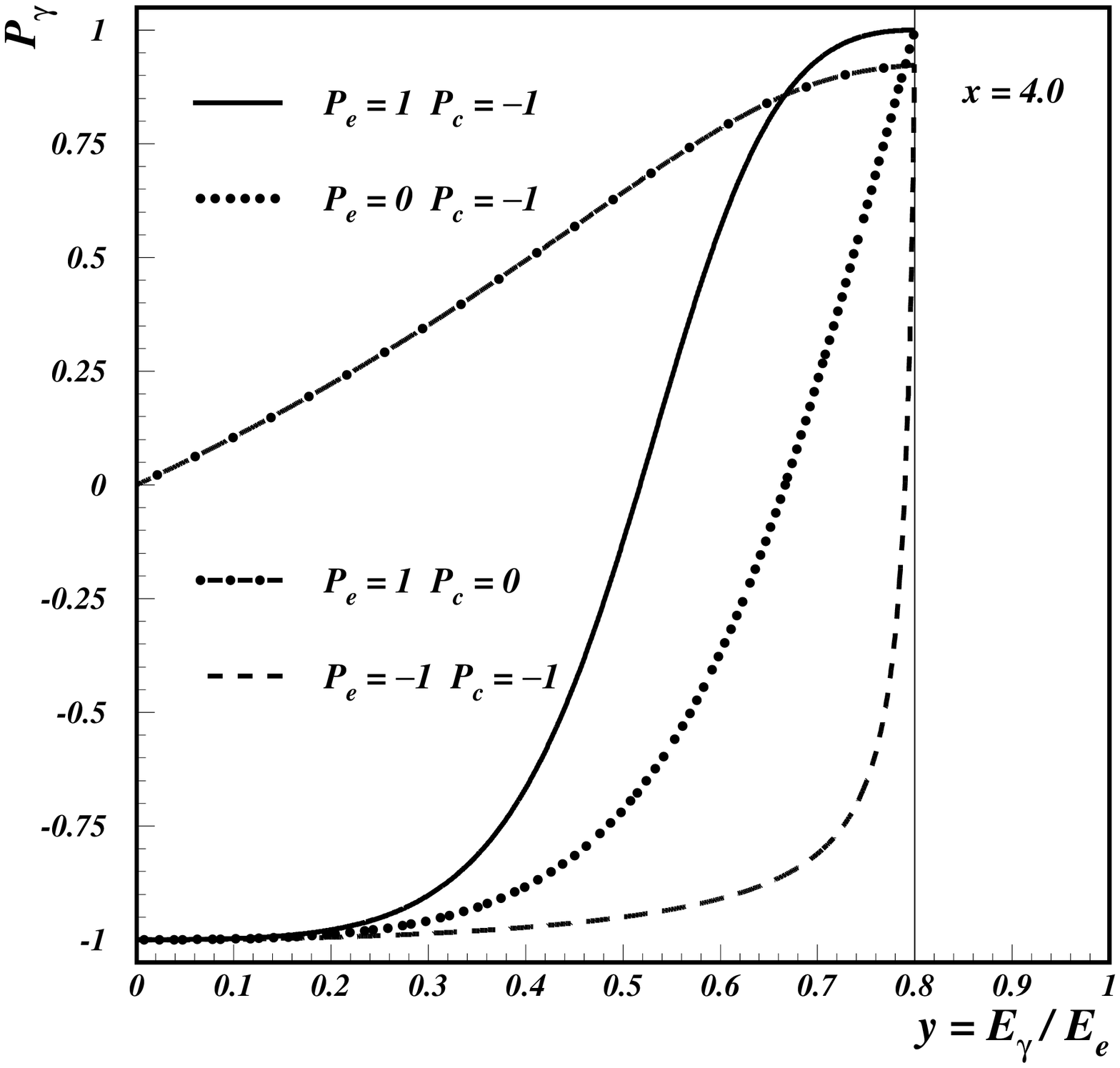}{fig:compton_pol_gamma}{
Circular polarization of scattered high energy photons, $P_{\ga}$,
as a function of relative photon energy, $y = E_{\ga}/E_{e}$.
Results for various combinations of laser and electron polarizations, $P_{c}$ and $P_{e}$, are shown.}

\pnfig{!bt}{\twofigheight}{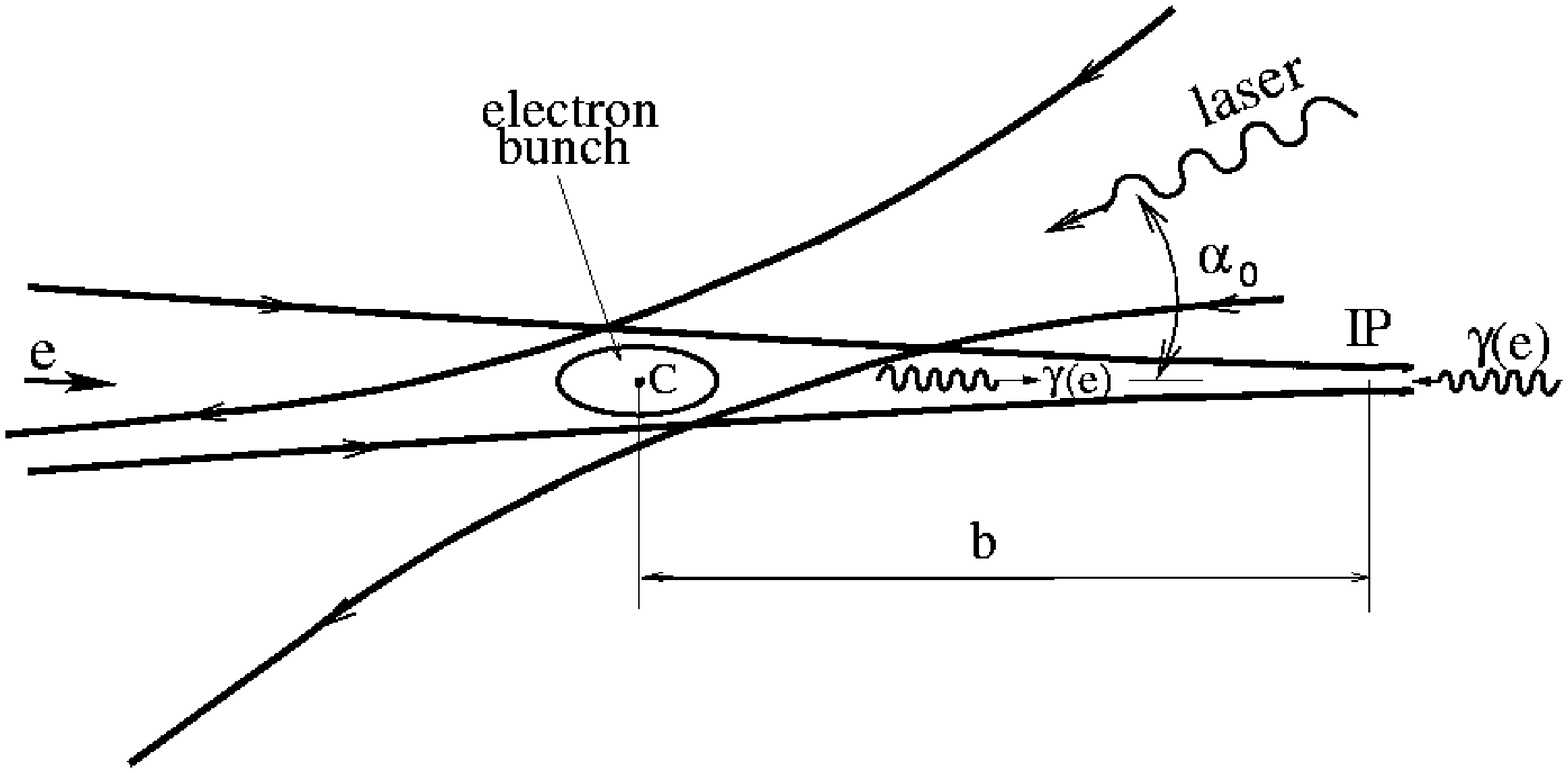}{fig:PCscheme}{
A basic scheme of the photon collider \cite{V.TelnovPlot}.
Primary electrons scatter on the laser photons (region C) at the distance $b$
from the interaction point (IP).
High-energy photons produced in the Compton back-scattering
follow the electron-beam direction 
and  collide  with photons  coming from the opposite side. 
}

%
In the laboratory frame photons from Compton back-scattering
are strongly boosted in the electron-beam direction.
The angular distribution has the width corresponding to 
the characteristic angle $\theta_{ch} \approx 1/\ga_{e}$ 
where $\ga_{e}$ is the relativistic factor for the beam electron, $\ga_{e}=E_e/m_e$.
For $E_e=250$~GeV one obtains $\theta_{ch} \approx 2$ $\mu$rad,
\ie the photon beam is strongly collimated along the incident electron beam direction.
Therefore, two high energy photon beams produced in Compton back-scattering
can be collided head on in the setup shown in Fig.\ \ref{fig:PCscheme}.
As the distance between conversion point (CP) and interaction point (IP), $b$, 
will be of the order of 3 mm, one can see that the additional spread of the photon bunch  
will be only about 6 nm
which is much smaller than the transverse size of the $e^{-}$-bunch in the $x$ direction 
and comparable with the $e^{-}$-bunch size in the $y$ direction.
Therefore, the $\gaga$-luminosity will be of the same order of magnitude 
as the $\emem$ geometrical luminosity.
Studies of the effects present in the conversion and interaction points revealed
that also the following corrections should be included in a description of $\gaga$-luminosity spectra:
\begin{enumerate}

\item 
Correlations between energy and scattering angle of Compton photons.
As more energetic photons scatter with smaller angles, 
high-energy photons in the 'core' of the beam
collide with high-energy photons of the opposite beam
with greater probability
than with low-energy photons forming beam 'halo'.
Thanks to this effect high-$\Wgaga$ part of $\gaga$-luminosity is enhanced
in comparison to the simple convolution of both spectra \cite{GinzburgFromCompAZ}.
\item 
An effective increase of the electron mass due to its transverse motion 
in the strong electromagnetic field of the very intense laser beam: 
$m_{e}^{2} \ar m_{eff}=m_{e}^{2}(1+\xi^{2})$. 
Here the $\xi$ parameter is related to the strength of the electromagnetic field 
in the conversion region and is used to describe nonlinear effects.     	

\item 
Scattering of electrons on two laser photons: $e^{-}+2\ga_{L} \ar e^{-}+\ga$. 
Interactions with three and more laser photons are supposed to be negligible.

\item 
 Interactions of laser photons with electrons which  already scattered one or more times.

\item 
Nonlinear $\epem$ pair creation $\ga+n\ga_{L} \ar \epem$ which should be taken into account even 
for $x < x_{thr}$.

\item 
Coherent $\epem$ pair creation by a high energy photon in the electromagnetic field 
which is present in the interaction point. 

%

\end{enumerate}

%
Before  the interaction point, 
to minimize some of aforementioned effects,
the possibility was studied to remove
electrons 
from the beam %
with special magnets.
For a design with $b\approx 1$ cm a small magnet was foreseen 
with magnetic field $B = 5$ kG deflecting electrons before the IP.
However, in the current design, optimized for highest luminosity,
this is no longer possible due to the short distance of 3 mm between conversion point and IP.

 The finite beams-crossing angle 
at the interaction point, 
with ``crab-wise'' tilted electron bunches \cite{Palmer}
has been recently accepted as the solution for the linear collider.
This is a preferred scheme for a photon collider 
because the removal of high-energy-photon bunches 
after the interaction would be very difficult with collinear beams.
Crab-crossing solution preserves the same luminosity as for head-on collisions.
%
%
%
%
However,  electromagnetic interaction between beams must be included    
in the full simulation of $\gaga$-luminosity because primary electrons 
are traversing through interaction point. 
The more complete description of processes outlined here and other effects 
influencing $\gaga$-luminosity spectra
can be found in \cite{V.TelnovPrinciples}.

\section{The Photon Collider at TESLA}

According to the  current design of the Photon Collider at TESLA \cite{TDR}, 
the energy of the laser photons is assumed to be fixed
for all electron-beam energies.
Laser photons are assumed to have circular polarization $P_{c} = -1$,
while longitudinal polarization of electrons is $P_{e} = 0.85$. 
This configuration of polarizations corresponds to the energy spectra of back-scattered photons
peaked at high energy (see Fig.\ \ref{fig:compton_y}).
With the same choice of parameters for each beam we maximize probability 
that two high-energy photons will collide  with the same polarization,
\ie in the state with total angular momentum, $J$, equal to zero,
so a spinless resonance can be produced.
To profit from the peaked $\gaga$-luminosity spectra 
the energy of primary electrons has to be adjusted 
in order to  enhance the resonance production signal at a particular mass.
The use of a by-pass for electron beams is considered if the  energy much lower 
than the nominal one is required. 
In this case the luminosity will be approximately proportional to the beam energy, $L \propto E_{e}$.
%
%

%

\subsection{Photon--photon luminosity spectra \label{ss_photon_luminosity}}

\pnfig{h}{\figheight}{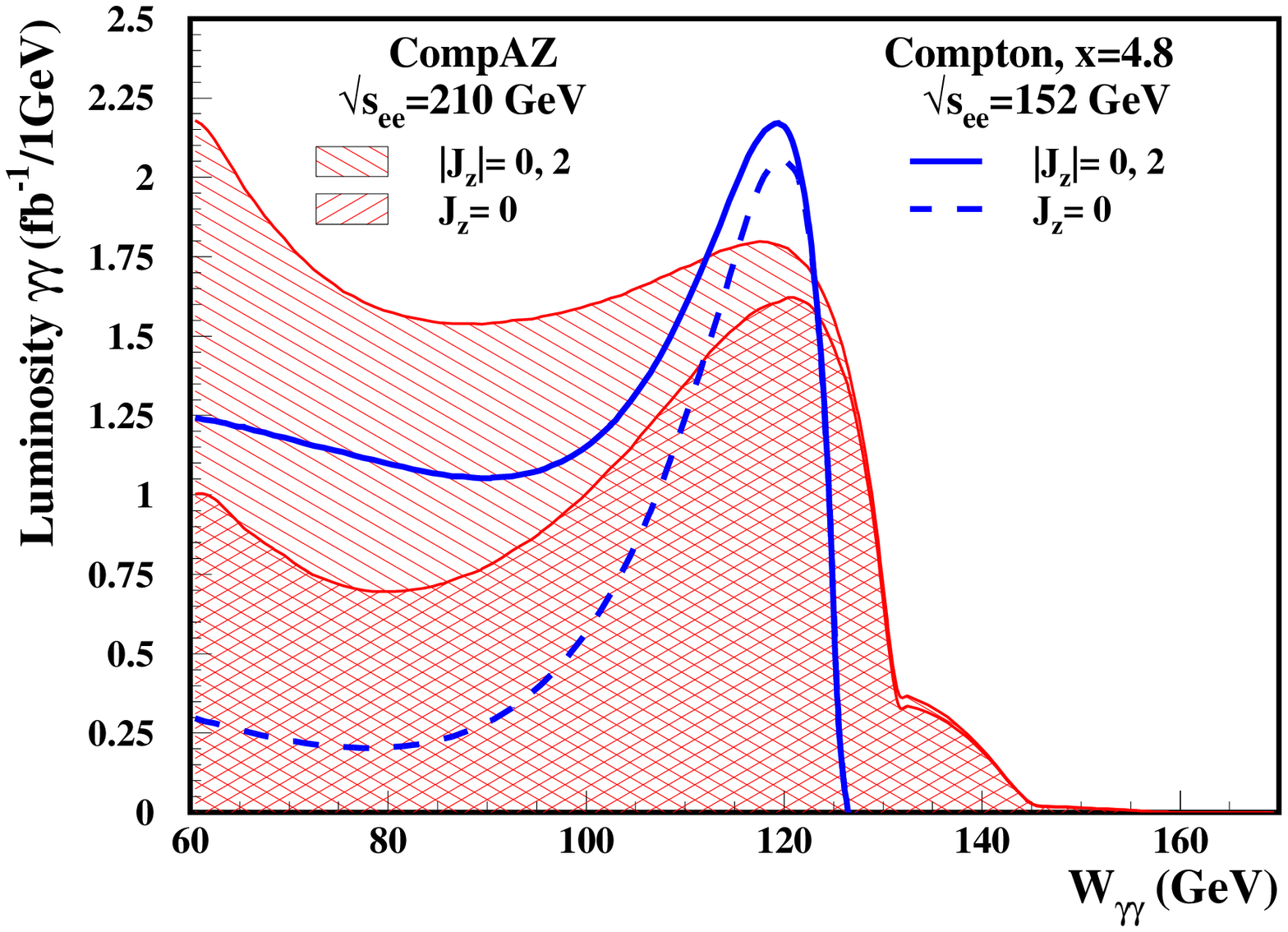}{fig:lumi_comparison_SRJ_CompAZ}{
Photon--photon luminosity spectra used in the analysis of the SM Higgs-boson production with mass \Mheq 120~GeV,
as a function of the invariant mass of the two colliding photons $\Wgaga$.
The spectrum 
 obtained from CompAZ parametrization 
is compared with 
the lowest order QED predictions 
for the Compton scattering, used in the earlier analysis (lines). 
%
The total luminosity distribution ($J_{z}=0, \pm 2$) and the $J_{z}=0$ contribution are shown, separately.}

As described in \ref{pc_as_ext_of_lc}, 
the $\gaga$-luminosity spectrum is influenced by many various effects.
To take them properly into account, 
a dedicated program for detailed beam simulation for the Photon Collider at TESLA has been developed \cite{V.TelnovSpectra}.
Large samples of $\gaga$ events were generated at selected energies and are
available for further analysis. 
The simulated photon-photon events were directly used in this analysis
when a proper description the low energy tail of the spectrum was crucial, \eg for the so-called \emph{overlaying events}%
\footnote{ See Section \ref{sec_overlaying_events} in Chapter \ref{ch_signalbackground}.}.
However, in the high-energy part of the $\gaga$ spectrum,
\ie for $\Wgaga > 0.3 \, \Wgagamf$, where $\Wgagamf=2 \Emf$, 
the results of the full simulation are well described by the \CompAZ{} parametrization \cite{CompAZ}.
The subroutines of the \CompAZ{} package were used when ``continuous'' description
was necessary. 
For example, in case of a very narrow resonance production
the full simulation provides only a few $\gaga$ events
in the region of interest per one million of simulated photon-photon collisions. 
Hence, analytical approach is much more efficient.
Also the NLO QCD program used for generating  $\gagaQQg$ events required
a functional description of the luminosity spectrum
for a proper calculation of the cross section.
This analysis is based on the realistic $\gaga$ luminosity simulation 
for the Photon Collider at TESLA \cite{TDR}. 
Some earlier studies of Higgs-boson production in the process $\gagahiggsbb$ 
assumed other laser parameters and/or ``ideal'' $\gaga$-luminosity spectrum
(\ie spectrum corresponding to the LO Compton cross-section formula) \cite{MMuhlleitner_Thesis,JikiaAndSoldner,MMuhlleitner}.
The ``ideal'' spectrum, used in  \cite{JikiaAndSoldner},
is compared with \CompAZ{} parametrization of the realistic beam simulation results
in Fig.~\ref{fig:lumi_comparison_SRJ_CompAZ}. 
As  can be seen, the ``ideal'' spectrum would be more advantageous 
for the narrow-resonance production.
Additional effects, which have to be taken into account in the realistic
study, increase the contribution of low energy $\gaga$ collisions
and make the high energy peak wider.
Moreover, the leading order results for Compton process assume $x=4.8$ 
whereas  fixed laser wave length 
is assumed in the present design for the whole energy range of electron beams,
resulting in $x$ parameter values smaller than the optimum value used in an ``ideal'' spectrum.
Therefore, results obtained in this analysis, using the realistic spectra description,
should not be directly compared to results obtained with ``ideal'' spectra.
 Details of 
the $\Wgaga$-spectrum obtained with \CompAZ{} for \sqrtseeeq 210~GeV are shown
in Fig.~\ref{fig:gaga_w_spectrum_see210_wgaga_gt_60}.
Contributions from two polarization combinations are indicated separately, \ie $J_z=0$ and $|J_z|=2$
where $J_z$ is the total $\gaga$ angular momentum projected  on a collision ($z$) axis.%
\footnote{
The coordinate system used in this document is a right handed system,
with the $z$-axis pointing in the direction of the electron beam in the $\epem$ mode, 
and the $y$-axis pointing upwards. 
The polar angle $\theta$  and the azimuthal angle $\phi$  are defined 
with respect to $z$ and $x$, respectively, 
while $r$ is the distance from the $z$-axis.
When describing selection procedure the angle with respect to the beam direction
is limited to $0 \leq \theta \leq \pi/2$.
} 
The suppression of $|J_z|=2$ luminosity can be clearly seen in the high-$\Wgaga$ part
of the spectrum.
The threshold at $\Wgagamf = $ 131~GeV, expected
for collisions of two photons 
produced in the lowest order Compton scattering, is not sharp.
There is a tail of collisions involving two photons 
from the second order process $e^{-} + 2 \ga_{L} \ar e^{-} + \ga $ 
for which the highest possible energy is around 161~GeV.
The intermediate structure emerges from ``mixed'' collisions with photons
originating from different scattering  processes.
If not stated explicitly otherwise, 
the results presented in this work are obtained for an integrated luminosity
expected after one year of the TESLA Photon Collider running \cite{V.TelnovSpectra}.
In Table \ref{tab:PLCluminosity} the total photon-photon luminosity per year, $\Lgaga$, 
is shown for different electron beam energies. 
Also shown are: the Higgs-boson mass  corresponding to the maximum of $J_z=0$ luminosity spectrum 
for given beam energy, 
and the expected luminosity in the high energy part of the spectrum, 
\ie for $\Wgaga > \Wgagamin$ where $\Wgagamin=0.5 \, \Wgagamf$.

\begin{table}[!b]
\label{tab:PLCluminosity}
\begin{tabular}{|c|c|c|c|c|c|}
\hline
$\sqrt{s_{ee}}$ [GeV] & $\Mhiggs$ [GeV] & $L_{\gaga}$ [fb$^{-1}$] & $\Wgaga^{\min}$ [GeV] & $L_{\gaga}(\Wgaga>\Wgaga^{\min})$ [fb$^{-1}$]  \\
\hline
 211    &   120 &  410  &     65  &   111  \\
 222    &   130 &  427  &     70  &   116  \\
 234    &   140 &  447  &     75  &   121  \\
 247    &   150 &  468  &     81  &   126  \\
 260    &   160 &  489  &     86  &   132  \\
 305    &   200 &  570  &    106  &   150  \\
 362    &   250 &  683  &    131  &   173  \\ 
 419    &   300 &  808  &    157  &   196  \\
 473    &   350 &  937  &    182  &   216  \\
\hline
\end{tabular}
\caption{Luminosity per year expected for the Photon Collider running at given $\sqrt{s_{ee}}$. 
 One accelerator year [y] is assumed to be equal to 10$^7$ s.
 Also shown are: the Higgs-boson mass, $\Mhiggs$, corresponding to the maximum of  the  $J_z=0$  luminosity,
and the expected luminosity in the high energy part of the spectrum, 
\ie for $\Wgaga > \Wgagamin$ where $\Wgagamin=0.5 \, \Wgagamf$.
}
\end{table}

%

%

%

\pnfig{t}{\figheight}{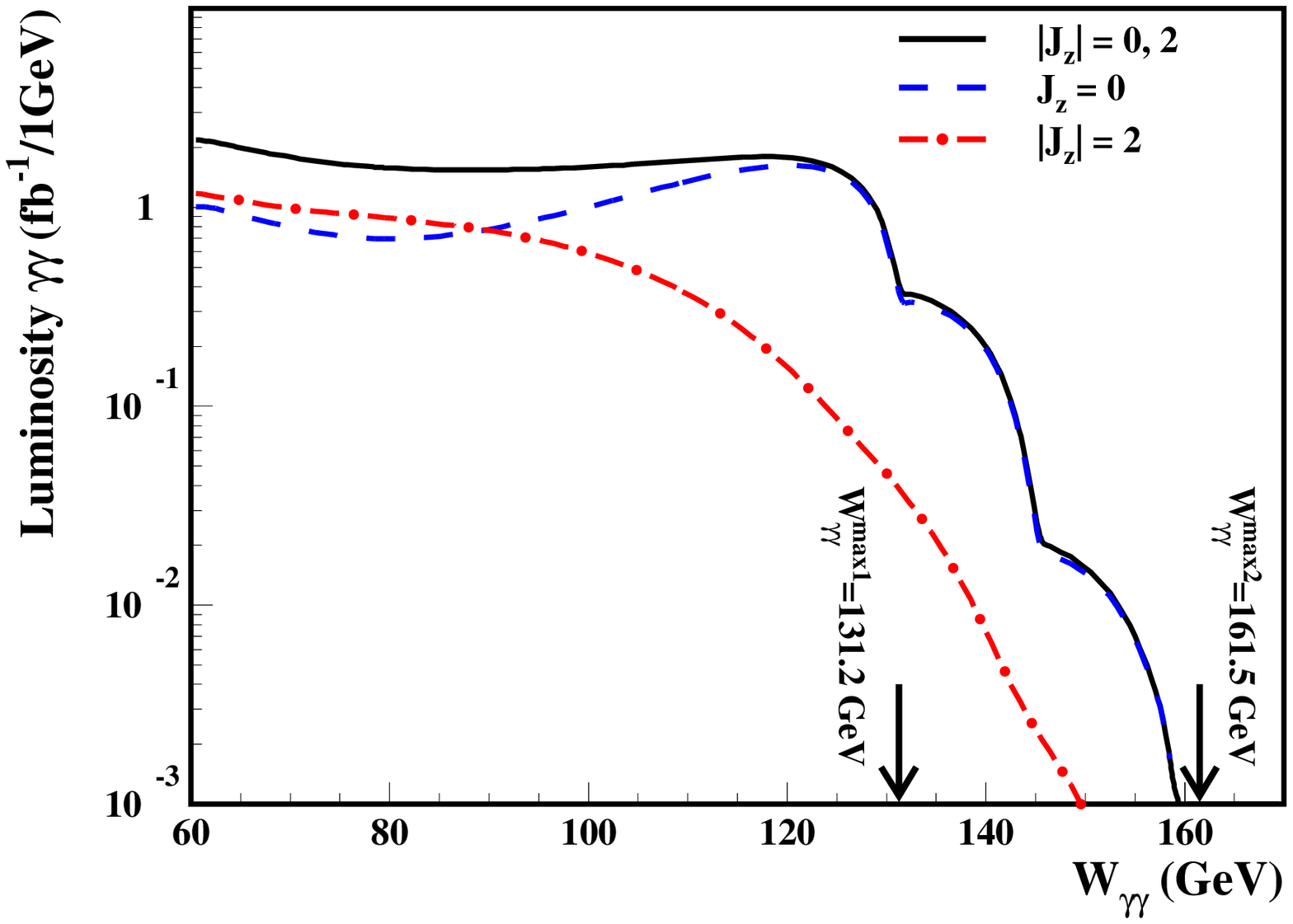}{fig:gaga_w_spectrum_see210_wgaga_gt_60}{
\CompAZ{} parametrization of the photon--photon luminosity spectra for \sqrtseeeq 210~GeV
as a function of the invariant mass of two colliding photons $\Wgaga$.
The total luminosity and contributions  
with $J_{z}$=0 and $J_{z}=\pm 2$
are shown.}

%
\subsection{Collision region}

The $\emem$ beams used in the photon collider will have similar geometrical parameters
as the $\epem$ beams of LC (they will be produced in the same damping rings, 
compressed by the same compression system etc.).
However, 
beamstrahlung due to beam-beam interactions is not present 
and the $\emem$ beams can be focused on a smaller area at the interaction point (IP).
For \sqrtseeeq 200~GeV electron bunches are assumed to have: 
$\sigma_{x}=140$~nm, $\sigma_{y}=6.8$~nm and $\sigma_{z}=0.3$~mm.
The longitudinal  ($z$)  photon-beam bunch size at the photon collider is approximately 
the same as the corresponding size of an electron bunch.
However, as a distance of $b = $ 2.6~mm is foreseen between CP and IP,
transverse sizes of the photon bunch will be greater than that of the electron bunch
due to 
the angular spread of the Compton scattering.
This affects distribution in the $y$ direction, 
as an additional spread  is of the order of $\sigma_{y}$,
but it does not influence the $x$-size of the bunch ($\sigma_{x} \gg \sigma_{y}$). 
For two head-on colliding bunches, which have Gaussian distribution with equal variances
(and the same speed),
the spacial distribution of collision probability follows 3-dimensional Gaussian 
distribution
with all three variances two times smaller than corresponding bunch parameters,
\ie $\sigma_{a}'^2=\sigma_{a}^2/2$ ($a=x,y,z$). 
%
%

%
Transverse dimensions of a photon bunch decrease slightly 
with $\sqrtsee$.
For \sqrtseeeq 200-800~GeV $x$ and $y$ dispersions of the photon bunch 
are 140-70~nm  and 15-5~nm, respectively. 
%
%
%
%
%
Vertical dimension of IP density, $\sigma_{y}'=\sigma_{y}/\sqrt{2}$, is about 10~nm or smaller,
so distribution in this direction is too narrow to influence
the event reconstruction and can be safely neglected.
So would be the horizontal dimension $\sigma_{x}'$ 
if the beams collided head-on.
However, the crab crossing scheme results in modified
 collision density in the horizontal direction.  
This effect is schematically shown in Fig.\ \ref{fig:crab_crossing}.
Assuming that beams collide with relative angle $\alpha_{c}=34$ mrad, 
the $x$-size of collision region is given by the following formula:
\[
 \sigma_{x}' = \sqrt{ \frac{1}{2}(\sigma_{x}^2 + \sigma_{z}^2 \tan^2 \frac{\alpha_{c}}{2}) }
\] 
%
%
This gives,  for all considered collider energies, 
$\sigma_{x}' \simeq 3.6$ $\mu$m. 
This  value is around 36 times greater than the spread $\sigma_{x}/\sqrt{2}$
expected in case of collinear beams
and comparable to the precision expected in the vertex position reconstruction.
Therefore horizontal spread of the interaction point position cannot be neglected.
For the analysis presented in this thesis the longitudinal size 
of the collision region  is most important.
As this is of the order of 100 $\mu$m, we can expect 
that additional tracks and clusters due to \emph{overlaying events}
(resulting in additional vertexes, changed jet characteristics etc.)
can influence the flavour-tagging algorithm
and affect the event selection.
Therefore, generation of all event samples used in the described analysis 
took into account the Gaussian smearing of primary vertex with 
 $\sigma_{x}'=3.6$ $\mu$m, $\sigma_{y}'=11$~nm and $\sigma_{z}'=0.21$~mm,
and the beams crossing angle in horizontal plane, $\alpha_{c}=34$ mrad. 
%

\pnfig{t}{\figheightsmall}{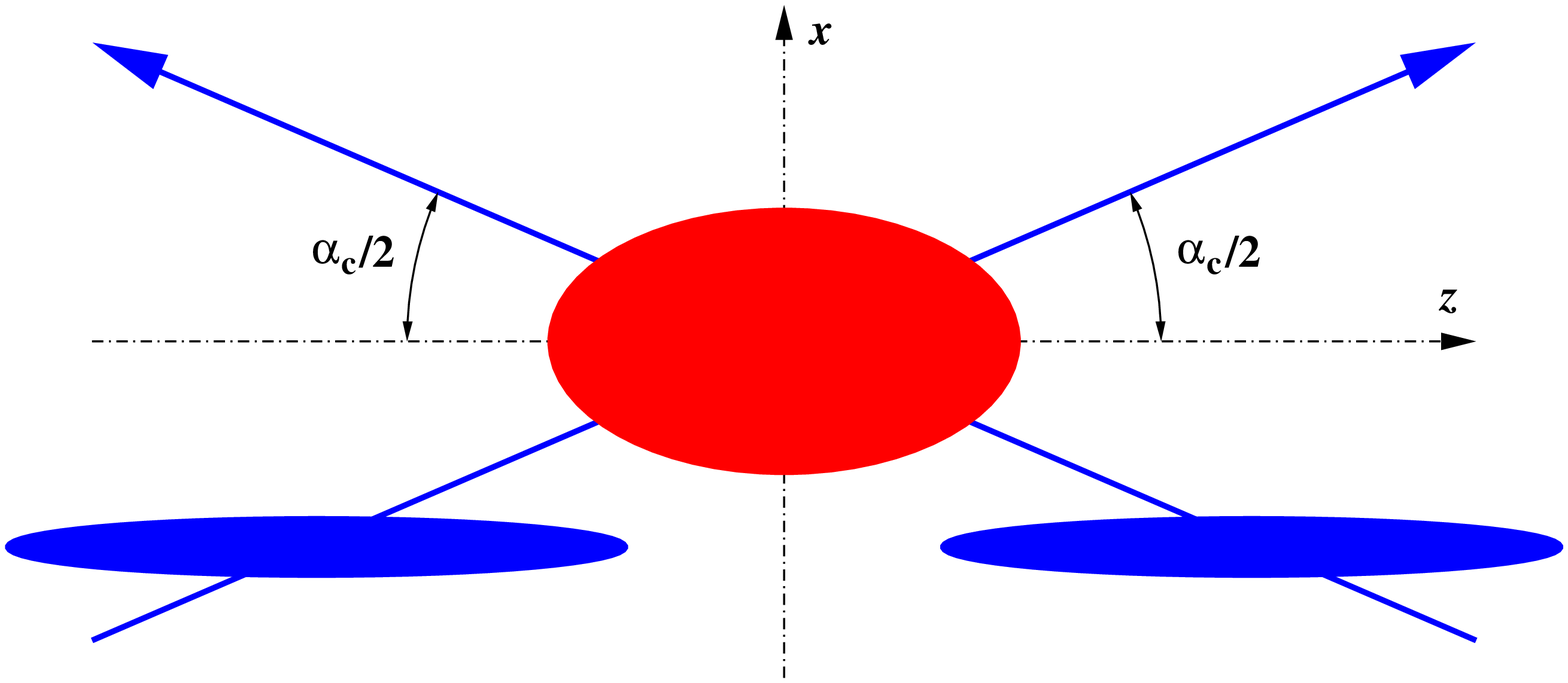}{fig:crab_crossing}{
A scheme of the crab-wise bunch--bunch collision
with the relative angle $\alpha_{c}$ between beams.
Bunches (the blue ellipses) are tilted to preserve the luminosity. 
The resulting primary vertex distribution, 
with $\sigma_{z}' < \sigma_{z}$ and $\sigma_{x}' > \sigma_{x}$, is shown (the red ellipse).
}

%

\section{The detector at TESLA}

The basic design of the TESLA detector for the Photon Collider
is the same as for the $\epem$ TESLA mode.
However, some modifications are needed due to the 
more complicated beam delivery system,
including optical system
which guides the laser beams to the conversion point.
To protect detector components against the high-intensity
low-angle radiation%
\footnote{
Background arises from synchrotron radiation and 
from upstream or downstream sources of $\ga$, $e^+$, $e^-$ and $n$.}
the tungsten mask is placed between the beam system and the detector.
In case of the $\epem$ collider the opening angle of the mask is $\thetamask = 83$ mrad.
Particles produced at smaller angles will not enter the detector.
In case of the Photon Collider the value of $\thetamask = 130$ mrad (7.5$^\circ$) has been chosen
as more space is required for optical system and beams removal.
This results in  moderate loss of hermeticity in comparison with the $\epem$-detector.
Moreover, in case of the $\epem$-detector two forward calorimeters
(Low Angle Tagger and Low Angle Calorimeter)
are foreseen which together cover the region down to around 5 mrad.
These components will not be installed in the Photon Collider option.
In the following, main components of the detector for the Photon Collider are described.
The description is based on the TESLA Technical Design Report (TDR) \cite{TESLATDR_part4} 
and the manual for the fast-simulation program \Simdet{} \cite{SIMDET401}.
For many detector components different choices of  technology 
and/or design were considered in the TDR.
We discuss only these solutions which have been implemented in the \Simdet{} program
and can be used to simulate the response of the detector.
Because all proposed designs are expected to fulfill performance standards
described in the TDR 
one can assume that our physical results should not worsen 
if alternative designs of the considered subdetectors are included in the final project.
%

\pnfig{!tb}{\figheight}{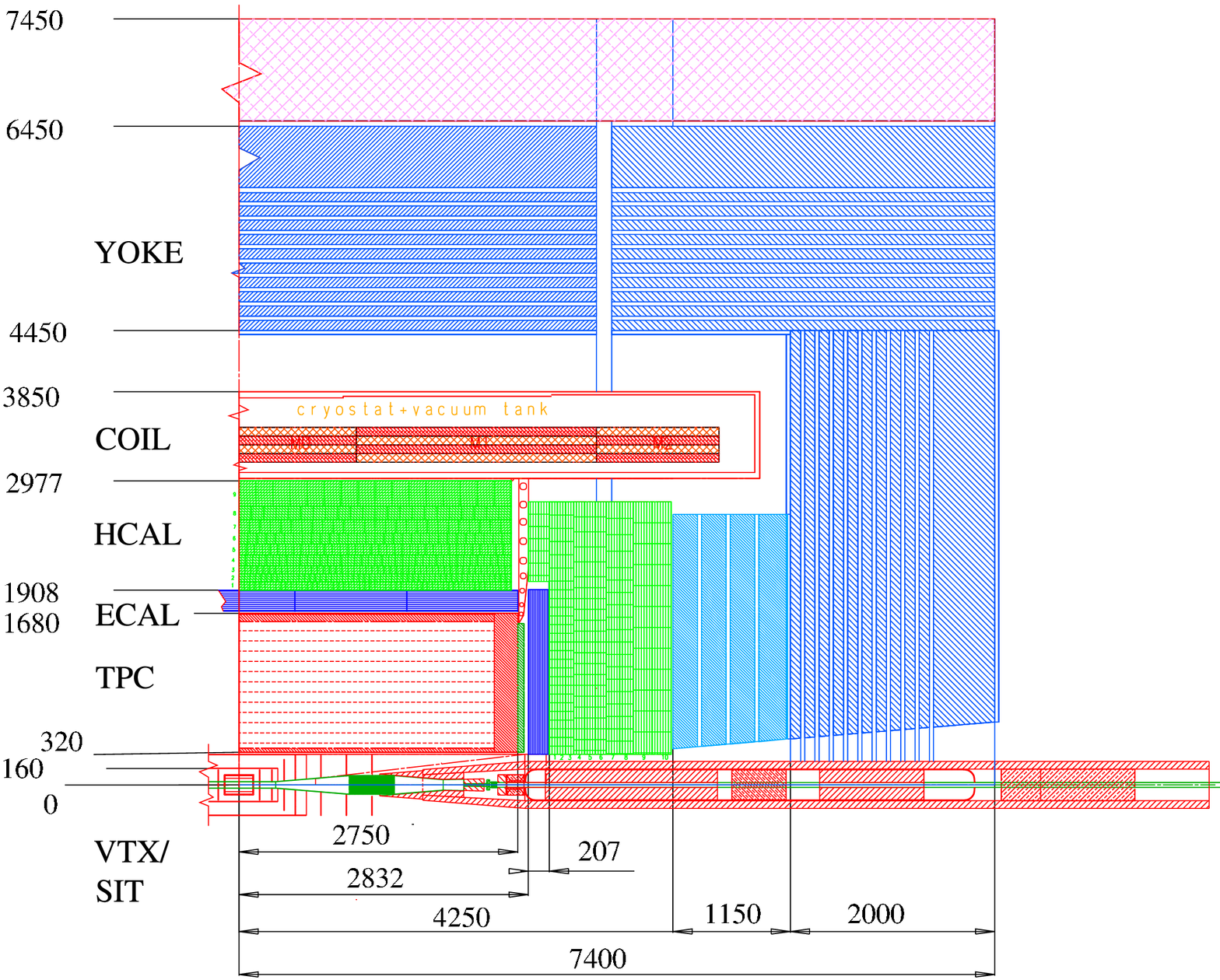}{fig:Detector}{
A scheme of one quadrant of the TESLA Detector \cite{TESLATDR_part4}. Dimensions are in mm. 
}

The schematic view of the TESLA detector is shown in Fig.\ \ref{fig:Detector}.
The detector closest to the interaction point is 
a multi-layer microvertex detector with a total length of around 30 cm.
Currently at least two technologies are considered for this detector:
charge-coupled devices (CCD) and active pixel sensors (APS).
In this work the CCD-based option is used. %
It has well-defined geometry, material budget
and the highest established performance in terms of precision over 
a wide range of incident angles
(for devices of the dimensions needed for this application, \ie tens of cm$^2$),
and only for this version the fast-simulation program provides a parametrized track covariance
matrix which is crucial for the realistic flavour tagging simulation. 
 With this design  precision of the position measurement of 3.5 $\mu$m can be achieved.
\footnote{
CCD vertex design implemented in \Simdet{} assumes  the radius of the innermost layer  of  1.5 cm, 
which is the optimum choice for $\epem$ based on the  background considerations.
In case of the Photon Collider the inner radius of the vertex detector 
will probably have to be increased to about 2 cm.%
}
To obtain a good reconstruction efficiency at least three detector layers 
are proposed so that, 
together with the silicon tracking subsystem (SIT), 
at least  five silicon layers inside the TPC are available.
In addition to the vertex detector
a tracker system consists of intermediate silicon tracking detectors (SIT),
a large Time Projection Chamber (TPC) and forward chambers.
Silicon tracking subsystem includes cylinders in the barrel 
(SIT) and disks in the forward region (FTD).
In the barrel region two layers of silicon strip detectors cover the region down to $\theta = 25^\circ$.
One of the cylinders, at $r$ = 16 cm, improves the track reconstruction efficiency
mostly for long-lived particles which decay outside  
the vertex detector.
Three pixel and four strip silicon detectors 
with point resolutions of 10 $\mu$m and 50 $\mu$m, respectively 
are placed on each side, in the forward region (the endcaps).
The main role of these detectors is to improve the momentum resolution for tracks
by adding a few very precise space points at comparatively large
distance from the primary interaction point, and to help the pattern recognition 
in linking the tracks found in the TPC
with tracks found in the vertex detector.
%
%

%
The central tracking system consists of two gas-filled chambers:
a large volume time projection chamber (TPC)
and a precise forward tracking chamber (FCH) located between the TPC endplate and the endcap calorimeter.
The TPC, with 200 readout points in the radial direction ($r = $ 32--170 cm),
provides a very precise measurement of a track curvature, which is used in the determination
of particle momentum.
Because of the high magnetic field of 4 T the minimal transverse momenta of a particle required
to enter and traverse the TPC are around 200 MeV and 1~GeV, respectively,
if the particle charge is equal to the electron charge.
Precise measurement of the specific energy loss in the TPC can be also used for particle identification. 
For tracks traversing  the TPC at large
polar angles  the expected errors on the transverse momentum and the energy loss measurements are
$\sigma (1/p_{T}) \approx 0.02\%/$GeV and $\sigma(\frac{dE}{dx})/\frac{dE}{dx} \approx 5 \%$, respectively.
%
%
%
%
For example, in case of a charged particle with  energy of 20~GeV at
the polar angle of 90$^\circ$
 the momentum resolution is about 80 MeV.
From ionisation losses the separation of kaons and pions should be possible in the  momentum range from 2 to 20~GeV.
Electron identification will be improved compared to 
what can be done with calorimeters alone, especially for low momenta ($p\lesssim$ 3~GeV)
where calorimetric identification is difficult.
%
%

%
Overall tracking system performance,
when the track parameters are determined from combining vertex detector, SIT and TPC measurements,
shows a very high precision of transverse
momentum determination $\sigma (1/p_{T}) \le 0.005 \%/$GeV
if systematic errors $\le 10$ $\mu$m for point position measurements are achieved.
It is worth noticing that the overall momentum-resolution has been improved by about 30\% 
by adding a cylindrical silicon detector (SIT) inside the TPC, \ie at $r$ = 30 cm.

A tracking electromagnetic calorimeter (ECAL), 
build of tungsten absorber plates and thin silicon sensors,
is placed behind the TPC.
%
%
%
The expected energy resolution $\sigma_{E}/E $ is around 11--14\%$/\sqrt{E / \mathrm{GeV}}$,
depending slightly on the energy.
The project assumes very high 3D granularity of this detector, allowing 
measurement of the particle momentum direction.
A hadronic calorimeter (HCAL) is an iron/scintillating tile calorimeter 
with fine transverse and longitudinal segmentations.
The energy resolution for single hadrons,
estimated from simulation of hadronic showers in both calorimeters (HCAL+ECAL),
is%
\footnote{The operator $\oplus$ means ``adding in quadrature'': $a \oplus b \equiv \sqrt{ a^2 + b^2}$}
$\sigma_E/E = 35\%/\sqrt{E / \mathrm{GeV}} \oplus  3\%$.
A large superconducting coil, 6 m in diameter,
produces a field of 4 T with very high uniformity ($\le 10^{-3}$).
The coil is placed behind calorimeters to preserve high
precision of energy measurement, 
reducing the amount of inactive material in front of the calorimeters.
The iron return yoke serves also as a muon ``separator'',
absorbing other particles escaping from the HCAL.

For muon chambers, which are placed inside the yoke and behind it,
resistive plate chamber (RPC) technology is considered.
Although the basic task for a muon detector is to identify muons,
it is also possible to use the inner muon chambers as the ``tail catcher'',
\ie to detect hadronic cascades which are not fully contained in the hadronic calorimeter.
%
%
Full efficiency of muon identification is reached for muons with energy above 5~GeV.
The total length and the diameter of the detector will be around 15 m each.
In general, the detector is designed to measure particles properties
with a very high accuracy in the collision energy range from about 90~GeV 
up to 1~TeV.
Electrons below 150~GeV, muons and charged hadrons 
are best measured in the tracking detectors.
Electrons above 150~GeV and photons by the electromagnetic calorimeter 
and neutral long-living hadrons by the combined response 
of the electromagnetic and hadronic calorimeters.
In the event reconstruction the so-called energy-flow technique will be used 
which combines the information from tracking  system and calorimeter 
to obtain the optimal estimate of the energy flow of produced particles 
and of the original parton four-momenta. 
%
%
%
%
For the energy-flow objects an average energy resolution  
of $\sigma_E / E \simeq 30\% / \sqrt{E / \mathrm{GeV}} $ is expected.
Due to the sparse beam structure 
(a long time interval of 199 ms between two bunch trains,
a separation of two bunches inside a train by 337 ns,
a train length of 950 $\mu$s) no hardware trigger is foreseen.
A total data volume of roughly 300 TByte per year will be stored for physical analyses.
%

\subsection{Simulation setup  \label{subs_simulation_setup}}

%
The fast simulation program for the TESLA detector, 
  \Simdet{} version \Siver{} \cite{SIMDET401},
was used  to model the detector performance. 
%
All detector components are implemented in the program according to the TESLA TDR. 
Parametrizations based on the full simulation of  detector performance with the  \Brahms{} program \cite{BRAHMS} 
are used to describe energy and angular resolutions.
The track reconstruction efficiency and charge misinterpretation probability are momentum dependent.
An energy-flow algorithm is used to link information from tracking system and calorimeters.
In the first stage energy deposits in the calorimeters are joined into clusters.
Then energy flow objects are defined by linking
clusters with tracks reconstructed in the tracking system%
\footnote{
 In the current \Simdet{} version
 the idealised pattern recognition is still used,
 \ie clusters are linked with tracks
 relying on the information about originally generated particles.}%
.
For the reconstructed particle tracks 
a track covariance matrix is calculated 
in the base: $xy$, $z$, $\theta$, $\phi$, $1/p_T$.
Because two forward calorimeters,
Low Angle Tagger and Low Angle Calorimeter, 
cannot be installed in the detector at the PC, they are not used in our simulation setup.
Also the information about  energy loss measured in the TPC,
which is not properly simulated yet,
is not used for particle identification. 
Instead, an appropriate misidentification probability is assumed for each particle species.
Within the current  \Simdet{} version it is not yet possible to set
a wider opening angle of the forward mask as required for the Photon Collider.
To take the modified mask setup into account 
all generator-level particles are removed from the event record, before entering the detector simulation,
if their polar angle is less than $\thetamask = 130$ mrad.
%


\chapter{Signal and background \label{ch_signalbackground}}

In this Chapter the signal and background processes, and methods used in their simulation are described. 
The signal of the Higgs-boson production 
considered in this thesis
is the process $\gagahiggsbb$, 
whereas the main background processes are $\gagabb$ and $\gagacc$.
The $\bbbar$ pair production is an irreducible background which can be suppressed
only by kinematic cuts.
The $\ccbar$ pair production contributes to the background due to the finite 
probability of being tagged as the $\bbbar$ pair production.
Background contributions from light quark and tau pair production,
$\gaga \ar \qqbar / \tautau$  ($q=u,d,s$), are also considered.
The process \gagahad{} is described in detail due to its contribution to \emph{overlaying events}.
For heavy Higgs bosons, $\Mhiggs \gtrsim$ 160~GeV, 
also the process $\gagaWW$ is taken into account. 
%
%
The underlying statistical principles used to describe 
the possibility of having more than one collision 
in single bunch crossing, and used in generation of the overlaying events
are described in Appendix \ref{app_eventgeneration}.


\section{Signal processes  \label{sec_signal}}

%
%
As the Higgs boson is a spinless particle, the distribution of its two-body decay products
is uniform in the three-dimensional space (in the Higgs-boson rest frame).
In spherical coordinates 
the distribution is uniform in $\cos\theta^*$ and $\phi^*$,
where $\theta^*$ and $\phi^*$ are polar and azimuthal angle, respectively.
In the Photon Collider \higgs{} will be produced in  collisions of photons which
 will have, in general, different energies.
Thus, the center of mass system will be boosted with respect to the laboratory frame,
 resulting in a non-uniform distribution in $\cos\theta$,
where $\theta$ is the polar angle in the laboratory frame.
%
%
%
%
%
However, as already mentioned in section  \ref{ss_photon_luminosity},  energy of 
the electron beam is assumed to be tuned for the  highest resonance production  rate. 
It means that most collisions will involve two photons with high and similar energies 
as~in~all considered cases the resonance is narrow 
(the total width, $\Gamma_{\higgs}$, is much smaller than the width of the $\Wgaga$-distribution in the high energy part).
Consequently, the Higgs boson will have small longitudinal momentum in comparison to the mass,
%
%
and its decays will be nearly isotropic also in the laboratory frame.

%
Total widths and  branching ratios of the Higgs bosons were calculated with the program \Hdecay{} \cite{HDECAY} (version 3.0), 
where higher order QCD corrections are included. 
The mass of the top quark equal to 174~GeV was assumed.
The contributions from the decay $\higgsgg^* \ar g \bbbar$ were not added to the branching ratio $\Brhiggsbb$
as the kinematical characteristic of such events is different from the direct decay to the $\bbbar$ pair.%
\footnote{Inclusion of the decay $\higgsgg^* \ar g \bbbar$ does not change 
the results of this analysis as $\Brhbb$ increases only by around 1\% for the SM case,
and for considered MSSM parameters $\BrAOHOgg / \BrAOHObb \lesssim 1\%$.
However, if  events with only one $b$-tagged jet were  accepted,
one would have to  use inclusive branching ratio
$\Brhbbinclusive$, measured at the $\epem$ LC, to obtain final results for $\Ghgaga$.
}
For the considered mass  range between 120 and 160~GeV
the total width of the SM Higgs boson increases from about 3.6 to 77 MeV, 
and its branching ratios $\Brhgaga$ and $\Brhbb$ decrease
 from 0.22\% to 0.06\% and from 68\% to 4\%, respectively,
in the mass
Event generation for Higgs-boson production process was done 
with the \Pythia{} program \cite{PYTHIA}. 
%
A parton shower algorithm implemented in \Pythia{}
was used to generate the final-state partons. 
%
The fragmentation into hadrons was also performed using the \Pythia{} program,
both for Higgs-boson production and for all background event samples. 
%

%


%

\begin{table}[!t]
\label{tab:MSSMparsets}
\bc
\begin{tabular}{|c|c|c|c|c|}
\hline
Symbol & $\mu$ [GeV] & $M_2$ [GeV] & $A_{\widetilde{f}}$ [GeV]  & $M_{\widetilde{f}}$ [GeV] \\
\hline
 $I$     &  200        &  200        &  1500                   &  1000           \\
 $II$    & -150        &  200        &  1500                   &  1000           \\
 $III$   & -200        &  200        &  1500                   &  1000           \\
 $IV$    &  300        &  200        &  2450                   &  1000           \\
\hline
\end{tabular}
\ec
\caption{MSSM parameter sets used in the described analysis.} 
\end{table}
%

In case of the MSSM Higgs-boson production the analysis  has been developed assuming
MSSM parameters  similar to these used in \cite{MMuhlleitner},
\ie $ \tbseven , \, \mu = \pm 200 $~GeV  and $M_{2} = $ 200~GeV,
taking into account decays to and loops of supersymmetric particles. 
The parameter value  $ \tbseven $ is used in the event generation and obtained results 
are rescaled to the parameter range $\tanb = 3 - 20$.
In the following, parameter sets from  \cite{MMuhlleitner} 
will be denoted as $I$ and $III$, see Tab.\ \ref{tab:MSSMparsets}. 
Only the values 
of trilinear couplings are changed (from  $A_{\widetilde{f}} = $ 0 to $A_{\widetilde{f}} = $ 1500~GeV),
%
so that the mass of the lightest Higgs boson, instead of being around 105~GeV (for \tanbeq 4 and \MAOeq 300~GeV)
is above the current lower limit for the  SM Higgs boson, $\Mh > $ 114.4~GeV.
Results for heavy neutral Higgs bosons are the same 
as for parameter sets proposed in \cite{MMuhlleitner}.
The intermediate scenario $II$ with $\mu = -150 $~GeV is also considered.
For comparison with predictions presented by LHC experiments, the scenario $IV$
 used in \cite{CMSDiscovery} is also included.
In all cases the common sfermion mass equal to 1 TeV was assumed.
We have checked that all parameter sets imply masses of neutralinos, charginos, sleptons and squarks
higher than current experimantal limits.

%
%
%
For the wide range of parameter values the heavy neutral Higgs bosons, $\AO$ and $\HO$,
are nearly mass degenerate.
The mass difference $\MHO - \MAO$  decreases with increasing $\tanb$ and $\MAO$
and is similar for all considered parameter sets.  
For  \MAOeq 200~GeV the mass difference decreases from
$\MHO - \MAO \approx 12 $~GeV for \tanbeq 3 to 0.7~GeV for \tanbeq 15.
If \MAOeq 350~GeV, the corresponding values are 6~GeV and 0.3~GeV.
The mass difference is larger or comparable 
to the total widths of  $\AO$ and $\HO$ 
which vary between 50~MeV and 4~GeV.
%
The branching ratios relevant for this study change between 3\% and 90\% for $\BrAOHObb$,
and from $2\cdot 10^{-7} $  to $ 9\cdot 10^{-5}$ for $\BrAOHOgaga$.
Processes $\gagaHbb$ and $\gagaAbb$ do not interfere%
\footnote{Although in general case two processes with $\AO$ and $\HO$ in the intermediate state
(different $CP$ quantum numbers) can interfere.
For example, there is interference between chargino production processes 
in $\gaga$ collisions via $\AO$ and $\HO$: $\gagaAH \ar \widetilde{\chi}^{+}_{i} \widetilde{\chi}^{-}_{j}$ 
(\eg see \cite{MMuhlleitner_Thesis} eq.~3.59).} 
(\eg see \cite{MMuhlleitner_Thesis} eq.~3.15), 
and $h-\HO$ interference is negligible due to the large difference in masses 
and relatively small widths of the bosons.
Therefore total $\gagaAHbb$ production rate is equal to the sum of both contributions.
%
%
%
%
%
%
%
%
%
%
\pnfig{t}{\figheight}{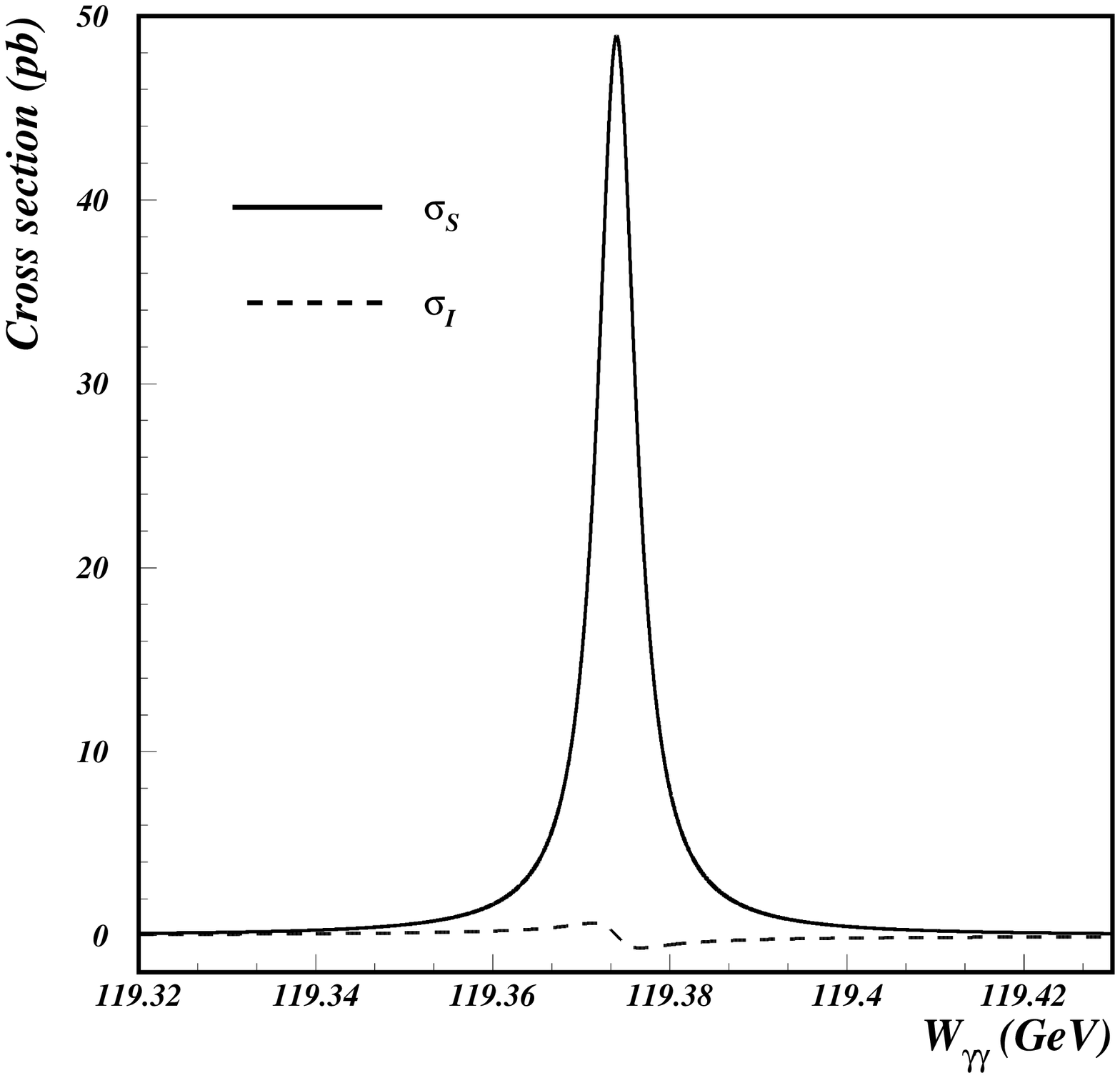}{fig:cross_section_signal_interference_l}
{The total cross section for process $\gagahbb$ for set I of the MSSM parameters values (see text) with \MAOeq 300~GeV and $\tbseven$.
 The signal only, $\xs_S$, and interference part, $\xs_I$, are shown separately. 
As the other neutral Higgs bosons are narrow and heavy, they do not influence the $h$ cross section.}
\pnfig{t}{\figheight}{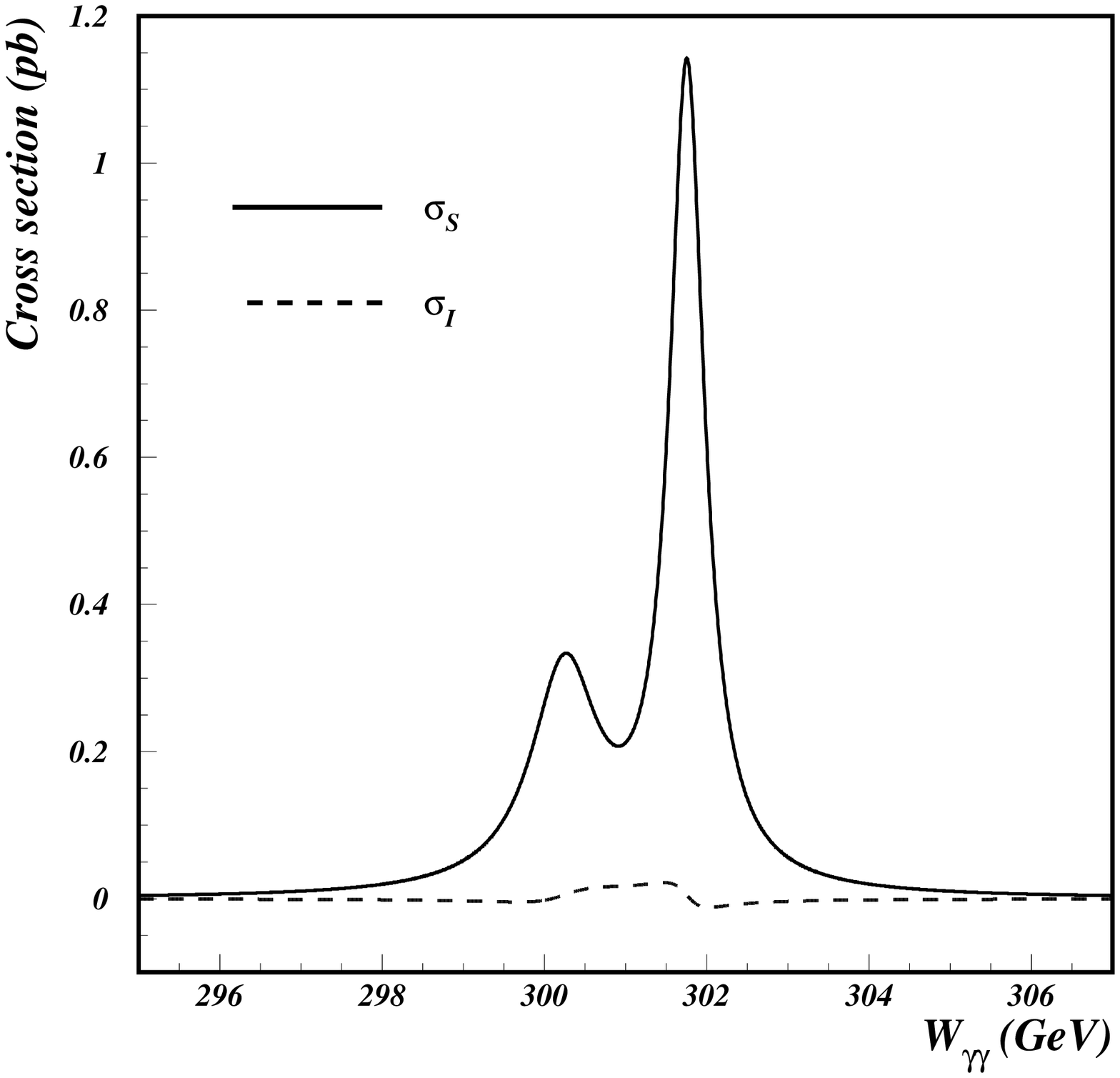}{fig:cross_section_signal_interference_h}
{The total cross section for process $\gagaAHbb$ for set I of the MSSM parameters values (see text) with \MAOeq 300~GeV and $\tbseven$.
 The signal only, $\xs_S$, and interference part, $\xs_I$, are shown separately. 
As the light Higgs boson, $h$, has mass of around 120~GeV and is narrow, it does not influence the $\AOHO$ cross section.}

However, for the complete description of the Higgs boson production we also have to consider
the interference  between $\gagahiggsbb$ and non-resonant $\gagabb$ production processes.
The LO interference terms for  $h$ and $\HO/\AO$ production  are shown 
in Fig.\ \ref{fig:cross_section_signal_interference_l} and \ref{fig:cross_section_signal_interference_h},
together with the signal cross sections  $\gagahbb$ and  $\gagaAHbb$.
For all considered cases
these terms are proportional to the real part of the propagator:
\[
 \Re \frac{M^{2}}{s - M^{2} + BM\Gamma} = \frac{M^{2}(s - M^{2})}{(s - M^{2})^{2} + (M\Gamma)^{2}}
\]
As on average (the integral  over $\Wgaga = \sqrt{s}$)  this expression is near to zero 
for small $\Gamma$  (of the order of $\Gamma/M$), 
the interference contribution can be safely neglected .
However, as discussed in the next section, the NLO corrections substantially modify 
predictions for heavy quark production.
The NLO corrections for interference term were calculated in \cite{MMuhlleitner_Thesis}.
As described in \cite{MMuhlleitner_Thesis},
after selection cuts the interference term was  below  the level of $10^{-3}$ of the signal.
Although selection cuts were different  than in our analysis, we can infer the total correction factor.
The cuts decrease interference contribution by the order of magnitude and the signal rate
by about 50\%.
Thus, one can estimate that the interference part contributes no more 
than 1\% of the total signal cross section.
Because this is smaller than other uncertainties, we neglect the interference term
in this analysis.


\section{Heavy quark production background \label{sec_heavy_quark_bkgd}}

The main background for the considered signal process, $\gagahiggsbb$, is the heavy quark-pair production.
An irreducible background consists of events with the $\bbg$ final state,
resulting from 'direct' nonresonant $\bbbar$ production,
$\gagabb$.
In LO approximation the cross section for $J_z=0$ is suppressed 
and the dominant contribution is due to the $|J_z|=2$ state.
This is very fortunate as the $\gaga$-luminosity spectrum is optimized to give
highest $J_z=0$ luminosity 
and the $|J_z|=2$ component is small in the \higgs-production region.
Unfortunately, NLO corrections compensate partially the $m_{Q}^2/s$-suppression 
and, after taking into account luminosity spectra,  both contributions
 (for $J_z=0$ and $|J_z|=2$) become comparable. 
The extensive comparison of NLO and LO results can be found for example in our work \cite{NZKhbbm120appb}.
The other processes $\gaga \ar q\bar{q} (g)$, where $q=u,d,s,c$, 
contribute to the reducible background.
However,  one has to consider these processes due to the non-zero probability 
of wrong flavour assignment in reconstruction
(impurity of flavour-tagging).
Events with $\ccg$ in the final state have the highest mistagging probability.
In comparison to the $\gagabb$ process there is an enhancement factor of $(e_{c}/e_{b})^4=16$ 
in the $\gagacc$ cross section.
 It turns out that after flavour tagging both processes give similar contribution to the background. 
%

%
The background events due to processes $\gagabbgccg$
were  generated using the program written by G.~Jikia \cite{JikiaAndSoldner},
where a complete  NLO QCD  calculation for the production of  massive quarks is performed 
in the massive-quark scheme. 
The program includes exact one-loop QCD corrections to the lowest order processes
$\gagabbcc$ \cite{JikiaAndTkabladze}, 
and the non-Sudakov form factor in the double-logarithmic approximation, 
calculated up to four loops \cite{MellesStirlingKhoze}.
%
Events generated with NLO QCD program were transfered to \Pythia{} program
for hadronisation.
To avoid double-counting of corrections due to real gluon emission,
the parton shower algorithm  was not applied.
However, to estimate the influence of higher order corrections on the event selection efficiency
we also prepared dedicated samples of $\gagabbgccg$ events with parton shower included.
Results of this comparison are presented in Appendix \ref{app_uncert_gagaQQg}.
%

%



%



\section{Other background processes}


In cases of the SM Higgs-boson production for \Mheq 150 and 160~GeV, and in the analysis  
of heavy neutral Higgs-bosons in the MSSM
also the pair production of $W$ bosons, $\gagaWW$, is considered
as a possible background.
The cross section for this process is very high for large $\Wgaga$,
and it can contribute to the background 
if the event is clustered to two or three jets
and
at least one of these jets is $b$-tagged. 
This  can be the case if two jets from hadronic $W$ decays 
are merged together by the jet-clustering algorithm,
or if some jets are ignored in the analysis because they are too close to the beam pipe.
For generation of $\gagaWW$ events the \Pythia{} program is used.
However, as only unpolarized cross section for this  process
is implemented in \Pythia{},
we use polarized differential cross section formulae from \cite{gagaWWpolarized} to obtain
correct distributions for $J_z=0$ and $|J_z|=2$ contributions. 

As there is non-zero probability of mistagging a light-quark jet as a $b$-jet,
the process  $\gaga \ar q\bar{q}$, where $q=u,d,s$,
is also taken into account as a possible background.
Due to the strong dependence of the cross section for this process on the fermion charge ($\xs \sim e_f^4$),
the $\gaga \ar \uubar$ contribution dominates.
The event generation is performed with \Pythia{} using unpolarized LO cross section.
We known that for $J_z=0$ the LO cross section for the process $\gagaqq$  
 is equal to zero for massless quarks.
By convoluting the cross section for $|J_z|=2$  with the total
luminosity spectrum (modulo factor 2)
we overestimate the light quark production background.
Comparing results for $\gagaccg$ and $\gaga \ar \uubar$ we have determined 
that the number of events with light quark-pair production
is  overestimated by a factor of about 2.6 for our SM analysis, 
and by a factor of about 4 for our MSSM analysis.
We do not apply any corrections to decrease this effect.
In the analysis of SM Higgs-boson production the light-quark contribution
is negligible and does not change the results.
In case of MSSM Higgs-boson production we use the overestimated  contribution
of $\gagaqq$, $\quds$, events to effectively take into account the  contribution of
\gagahad{} events (without \emph{direct$\times$direct} interactions)
from which no generated events passed all selection cuts.
We estimated that in the mass window optimal for the cross-section measurement 
(see Chapters \ref{ch_sm_analysis} and \ref{ch_mssm_analysis}) 
the contribution of \gagahad{}  events  would correspond to about 50\%
of the (overestimated) light-quark  contribution. 

 A large number of $\tau$-pair production events will also be observed in the Photon Collider
(the cross section about 1.7 times higher than the LO cross section for $\gagacc$).
Thus, even small $\tau$ mistagging probability could in principle significantly influence our results.
Fortunately, the $J_z=0$ contribution is negligible in this case  due to the $m_{\tau}^2/s$-suppression
(QED higher order corrections are small).
In addition,  considerable amount of energy is carried out by neutrinos.
Consequently, the number of $\tautau$  events 
reconstructed as two-jet events with  high invariant mass should be very small.
For this study  $\gagatautau$ production events were generated with \Pythia{} using polarized cross section.




%

\section{Overlaying events {\protect \gagahad{}} \label{sec_overlaying_events}}

Because of the large cross section and huge $\gaga$-luminosity at low $\Wgaga$,
from one to two \gagahad{} events%
\footnote{%
           For technical reasons we consider only photon--photon events with $\Wgaga >4 $~GeV.
	   However, events with lower $\Wgaga$ are mostly produced with high boost
	   and particles going at very small angles do not enter the detector.
	   For further detailed discussion see Appendix \ref{app_thetatc}.
}
are expected 
at the TESLA Photon Collider per bunch crossing.
These events hardly contribute to the background on their own.
However, they can have a great impact on the reconstruction of other events
produced in the same bunch crossing,
by changing their kinematical
and topological characteristics.
We generate \gagahad{} events with \Pythia{},
%
using the luminosity spectra from a full simulation of the photon-photon 
collisions \cite{V.TelnovSpectra}, 
rescaled 
to the chosen beam energy.
For each considered $\emem$ energy, $\sqrtsee$, 
an average number of the \gagahad{} events 
per bunch crossing is calculated.  
Then, for every signal 
or background event, 
the \gagahad{} events are overlaid (added to the event record)
according to the Poisson distribution.  
In Appendix \ref{app_eventgeneration} principles of  event generation 
and approximations used are discussed in detail.
The \gagahad{} processes are classified according to the type of photon(s) interaction.
If both photons interact as point-like, as described by QED, 
then we call the process \emph{direct$\times$direct}.
However, if one or two  photons interact as a hadronic state 
(vector meson or quantum fluctuation with gluons), 
then they are denoted as \emph{hadron-like$\times$direct} or \emph{hadron-like$\times$hadron-like},
respectively.
As shown in Fig.\ \ref{fig:xsec_gagahad_py} the biggest contribution to the cross section 
is due to processes with  \emph{hadron-like$\times$hadron-like} photons.
Fortunately, the  \gagahad{} cross section is very forward-peaked as seen in Fig.\ \ref{fig:gagahad_3contr_col4}.
A cut on the polar angle of tracks and clusters measured in the detector 
should greatly reduce contribution of 
particles from \gagahad{} processes to selected events.
%
%
%
%
%
Events with  \gagahad{} interactions only, without \emph{direct$\times$direct} interactions,  are  simulated as well.
Such events can mimic signal if two or more of them are overlaid.
As the generation of high-energy \gagahad{} events with significant transverse energy
is very inefficient,
these events are included only in the SM analysis.
In case of MSSM analysis we have estimated the contribution of \gagahad{} events,
and include it effectively as the part of light quark-pair production (see previous Section).
%
%

For more details concerning  \gagahad{} overlaying events and their influence on 
the reconstruction see Appendix~\ref{app_thetatc}.
The package \Orlop{}, created for including the \gagahad{} events 
by the generation of hard $\gaga$ scattering processes,
is described in Appendix~\ref{app_orlop}.


\pnfig{!}{\figheight}{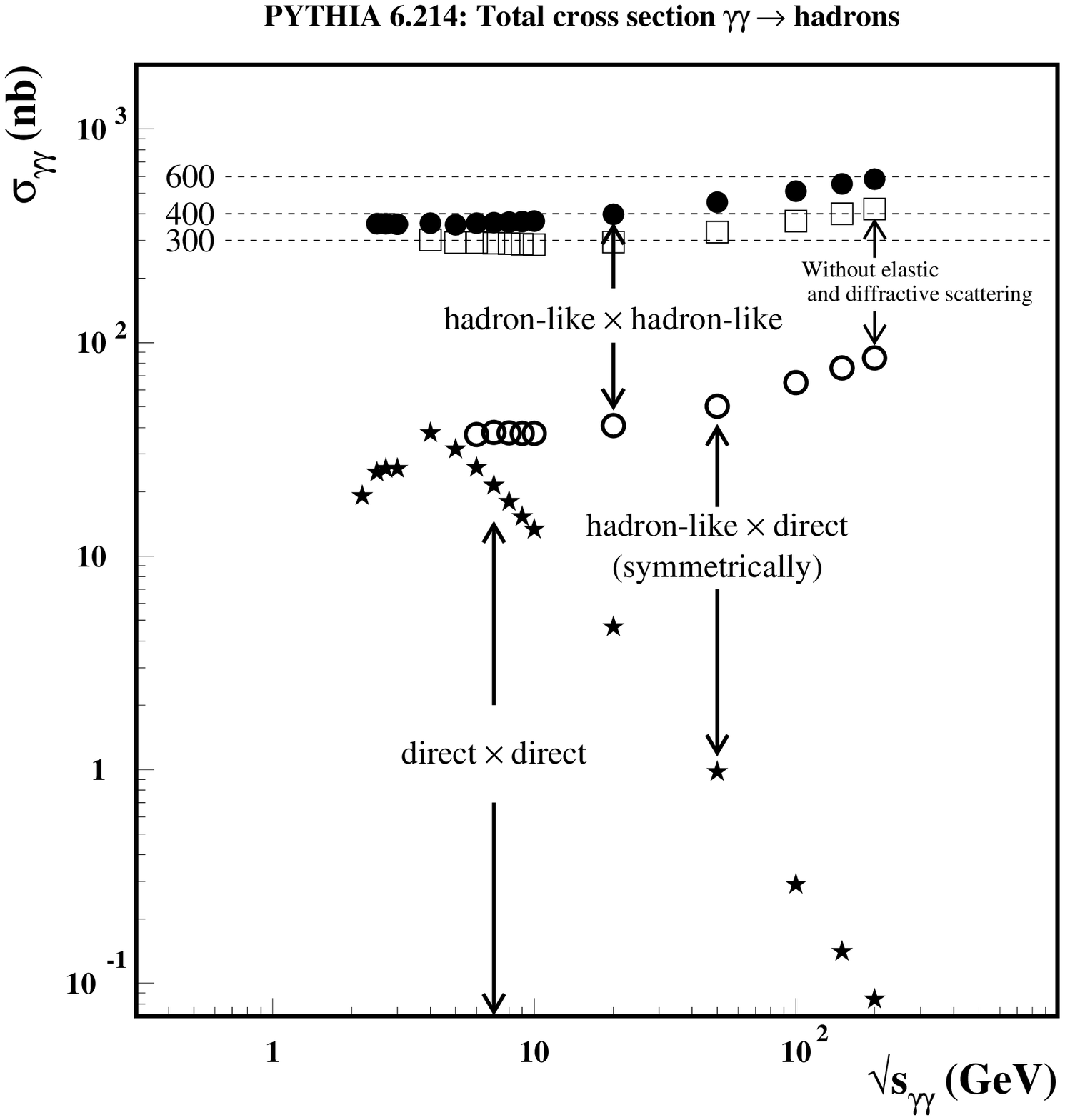}{fig:xsec_gagahad_py}{
Cross sections for the process \gagahad{} obtained with the \Pythia{} program.
Contributions from various event classes are indicated.
Also the magnitude of the cross section for elastic and diffractive processes is indicated.  
}

\pnfig{!}{\figheight}{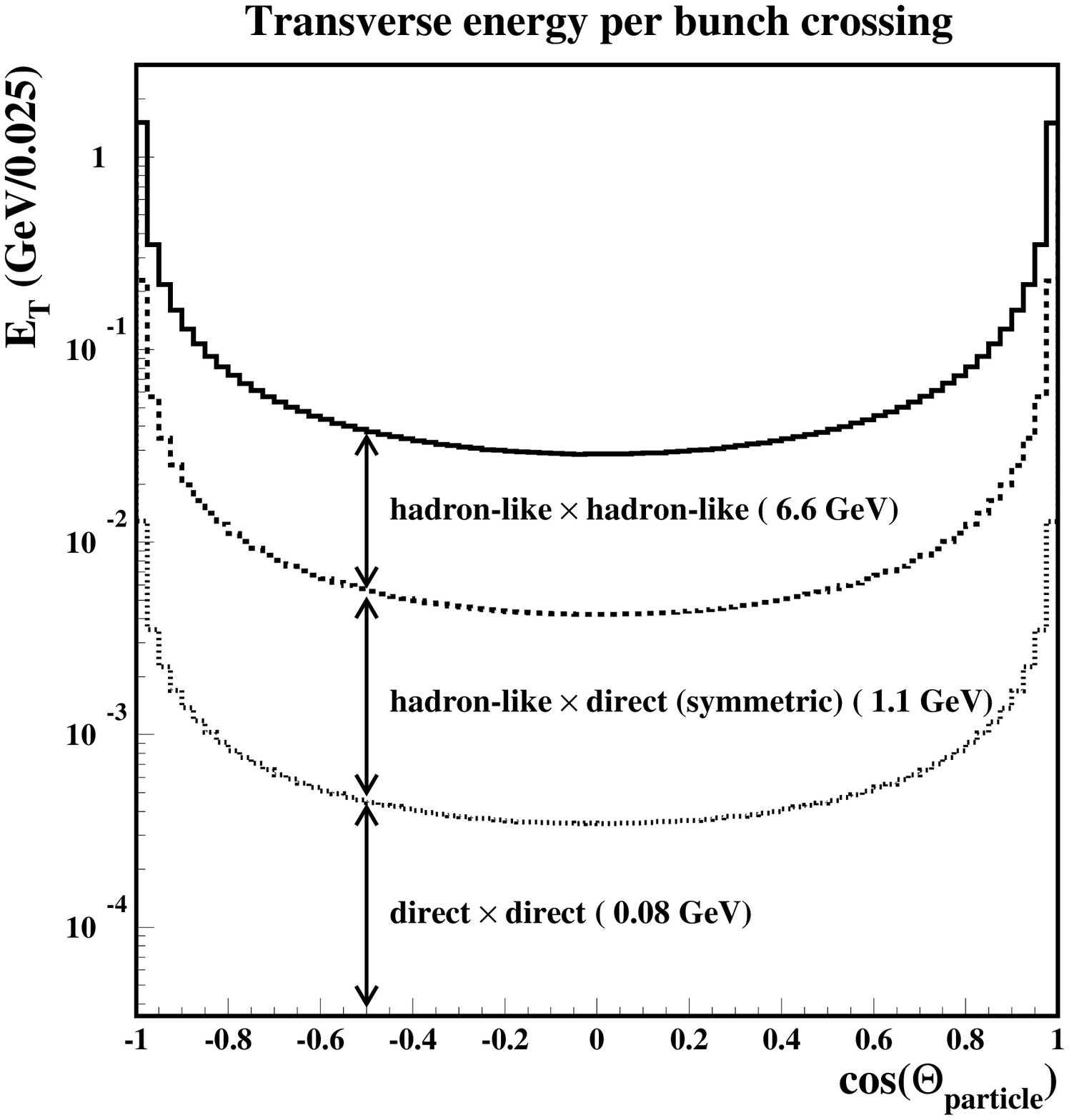}{fig:gagahad_3contr_col4}{
Angular distributions of the transverse energy flow,  \( E_T \), 
for overlaying events \gagahad{} per bunch crossing.
Various components  and their total contributions (numbers in parentheses) are indicated. 
\Pythia{} generation results with luminosity spectrum for \sqrtseeeq 210.5~GeV. 
}


\
\newpage 
\
\newpage 

\chapter{ Standard Model Higgs-boson production  \label{ch_sm_analysis}}

In this Chapter the measurement of the Standard Model Higgs-boson production cross section
at the TESLA Photon Collider is discussed. 
Following steps of the analysis are described:
selection of energy-flow objects,
jet reconstruction,
%
%
 kinematical and topological selection cuts optimized for cross section measurement,
and the role of \btagging{} in selection of signal events.
%
%
To simplify the description, the analysis is presented in detail for
the Higgs-boson mass \Mheq 120~GeV.
%
For other considered masses of the Higgs boson, 
\Mheq 130, 140, 150 and 160~GeV, 
the same procedure was performed with independent optimization of  selection thresholds.
The cuts dedicated to suppress $\gagaWW$ background,
which is not relevant for lower Higgs-boson masses, are described for \Mheq 160~GeV.
%
%
Expected precisions of the measurement obtained in this analysis are compared with results of our earlier works 
in which some of experimental aspects and background contributions 
considered here
were not yet taken into account.

\section{Preselection of energy-flow objects and  jet reconstruction} 

%
%
In the  energy range \sqrtseeeq 210--260~GeV about one \gagahad{} event 
takes place on average at each bunch crossing.
The contribution from these overlaying events 
is expected to affect observed particle and energy flow mainly at low polar angles (see Section \ref{sec_overlaying_events}).
Therefore, we introduce an angle $\thetamindet$ defining the region strongly contaminated by this contribution;
tracks and clusters with polar angle less than $\thetamindet$  
are not taken into account when  applying energy-flow algorithm.
In spite of that  energy-flow objects  with
polar angle less than $\thetamindet$ can still be formed; they are also ignored in further steps of analysis.
A few values of $\thetamindet$ were considered in the analysis
as discussed in detail in Appendix \ref{app_thetatc}.
 We decided to use the value  \thetamindeteq 0.85 as with this choice almost the whole
contribution from hadron-like photon interactions is suppressed
and distributions of jet transverse momentum and jet mass 
are similar
to those obtained without overlaying events and without $\thetamindet$ cut.
It was checked that \thetamindeteq 0.85 results also in the best 
final cross section measurement precision.
For the signal process considered in this analysis 
we expect that
the produced partonic state
is well reproduced
by jets reconstructed from energy-flow objects.
In the presented study  jets are reconstructed using the Durham algorithm \cite{Durham}
where the distance measure between two jets, $i$ and $j$,
is defined as
 \[ 
    y_{ij}=2\min (E^{2}_{i},E^{2}_{j})(1-\cos\theta_{ij})/E^{2}; 
 \]
$E_{i}$ and $E_{j}$ are energies of jets, $\theta_{ij}$ is the relative angle
between jets
and $E$ is the total energy measured in the detector.
The list of energy-flow objects reconstructed in the detector
is used as the input to
the algorithm,  assuming that each energy-flow object is a jet.
In following steps a pair of jets which has the smallest value of $y_{ij}$ 
is searched for and  these two jets are merged into one jet. 
The algorithm terminates when all possible values of $y_{ij}$ are greater than 
the value of the cut-off parameter, $y_{cut}$. 
%
%
The choice of the distance measure $y_{ij}$ and of the parameter $y_{cut}$ value
used in this analysis is based on the approach adopted in the NLO QCD \cite{JikiaAndSoldner} program 
which is used
for generation of background events $\gagaQQg$.
In this program the real gluon emission is considered only for $y_{qg} >$ 0.01.
Soft gluon emissions, \ie emissions with $y_{qg} \le$ 0.01, are absorbed in 
the cross-section  calculation for $\QQbar$ final state.
%
%
For consistency with this approach jets have to  be reconstructed with $y_{cut} >$ 0.01,
as for lower $y_{cut}$ values additional jets expected from soft-gluons  emission
would not be described by the generator.
Moreover, the distance measure used in the NLO generator  is calculated using true values of kinematic variables
and is inversely proportional to the $\gaga$ invariant mass
squared, $y_{ij}^{gen} \propto 1/s_{\gaga}$, whereas  the visible
energy is used in the jet reconstruction, $y_{ij} \propto 1/E^{2}$.
In the significant fraction of events we expect that due to detector acceptance $E^{2} < s_{\gaga}$.
Therefore, the value $y_{cut}=0.02$, two times larger than the one used in generator, 
has been chosen.
With this value, reconstructed jets can be relatively wide.
For example,  two perpendicular jets will be joined together 
if one of them has energy of 12~GeV or below (assuming $E \approx 120$~GeV, most probable value  for \Mheq 120~GeV).
As the NLO QCD generator does not include additional  gluon emissions due to higher order corrections, 
we study 
the resulting systematic uncertainty of the result, by applying the parton shower
algorithm to the NLO heavy quark background events.
Although some gluon contributions are double counted in such procedure,
it allows us to determine the sensitivity of the analysis to the higher order corrections.
Results are presented in Appendix \ref{app_uncert_gagaQQg}. 
%

To correct for the non-zero beam crossing angle, all reconstructed jets are transformed 
from the laboratory frame
to the frame moving 
with the speed factor $\beta=\sin\frac{\alpha_{c}}{2} \approx \frac{\alpha_{c}}{2} = 0.017$
in the $x$ direction, where $\alpha_{c}$ is the beam crossing angle.
After this correction the average value of the measured transverse momentum in the  horizontal direction, $P_x$, 
is zero.

\section{Kinematical and topological cuts}

The first cut applied after detector simulation
is introduced to exclude possible
  influence of 
the cut $\Wgaga^{\min}$, the lower limit on the  $\gaga$ invariant mass, used in the event generation.
Therefore, the condition $W_{rec} > 1.2 \: \Wgaga^{\min}$ is imposed for all considered events,
where $W_{rec}$ is the total reconstructed invariant mass of the event
(calculated from all energy-flow objects above $\thetamindet$).
Higgs-boson decay events are expected to consist mainly of 
two $b$-tagged jets with large transverse momentum and nearly isotropic distribution 
of the jet directions.
The significant number of events ($\sim 25\%$) contains the third jet 
due to the real gluon emissions which are approximated in this analysis by the parton shower algorithm,
as implemented in the \Pythia.
The following cuts are used  to 
select properly reconstructed $\bbbar$ events coming from Higgs decay.
\begin{enumerate}
%

%
\item Number of selected jets should be 2 or 3.
      In addition to two $b$-quark jets we allow for one additional jet from hard gluon emission. %
      The  signal-to-background ratio is similar for both jet multiplicities.
      Moreover, the NLO QCD generator used for heavy-quark background generation does not
      include resummation of the so-called Sudakov logarithms which would be relevant
      if  2- and 3-jet events classes were considered separately.
\item The condition \( |\cos {\theta}_{jet}| < \Cct \)
      is imposed for all jets in the event where ${\theta}_{jet}$ 
      is the jet polar angle, \ie the angle between the jet axis and the beam line.
      This cut should improve signal-to-background ratio
      as the signal is almost uniform in $\cos\theta$, 
      while the background is peaked at $|\cos\theta|=1$. 
\item Since the Higgs bosons are expected to be produced almost at rest, 
      the ratio of the total longitudinal momentum calculated from all  jets in the event, $P_{z}$,
      to the total energy, $E$, should fulfill condition \( |P_{z}|/E < \Cpz \).

\end{enumerate}

To determine the cut parameter  values $\Cct$ and $\Cpz$ the corresponding distributions of the signal and 
heavy quark background events were compared
(other background contributions were not considered at this stage).
After cut 1  the optimal value of parameter $\Cct$ (cut 2) is found
as  the one which minimizes the estimated statistical uncertainty of the measurement:
\[
\frac{\Delta \sgagahbb}{\sgagahbb} =
\frac{\sqrt{\mu_{S}+\mu_{B}}}{\mu_{S}},
\]
%
where $\mu_{S}$ and $\mu_{B}$ are the numbers 
of expected signal and background events after the cut, respectively.
With optimized  cut 2 the same procedure is repeated for parameter $\Cpz$ (cut 3).
The expected event distributions for  $|\cos {\theta}_{jet}|^{\max}$
(the maximum value of   $|\cos {\theta}_{jet}|$ over all jets in the event)
 and $|P_{z}|/E$ 
are shown in Fig.~\ref{fig:max_abs_costhjet_mh120} and \ref{fig:pz_to_evis_ratio_mh120}, respectively. 
For simplicity only $\gagabbg$ background contribution is shown.
The $\gagaccg$ contribution, which is around 16 times larger, has a very similar shape.
Both background contributions are taken into account in  cut optimization.
For \Mheq 120~GeV the
optimized cut values are  $\Cct = 0.725$ and $\Cpz = 0.1$, as indicated  in the figures (vertical arrows).
The measurement precision is estimated to be around 7\% and 5\% after
the $|\cos {\theta}^{jet}|$ cut and after the $|P_{z}|/E$  cut, respectively.
Angular cuts used in the event selection procedure are compared in Fig.\ \ref{fig:angles}. 


%
\pnfig{tb}{\figheight}{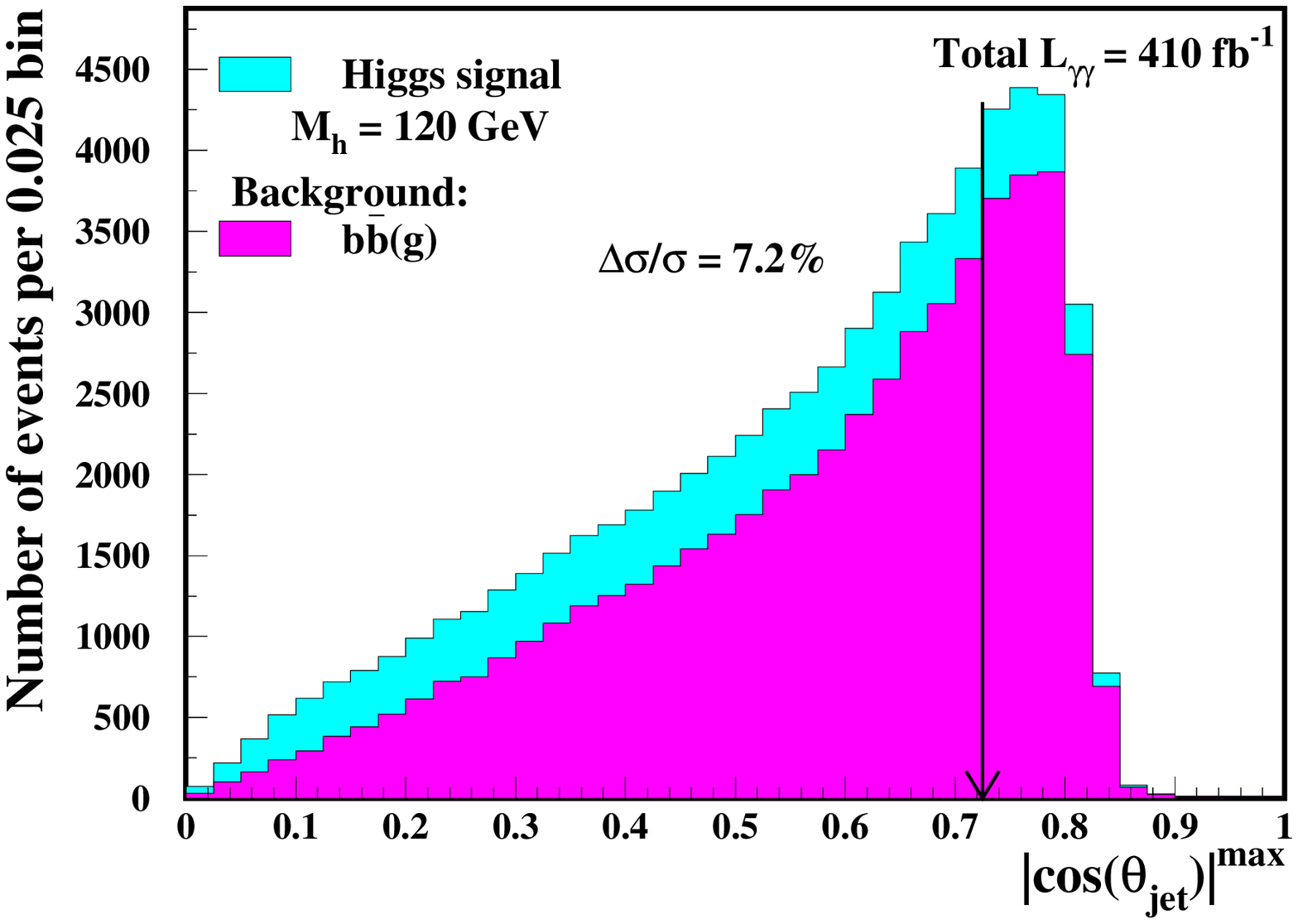}{fig:max_abs_costhjet_mh120}
{
 Distributions of $|\cos {\theta}_{ jet}|^{\max}$ (maximal value of   $|\cos {\theta}_{jet}|$ over all jets) 
 for signal and background events. 
  For background only $\gagabbg$ events are shown.
 The signal measurement precision  $\Delta\xs/\xs$ of about 7\% is obtained 
 for $ |\cos {\theta}_{jet}|^{\max} < \Cct = 0.725$. 
}


%
\pnfig{tb}{\figheight}{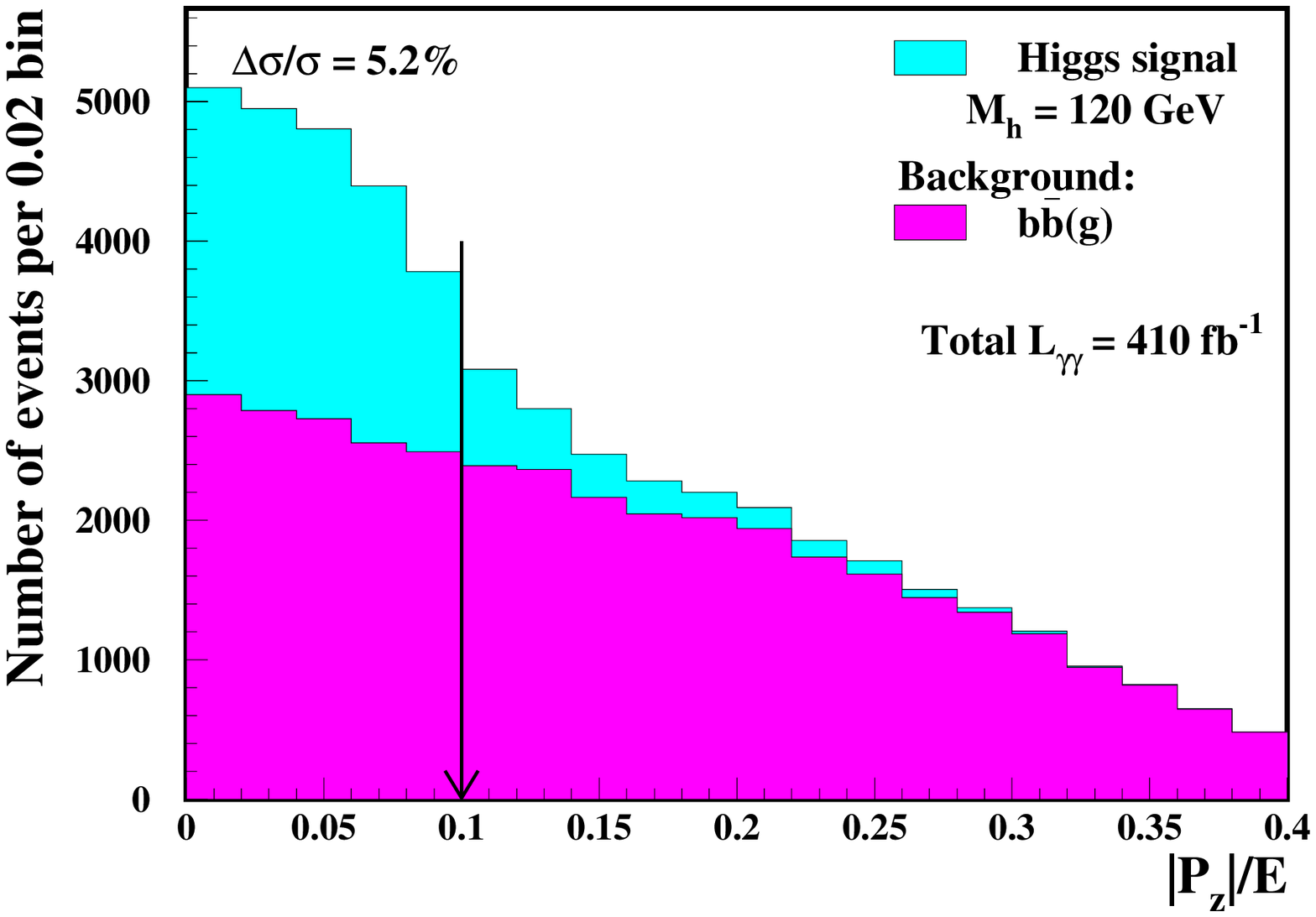}{fig:pz_to_evis_ratio_mh120}
{Distributions of $|P_{z}|/E$ for signal and background events. 
  For background only $\gagabbg$ events are shown.
 The signal measurement precision  $\Delta\xs/\xs$  of about 5\% is obtained  
  for $  |P_{z}|/E < \Cpz  = 0.1$.}

\pnfig{htb}{\figheightsmall}{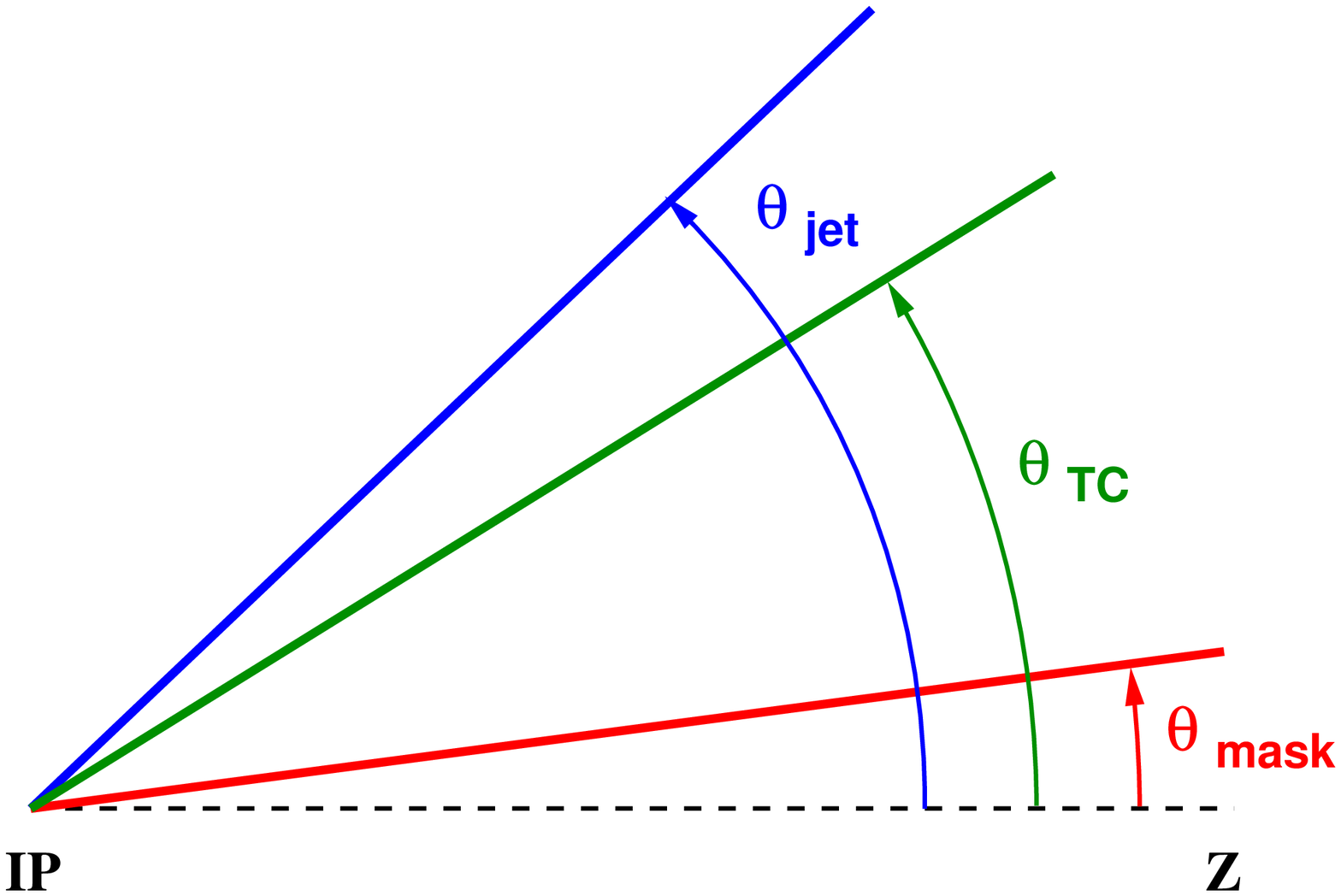}{fig:angles}{
 Comparison of the angular cuts used in the selection procedure of the Higgs-boson production events for \Mheq 120~GeV. 
}


\section{ \protect\btagging{} algorithm}
For  \btagging{} 
the \ZBHT{}  package prepared for the TESLA project was used \cite{HawkingBT,XellaBT,Btagging}.
%
The flavour tagging algorithm is based primarily on ZVTOP, 
the topological vertex finding procedure developed at SLD \cite{ZVTOP}. 
%
In addition to the ZVTOP results, a one-prong charm tag \cite{HawkingBT} 
and an impact parameter joint probability tag \cite{ALEPH_IMPP_JP} 
outputs are used to train a neural net. 
%
Following parameters are given as an input to the neural-network algorithm (for all tracks or all vertices):
\begin{enumerate}
\item Impact parameters in $r-\phi$ and $r-z$. 
Impact parameter in the $r-\phi$ plane is defined 
as the minimal distance between the track trajectory and the beam axis;  
impact parameter in $r-z$ plane is defined as the distance between the reconstructed primary vertex position
and the point on the beam axis nearest to the track trajectory.
\item Significance of the track impact parameters  -- the ratio of the impact parameters to their estimated errors.
\item Vertex decay length -- the distance between the primary vertex and the secondary or tertiary vertex.
\item Vertex decay length significance -- the ratio of the vertex decay length to its measurement error.
\item $p_t$-corrected mass of the secondary vertex  -- the invariant mass 
of particles coming from the vertex. As only charged particles (tracks) are considered, 
the correction for neutral particles is applied.
The correction is based on the assumption 
that the total momentum of all particles coming from the secondary vertex must
be parallel to the vector between the primary and secondary vertex positions. 
\item Vertex momentum -- the total momentum of all tracks belonging to the vertex.
\item Secondary vertex track multiplicity.
\item Secondary vertex probability -- the probability that all tracks assigned by the ZVTOP algorithm 
      to the secondary vertex belong to this one vertex. 
\end{enumerate}
The neural-network algorithm was trained on the $Z$ decays. 
For each jet the routine returns a ``$b$-tag'' value -- the number 
between 0 and 1 corresponding to ``$b$-jet'' likelihood.

%
In order to optimize the signal cross-section measurement,  
the two-dimensional cut on $b$-tag values is used.
In the signal events two jets with the highest transverse momentum
are most likely to originate from $b$ quarks. 
%
Therefore, all jets in the event are sorted according 
to the value of their transverse momentum. 
The distribution of $b$-tag values for 2 and 3-jet events is considered in the plane 
$b$-tag(jet$_1$)$\otimes$$b$-tag(jet$_2$) 
where indices 1 and 2 correspond to two jets with the highest transverse momenta.
The two-dimensional distributions of $b$-tag values 
for the signal,  $\gagahbb$, and for the background, $\gagabbg$,  events are shown in Fig.\ \ref{fig:btag_h_bbg}.
The corresponding distributions for other considered background contributions,
 $\gagaccg$ and $\gagaqq$ ($\quds$), are shown in Fig.\ \ref{fig:btag_ccg_uds}.
Events considered in  the \btagging{} studies 
 fulfill fore-mentioned, optimized selection cuts and an additional
cut $W_{rec}>0.7 \: \Mh$ which removes low-mass events not relevant for the final result
(this cut is used only for tagging optimization).
As expected, for processes $\gagahbb$ and  $\gagabbg$ most events populate the regions 
with high $b$-tag values (Fig.\ \ref{fig:btag_h_bbg}),
whereas most $\gagaccg$ and  $\gagaqq$  events have small $b$-tag values (Fig.\ \ref{fig:btag_ccg_uds}).
Nevertheless, significant fraction of $\gagaccg$ events 
populates the region of high $b$-tag values, and the event  distribution is more flat
than the one for   $\gagaqq$ events. 
The optimal \higgstagging{} cut is found by considering the value of 
the signal to background ratio $S/B$, 
where  $S$ and $B$ denote the expected numbers of events for the signal $\gagahbb$ 
and for the sum of background contributions from processes  $\gagaQQg$ ($\Qcb$)  and $\gagaqq$  ($\quds$),
respectively.
Obtained $S/B$ distribution in the $b$-tag(jet$_1$)$\otimes$$b$-tag(jet$_2$)
plane for Higgs-boson production with \Mheq 120~GeV is shown in Fig.\ \ref{fig:plot_btag2j_m120_modsm}.
 The selection criteria
which results in the best precision of the $\Ghgagahbb$ 
measurement corresponds to $S/B> 0.19$ as indicated  in the figure (stars).

\pnfiggeneral{t}{\twofigheight}{\includegraphics{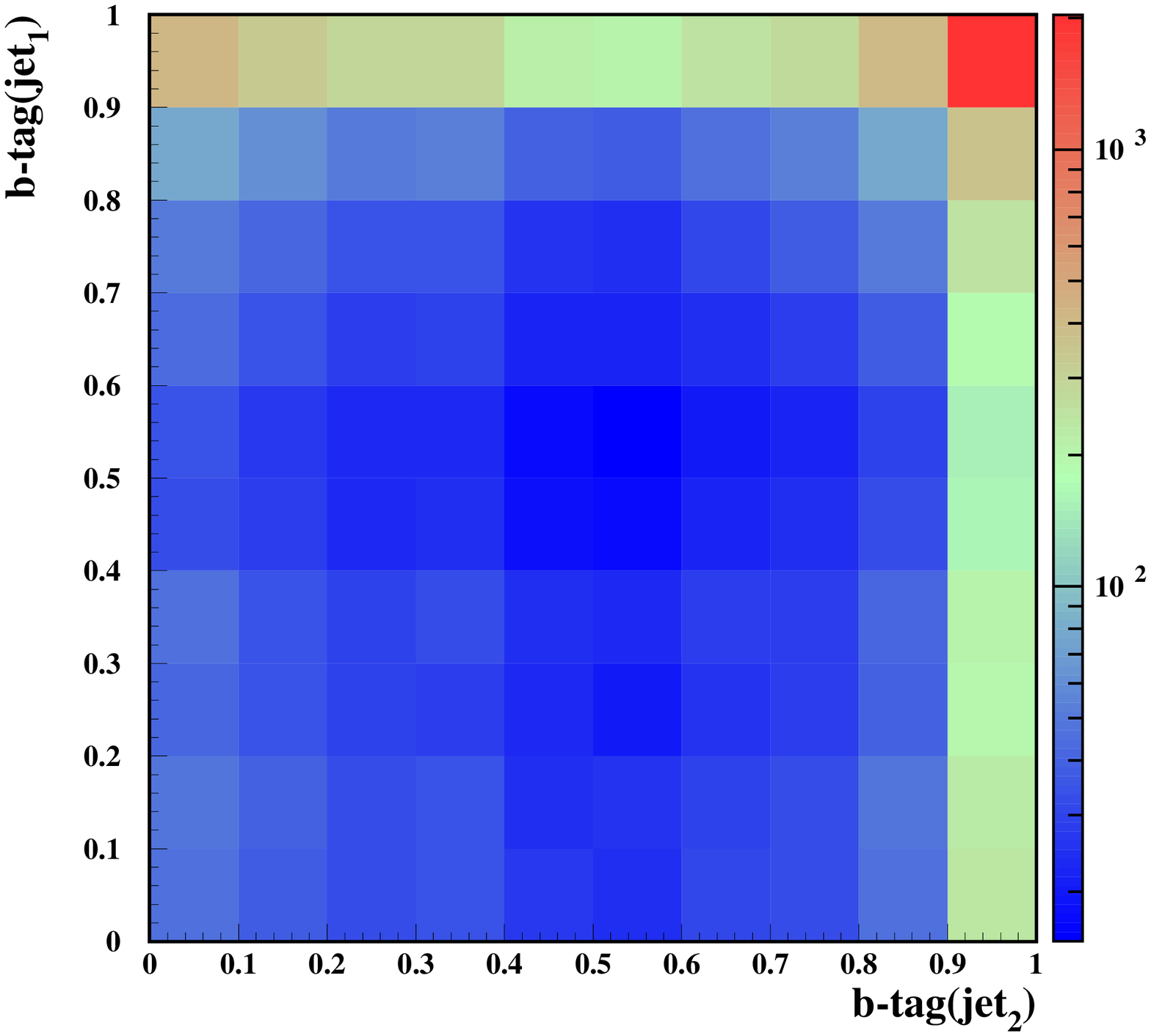} 
   \includegraphics{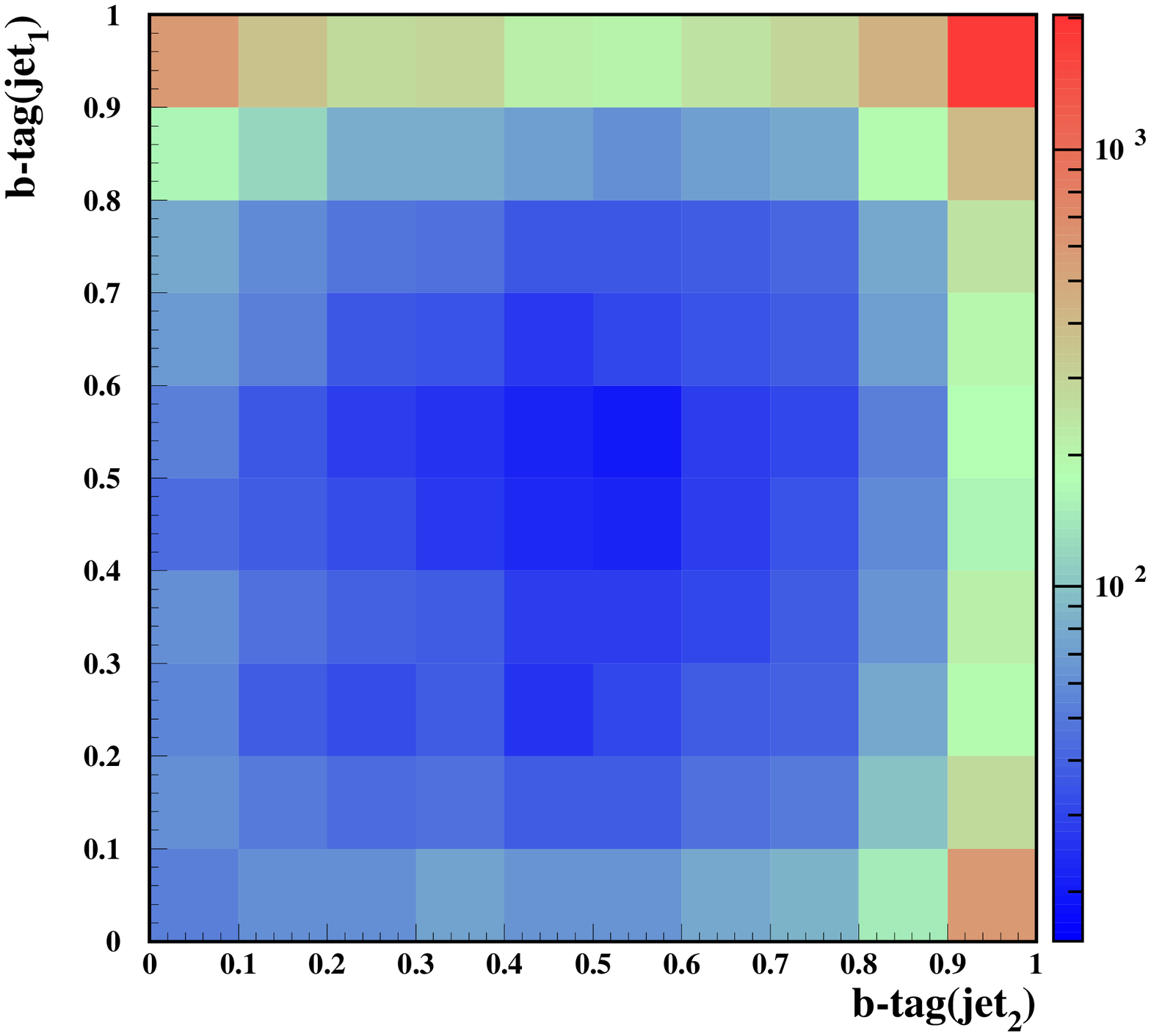}}{fig:btag_h_bbg}{
Distributions of $\gagahbb$ (left) and  $\gagabbg$ (right) events 
in the plane ${\rm btag}({\rm jet_{1}}) \otimes {\rm btag}({\rm jet_{2}})$.
}
\pnfiggeneral{t}{\twofigheight}{\includegraphics{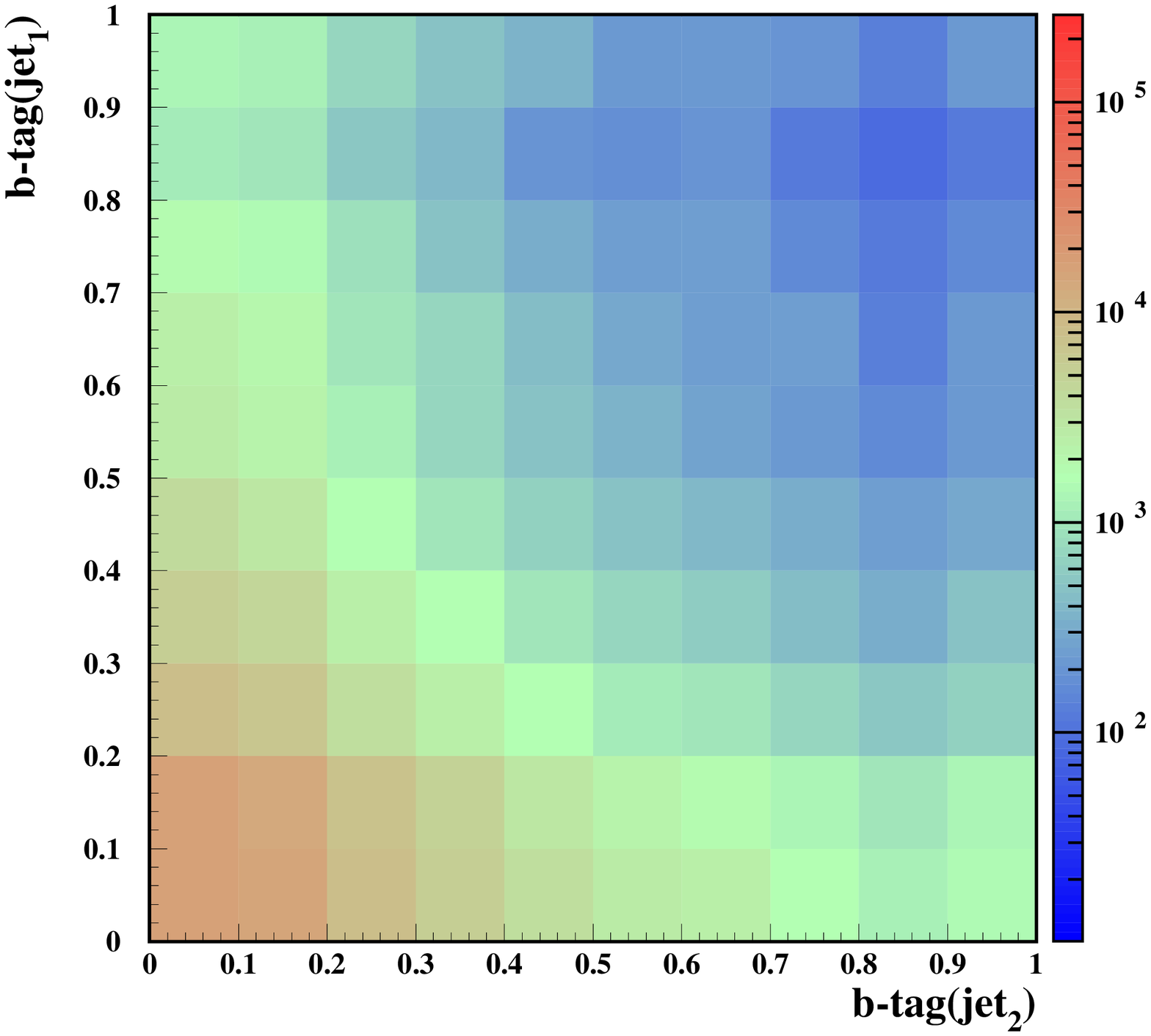} 
   \includegraphics{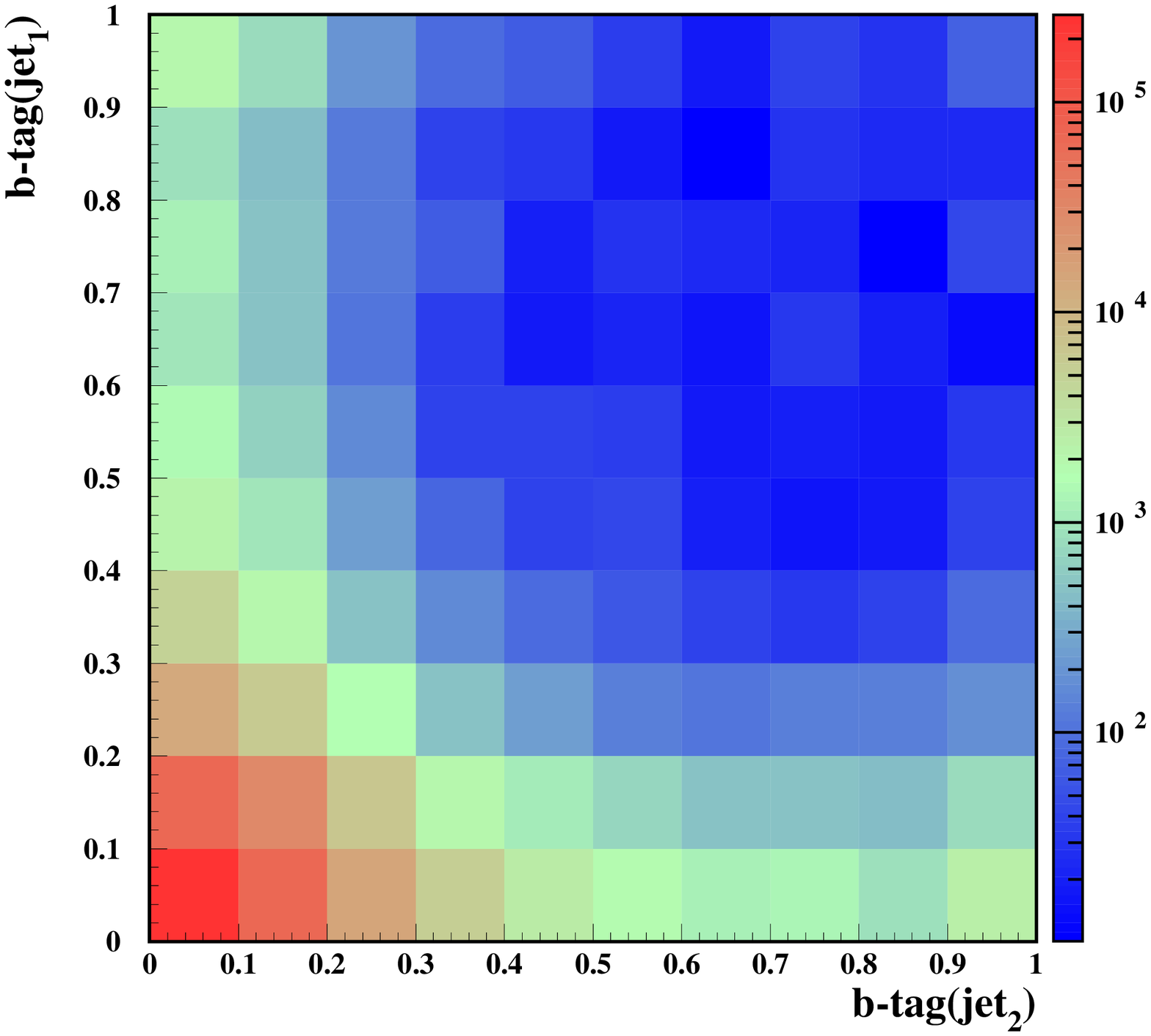}}{fig:btag_ccg_uds}{
Distributions of $\gagaccg$ (left) and  $\gagaqq$, $\quds$,  (right) events 
in the plane ${\rm btag}({\rm jet_{1}}) \otimes {\rm btag}({\rm jet_{2}})$.
}
\pnfig{htb}{\twofigheight}{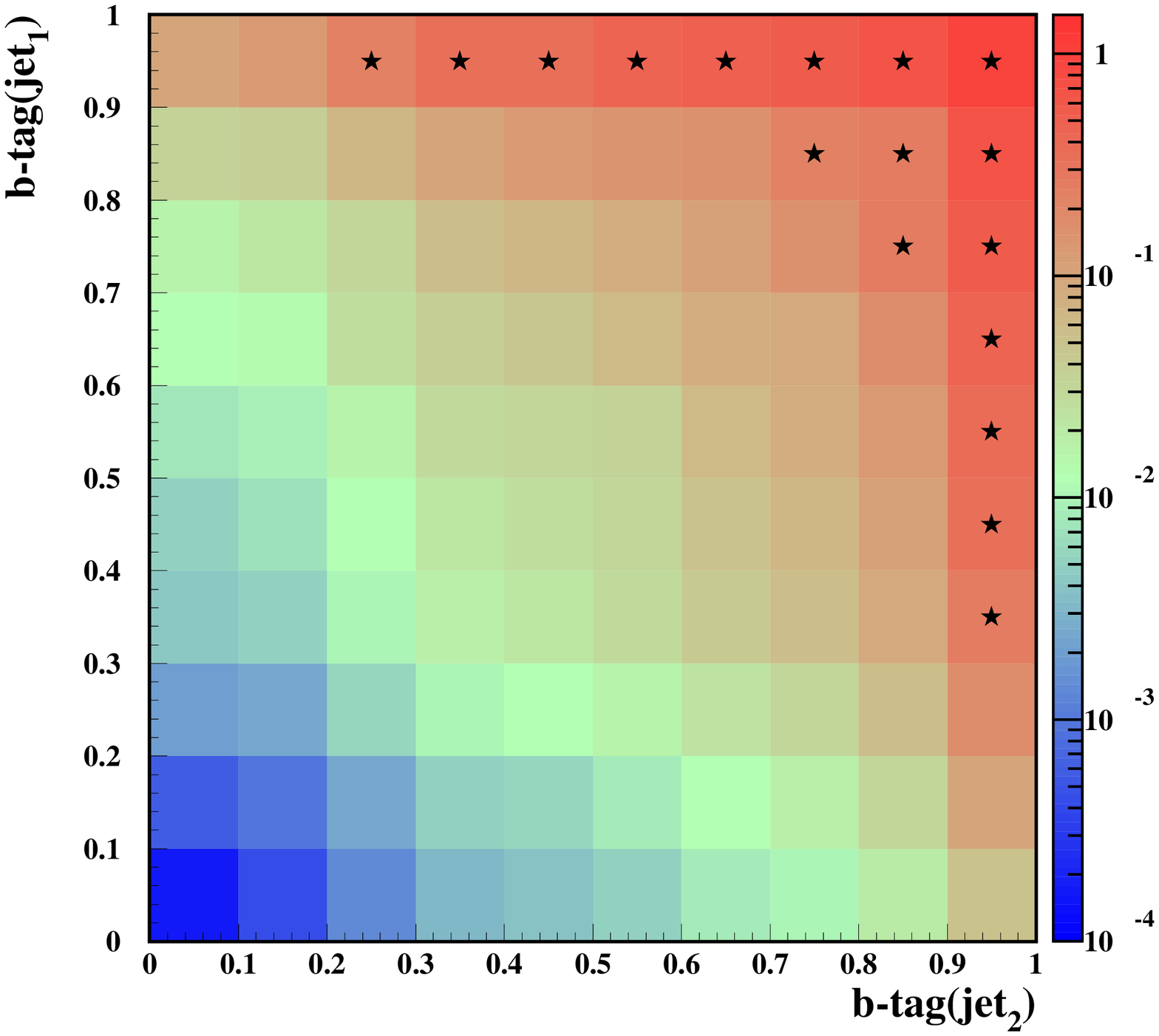}{fig:plot_btag2j_m120_modsm}
{
The expected ratio of signal ($\gagahbb$) to background ($\gagaQQg$, $\Qcb$, and $\gagaqq$, $\quds$) 
event distributions 
in the plane ${\rm btag}({\rm jet_{1}}) \otimes {\rm btag}({\rm jet_{2}})$. 
The region which results in the best precision measurement for the cross-section measurement is indicated by stars.} 




%
The obtained efficiencies for tagging \higgs{} events, 
$\bbbar$ background events, 
and the probabilities for mistagging of the $\ccbar$ and $\qqbar$ ($\quds$) events 
are $\varepsilon_{h}=58\%$, $\varepsilon_{bb}=50\%$,  $\varepsilon_{cc}=2.2\%$ and  $\varepsilon_{uds}=0.16\%$, respectively.
Similar efficiencies were obtained for other considered electron-beam energies.
We call this procedure '\higgstagging' as the efficiency for tagging signal 
events, $\varepsilon_{h}$, is significantly higher than the efficiency
for tagging $\bbg$  background events, $\varepsilon_{bb}$.
This is because large fraction of background events is reconstructed
as 3-jet events (LO contribution is suppressed for $J_z=0$)
in which the gluon jet is often one of the two jets with highest transverse momenta.
In the earlier analyses \cite{JikiaAndSoldner,NZKhbbm120appb}
a fixed \bbtagging{} efficiency, \( \varepsilon_{h}=\varepsilon _{bb}=70\% \), 
and a fixed \ccmistagging{} efficiency, \( \varepsilon _{cc}=3.5\% \), 
were assumed. 
Although the efficiencies resulting from the optimized selection
are much lower,
signal to background ratio  $\varepsilon_{h}/\varepsilon_{cc}$ improves significantly.
Particles from \gagahad{} overlaying events can significantly change properties
of the jet to which they are assigned by the jet clustering algorithm.
For example, the invariant mass of the jet increases on average by 3~GeV, 
if the angular cut is not applied (\ie parameter $\thetamindet = \thetamask$; 
see Fig.~\ref{fig:ptjet_mjet_120} in Appendix~\ref{app_thetatc}).
Although  the average invariant mass of the
jet after the cut corresponding to \costhmindeteq 0.85
is similar to the jet mass without overlaying-events contribution,
 the jet structure can still be affected by the remaining particles from \gagahad{} interactions,
and by rejection of some particles coming from the signal process.
These effects influence also the flavour tagging algorithm 
and   cause a significant change
in the results of the \bbtagging{} optimization.
%
%
To quantify the influence of overlaying events we repeated
the optimization procedure,
for production of the SM Higgs boson with \Mheq 120~GeV,
without overlaying events and with \costhmindeteq 0.99.
 The resulting  efficiencies, 
corresponding to optimal cut $S/B>0.16$,
are
$\varepsilon_{h}=71\%$, $\varepsilon_{bb}=64\%$,   
$\varepsilon_{cc}=2.9\%$ and  $\varepsilon_{uds}=0.11\%$.
The corresponding selection region in the $b$-tag(jet$_1$)$\otimes$$b$-tag(jet$_2$) plane 
is significantly wider than for the nominal analysis,
but the $\ccbar$ background suppression factor $\varepsilon_{bb}/\varepsilon_{cc}$
is similar.

Although the  selection region is smaller when overlaying events are taken into account,
the efficiency for $\gagaqq$, $\quds$, is greater by about 50\%. 
Jets coming from light-quark production can be significantly modified by \gagahad{} events,
and are more likely to be recognized as $b$-jets. 
%
One of important reasons is that the primary vertex of the overlaying
event is usually shifted with respect to the primary vertex of the hard interaction.
If particles from both interactions are combined in the reconstructed hadronic jet,
the vertex finding algorithm is likely to reconstruct two vertices
treating one of them as a primary vertex, 
and the second one as the vertex resulting from $b$ decay.
The effect of tagging deterioration is even stronger for higher beam energies as
more overlaying events per bunch crossing are produced  
due to higher luminosity and cross section.
The influence of overlaying events on the event reconstruction
is clearly seen also in case of the heavy MSSM Higgs-bosons production
as will be discussed in Chapter \ref{ch_mssm_analysis}.
%




The influence of the optimized \higgstagging{} criteria on 
the reconstructed invariant-mass, \( W_{rec} \), distribution  is shown in Fig.\ \ref{fig:plot_ptet_m120_modsm_var1}
for signal events $\gagahbb$  after all described selection cuts. 
These results were obtained without overlaying events \gagahad{}. 
The tail towards low masses is due to  events with energetic neutrinos 
coming from semileptonic decays of $D$ and $B$ mesons (see \cite{NZKhbbm120appb} for more details). 
Contribution of these events can be suppressed by an additional cut 
 \( P_{T}/E_{T} < 0.04 \), where
\( P_{T} \) and \( E_{T} \) are the absolute values of the total transverse
momentum of an event, $\vec{P}_{T}$, and the total transverse energy, 
respectively. 
We see that the efficiency of  \btagging{} is similar for events with and without 
energetic neutrinos as the algorithm does not significantly influence
the shape of the distributions.
%

\pnfig{tb}{\twofigheight}{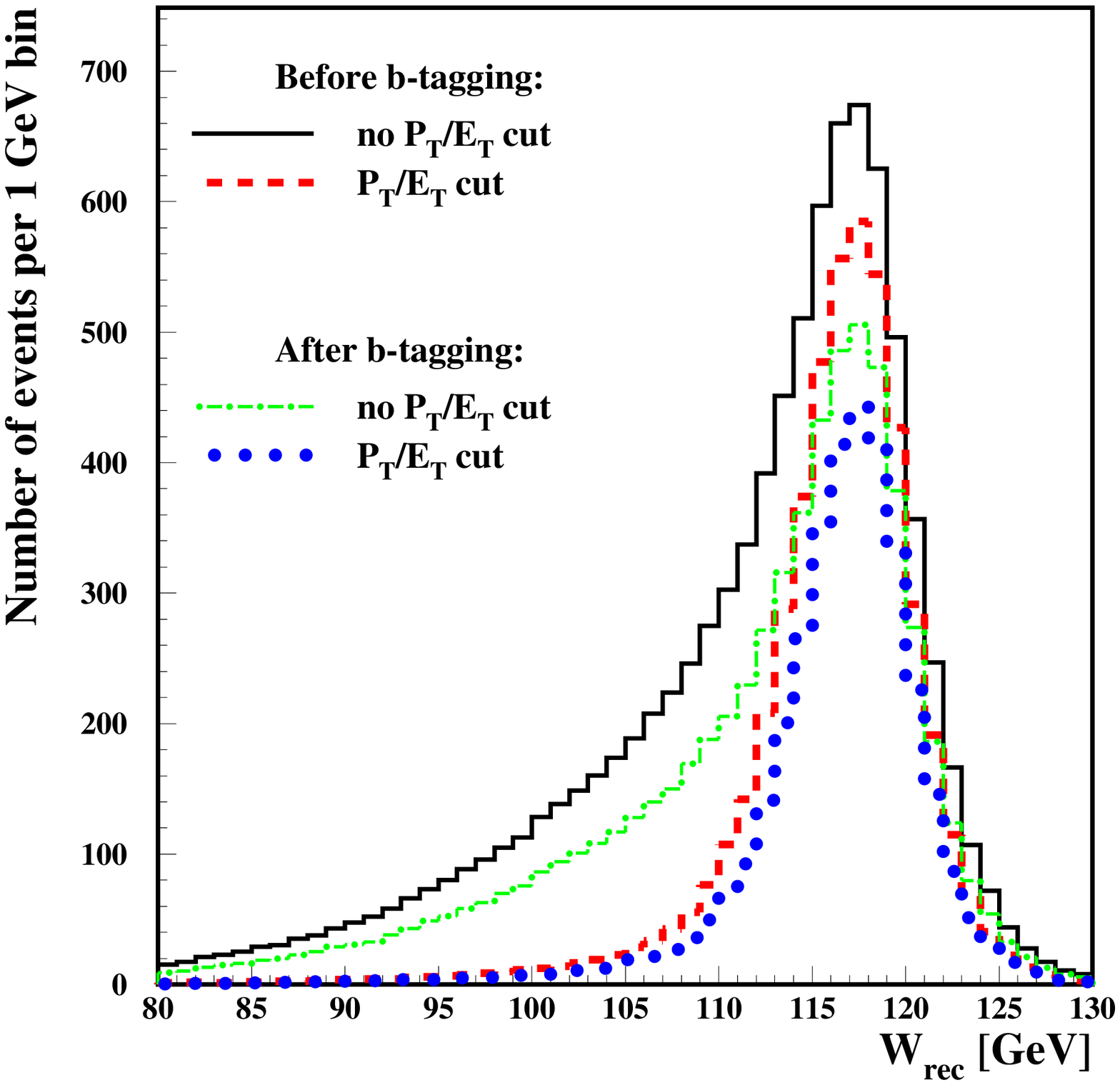}{fig:plot_ptet_m120_modsm_var1}{
Distributions of the reconstructed invariant mass, $W_{rec}$,
for selected $\gagahbb$ events, for \Mheq 120~GeV.
Distributions obtained before and after applying the \btagging algorithm, 
without and with the additional $P_{T}/E_{T} < 0.04$ cut are compared. 
 Overlaying events are not included.
}

\section{Results}

The invariant-mass distributions 
for signal events passing all optimized selection cuts,
before and after taking into account the overlaying events \gagahad{}
are compared in Fig.\ \ref{fig:wrec_h_oe01_m120} (left). 
Mass resolution, derived from the 
Gaussian fit in the region from \( \mu - 1.3\sigma  \)
to \( \mu + 1.3 \sigma  \), is about 4 and 6~GeV, respectively. 
%
The overlaying events and cuts suppressing their contribution 
significantly influence the mass reconstruction 
and result in  an increase of distribution width by about 2~GeV, 
and in a shift of the mean value, $\mu$, by about 3~GeV.
%
%
%
%
%
%
\pnfiggeneral{tb}{\twofigheight}{\includegraphics{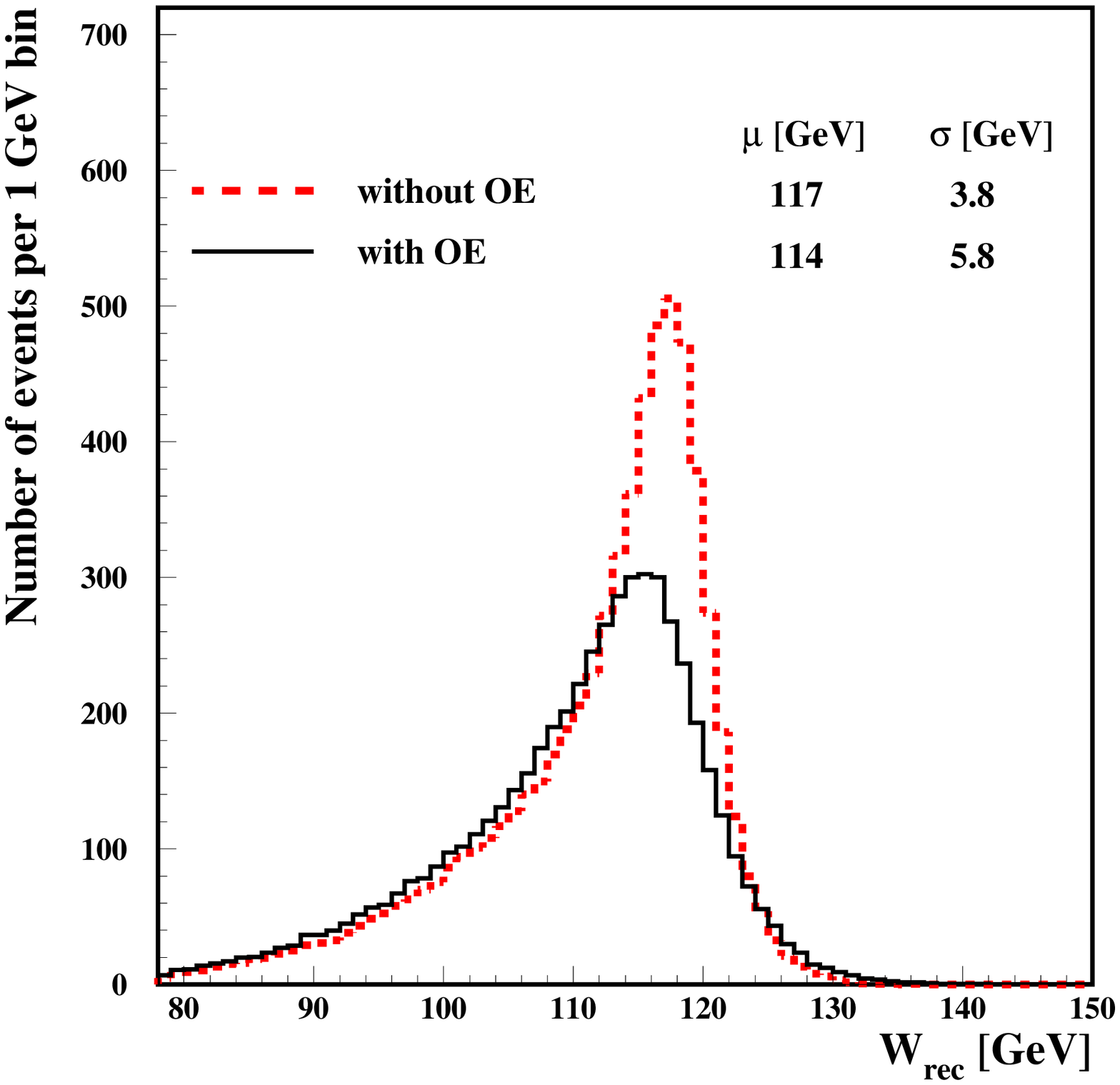}
\includegraphics{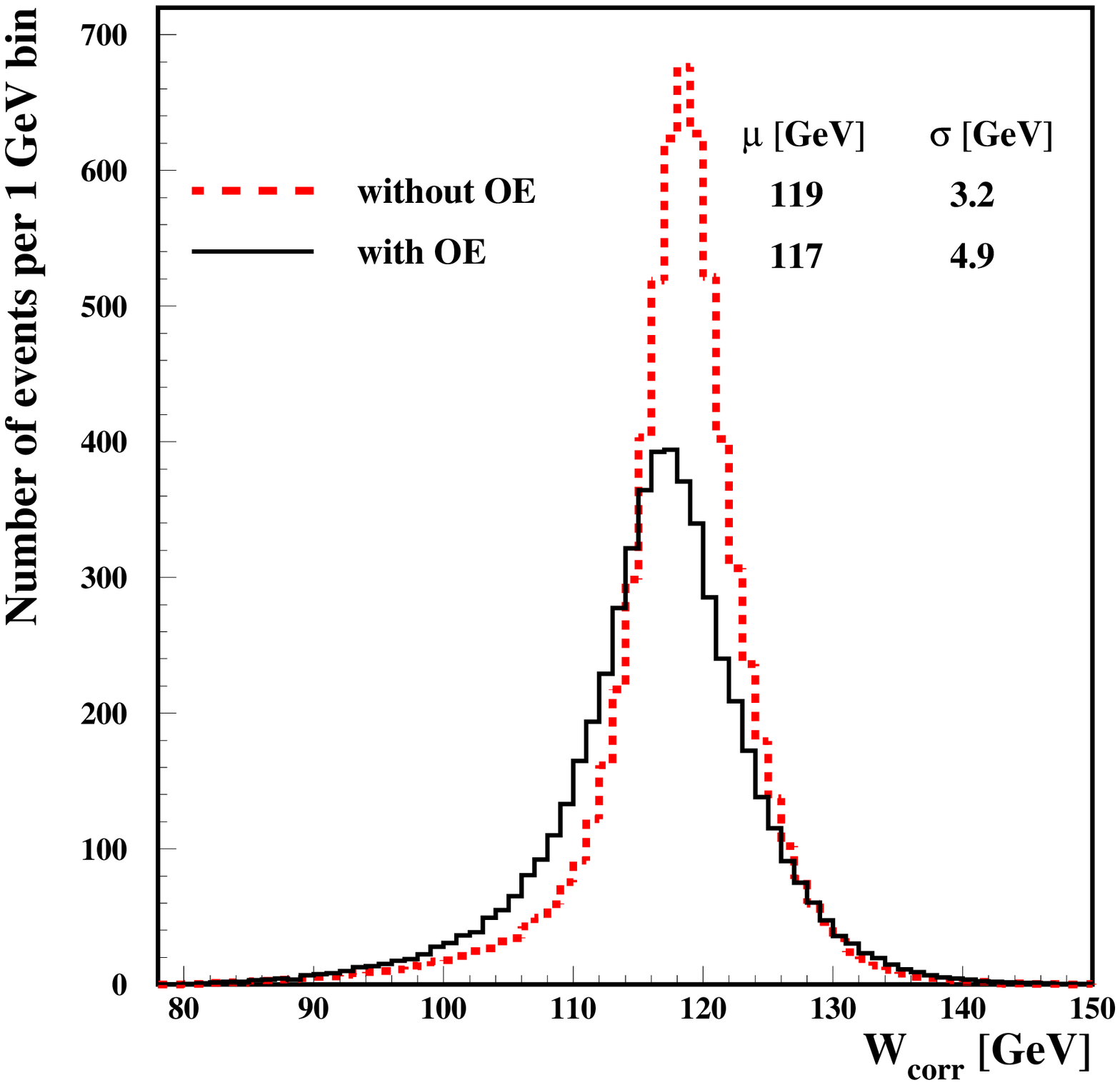}}{fig:wrec_h_oe01_m120}
{Reconstructed invariant-mass , $W_{rec}$, (left)
and corrected invariant-mass , $W_{corr}$, (right)
distributions for selected $\gagahbb$ events, for \Mheq 120~GeV.
Distributions obtained without and with overlaying events (OE) are compared.
Results for the mean $\mu$ and dispersion $\sigma$ from the Gaussian fit in the region from \( \mu - 1.3\sigma  \)
to \( \mu + 1.3 \sigma  \), are also shown. 
}
%
A drop in a selection efficiency, 
resulting in the reduced number of events expected after selection cuts 
(from about 6450 to 5530 events), 
is also observed. 
This is because tighter \btagging{} cuts have to be imposed to reduce influence 
of overlaying events.
More events are also rejected by the cut on the longitudinal momentum.
%
Some drop in the selection efficiency is also due
to the fact 
that the energy deposits from the \gagahad{} processes,
remaining after the $\thetamindet$ cut, ``shift'' 
jets nearer to the beam axis and the event 
can be rejected by the jet-angle cut.
%
%
%
%


\pnfig{tb}{\figheight}{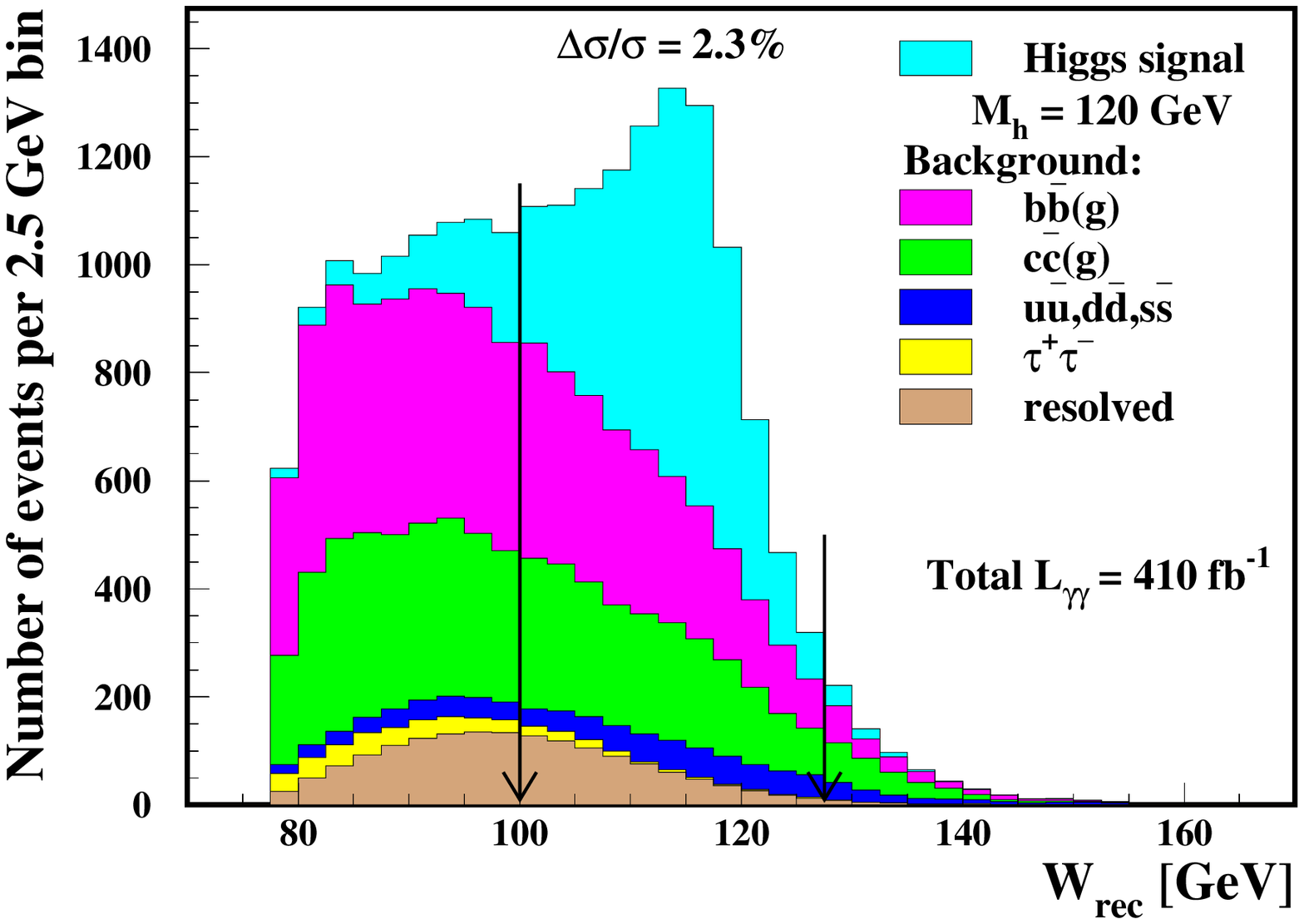}{fig:m120_modsm_var31_oe1}{
Distributions  of the reconstructed invariant mass, $W_{rec}$,
for selected $\bbbar$ events.
Contributions of the signal, for \Mheq  120~GeV, 
and of the  background processes, \ie
$\gagaqq$ for $\quds$, $\gagatautau$,
and \gagahad{} (as a separate contribution with \emph{hadron-like$\times$hadron-like} interactions only,
indicated as 'resolved'),
are shown separately.
Arrows indicate the mass window, 100 to 127.5~GeV, optimized for the measurement of the 
$\Ghgagahbb$, which leads to the statistical precision of 2.3\%.
}

Distributions of the reconstructed invariant mass, $W_{rec}$, 
expected  after applying all selection cuts and \btagging{} algorithm, 
%
%
%
%
for the signal ($\gagahbb$) and all considered background contributions
are shown in Fig.\ \ref{fig:m120_modsm_var31_oe1}.
Beside the heavy quark production,  $\gagabbg$ and $\gagaccg$, all background contributions are small,
of the order of 30\%, 15\% and 5\% of the signal
for \gagahad{}, $\gagaqq$ and $\gagatautau$, respectively.
We assume that the number of observed   Higgs-boson production events 
will be extracted
by counting the number of \( \bbbar \) events in the mass window 
around the Higgs-boson mass peak, $N_{obs}$, 
and subtracting the expected contribution of background events, $\mu_{B}$.
The relative statistical error expected in  the measurement of the Higgs-boson production cross section 
\( \sgagahbb  \),
or of the partial width multiplied by the branching ratio \( \Ghgagahbb  \),
can be estimated from the following formula:
\begin{equation}
\label{eq:prec1}
\frac{\Delta \sgagahbb}{\sgagahbb} =
\frac{\Delta \left[ \Ghgagahbb \right] }{ \Ghgagahbb }=
\frac{\sqrt{\mu_{S}+\mu_{B}}}{\mu_{S}}
\end{equation}
where $\mu_s$ is  expected number of signal events,
and the expected number of observed events is $\left< N_{obs} \right> = \mu_{S}+\mu_{B}$.
The accuracy obtained 
from  the reconstructed invariant-mass distribution for \Mheq 120~GeV 
is equal to 2.3\%. 
The mass window used to calculate the signal measurement precision
is again optimized to obtain the lowest relative error.
For \Mheq 120~GeV the selected mass window is 100 to 127.5~GeV 
as indicated in  Fig.\ \ref{fig:m120_modsm_var31_oe1} (vertical arrows).
The obtained result  is consistent with 
 results of our  previous analyses \cite{NZKhbbm120appb,NZKSMeps2003}
which however did not take into account many aspects of the measurement considered here.
In the current analysis additional background contributions
deteriorate  the precision of the cross section measurement.
However, the effect  is partially compensated by the performed optimization of selection procedure.
%
%

  Significant part of the energy in the signal events can be lost due to
escaping neutrinos.
As shown in Fig.\ \ref{fig:plot_ptet_m120_modsm_var1}, 
this effect worsens the mass resolution and, as a result, the cross section measurement precision.
We have looked for a correction method which would improve the mass resolution without reducing the event statistics.
Unfortunately, due to a large spread of the photon beam energy, no unambiguous constraint can
be imposed on the reconstructed longitudinal momentum.
\pnfig{tb}{\figheightsmall}{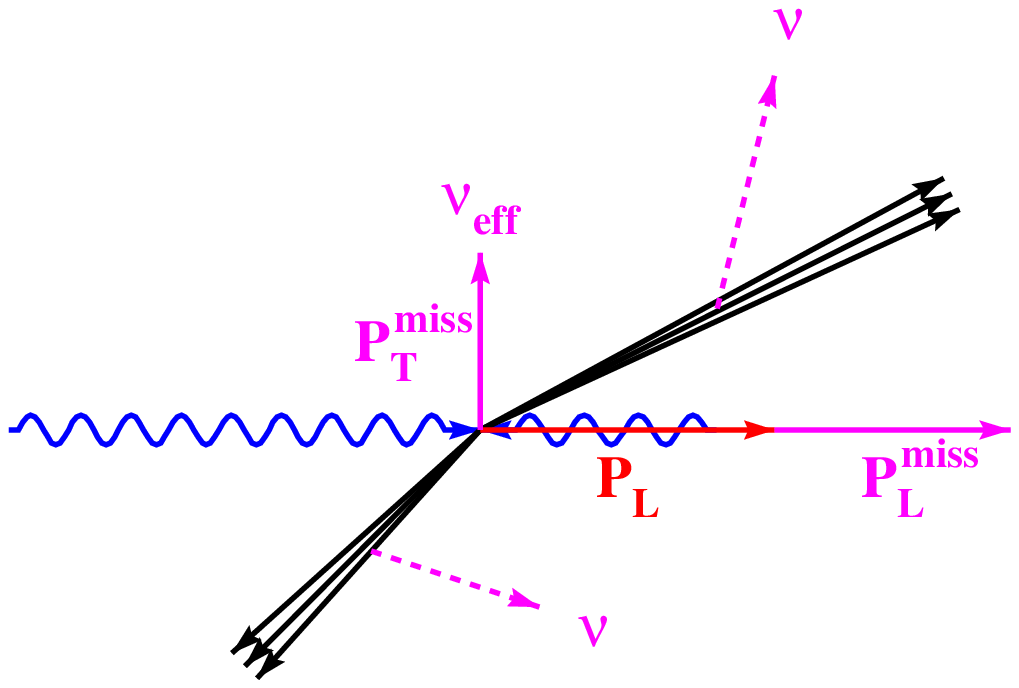}{fig:diagram_gagaptpl_3}
{
	 The 2-jet $\gagahbb$ event in which part of energy is carried out by neutrinos.
	 As the total longitudinal momentum, $P_L$, is unknown, the missing
	 longitudinal momentum, $P_L^{miss}$, cannot be estimated.
	 However, the total transverse momentum should be balanced
	 and the missing transverse momentum, $P_T^{miss}$, can be attributed
	 to the 'effective' neutrino, $\nu_{eff}$.
}
We have considered four methods of correcting the measured invariant 
mass:
\begin{enumerate}
  \item  The value of the measured transverse momentum is added 
         to the total energy and the transverse momentum is balanced. 
         This is equivalent to the assumption that the missing transverse
	 momentum is due to a single neutrino emitted perpendicularly to the beam line. 
	 The procedure is illustrated in Fig.\ \ref{fig:diagram_gagaptpl_3} 
	 where the 2-jet $\gagahbb$ event is shown
	 with energy and momentum carried out by two neutrinos.
	 As the total longitudinal momentum, $P_L$, is not constrained, the missing
	 longitudinal momentum, $P_L^{miss}$, cannot be estimated.
	 However, the total transverse momentum should be balanced.
	 Thus, the missing transverse momentum, $P_T^{miss}$, can be attributed
	 to the 'effective' neutrino.
	 This correction was introduced in our earlier analysis \cite{NZKhbbm120appb}.
  \item  The transverse momentum, $p_T$, of the jet with lowest $p_T$ is increased by 
         a value of the total missing transverse momentum
	 and the longitudinal jet momentum is rescaled to preserve its original direction.
	 This method is applied for 2-jet events 
	 and assumes that the missing $P_T$ is due to the single neutrino
	 emitted along the jet with lower $p_T$. 
	 In general the total transverse momentum is still unbalanced after this correction.
  \item  The transverse momentum of 2-jet event is balanced 
         under assumption that the missing $P_T$
         is due to the neutrino emitted under the polar angle equal to the polar angle 
	 of the jet with lowest $p_T$. 
  \item  The transverse momentum of 3-jet events is balanced by rescaling momenta of two jets.
         All combinations are checked and the most ``reasonable'' one is chosen,
	 \ie the one satisfying the requirement that each of two rescaling factors is greater than 1 and less than 1.3.
\end{enumerate}
Approaches combining  these methods were also taken into account.
Moreover, we also considered the algorithms where  the correction were limited to events with transverse momentum
greater than some threshold value, which was varied to obtain the best mass resolution.
Surprisingly, the first (and simplest) procedure proved to be the best one when applied to all events. 
Other correction methods introduce 
systematic bias in the corrected invariant
mass distributions and sometimes result even 
in the  deterioration of the final cross-section measurement  precision.
Thus, the corrected reconstructed invariant mass used for the final analysis was defined as \cite{NZKhbbm120appb}: 
\begin{equation}
W_{corr} \equiv \sqrt{W_{rec}^{2}+2P_{T}(E+P_{T})}.
\end{equation}
%

%
In Fig.\ \ref{fig:wrec_h_oe01_m120} (right) the distributions of \( W_{corr} \)
for the selected signal events, without and with overlaying events, 
are presented. 
The tail of events with invariant masses below $\sim 110$~GeV 
is much smaller than for the $W_{rec}$ distributions (compare with the left figure).
The mass resolutions, 
derived from the Gaussian fits to the $W_{corr}$ distributions 
in the region from \( \mu - 1.3\sigma  \) to \( \mu + 1.3\sigma  \),
are equal to 3.2 and 4.9~GeV, 
without and with overlaying events,  respectively. 
Also the mean values obtained from the fit  are closer to $\Mh$ than the values when $W_{rec}$ was used.
%



\pnfig{tb}{\figheight}{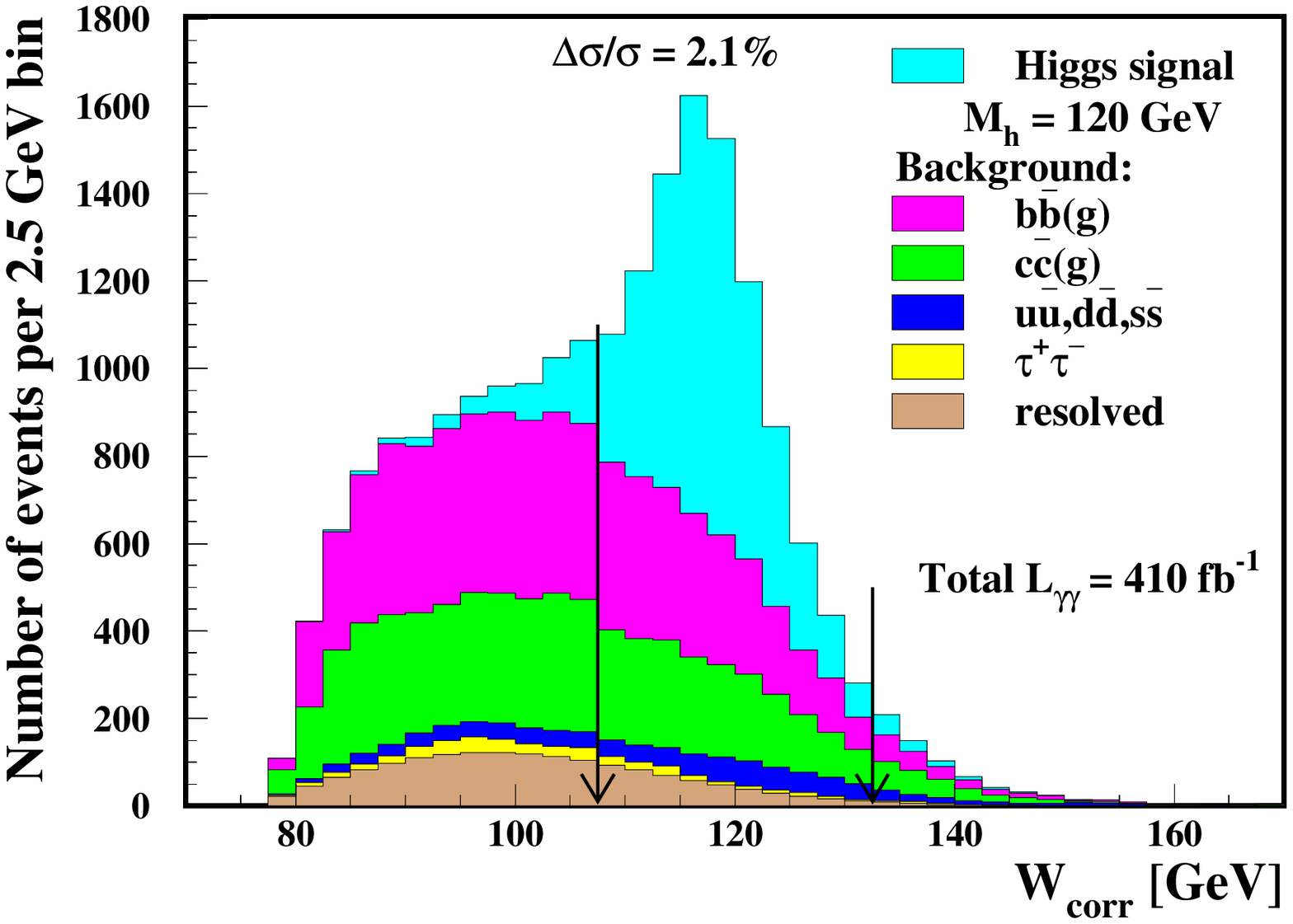}{fig:m120_modsm_var34_oe1}{
As in Fig.\ \ref{fig:m120_modsm_var31_oe1}, for the corrected invariant mass, $W_{corr}$,
distributions.
Arrows indicate the mass window, 107.5 to 132.5~GeV, optimized for the measurement of the 
$\Ghgagahbb$, which leads to the statistical precision of 2.1\%.
}

The final  \( W_{corr} \) distributions  for the  signal and
background events (with overlaying events included)  are shown in Fig.\ \ref{fig:m120_modsm_var34_oe1}. 
For \Mheq 120~GeV the most precise measurement of the Higgs-boson production cross section
is obtained for the  mass window 
between  108 and  133~GeV, as indicated by arrows.
In the selected \( W_{corr} \) region one expects, after one year of
the Photon Collider running at nominal luminosity,
about 4900  reconstructed signal
events and  5400 background events  (\ie \( \mu_S/\mu_B \approx 0.9 \)).
This corresponds to the statistical precision of:
\[
\frac{\Delta \left[ \Ghgagahbb \right] }{ \Ghgagahbb }=2.1\%. 
\]
The statistical precision calculated with the formula \ref{eq:prec1} 
should be considered a conservative estimate as it
does not take into account our 
knowledge of the shape of the  signal and background contributions.
To  exploit this additional information we can determine the measured number 
of signal events by the maximum likelihood method.
In this procedure the background contribution is assumed to be fixed
and the likelihood function depends only on the total number of signal events.
The Poissonian distribution of events in each bin is assumed.
For the production of the Higgs boson with \Mheq 120~GeV
we find that 
precision of the signal cross-section determination
from the measured $W_{corr}$ distribution
(without mass window cut)
  is 2.0\%.
The maximum likelihood method can also be used to estimate the systematic uncertainty
of the cross section measurement. 
With this approach the systematic uncertainties of 
the estimated total background contribution 
and of the luminosity determination are included.
We assume that both uncertainties are described by 
Gaussian distributions. 
%
The systematic uncertainty of the total background contribution 
is estimated to be about 2\%.
This is based
on the assumption that the background can be constrained from the dedicated run of the Photon Collider
at lower value of $\sqrtsee$.
If the center-of-mass-system energy is reduced by about 10~GeV, 
then only background events are measured  (the Higgs boson will not be produced 
due to very low luminosity at $\Wgaga = \Mh$)
while the detector performance remains almost unchanged.
During half a year of the Photon Collider running 
 about 3000 events can be  selected
 in the invariant-mass range corresponding to 
the optimal mass window for the Higgs-boson production measurement.
The statistical uncertainty of 2\% on the background contribution 
can be obtained
which could be reduced by in a longer run if required. 
This uncertainty will result in the corresponding  systematic error
in the (independent) Higgs-boson production measurement.
The $J_z=0$ luminosity contribution will be measured with precision of around 1\% \cite{KMonigLumi}
and  this is assumed to be the expected uncertainty for the total luminosity
(the $|J_z|=2$ luminosity contribution will be known with much better precision
but it is small in the Higgs-resonance region).
%
Using maximal likelihood procedure with assumed systematic uncertainties 
we obtain precision of 2.7\% for $\sgagahbb$ measurement at \Mheq 120~GeV.
Therefore we can conclude that the systematic error of the measurement is of the order of 1.8\%. 

The final result for $\Ghgagahbb$ should be extracted from the measured event rate
by applying correction for the selection efficiency.
In our analysis the total efficiency for signal events  is only about 30\%.
The significant reduction of the signal events is due to the \btagging{} cut.
Therefore a very precise determination of flavour tagging efficiency will be crucial.
To minimize influence of the uncertainties
resulting from the Monte Carlo description of the detector performance
we propose to use hadronic decays of $Z$ bosons for \btagging{} studies.
After  one year of the Photon Collider running at \sqrtseeeq 419~GeV, 
the sample of about 26000  $\gaga \ar ZZ$ events
will be collected \cite{wwzz}.
Taking into account branching ratios and the selection efficiency of about 80\%,
the expected number of $Z \ar \bbbar$ decays
will be about 5000.
This sample will allow us to determine the flavour tagging efficiency for $b$-jets
with statistical precision of about 1.4\%.
There is also 
an alternative solution: 
taking $\epem$  data  at \sqrtseeeq $M_Z$
for two to three months (the so-called GigaZ project),
using the same detector and selection procedure as described above.
Having about $10^9$ events of $Z$-boson production we could determine flavour-tagging efficiencies
with exceptional statistical precision of about $10^{-4}$.
Also other systematic uncertainties could be significantly reduced with such a large sample of $Z$ decays.
Therefore we would like to stress that
the possibility of $\epem$ operation in the collision point designed for the $\gaga$ mode should be guaranteed
to make full use of the physics potential of the Photon Collider.
For all methods, results for systematic uncertainties  will require some extrapolation.
However, we estimate that after a year of  additional running,   
the total systematic uncertainty of the $\Ghgagahbb$ measurement can be reduced to about 2\%.
The background contribution due to $\gagaWW$ production is included in the analysis for \Mheq 150 and 160~GeV.
Hadronic decays of $\WW$ pairs  result in 4-jet final state.
However, in significant fraction of events only 3 or 2 jets are reconstructed.
As cross section for $\WW$ production is very high,
additional cuts, dedicated to suppress $\gagaWW$ background are introduced:
\begin{enumerate}
%
%
\item Events 
      are rejected 
      if $M_{jet}^{\max} > \Cmj$,
      where $M_{jet}^{\max}$ is the highest invariant mass of the jet in the event.
%
%
\item Total energy measured below $\thetamindet$, $E_{TC}$, should be less than $\Cec$. 
\item Each jet in the event should contain at least $ \Cnt $ tracks, \ie $N_{trk}^{\min} \geq \Cnt $
      where $N_{trk}^{\min}$ is the minimal number of tracks in jet for a given event. 
\end{enumerate}
The first cut removes the events in which two jets from decay of one $W$ boson have been joined into one jet
by the jet-clustering algorithm.
The second condition rejects events with substantial energy in the forward region, 
below $\thetamindet$,   where one or two jets
from $\WW$ decay could deposit their energy.
The last cut, with the parameter value  $\Cnt=4$, suppresses leptonic decays of $W$.
For \Mheq 160~GeV the  optimal values for the two other  thresholds   are  $\Cmj=70$~GeV and  $\Cec=90$~GeV.

%
We have performed a full simulation of signal and background events also
for \Mheq  130, 140, 150 and 160~GeV
choosing optimal $\emem$ beam energy for each Higgs-boson mass.
Statistical precision of $\Ghgagahbb$ measurement was estimated in each case.
It is equal to 2.1\%, 2.5\%, 3.4\% and 7.7\%, respectively.
These results, together with the result for \Mheq 120~GeV described above, 
are presented in Fig.\ \ref{fig:plot_precision_summary_modsm}. 
For comparison, our earlier results 
obtained without overlaying events,
without various background contributions or
without distribution of interaction point 
are also shown.
For \Mheq 160~GeV, after the full optimization of the selection cuts,  
better precision is obtained 
than in earlier analyses, which did not take into account some background contributions.


\pnfig{tb}{\figheight}{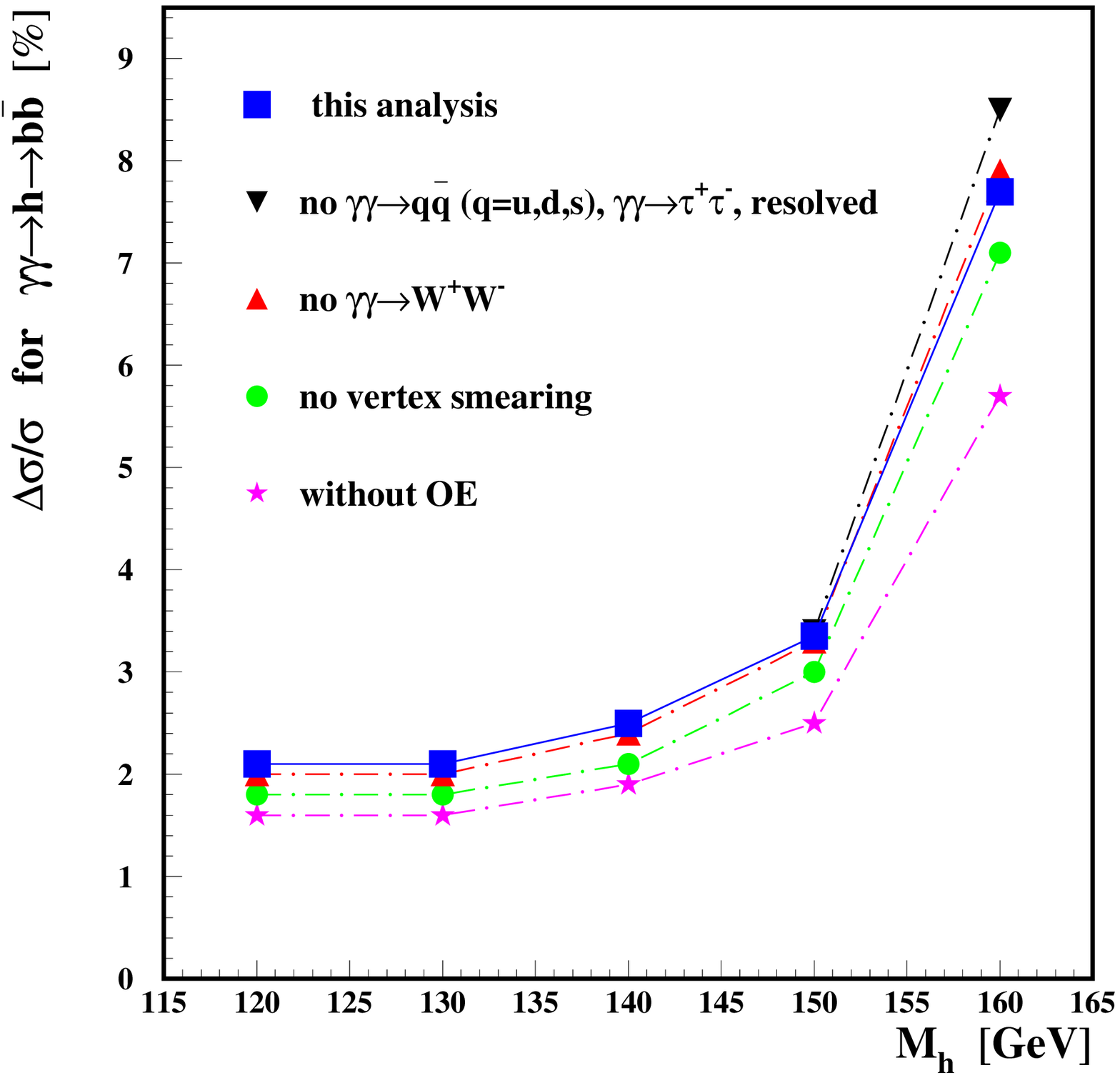}{fig:plot_precision_summary_modsm}{
Statistical precision of $\Ghgagahbb$ measurement for the SM Higgs boson with mass 120--160~GeV.
%
Final results of this analysis are compared with our earlier results,
which did not take into account  some of the
background contributions, 
distribution of the interaction point or overlaying events.
The lines are drawn to guide the eye. 
}

\
\newpage 
\
\newpage 
\chapter{  Production of heavy neutral Higgs bosons in the MSSM  \label{ch_mssm_analysis}}
The analysis of the MSSM Higgs-bosons production
closely follows the SM study described in Chapter \ref{ch_sm_analysis}.
Therefore only these parts of the analysis which differ from the SM case
are discussed in detail below.
The analysis was developed for  set \textit{I} of the MSSM parameters,  
as defined in Section \ref{sec_signal},  with $\tbseven$.
To simplify the description the procedure is first presented for \MAOeq 300~GeV, 
and later extended to \MAOeq 200, 250 and 350~GeV.
In each case the beam energy was chosen in such a way,
as to provide
the highest luminosity for $\gaga$ collisions with $J_z=0$ at $\Wgaga = \MAO$.
For detailed information about beam energies and $\gaga$-luminosities, 
for the considered Higgs-boson masses, 
see Tab.\ \ref{tab:PLCluminosity}.


For $\sqrtsee \approx$ 400~GeV about two \gagahad{} events are expected on average 
in each bunch crossing.
To suppress the influence of these events optimization of the $\thetamindet$ cut was repeated,
as described in Appendix \ref{app_thetatc}
and the value  \thetamindeteq 0.85, used in the SM analysis,
turned out to be optimal also for the MSSM case.
Selection criteria were optimized for the measurement of the \higgs{} production cross section.
The event distributions for  $|\cos {\theta}^{jet}|^{\max}$ and $|P_{z}|/E$ 
are shown in Fig.\ \ref{fig:max_abs_costhjet_mh300} and \ref{fig:pz_to_evis_ratio_mh300}, respectively. 
For \MAOeq 300~GeV the
optimized cut values are $\Cct = 0.65$ and $\Cpz = 0.06$  (indicated by arrows).
The measurement precision is estimated to be around 32\% and 22\%,
 after the $|\cos {\theta}^{jet}|$ cut and after the $|P_{z}|/E$  cut, respectively.

%
\pnfig{!}{\figheight}{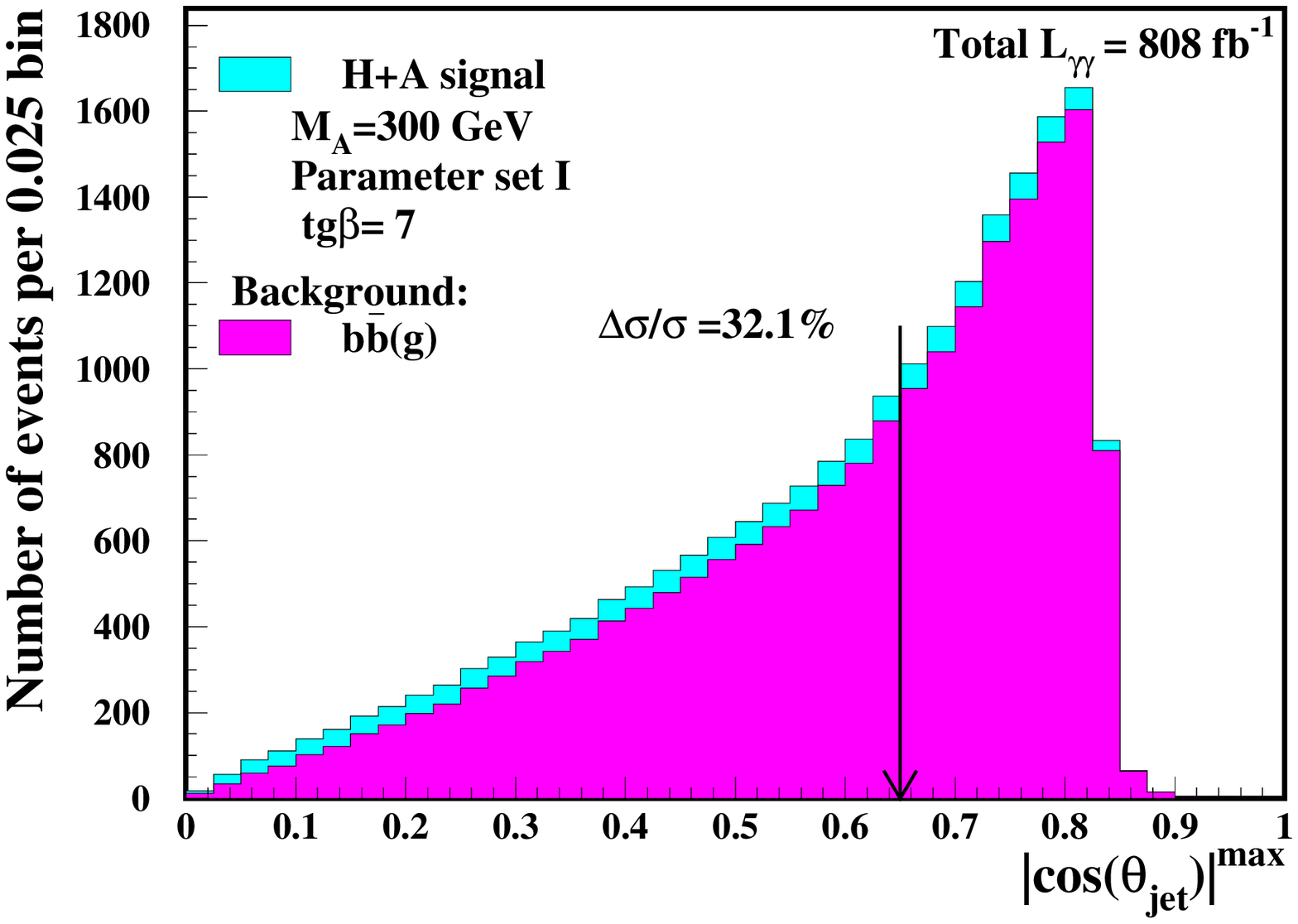}{fig:max_abs_costhjet_mh300}
{
Distributions of $|\cos {\theta}_{ jet}|^{\max}$ for signal and background events. 
Only $\gagabbg$ events are shown for the background.
 The signal measurement precision for events fulfilling the cut $ |\cos {\theta}_{jet}|^{\max} < \Cct = 0.65$, $\Delta\xs/\xs$, 
is around 32\%.
}

%
\pnfig{!}{\figheight}{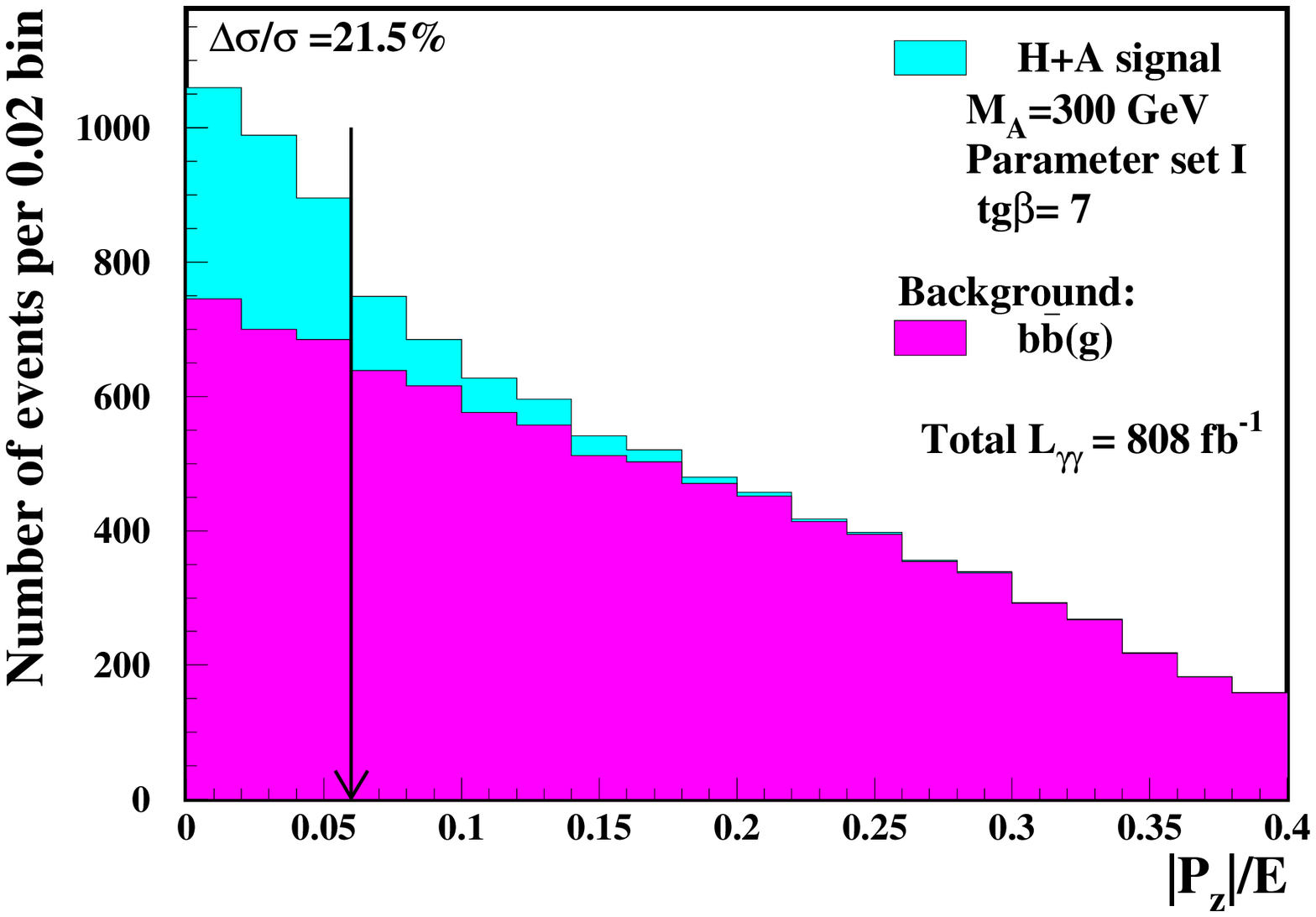}{fig:pz_to_evis_ratio_mh300}
{
Distributions of $|P_{z}|/E$ for signal and background events. Only $\gagabbg$ events are shown for the background.
 The signal measurement precision for events fulfilling the cut $  |P_{z}|/E < \Cpz  = 0.06$, $\Delta\xs/\xs$, is around 22\%.
}

As in the SM case, $\bbbar$ events were selected by considering $b$-tag values
for two jets with highest transverse momentum.
 For \MAOeq 300~GeV  the selection region in the  ${\rm btag}({\rm jet_{1}}) \otimes {\rm btag}({\rm jet_{2}})$ plane
which gives the best precision of the $\sgagaAHbb$
measurement is shown  in Fig.\ \ref{fig:plot_btag2j_m300_modmssm}. 
Optimal $\bbbar$ selection,
when  overlaying events are taken into account,   
corresponds to the efficiencies
 $\varepsilon_{h}=53\%$, $\varepsilon_{bb}=47\%$, $\varepsilon_{cc}=2.9\%$, and $\varepsilon_{uds}=0.5\%$,
\ie $\ccbar$  background suppression 
by a factor of $\varepsilon_{bb}/\varepsilon_{cc} \approx 16$.
%
This should be compared to results obtained when overlaying events are not included 
-- then the optimized selection  
corresponds to the efficiencies 
 $\varepsilon_{h}=57\%$, $\varepsilon_{bb}=52\%$, $\varepsilon_{cc}=1.8\%$, and $\varepsilon_{uds}=0.1\%$,
\ie  $\varepsilon_{bb}/\varepsilon_{cc} \approx 29$.
This comparison shows that for heavy Higgs bosons overlaying events 
increase relative probability of \ccmistagging{} by a factor of about 2. 
Even larger effect is observed for light quark-pair production as we
observe  increase of  $\varepsilon_{uds}$ by a factor of  5
despite the tighter selection cut.
In Fig.\ \ref{fig:wrec_a0_oe01_m300} we compare the  ($W_{rec}  - \Wgaga $) distributions,  
for signal events $\gagaAbb$,
before and after taking into account the overlaying events.
Mass resolution, derived from the 
Gaussian fit in the region from \( \mu - 1.3 \sigma  \)
to \( \mu + 1.3 \sigma  \), is about 8 and 12~GeV, respectively. 
The overlaying events and cut $\thetamindet$ suppressing their contribution
significantly influence the reconstructed mass  distribution
and result in a deterioration of the mass resolution.
However, no significant drop in the selection efficiency 
is observed (the number of selected events is about 220 in both cases),  
in contrary to the SM case.
Corresponding distributions of  ($W_{corr}  - \Wgaga $) are compared in Fig.\ \ref{fig:wrec_a0_oe01_m300}.
The mass resolution for $\gagaAbb$ events, for \MAOeq 300~GeV,
is about 7~GeV without and about 10~GeV with overlaying events.
As the mass resolution is much bigger than 
the mass difference between heavy neutral Higgs bosons $\HO$ and $\AO$, $\MHO - \MAO \approx 1.5$~GeV,
the separation of these two contributions will be impossible for the considered parameter set.
Even for \tanbeq 3 the mass difference ($\MHO - \MAO \approx 6.8$~GeV) will be too small 
to resolve $\AO$ and $\HO$ contributions, 
see  Fig.\ \ref{fig:wcorr_a0hh0_ma300}.

\pnfig{tb}{\twofigheight}{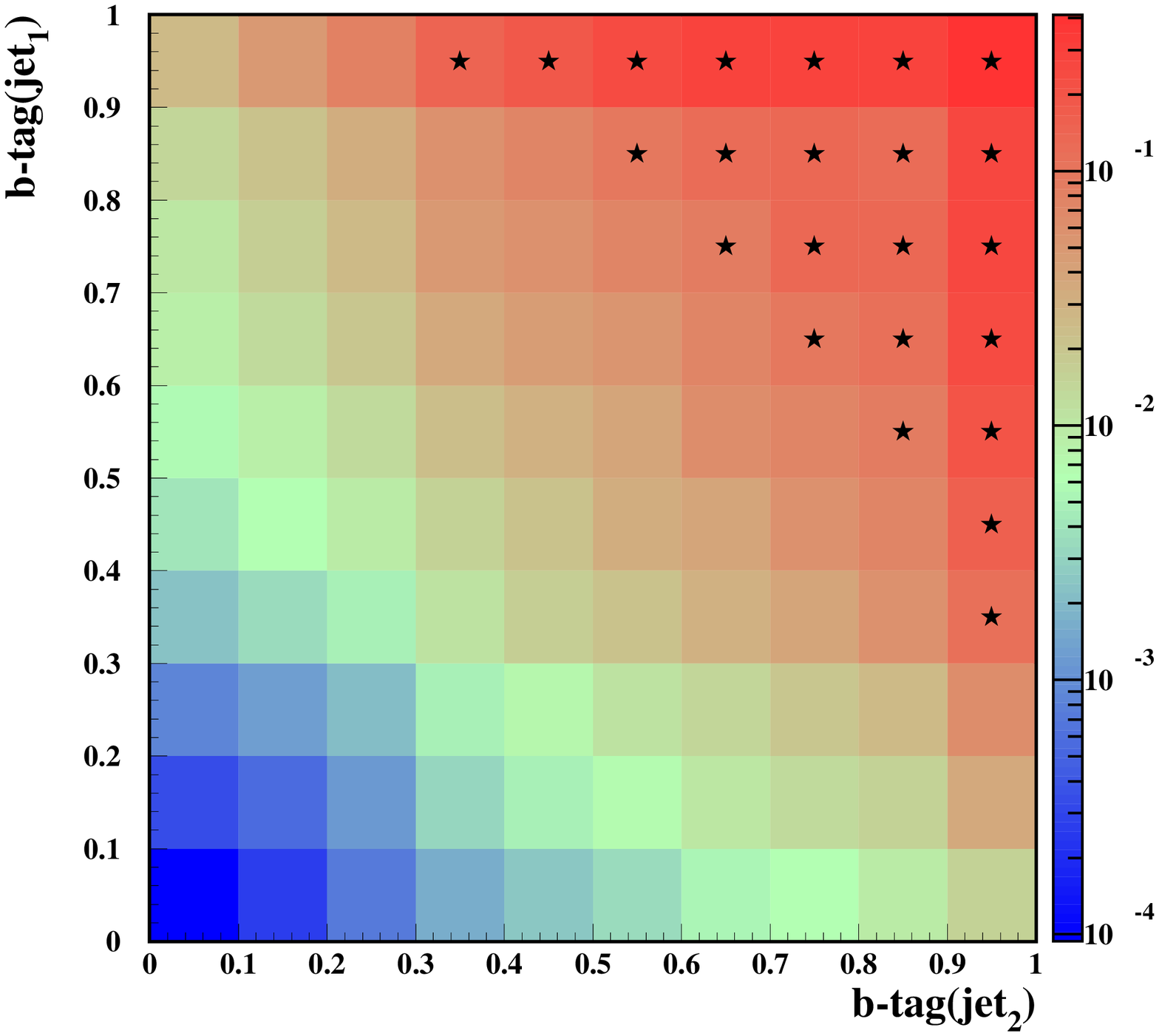}{fig:plot_btag2j_m300_modmssm}
{
Ratio of $\gagaAHbb$ events to $\gagabbg$, $\gagaccg$ and $\gagaqq$, $\quds$, events distributions 
in the plane ${\rm btag}({\rm jet_{1}}) \otimes {\rm btag}({\rm jet_{2}})$. 
The region is indicated by stars which gives the best precision measurement for $\sgagaAHbb$.} 

\pnfiggeneral{p}{\twofigheight}{\includegraphics{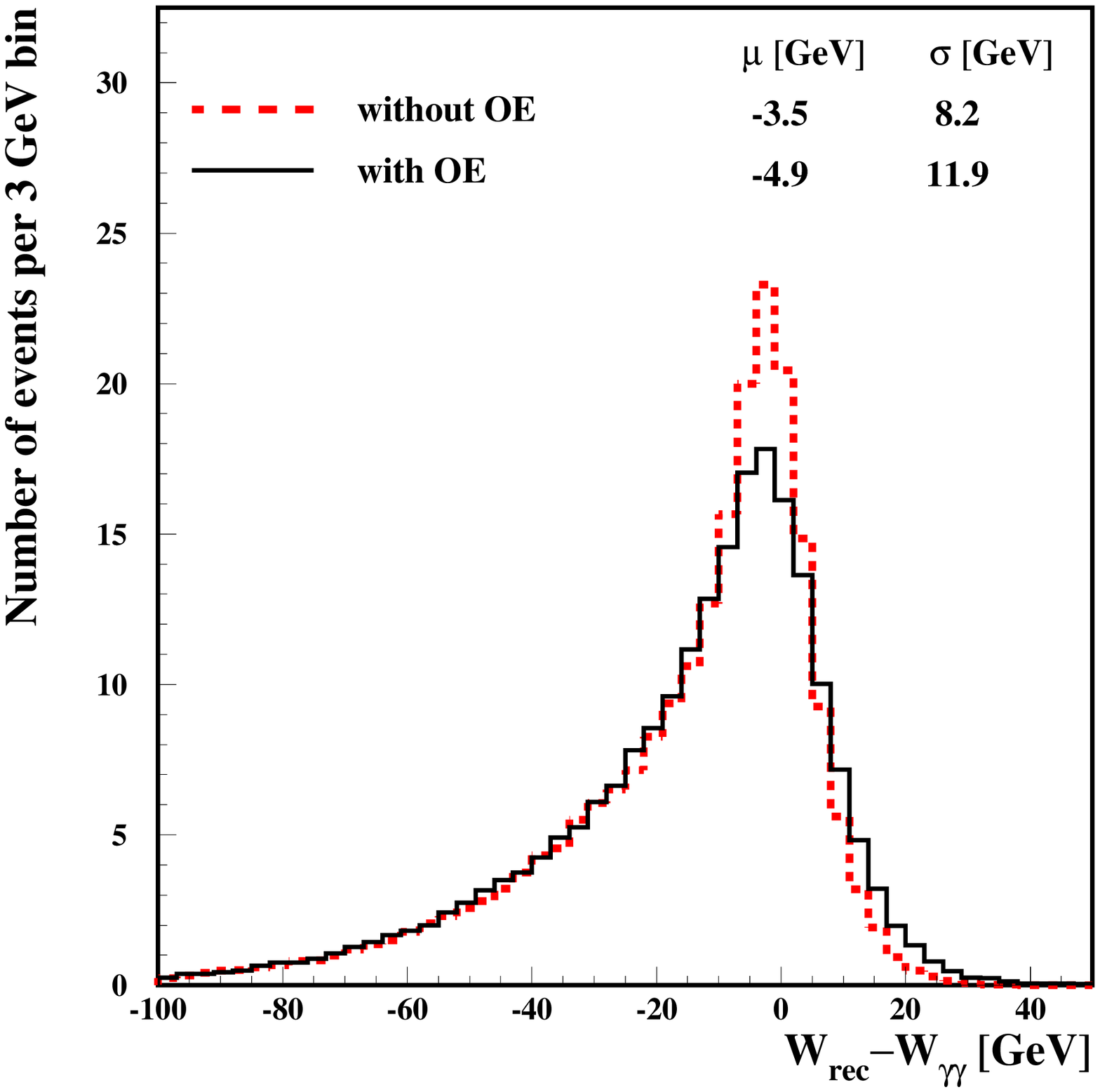}
\includegraphics{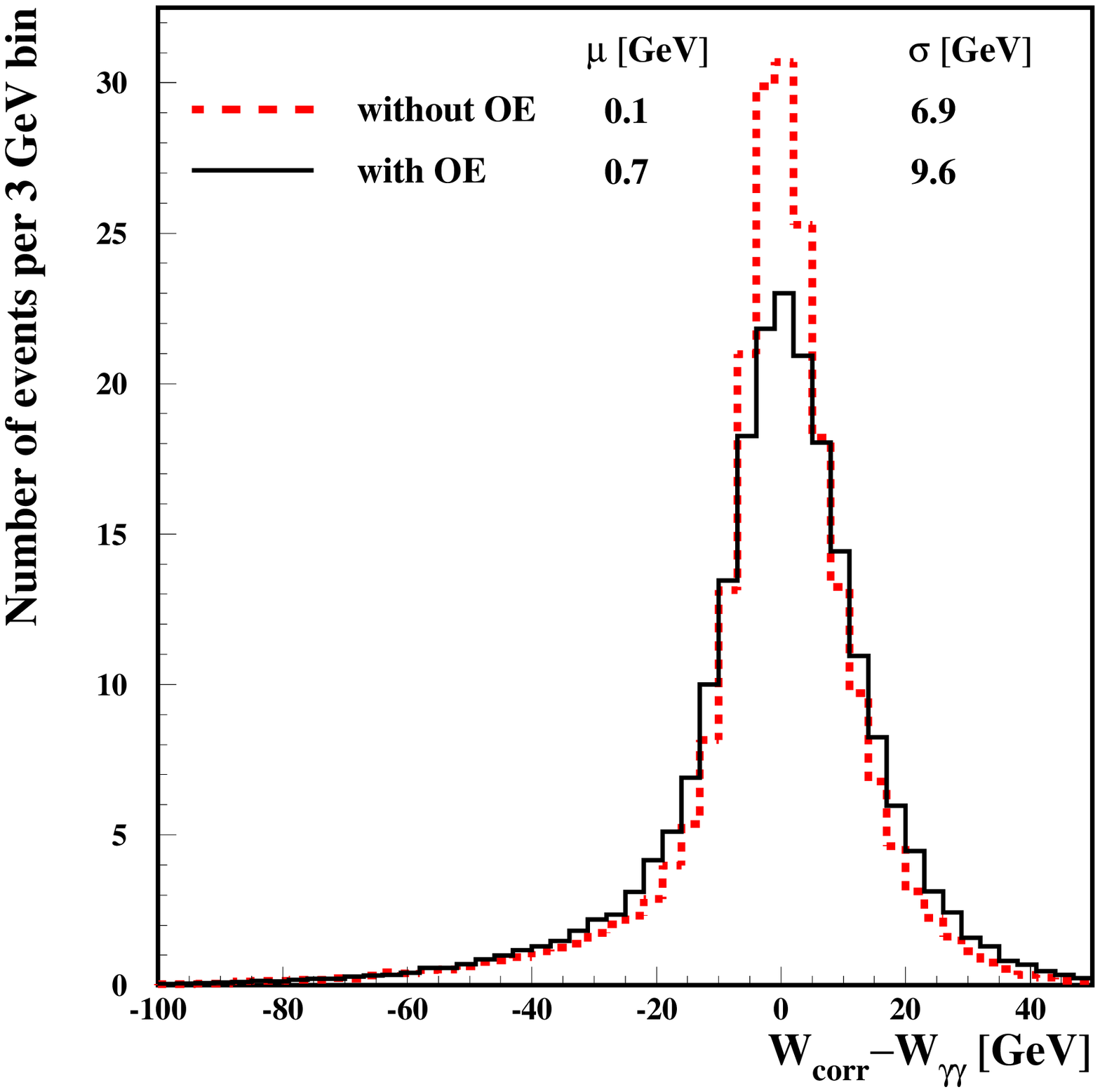} }{fig:wrec_a0_oe01_m300}{
Mass resolution of the Higgs-boson $\AO$ , $W_{rec}  - \Wgaga$ (left), and  $W_{corr}  - \Wgaga$ (right),
for the selected events for \MAOeq 300~GeV, 
obtained without and with overlaying events (OE).
Mean and dispersion values, $\mu$ and $\sigma$, from the Gaussian fit 
between $\mu - 1.3\sigma$ and  $\mu + 1.3\sigma$ are indicated.
}
\pnfiggeneral{p}{\twofigheight}{\includegraphics{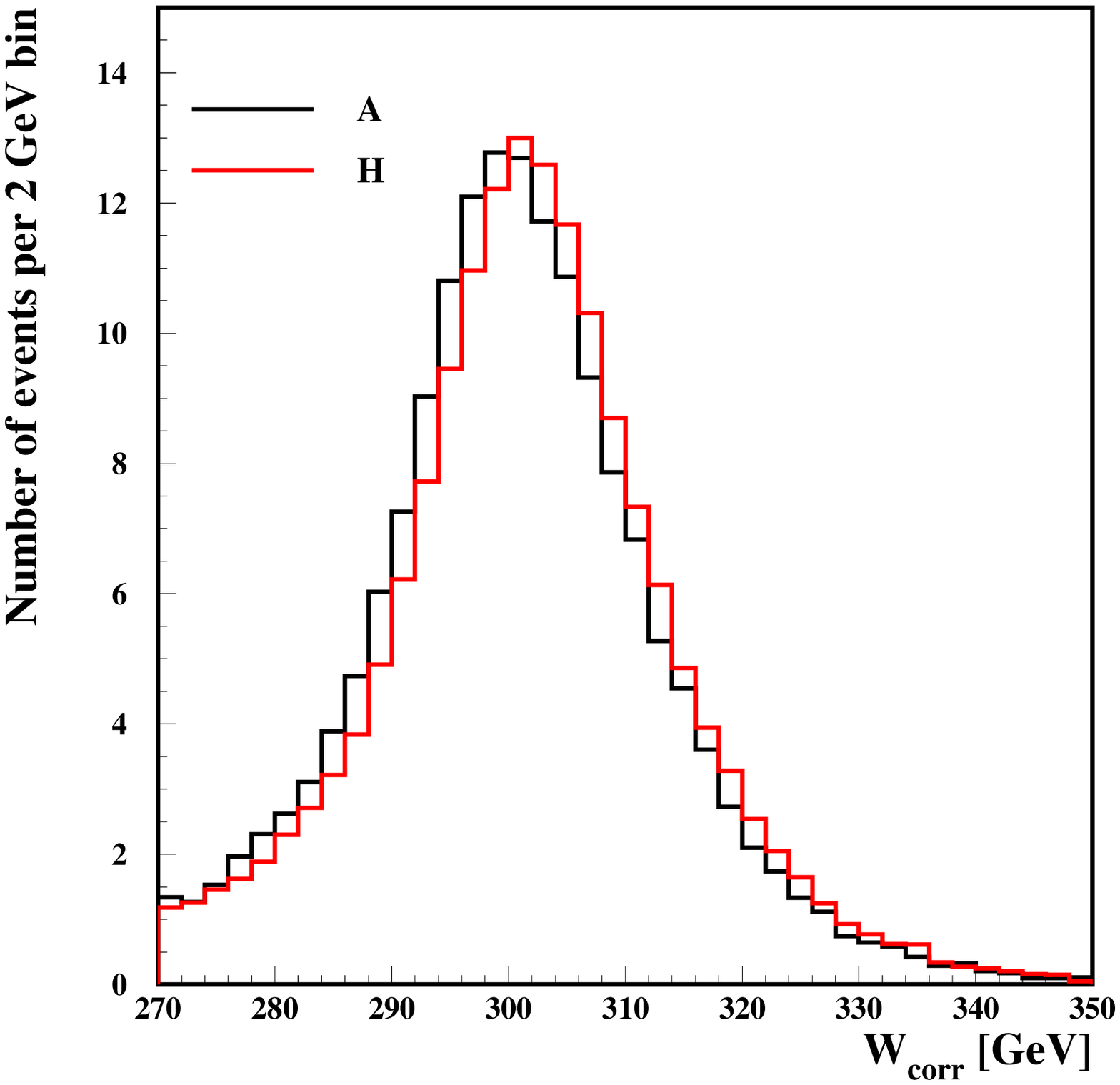}
\includegraphics{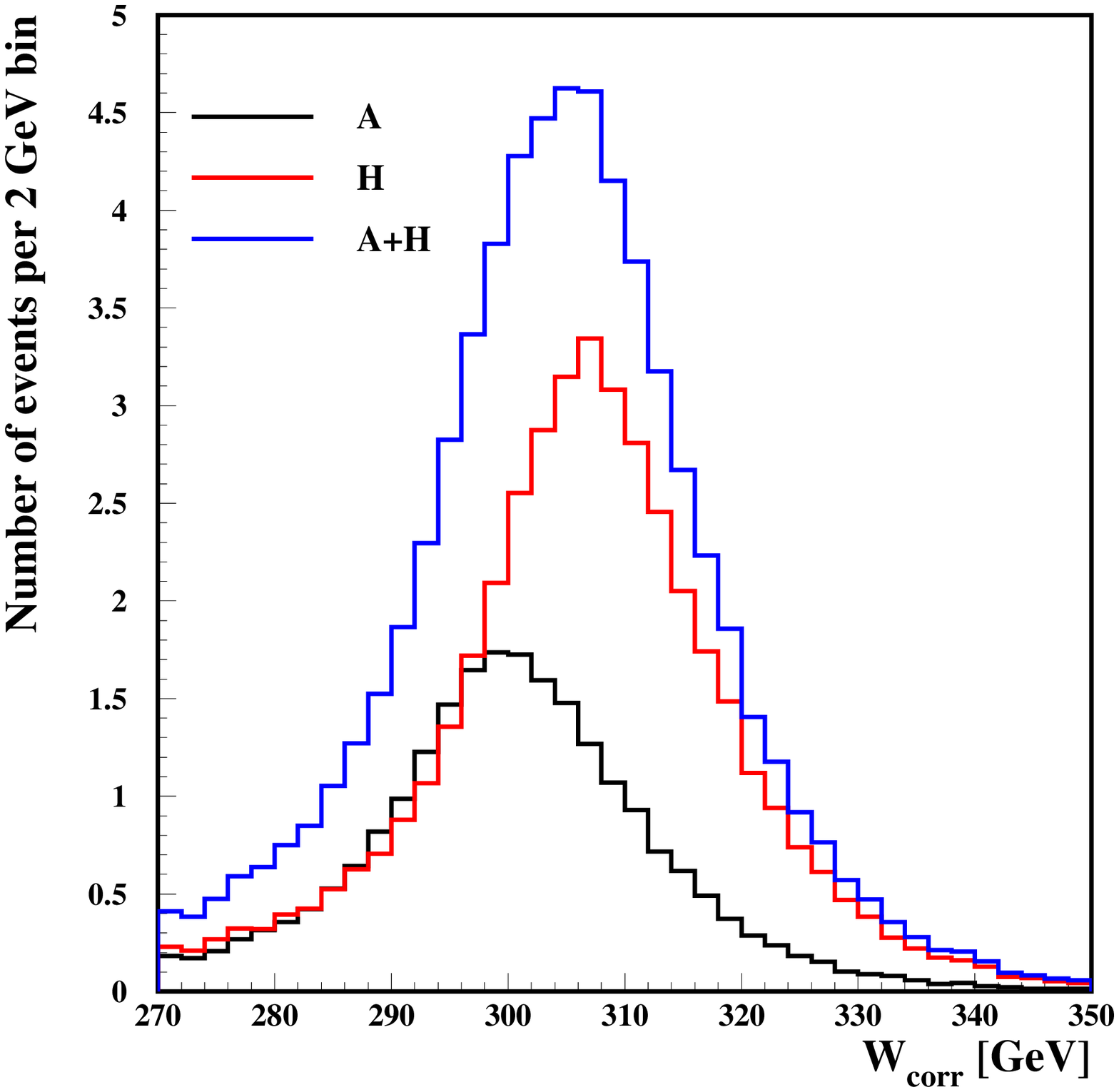}}{fig:wcorr_a0hh0_ma300}{
Corrected invariant mass, $W_{corr}$,
distributions for the selected $\gagaHbb$ and $\gagaAbb$ events  (\MAOeq 300~GeV), 
with overlaying events, for \tanbeq 7 (left) and \tanbeq 3 (right), respectively.
For  \tanbeq 3 also the sum of both contributions is shown.
}

\pnfiggeneral{p}{0.39\textwidth}{\includegraphics{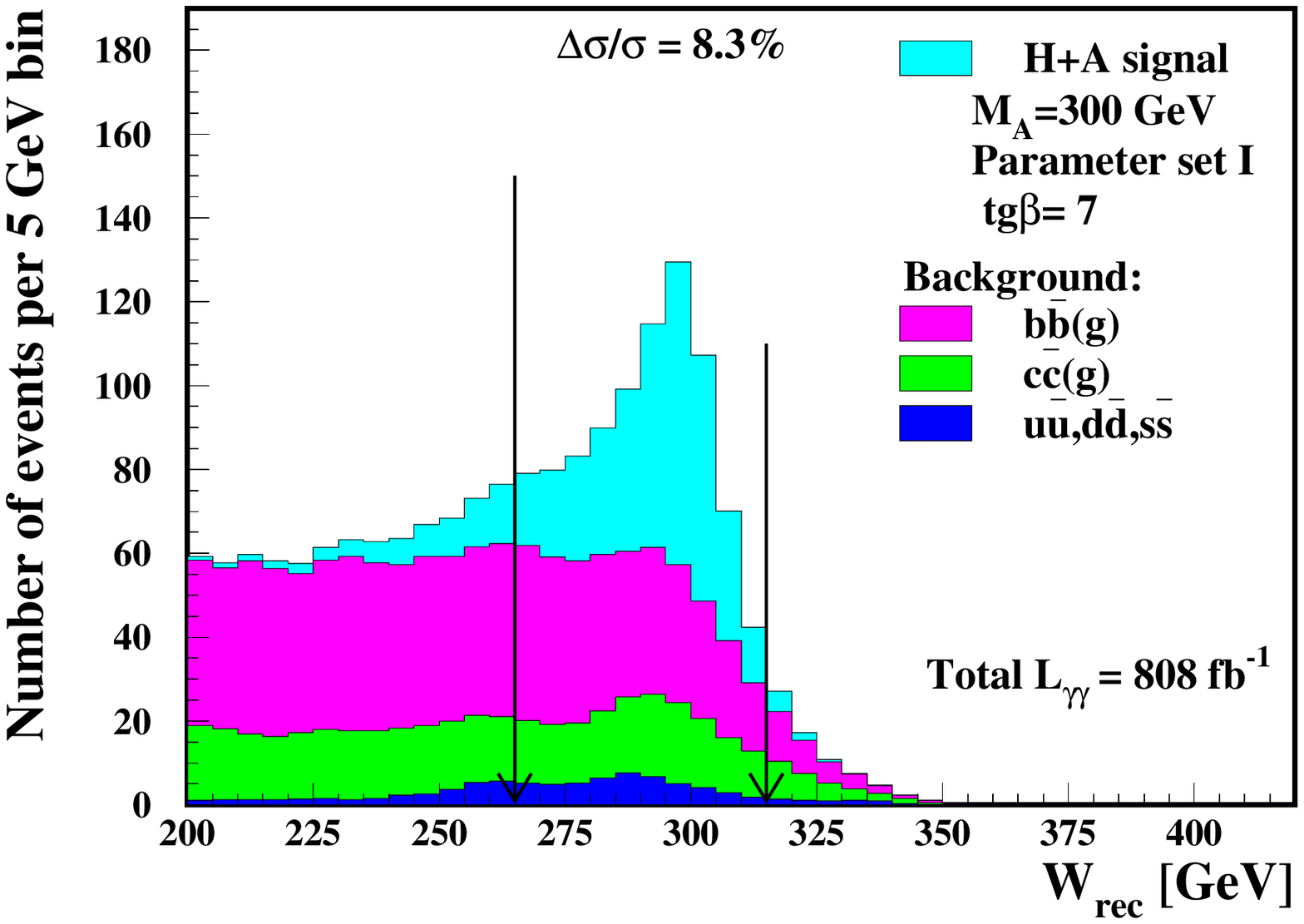} 
                       \includegraphics{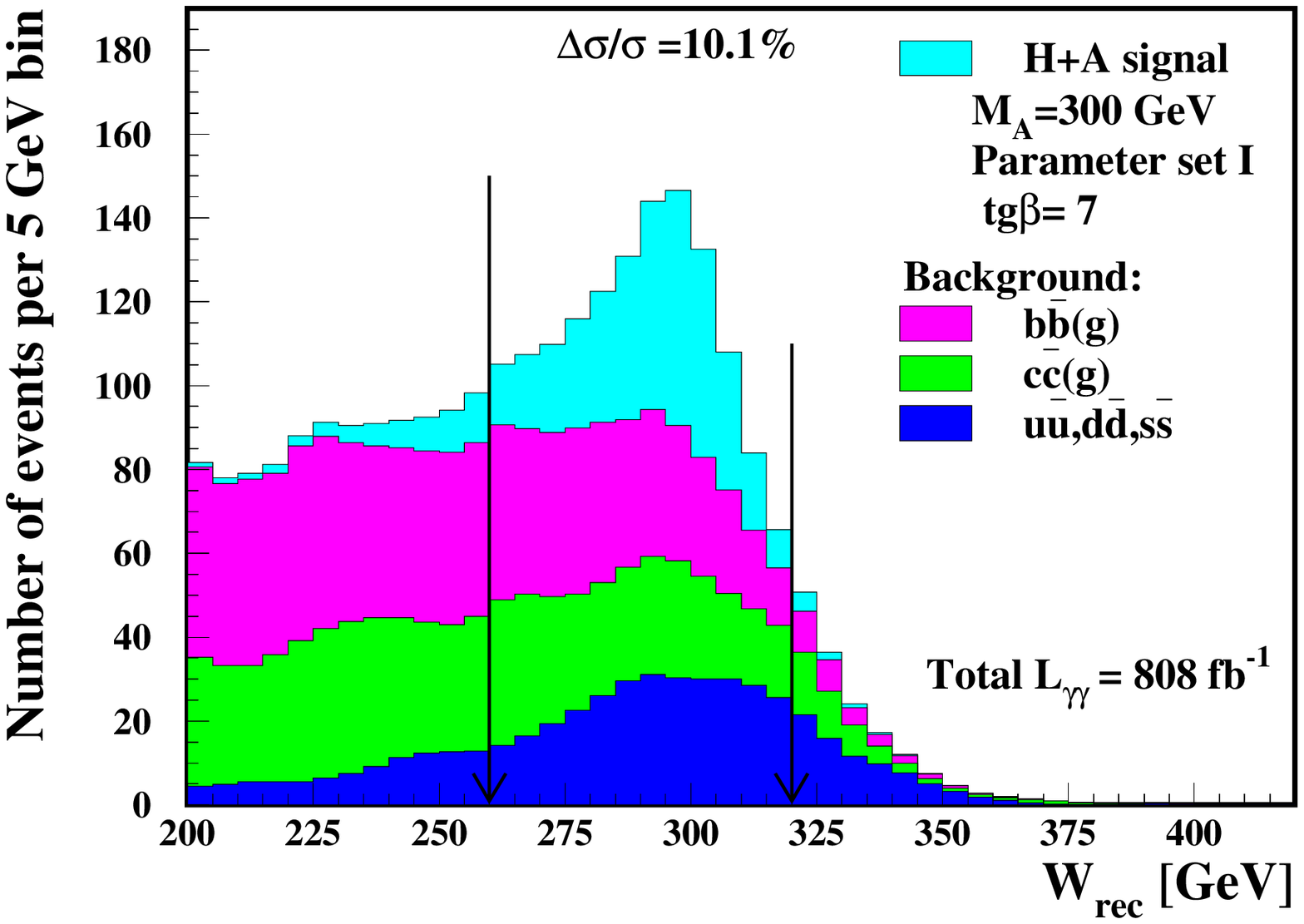}}
 {fig:plot_m300_modmssm_var1_oe01}{
Distribution of the reconstructed invariant mass, 
$W_{rec}$, for $\bbbar$ events  (\MAOeq 300~GeV)  selected
before (left plot) and after (right plot) taking into account 
overlaying events. 
Contributions of the $\HO$ and $\AO$ signal and of the quark-pair backgrounds are shown separately. 
Arrows indicate the mass windows optimized for the measurement of $\sgagaAHbb$.
}
%
\pnfiggeneral{p}{0.39\textwidth}{\includegraphics{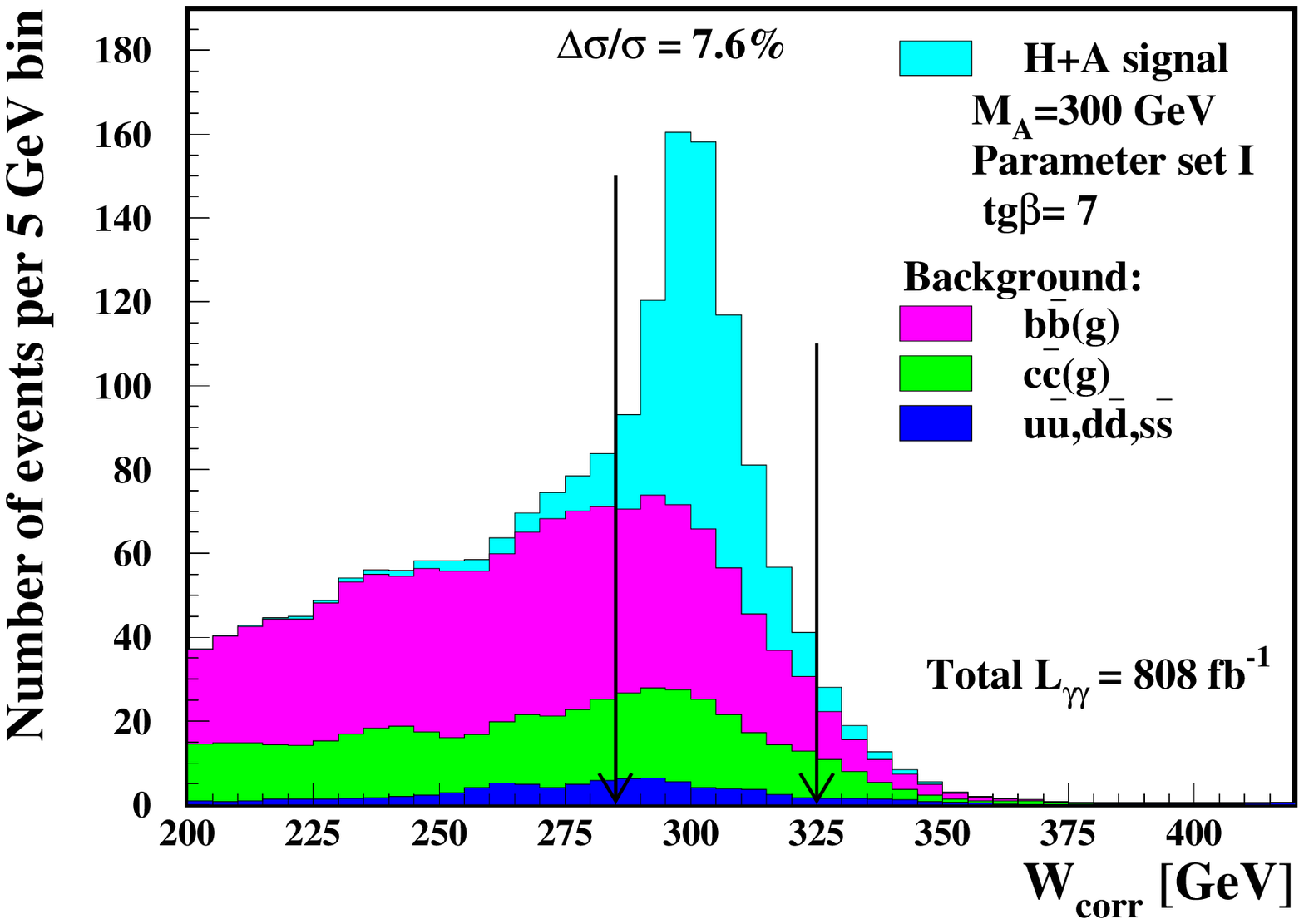} 
                       \includegraphics{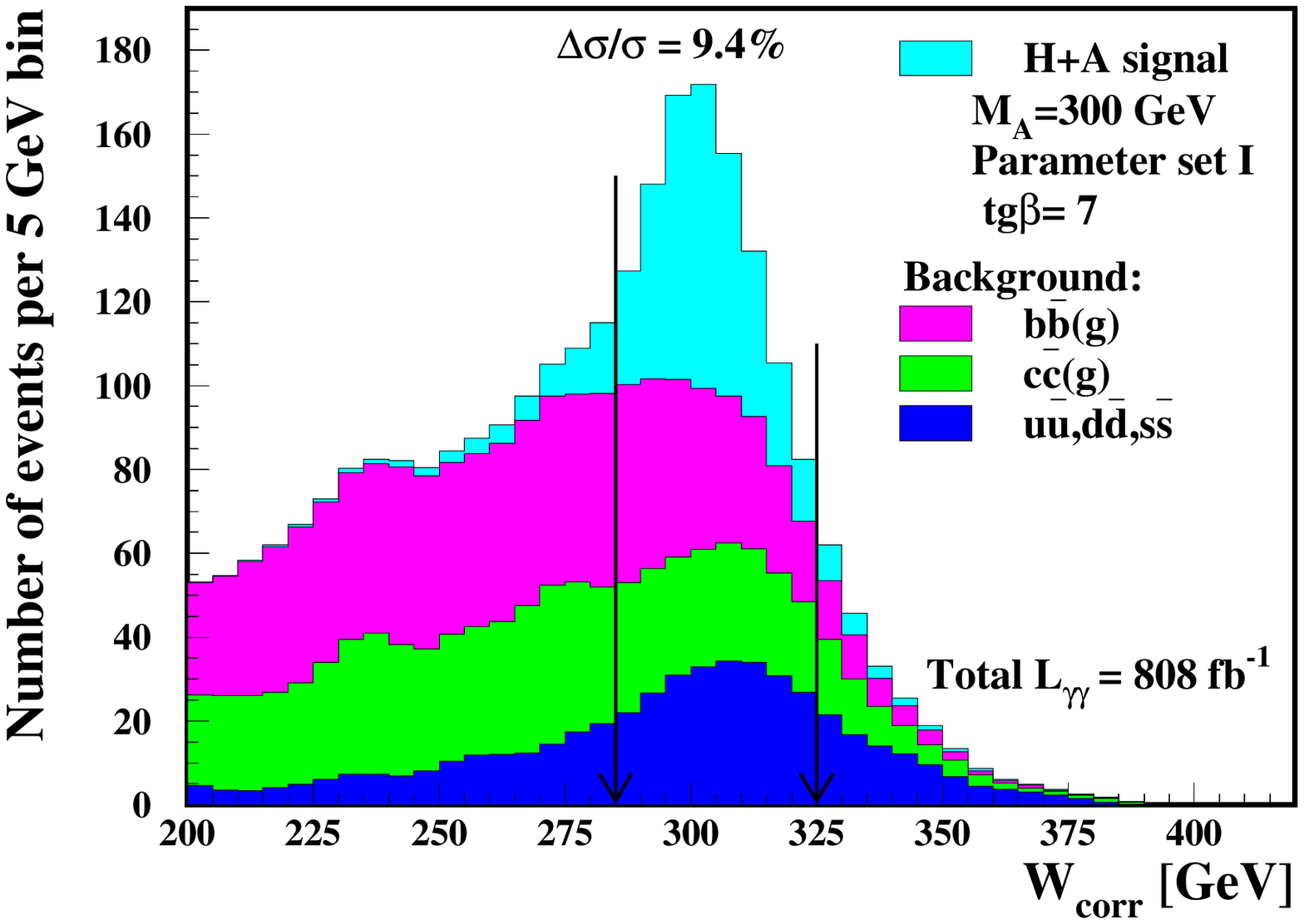}}
{fig:plot_m300_modmssm_var4_oe01}{
As in Figs.\ \ref{fig:plot_m300_modmssm_var1_oe01}, 
for the corrected invariant-mass, $W_{corr}$, distributions.
The statistical precision is 7.6\% without overlaying events (left plot)  
and 9.4\% with overlaying events (right plot).
\vspace{3cm}
}

The influence of overlaying events on the Higgs-boson production measurement
is also illustrated in Fig.\ \ref{fig:plot_m300_modmssm_var1_oe01} and \ref{fig:plot_m300_modmssm_var4_oe01}
where
the $W_{rec}$ and $W_{corr}$ distributions, respectively,
are shown for signal and background  events (for this comparison only quark-pair production is taken as the background).
In both cases distributions obtained without and with overlaying events are compared. 
We conclude that overlaying events significantly deteriorate  the cross section measurement
by increasing the selection efficiency of background contributions.
As in the SM case,  better estimates for measurement error $\DssgagaAHbb$ 
are obtained when using the variable $W_{corr}$.

For all considered heavy Higgs-boson masses in the MSSM,
additional cuts to suppress $\gagaWW$ background were also applied,
as described for the SM Higgs boson with \Mheq 160~GeV.
For \MAOeq 300~GeV the  optimal threshold values  are  $\Cmj=65$~GeV and $\Cec=80$~GeV.
The corresponding distributions of events are shown
in Fig.\ \ref{fig:plot_var420_m300_modmssm_oe1_costhtc0.85} and \ref{fig:plot_var430_m300_modmssm_oe1_costhtc0.85}.
The resulting estimate of the cross-section measurement precision is about 14\%.
After the additional cut on the number of tracks in each jet ($\Cnt=4$), 
the final result for \MAOeq 300~GeV is shown in Fig.\ \ref{fig:plot_var34_m300_modmssm_oe1_costhtc0.85}.
For the optimized $W_{corr}$ window the cross-section measurement  precision of 11\% is obtained.

\pnfig{tb}{\figheight}{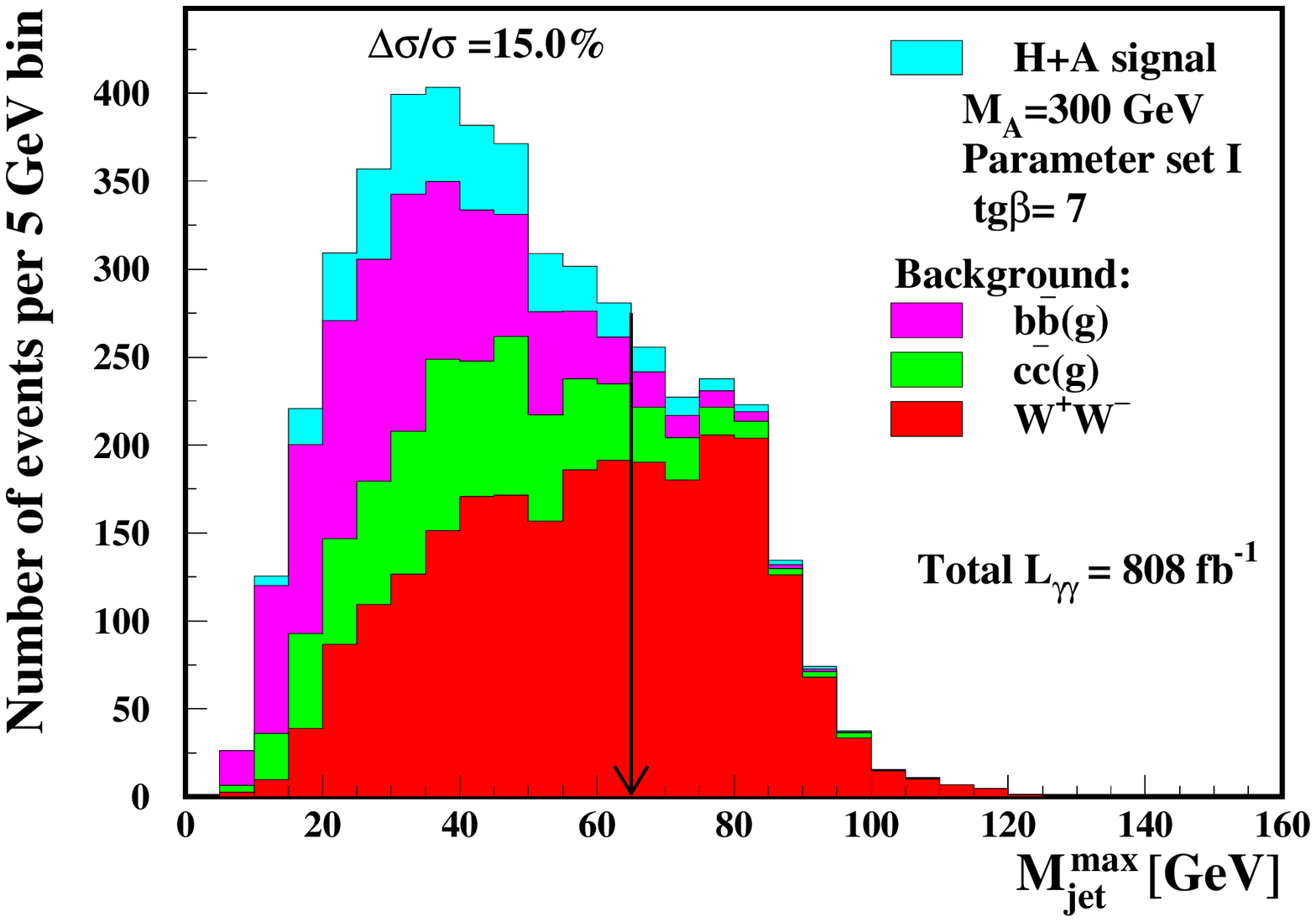}{fig:plot_var420_m300_modmssm_oe1_costhtc0.85}{
Distributions of the maximal jet mass, $M_{jet}^{\max}$, expected for signal and background contributions (overlaying events included).
}

\pnfig{tb}{\figheight}{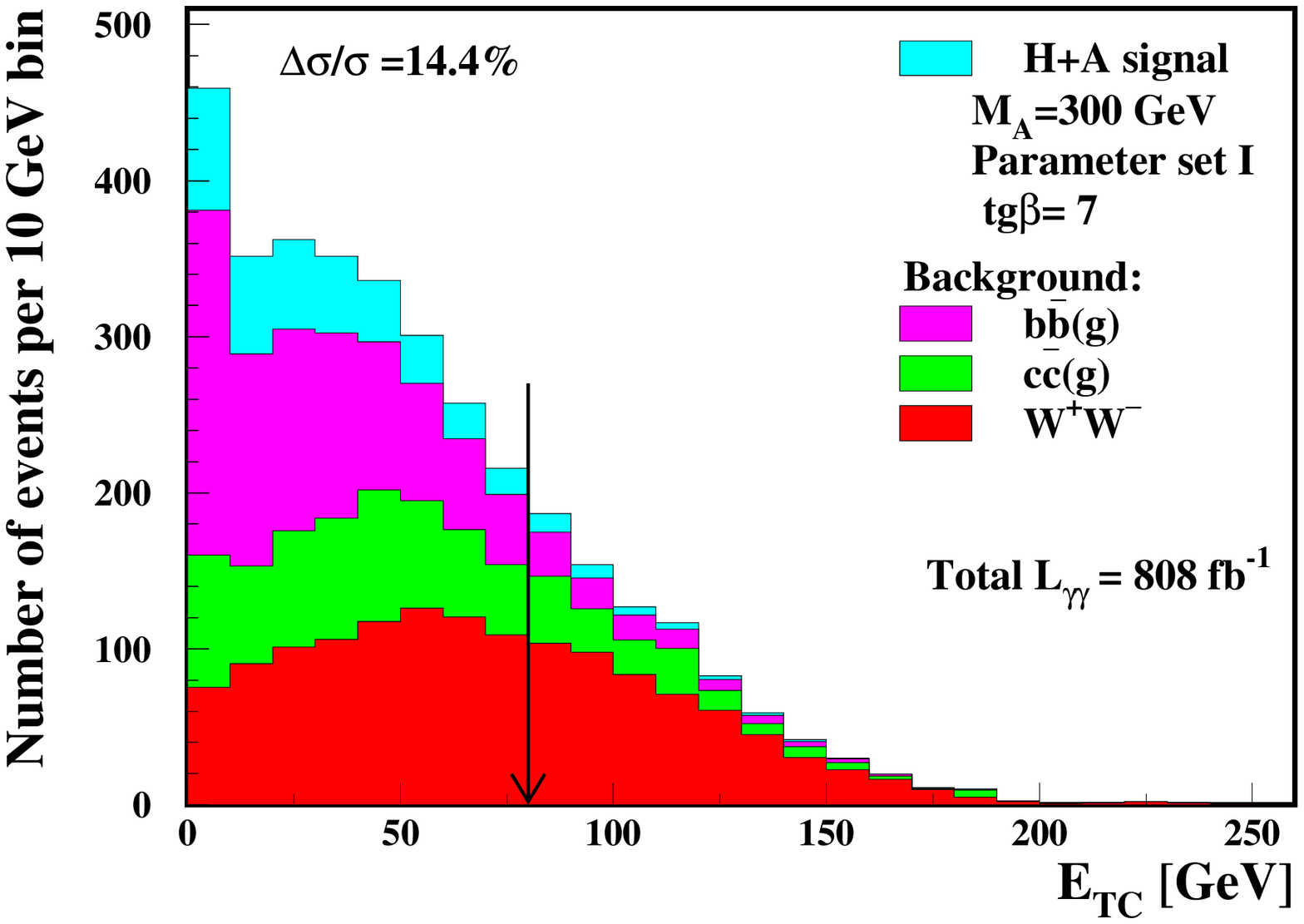}{fig:plot_var430_m300_modmssm_oe1_costhtc0.85}{
Distributions of the energy below $\thetamindet$ measured in the calorimeters, $E_{CAL}$, for signal and background contributions,
with  overlaying events included.
}

\pnfig{tb}{\figheight}{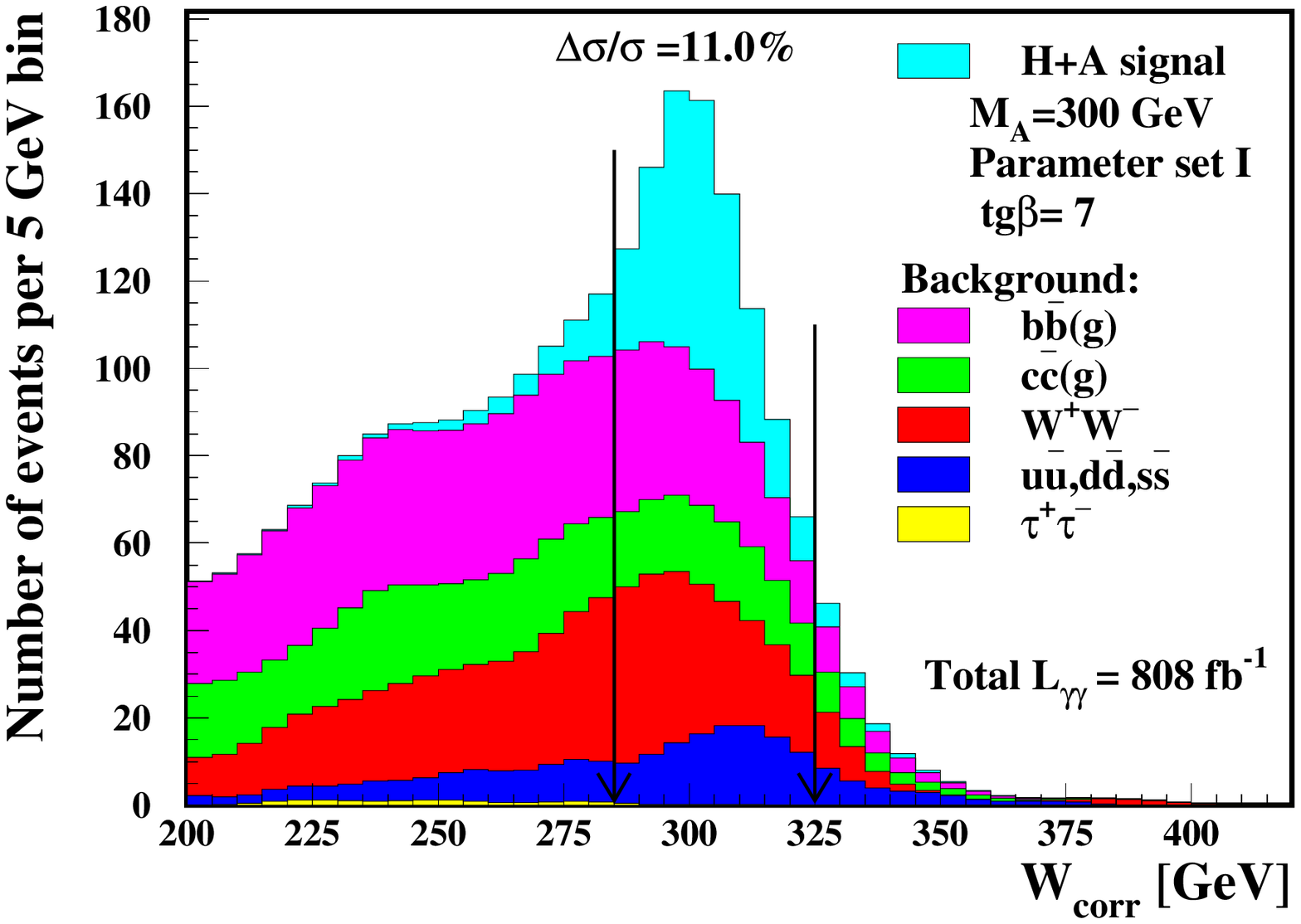}{fig:plot_var34_m300_modmssm_oe1_costhtc0.85}{
Distributions of the corrected invariant mass, $W_{corr}$, for signal and all considered background contributions,
with  overlaying events included.
The best precision of 11\% for $\gagaAHbb$ cross section measurement 
is achieved in the $W_{corr}$ window between   285 and  325~GeV.
}

After discovery or a 'hint' of the resonant-like excess of events 
at LHC or ILC
the~Photon Collider can be used
to confirm the existence and to measure the cross section
for~production of the new state.
In this case the beam energy will also be optimized for the production of the observed resonance.
Under this assumption,
we estimate precision expected  for $\gagaAHbb$ cross section measurement
in the considered MSSM scenario
after one year of experimentation.
In Fig.\ \ref{fig:plot_precision_summary_modmssm} the precisions for all considered values of $\MAO$ are shown
and compared with our previous results.
Thanks to the cuts optimization better estimated precision  for \MAOeq 200~GeV has been obtained;
earlier results were obtained with cuts optimized for Higgs-bosons production for \MAOeq 300~GeV.  


\pnfig{tb}{\figheight}{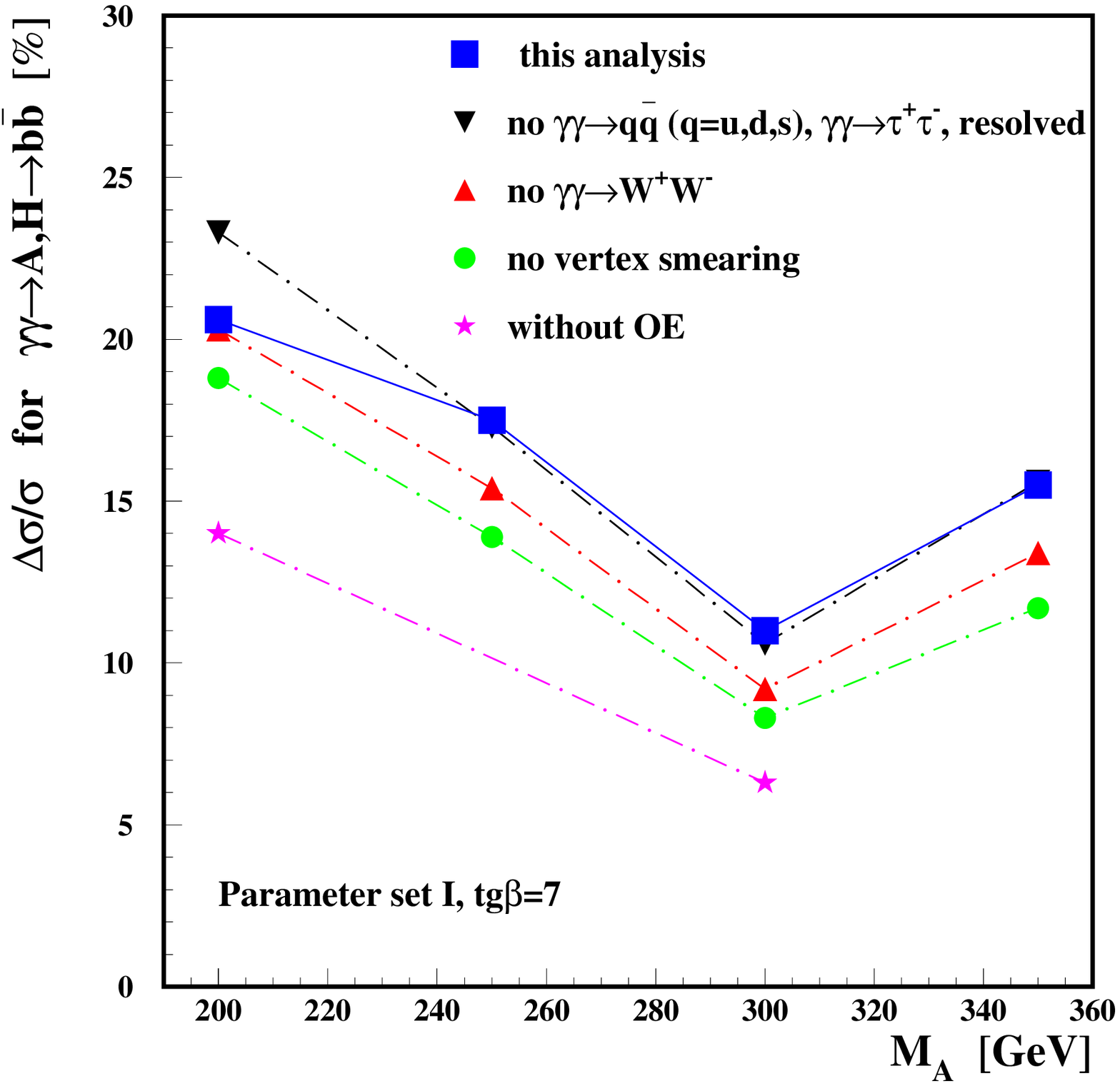}{fig:plot_precision_summary_modmssm}{
Precisions of $\sgagaAHbb$ measurement for MSSM parameter set $I$, for \MAOeq 200--350~GeV  and \tanbeq 7.
Final results of this analysis are compared with our earlier results,
which did not take into account  some of the
background contributions, 
$\gagaWW$,
distribution of primary  vertex or overlaying events.
The lines are drawn to guide the eye. 
}

Presented results were also used to estimate precision of the cross-section measurement at the Photon Collider for 
other parameter sets described in Section \ref{sec_signal}.
For all sets the 
total widths of the Higgs bosons $\AO$ and $\HO$ are comparable or smaller than for the set \textit{I} 
and 
their masses are nearly degenerated (the mass difference is smaller or comparable with detector resolution).
Precision estimates for $\sgagaAHbb$ measurement for all considered  parameter sets,
assuming  \tanbeq 7 are compared
in  Fig.\ \ref{fig:plot_mssm_precision_vs_ma_tgb7}.
The most precise measurement is expected  
for parameter sets $II$ and $III$  --- precision hardly depends on $\MAO$
and is about 10\%.
The worst measurement is expected for scenario $IV$, \ie the one considered by the CMS collaboration.

\pnfig{p}{\figheight}{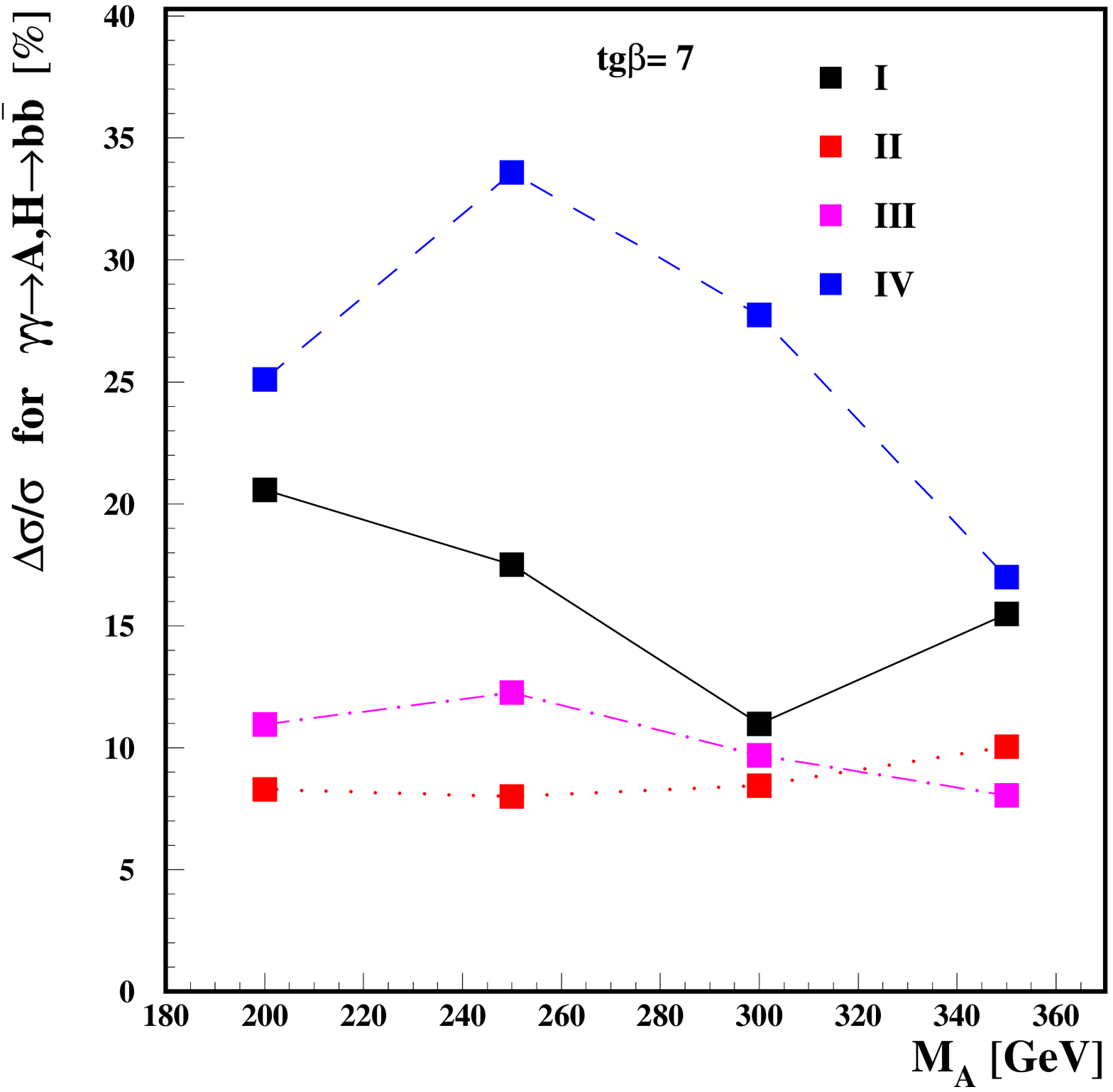}{fig:plot_mssm_precision_vs_ma_tgb7}{
Precisions of $\sgagaAHbb$ measurement are shown  for MSSM parameter sets $I$-$IV$, for \MAOeq 200--350~GeV and   \tanbeq 7.
The lines are drawn to guide the eye. 
}

\begin{figure}[p]
{
 \begin{minipage}[b]{0.5\textwidth}
   \includegraphics[width=\textwidth]{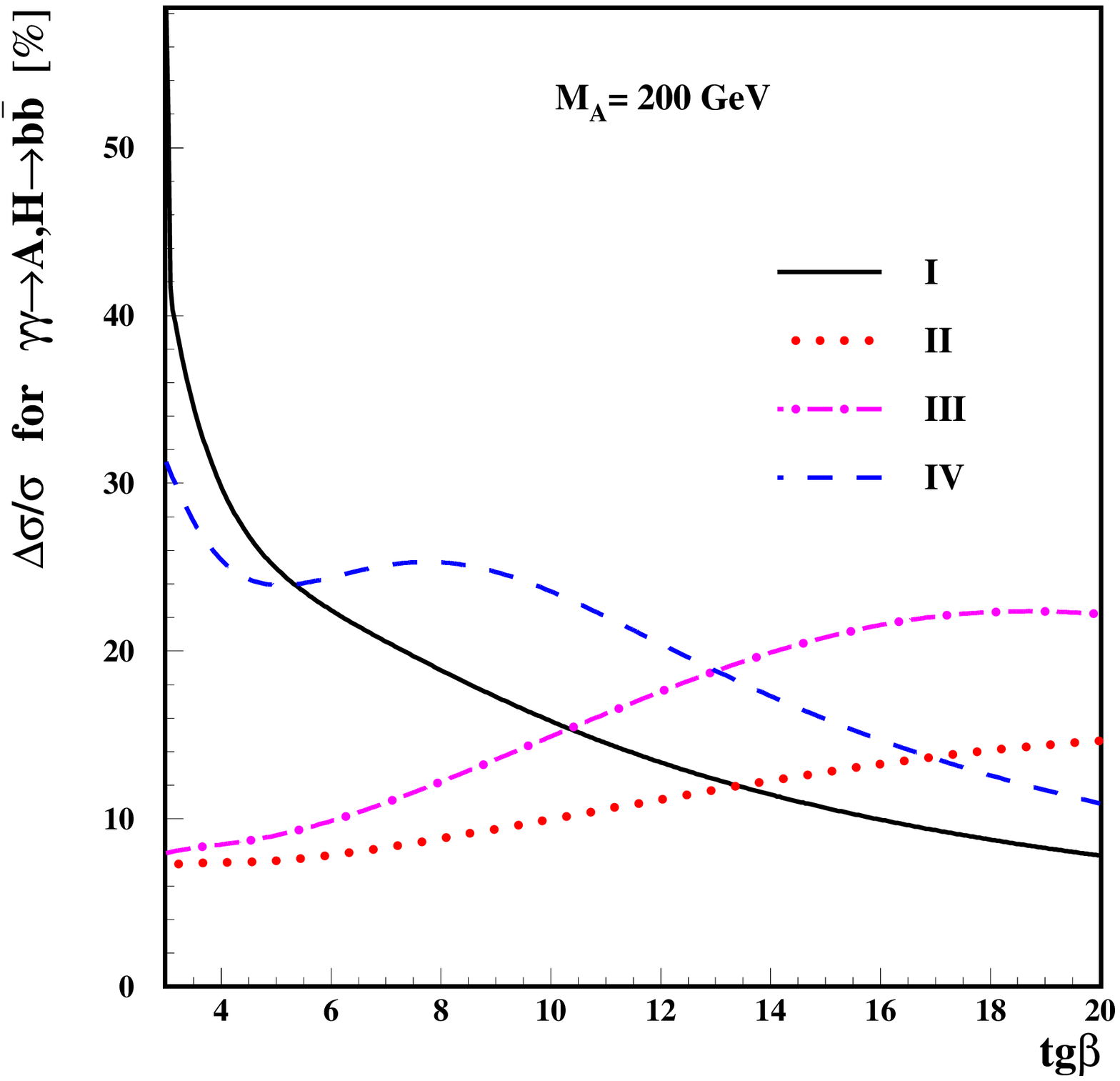}  
 \end{minipage} 
 \begin{minipage}[b]{0.5\textwidth}
   \includegraphics[width=\textwidth]{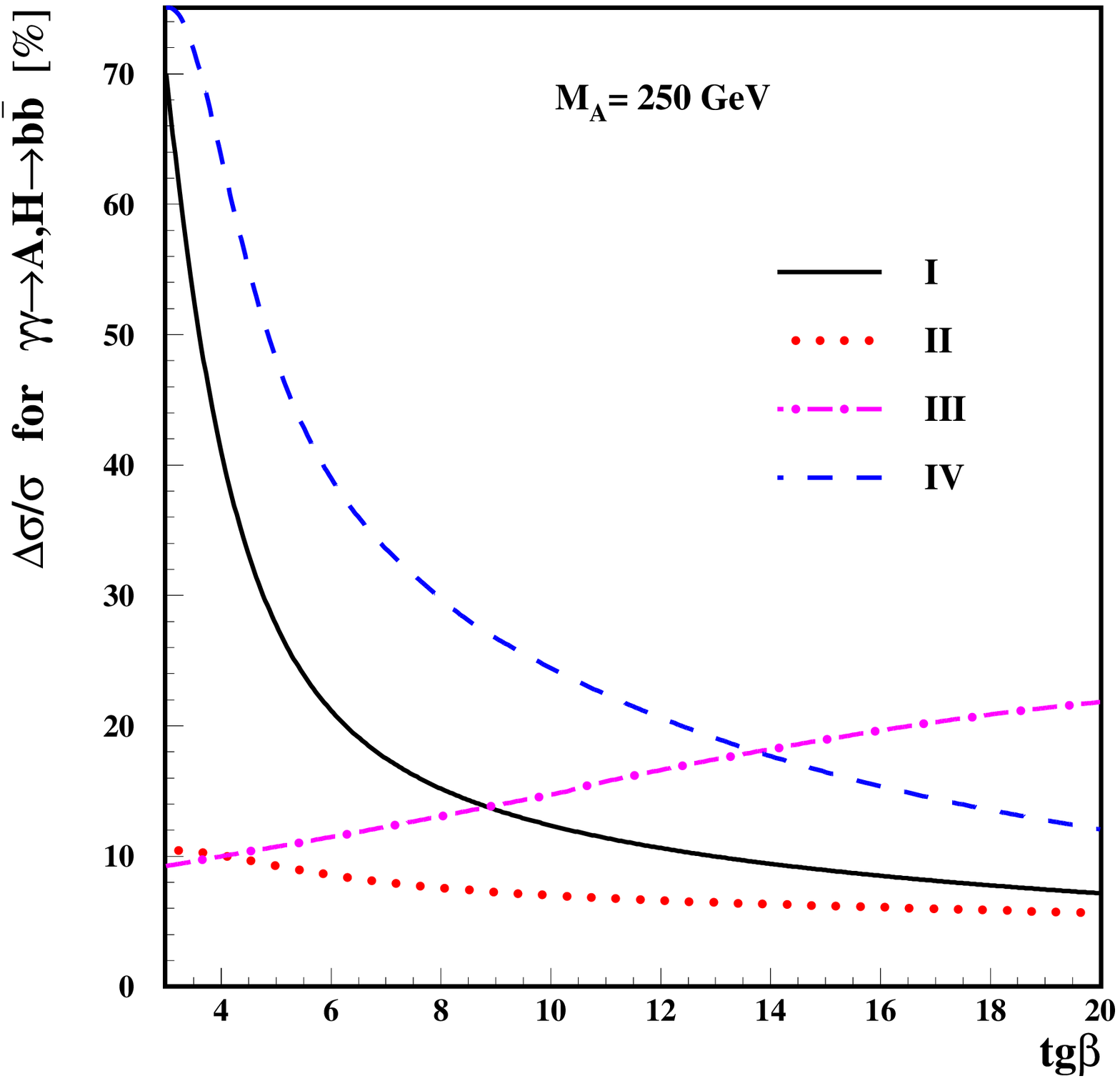} 
 \end{minipage} \\
 \begin{minipage}[b]{0.5\textwidth}
   \includegraphics[width=\textwidth]{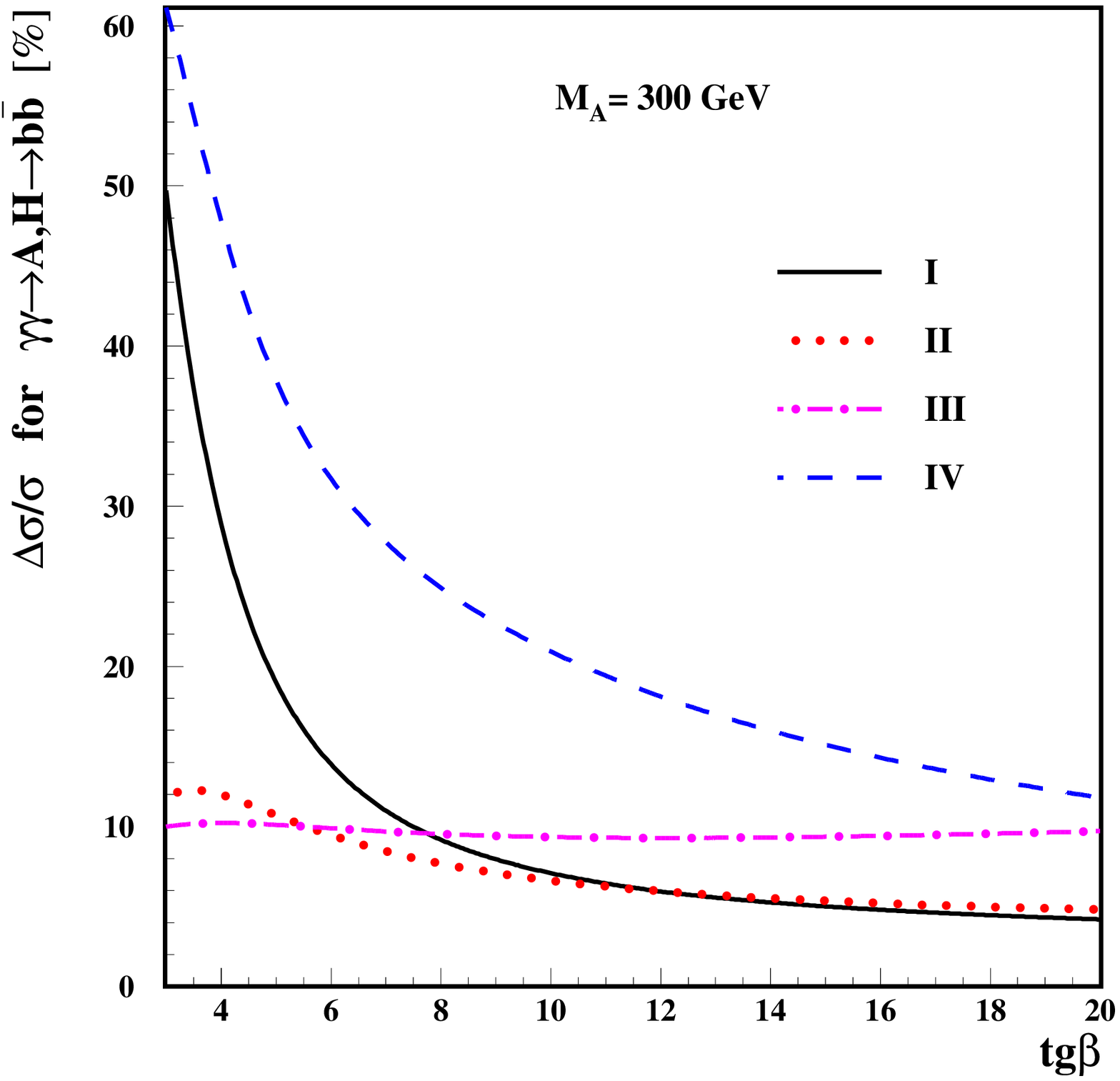} 
 \end{minipage} 
 \begin{minipage}[b]{0.5\textwidth}
   \includegraphics[width=\textwidth]{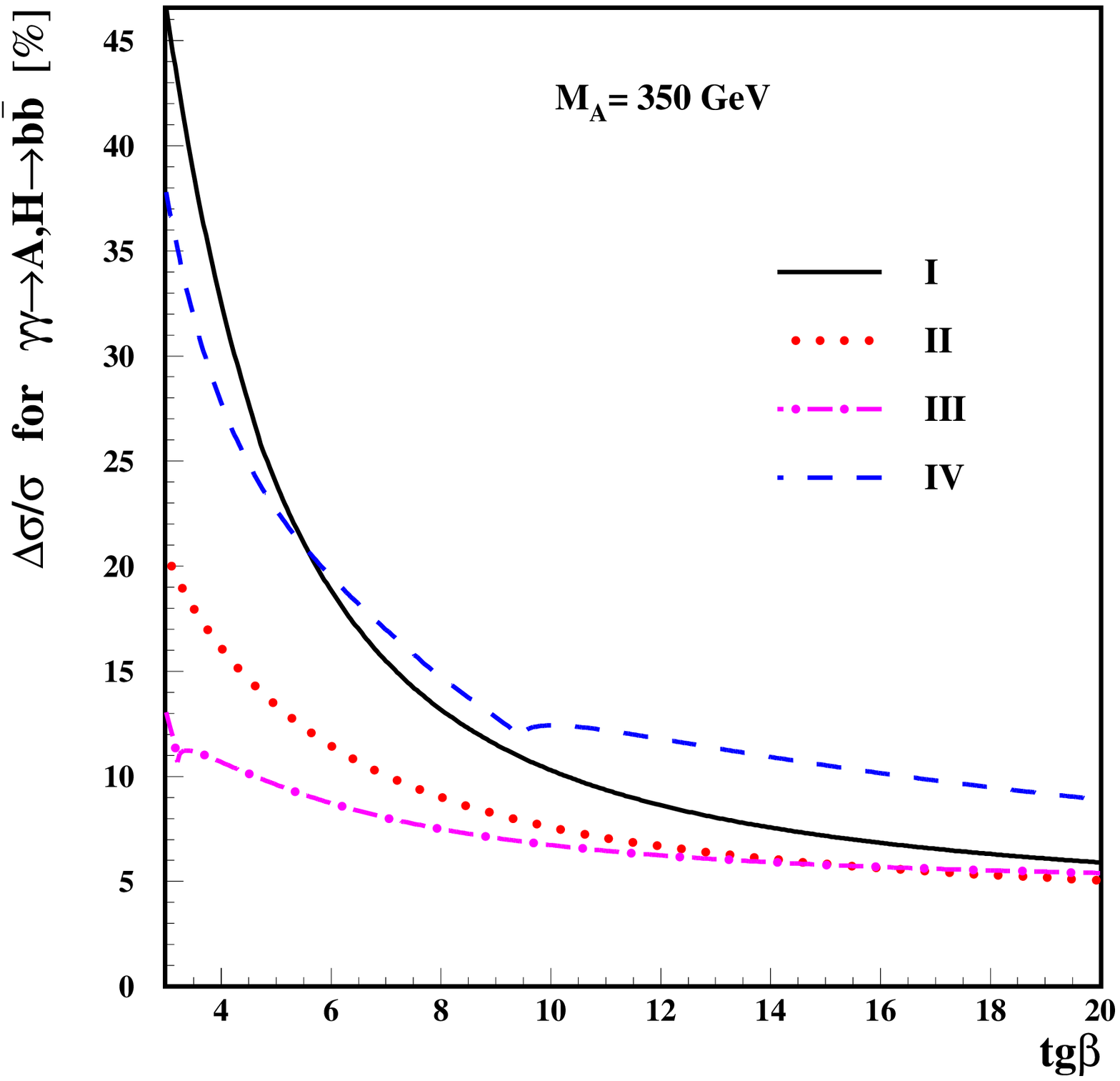}
 \end{minipage} 
}
\caption{\label{fig:mssm_precision_vs_tgb_ma200_350}
Precisions of $\sgagaAHbb$ measurement are shown  for \MAOeq 200, 250, 300 and 350~GeV, for MSSM parameter sets $I$-$IV$ with \tanbeq 3--20.
}
\end{figure}

For all considered values of $\MAO$ the dependence of the measurement precision 
on $\tanb$ was studied
and the results are  shown in Fig.\ \ref{fig:mssm_precision_vs_tgb_ma200_350}.
The precision  weakly  depends on  $\tanb$ 
if parameter sets $II$ or $III$ are  considered.
In case of parameter sets  $I$ or $IV$ the precise measurement will not be possible
for low  $\tanb$ values,   $\tanb \lesssim 5 $.

As already mentioned in Chapter \ref{ch_motivation} considered range of MSSM parameters
covers the region of so-called LHC wedge
where the expected number of reconstructed $\AO$ and $\HO$ production events 
will be too small to claim 5$\sigma$ discovery.
Also the heavy MSSM Higgs-boson production signal at the ILC 
may not be sufficient. 
In such a case
the Photon Collider was considered as the only machine which will be able to 
confirm the existence of the heavy neutral \higgses.
For \MAOeq 300~GeV, assuming parameter set $I$ and \tanbeq 7, 
the expected statistical significance $\delta$ of the $\AO+\HO$ signal measurement 
at the Photon Collider is 
$\delta=\mu_S/\sqrt{\mu_B} = 7.2$, in the optimal mass window.
The significance of the Higgs-bosons measurement, $\delta$, 
for different parameter sets considered in this analysis
is shown in Fig.\ \ref{fig:plot_mssm_significance_vs_ma_tgb7}.
The bands widths  indicate the level of possible  statistical fluctuations of the actual measurement:
\[
  \delta = \frac{\mu_S}{\sqrt{\mu_B}} \pm \sqrt{1 + \frac{\mu_S}{\mu_B}}
\]
In three from the four considered MSSM scenarios, assuming $\tbseven$, 
5$\sigma$ discovery of heavy, neutral \higgses{} will be possible  
in the whole range of $\MAO$,
if only the optimum beam energy is chosen
(some hint for $\MAO$ value is expected from LHC and ILC measurements).

\pnfig{p}{\figheight}{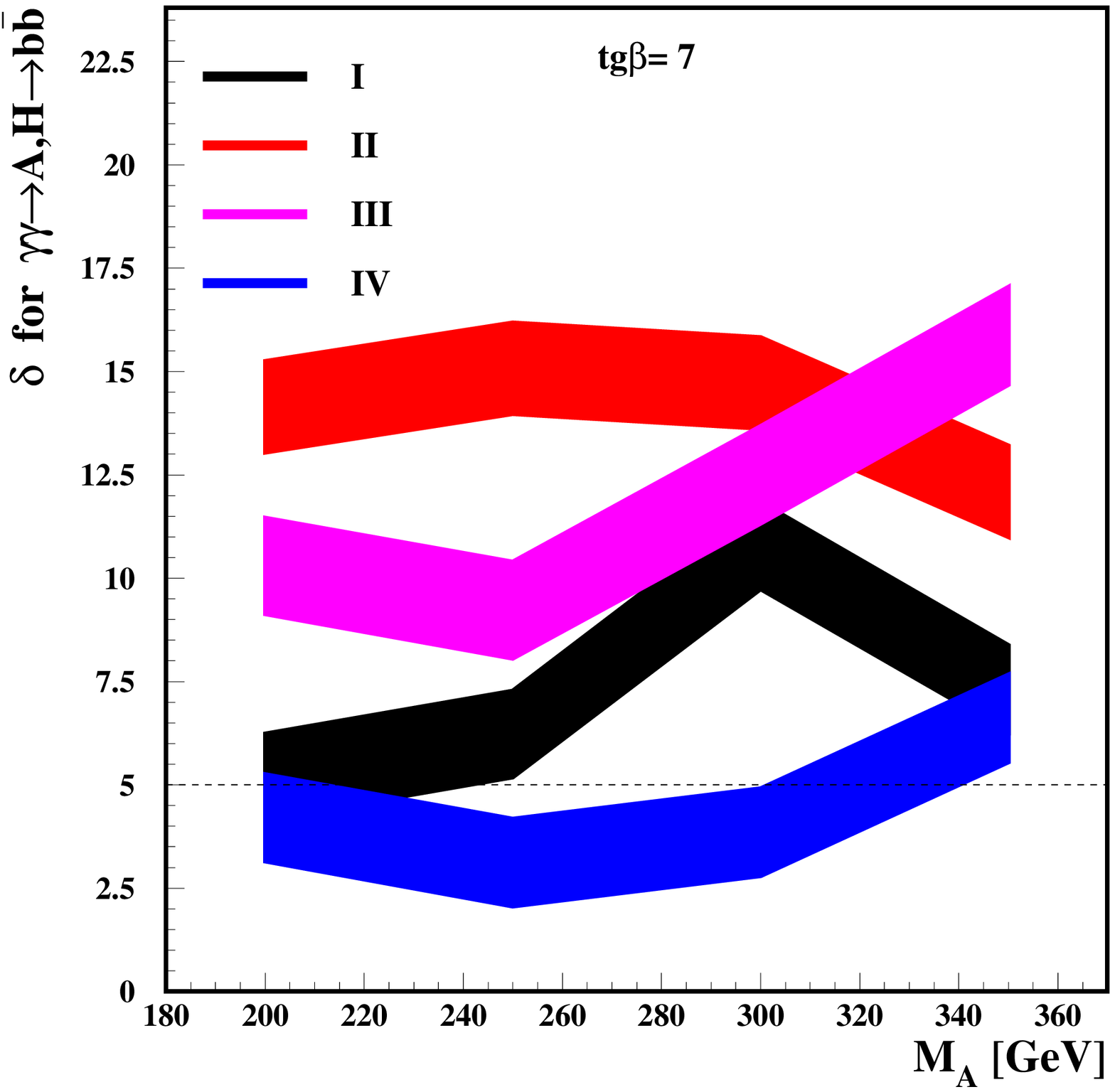}{fig:plot_mssm_significance_vs_ma_tgb7}{
Significances of $\sgagaAHbb$ measurement for MSSM parameter sets   $I$-$IV$, for \MAOeq 200--350~GeV and \tanbeq 7.
The band widths indicate the level of possible  statistical fluctuations of the actual measurement.
}

\begin{figure}[p]
{
 \begin{minipage}[b]{0.5\textwidth}
   \centering
   \includegraphics[width=\textwidth]{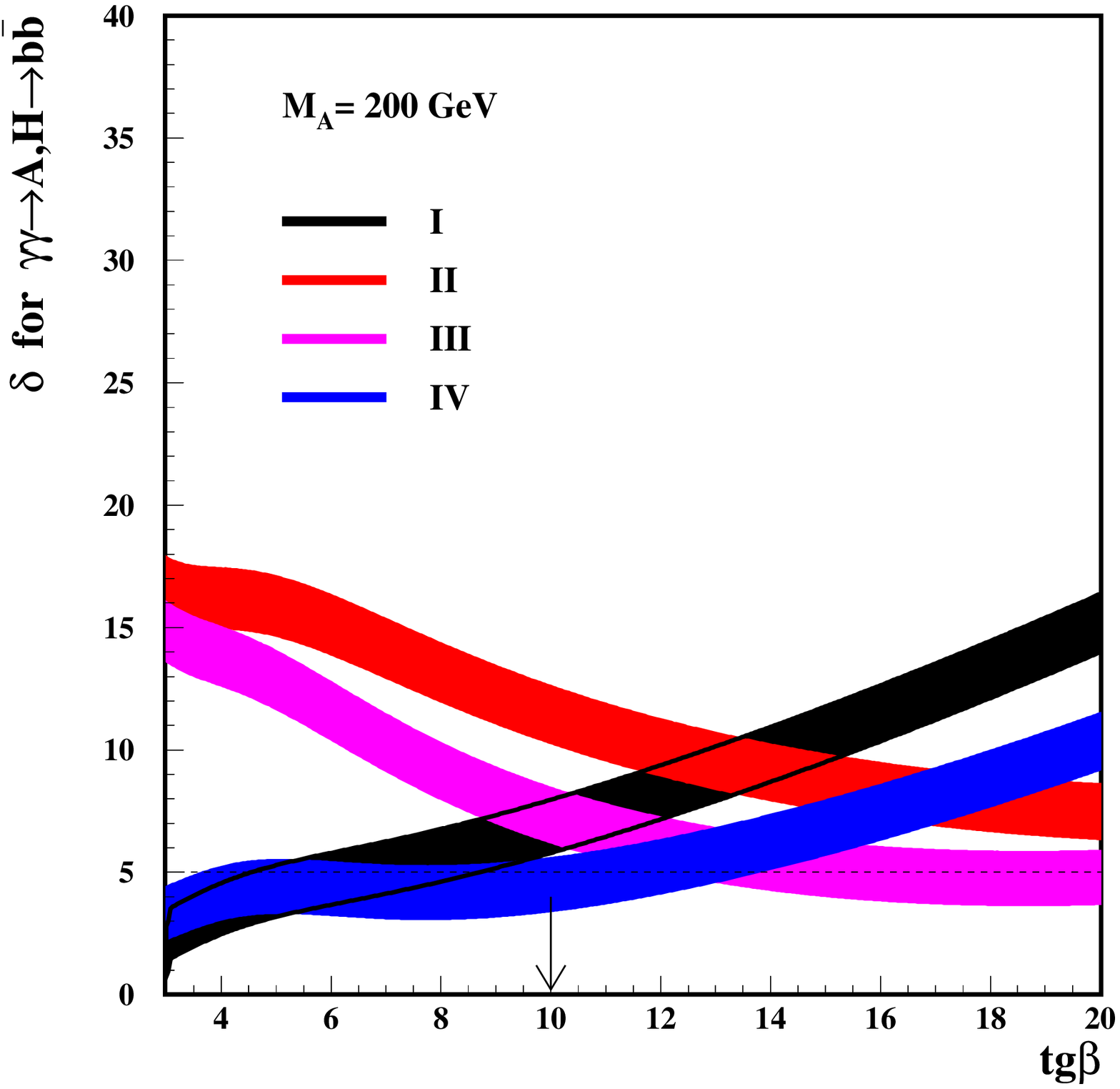}  
 \end{minipage} 
 \begin{minipage}[b]{0.5\textwidth}
   \centering
   \includegraphics[width=\textwidth]{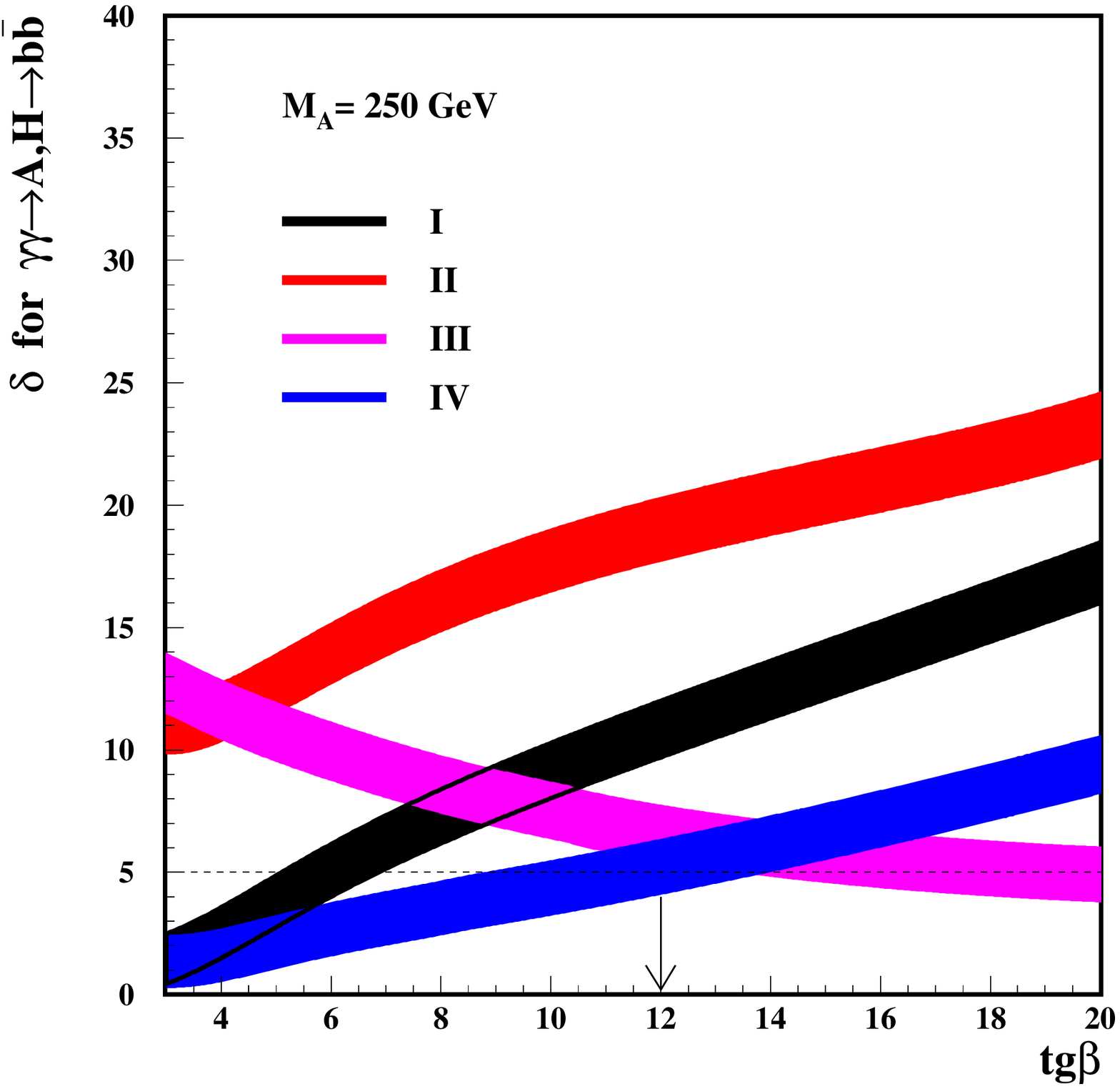} 
 \end{minipage} \\
 \begin{minipage}[b]{0.5\textwidth}
   \centering
   \includegraphics[width=\textwidth]{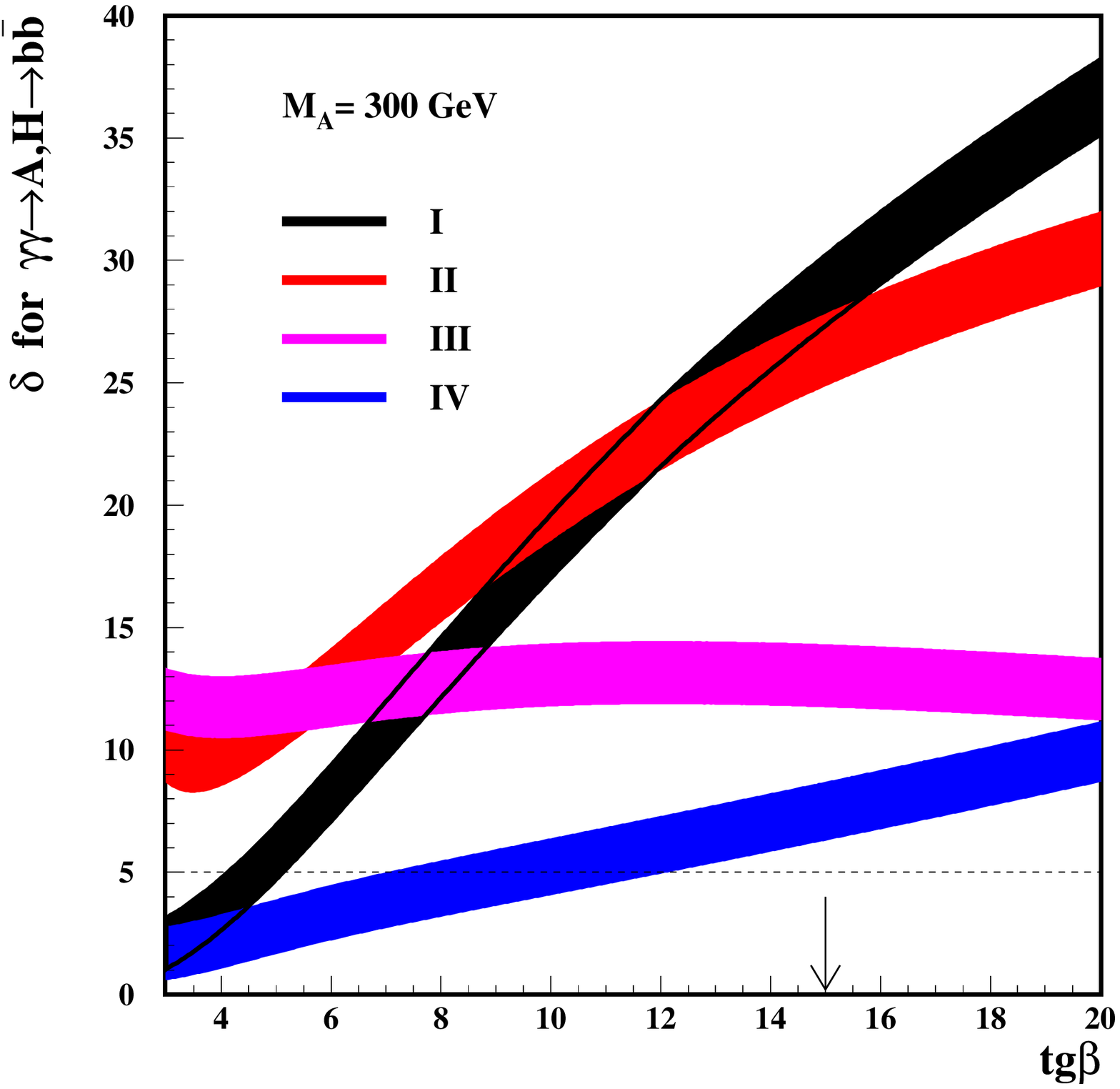} 
 \end{minipage} 
 \begin{minipage}[b]{0.5\textwidth}
   \centering
   \includegraphics[width=\textwidth]{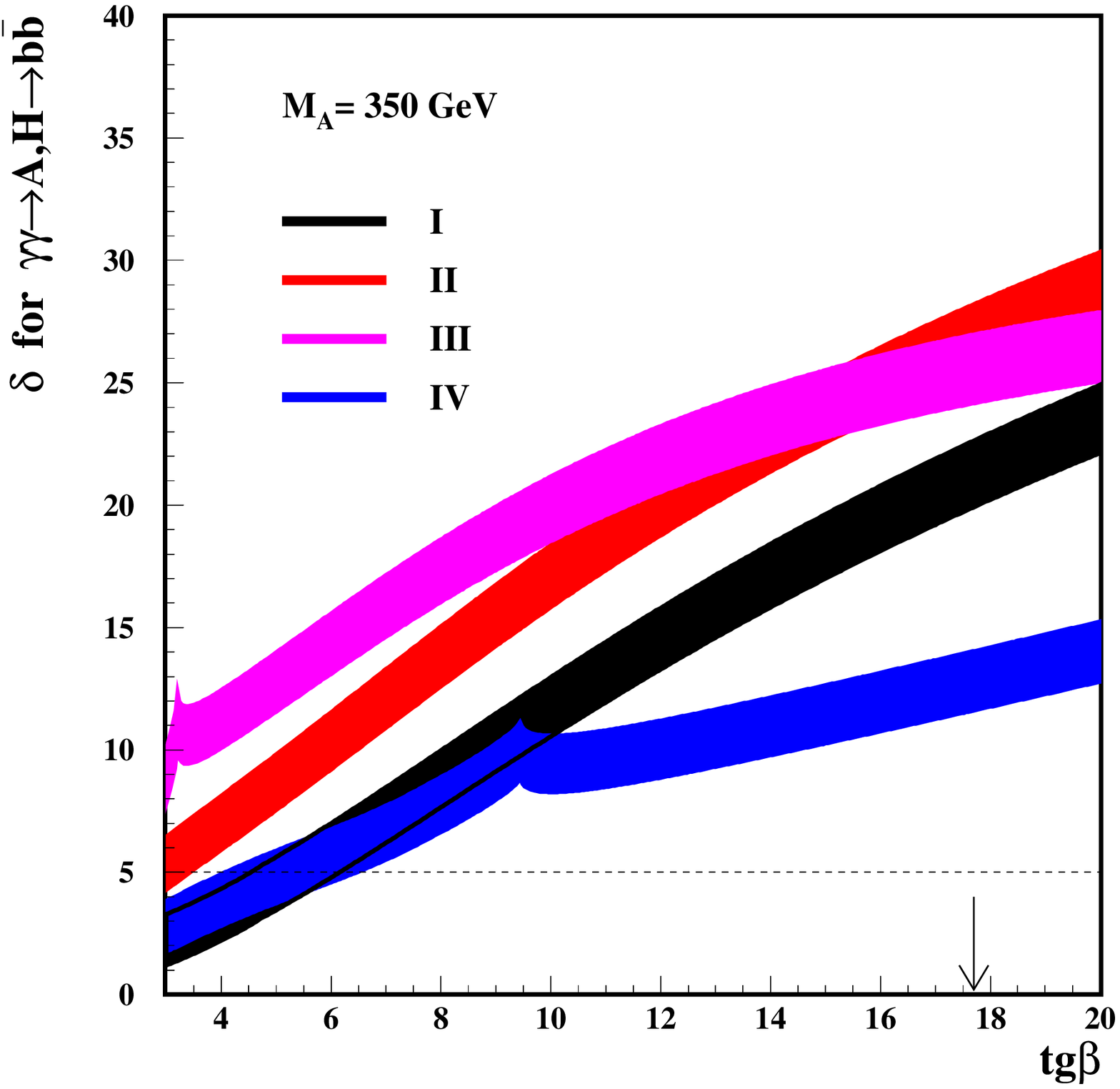}
 \end{minipage} 
}
\caption{\label{fig:mssm_significance_vs_tgb_ma200_350}
Significances of $\sgagaAHbb$ measurement for MSSM parameter sets $I$-$IV$,
   for \MAOeq 200, 250, 300 and 350~GeV,  and  \tanbeq 3--20.
The band widths indicate the level of possible  statistical fluctuations of the actual measurement.
}
\end{figure}

For all considered values of $\MAO$ we also studied  the significance
of  signal measurement  
as a function of  $\tanb$, for \tanbeq 3--20.
Results obtained for different parameter sets are compared 
in Fig.\ \ref{fig:mssm_significance_vs_tgb_ma200_350}.
The estimated lower limit of the discovery region of LHC experiments
(as presented by CMS collaboration \cite{CMSDiscovery}) is indicated by arrows.
For all parameter sets the expected statistics of signal events for \MAOeq 200--350~GeV will be sufficient 
to cover most of  the considered MSSM parameters space.
We can conclude that for $\MAO \gtrsim 300 $~GeV the Photon Collider 
should be able to discover Higgs bosons
for much lower values of $\tanb$ than experiments at the LHC.

\chapter{Summary \label{ch_summary}}

Search for the Higgs boson and precise measurement of its properties are
among most important tasks of all future colliders.
One of the measurements crucial for understanding of the Higgs sector
and for the verification of the particle physics models
is the measurement of $\Ghgaga$.
The Photon Collider, which has been proposed as an extension of the $\epem$ linear
collider project, is considered the best place to do this measurement.
In this thesis the first fully realistic estimates for the precision of $\gagahiggsbb$ 
cross-section measurement at the Photon Collider are obtained.
Assumed photon collider parameters  correspond to 
the  superconducting Linear Collider project TESLA.
The analysis is based on the  realistic $\gaga$-luminosity 
spectrum simulation. 
Due to the high beam intensity,  resulting in high $\gaga$-luminosity 
per bunch crossing, the contribution of overlaying events \gagahad{}
turns out to be sizable and affects event reconstruction.
Crossing angle between beams
resulting in the significant broadening
of the interaction region is also taken into account.
These two factors significantly affect the performance of the  \btagging{} algorithm.
It is shown that 
 the contamination of \gagahad{} overlaying events in the signal
can be reduced by
rejecting low-angle tracks and clusters in the event.
Additional cuts are proposed to suppress  contributions from other background sources
which are taken into account.
For the realistic estimate of the heavy quark production  background
which is considered to be the most important one, 
the NLO QCD results are used.
Other processes are taken into account using the LO cross sections implemented in \Pythia{}: $\tau$-pair production,
light-quark production $\gagaqq$, \gagahad{} events 
(without \emph{direct$\times$direct} processes and without other events in the bunch crossing),
and $\gagaWW$ (with modified cross section to take into account polarization). 
After optimizing selection cuts and applying
correction for escaping neutrinos  from $D$- and $B$-meson decays 
the quantity $\Ghgagahbb$ for the SM Higgs boson with mass around 120~GeV 
can be measured with the precision of about 2\% already after one year of the Photon Collider running.
The systematic uncertainties of the measurement are estimated to be of the order of 2\%.
The statistical precision of the measurement decreases
up to 7.7\% for the SM Higgs boson with  mass  \Mheq 160~GeV.
For  higher
masses of the SM Higgs boson 
other decay channels are expected to give better precision of $\Ghgaga$ measurement,
see \eg \cite{wwzz}.
The measurement discussed in this paper can be used to derive the partial width
$\Ghgaga$, taking 
$\Brhbb$ value from precise 
measurement at the $\epem$ International Linear Collider. 
With 2\% accuracy on 
$\Ghgagahbb$, 
as obtained in this analysis, and assuming $ \Brhbb$
will be measured to 1.5\% \cite{BRhbb}, Higgs-boson partial width 
$\Ghgaga$ can be extracted with accuracy of about 2.5\%.
With this precision the measurement will be sensitive to the deviations
from the SM coming from loop contributions
of new heavy charged particles.
For example, heavy charged \higgs{} contribution in the SM-like 2HDM
is expected to change $\Ghgaga$ by 5--10\% \cite{2HDM}.
Using in addition the result from the $\epem$ Linear Collider for 
$\Brhgaga$
\cite{BRhgaga}, one can also extract 
$\Gh$ with precision of 10\%.


%

The presented analysis has been extended to the MSSM case by considering signal 
of heavy MSSM Higgs bosons $\AO$ and $\HO$ production with decays into $\bbbar$. 
We study masses \MAOeq 200, 250, 300 and 350~GeV,
and for each mass we choose an optimal $\emem$ beam energy
and selection cuts, as in SM case. 
Following \cite{MMuhlleitner}, we study 
MSSM parameter range which corresponds to the so-called  ``LHC wedge'',
where $\AO$ and $\HO$ are almost   degenerate in mass.
%
%

Our analysis shows that, for  $\MAO \sim$ 300~GeV,
the cross section for the MSSM Higgs-bosons production $\sgagaAHbb$ can be measured
with a statistical precision of about 11\% already after one year of Photon Collider running.
For other considered values of $\MAO$   it turns out to be  lower -- from 16\% to 21\%.
Although this result is less optimistic than the earlier estimate \cite{MMuhlleitner},
  still the photon--photon collider gives opportunity
of a precision measurement of $\sgagaAHbb$, 
assuming that we know the mass of the Higgs boson(s). 
Unfortunately, separation of $\AO$ and $\HO$ will not 
be possible 
as the mass resolution is about 15~GeV.
A discovery of  MSSM Higgs-bosons requires energy scanning or a run with 
a broad luminosity spectrum, perhaps  followed by the run with a peaked one. 
We estimate the  significance  expected for the Higgs production measurement 
for four different  parameter sets and for \tanbeq 3--20.
The discovery reach of the Photon Collider is compared with the estimated reach of the CMS experiment.
 For the optimum energy choice the photon--photon collisions allow for the discovery of the Higgs bosons 
even for $\tanb$ values  lower than the expected reach of the LHC experiments.
Thus, at least partially, the Photon Collider will cover the so-called ``LHC wedge''.
For low $\tanb$ values the measurement 
could probably profit from use of additional channels: 
decays of $\AO$ and $\HO$ to charginos and neutralinos,
and in case of $\HO$ also decays to $hh$.

Presented analysis takes into account all corrections which were available to date.
However, some aspects of the analysis can probably be improved in the future.
The first one is the cut on the number of selected jets.
In the current analysis 2 or 3 jets are required.
These are the 'natural' jet multiplicities for the signal.
However, when 
full NLO generators with 'matrix element -- parton shower matching'
will become available for the heavy quark background, 
then this study should be repeated and  the cuts depending on the number of jets
should be considered.
Secondly, the jet-clustering algorithm with smaller $y_{cut}$ values could be 
used.
Narrower jets (possibly including tracks and clusters rejected by $\thetamindet$-cut)
could be used in the 4-jet kinematical fit 
to discriminate $\gagaWW$ background events  more effectively. 
It is also not excluded that further minimization of  the \gagahad{} overlaying-events contribution
is possible. 
We have not found any effect due to overlaying events on the spacial distributions  
of jet primary vertices,
probably because tracks with largest impact parameters are rejected by the vertex fit.
However, the cut on the impact parameter of single tracks  and rejection of accompanying clusters 
could still be considered.


\
\newpage 
\
\newpage 

\newpage

\section*{Acknowledgments}






First of all, I would like to thank my supervisor Aleksander Filip \Zarnecki{}
for the enormous amount of time and work he devoted to me.
His ingenuity has been  a constant inspiration and challenge for me.   
During our discussions and collaboration I have learnt a lot about physics
and started to believe that I can obtain important results.
  
%
I am very grateful to Krzysztof Doroba for his friendliness and support.
He introduced me to particle physics, especially to photon-photon collisions.
He also encouraged me to continue my studies when I had difficult time and considered giving up.
%

The subject of this thesis was suggested to me by Maria Krawczyk
who invited me to work on the Photon  Collider.
I thank her for this brave proposition, and for our  fruitful collaboration.

Thinking about my journey as a physicist,
I would like to thank my  teacher Bo\.zena Bugaj, 
who initiated me into the realm of experiments and equations,
for her warmth and for  my first joy from solved exercises
when she was organizing 'tajne komplety' without being paid for it.
I thank my mother and my sister  for their love, support and encouragement during my research work.
My special thanks go to my wife for her great love and endless  patience.
I am very grateful to her for taking over all my home duties during last months of my work.

%
%
This work was partially supported 
by the Polish Committee for Scientific Research, 
grant 
no.~2~P03B~128~25
and
no.~1~P03B~040~26.
%

\appendix
\
\newpage 
\
\newpage 

\chapter{Event generation with overlaying events \label{app_eventgeneration}}

\vspace*{-2cm}
In this Appendix we discuss the statistical principles of event generation,
with emphasis on the possibility of having multiple events in single bunch crossing.
We assume that for a given invariant mass of colliding photons, $\Wgaga$,
we know how to calculate the total cross section for the process $a$, $\xs_a(\Wgaga)$. %
%
%
In the general case, for the $\gaga$-luminosity spectrum (integrated over time), given by ${d L (\Wgaga)}/{d \Wgaga}$, 
the expected total number of events, $N_a$, is:
\[ N_a = \int \!\! d\Wgaga  \: \frac{d L(\Wgaga)}{\,d\Wgaga}  \: \xs_a(\Wgaga). \]   
If we take the luminosity spectrum corresponding to one bunch crossing only
then we obtain an average number of events of the process $a$
per one crossing, $\mu_a$.
In the realistic event generation one must 
take into account the possibility that two or more $\gaga$ collisions in channel $a$ take place in one crossing.
Therefore in each generated event, corresponding to one bunch crossing,
$k_a$ physical events of process $a$ should be included,
where $k_a$ is the random number from Poisson distribution with mean $\mu_a$:
\[ P(k_a,\mu_a)=\frac{\mu_a^{k_a}}{k_a!}e^{-\mu_a}.\]
In this analysis the exact algorithm based on the Poisson distribution 
is used to generate \gagahad{} events 
for which $\mu \sim 1$.
For small values of $\mu_a$, $\mu_a \ll 1$, 
an approximation is valid in which one generates $N_a$ bunch crossings
with one event of the process $a$ in each crossing,
\ie one assumes that the fraction of bunch crossings with exactly one $a$ event is $\mu_a$.
Actually, this fraction is smaller 
and is equal to $P(k_a=1,\mu_a) = \mu_a e^{- \mu_a }$.
Moreover, in this simple approach bunch crossings with more than one 
$a$ event are neglected; 
they contribute to the fraction $P(k_a>1,\mu_a)$ of  crossings.
%
Inaccuracy of this procedure 
is of the order of $\mu_a^2$.
Therefore it can be safely used for processes with $\mu_a \ll 1$.
 In this work this approach is used for following processes: 
$\gagahiggsbb$, $\gagaQQg$, $\gagaqq$, $\gagaWW$, $\gagatautau$
(for the process with the highest cross section, $\gagaQQg$, $\mu \sim 10^{-4}$).
In general one would like to consider many independent physical processes.
Here, to simplify the description, we assume that only three different processes
have to be taken into account: $a$, $b$ and $c$, with $\mu_a,\mu_b \le 0.01$ and $\mu_c \sim 1$.
In the exact (but very ineffective) generation algorithm 
one would have to include $k_a$, $k_b$ and $k_c$ events of corresponding processes
for each bunch crossing  
where $k_i$ are random numbers coming from Poisson distributions $P(k_i,\mu_i)$.
However, for small $\mu_a$ and $\mu_b$ one can neglect bunch crossings in which events 
of both process $a$ and process $b$ are present; 
the fraction of such crossings is $P(k_a>0,\mu_a)P(k_b>0,\mu_b) \approx \mu_a \mu_b \le 10^{-4}$.
Therefore only 
bunch crossings with $a$ and $c$, or $b$ and $c$ processes can be considered.
As already pointed out, only bunch crossings 
with $k_a=1$ and $k_b=1$  can be taken into account in generation.
In this case $c$ events, with $k_c$ given by the Poisson distribution, are called \emph{overlaying events}.
However, one has to include also events with $c$ process only. 
The fraction of bunch crossings with $k_c>0$ and $k_a=k_b=0$
is equal to $P(k_c>0,\mu_c)P(k_a=0,\mu_a)P(k_b=0,\mu_b)$.
As values of $\mu_a$ and $\mu_b$ are small $P(k_a=0,\mu_a)P(k_b=0,\mu_b) \approx 1$, 
and it is enough to generate a sample of $P(k_c>0,\mu_c) N_c / \mu_c$ 
bunch crossings with random number $k_c$ ($k_c>0$) of $c$ events in each crossing.
In this analysis processes
$\gagahiggsbb$, $\gagaWW$ and $\gagatautau$ 
(corresponding to processes $a$ or $b$) with overlaying events \gagahad{} (process $c$)
are generated according to this approach.%
\footnote{
When the \gagahad{} process is considered as a separate background contribution
(only  overlaying events)
the direct processes $\gagaqq$ with $q=u,d,s,c,b$ are excluded from
generation as they are taken into account individually ($\gagaQQg$ for $Q=b,c$ and $\gagaqq$ for $q=u,d,s$).
}
Slightly different procedure can be applied 
if process $a$ is a subclass of overlaying event process.
Let us first consider the generation of processes $a$ with overlaying events $c$
in more detail.
%
%
%
The fraction of bunch crossings in which there are two $a$ events
is $P(k_a=2,\mu_a)\approx \mu_{a}^2/2$.
%
Using simple approach for generation of $a$ events,
\ie generating $N_a$ bunch crossings with $k_a=1$, 
and overlaying a random number of $c$ events, $k_c$, 
one overestimates number of crossings 
with one $a$ event by about $\mu_a^2$
and -- by neglecting a possibility of having $k_a >1$ -- underestimates number of bunch crossings 
with two $a$ events by about $\mu_a^2/2$.
However, there is also an alternative version of the generation algorithm.
Instead of generating $k_c$ overlaying events from class $c$,
one can generate $k_{a+c}$ overlaying events from joined class $a+c$
where $k_{a+c}$ is a random number from distribution $P(k_{a+c},\mu_a+\mu_c)$. 
Then, 
one obtains almost correct fraction of bunch crossings with $k_a=1$ (about $\mu_a-\mu_a^2$)  
and 
overestimates a fraction of crossings with $k_a=2$ (by about $\mu_a^2/2$).
%
%
This approach is applied in this work for processes
$\gagaQQg$ and $\gagaqq$ (corresponding to the process $a$)
with overlaying events \gagahad{} ($a+c$)
because the first two processes are included in generation
of \gagahad{} events.%
\footnote{
The processes $\gagaQQg$  are included in \gagahad{} events generated with \Pythia{} in the LO approximation
for the process $\gagaqq$ with $q=c,b$, and assuming unpolarized photon-photon collisions.
}

\
\newpage 

\chapter{Influence of higher order corrections  \label{app_uncert_gagaQQg}} 

In this Appendix we try to estimate the uncertainty of the presented results due to
 higher order corrections
to the NLO QCD background process $\gagaQQg$.
As explained in Section \ref{sec_heavy_quark_bkgd}, the parton shower algorithm
was not applied in the generation of heavy-quark background events 
as it would lead to double-counting of some gluon emissions.
However, we can use it to estimate influence of higher order
QCD corrections which are not included in the NLO generating program.
The parton shower algorithm was applied to the $\QQbarg$ final state generated with the NLO code,
and the full analysis including cut optimization was repeated 
for the SM Higgs boson with mass of 120~GeV (\sqrtseeeq 210~GeV) 
and for the MSSM Higgs bosons $\HO$ and $\AO$ for \MAOeq 300~GeV (\sqrtseeeq 419~GeV).
However, here background contributions due to $\gagatautau$ and resolved \gagahad{} events
were not included.
The cut suppressing  overlaying event contribution was not varied: \costhmindeteq 0.85.
As some gluon emissions are neglected in the generation without parton shower (our standard analysis),
and  with parton shower some emissions are double counted,
we can expect that the actual event distributions  will be somewhere in between.
Thus, the difference between results obtained without and with parton shower can indicate 
the size of  effects
which are expected  after higher QCD corrections are taken into account.
In Table \ref{tab:ps0ps1} results obtained with and without parton shower are compared.
The precisions of cross section measurement for both cases are nearly the same.
After applying the parton shower algorithm the   $\gagabbg$ events 
have on average slightly lower $b$-tag values 
as  $B$-mesons are slower due to the additional gluon emissions.
It can be clearly seen in  the  analysis of SM Higgs-boson production 
when in both cases we obtain similar
optimal region of acceptance in the $b$-tag(jet$_1$)$\otimes$$b$-tag(jet$_2$) plane
(nearly the same efficiencies for \higgs{} signal)
but $\varepsilon_{bb}$ is smaller if the parton shower algorithm is applied.
Moreover, the increase of \ccmistagging{} probability is observed.
It is due to the fact that additional soft gluons 
broaden a $c$-jet and some tracks appear which are not present in events without parton shower.
These modifications can cause the $c$-jet being sometimes similar to the $b$-jet.
In case of the MSSM analysis the situation is more complicated 
as the wider optimal acceptance region in the $b$-tag(jet$_1$)$\otimes$$b$-tag(jet$_2$) plane 
is chosen if parton shower is included (greater  $\varepsilon_{h}$).
Then it is advantageous to take also signal events with jets having lower $b$-tag value
because the $\gagabbg$ background events have  slightly lower \bbtagging{} efficiency.


We conclude that our results are not sensitive to the higher order corrections
for the heavy-quark background process
under assumption than these corrections do not change significantly 
the overall normalization of the $\gagaQQg$ cross section.

\begin{table}[t]
\label{tab:ps0ps1}
\bc
\begin{tabular}{|c|c|c|c|c|c|c|}
\hline
$\sqrtsee$ [GeV] & Signal process & $\Dss$ [\%] & $\varepsilon_{h}$ [\%] &  $\varepsilon_{bb}$ [\%] &  $\varepsilon_{cc}$ [\%] &  $\varepsilon_{uds}$ [\%] \\
\hline
210              & $\gagahbb$     &  1.96 (2.03)  &  57 (58)               &  43 (50)                 &  2.8 (2.2)               &  0.16 (0.16)                \\
419              & $\gagaAHbb$    & 10.6 (11.0) &  59 (53)               &  50 (47)                 &  5.2 (2.9)               &  0.6 (0.5)                \\
\hline
\end{tabular}
\ec
\caption{ Results for precision measurement of signal $\gagahbb$ (\Mheq 120~GeV) and $\gagaAHbb$ (\MAOeq 300~GeV),
          and optimal tagging efficiencies  $\varepsilon_{h}$, $\varepsilon_{bb}$, $\varepsilon_{cc}$, and  $\varepsilon_{uds}$.
          Values obtained with parton shower algorithm applied to $\QQbarg$ parton system.
	  In parentheses values for generation without parton shower are shown (our standard analysis).
	  All cases were subject to the same procedure with cuts and \higgstagging{} optimization.
          }
\end{table}
\chapter{Studies on the influence of overlaying events  \label{app_thetatc}}

In this Appendix the contribution of the \gagahad{} overlaying events to the 
hard $\gaga$ scattering events in the detector is studied in detail.
To understand better properties of overlaying events
we first study particle distributions on the generator level.
The expected distributions of the particle, energy and transverse energy 
flows in  $\cos(\theta_{\mathit{particle}})$,
where $\theta_{\mathit{particle}}$ is the particle polar angle,
are shown in Fig.\ \ref{fig:gagahad_hadron_hadron}  for \emph{hadron-like$\times$hadron-like} events
at \sqrtseeeq 210~GeV.
The average number  of \gagahad{} events per bunch crossing, $\mu$, is about 0.9.
All distributions are strongly forward-peaked.
This behaviour, expected for soft hadron-hadron interactions,
is intensified in majority of events by the strong Lorentz boost of the $\gaga$
center of mass system.
As shown in Fig.\ \ref{fig:gagahad_hadron_direct} and \ref{fig:gagahad_direct_direct}, 
the same behaviour is observed for \emph{hadron-like$\times$direct} and \emph{direct$\times$direct} 
processes for which $\mu \approx 0.1$ and 0.01, respectively.
Therefore,  the cut on the particle polar angle seems to be the simplest way to 
reduce the contamination of particles coming from overlaying events.
We studied the average contribution of overlaying events
to the energy measured in the detector
 after a cut $\theta_{min}$ on the minimal polar angle of particle.
In Fig.\ \ref{fig:gagahad_e_dist} and \ref{fig:gagahad_et_dist} the probability distributions
for the energy and trasverse energy flows are shown for two $\theta_{min}$ values.
The significant decrease of the average energy and transverse energy is clearly observed:
the average energy changes from around 60~GeV without $\theta_{min}$ cut 
to 20 and 10~GeV for $\theta_{min} = $ 80 and 250 mrad, respectively.


\pnfiggeneral{p}{5.5cm}{\includegraphics{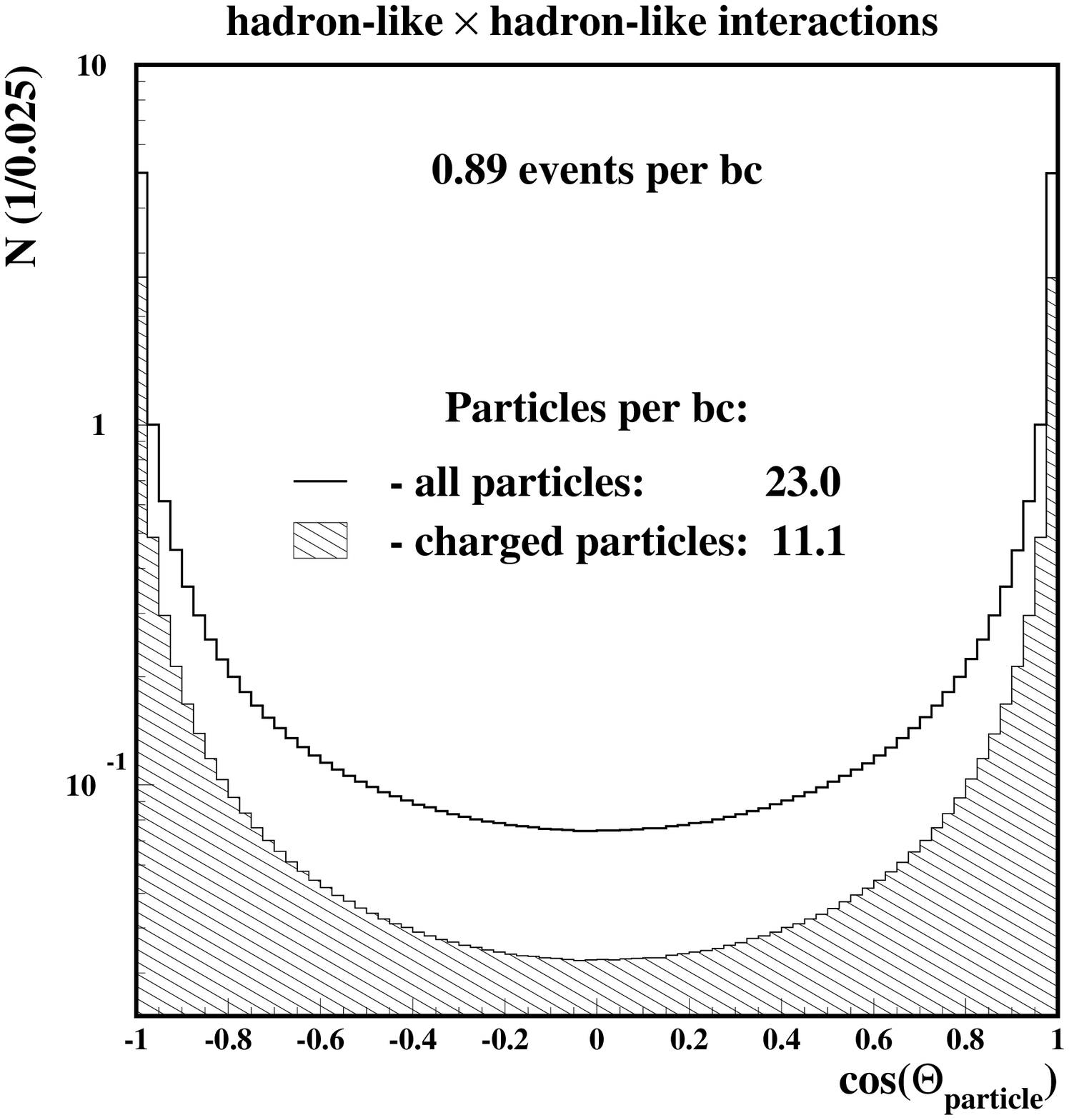} \includegraphics{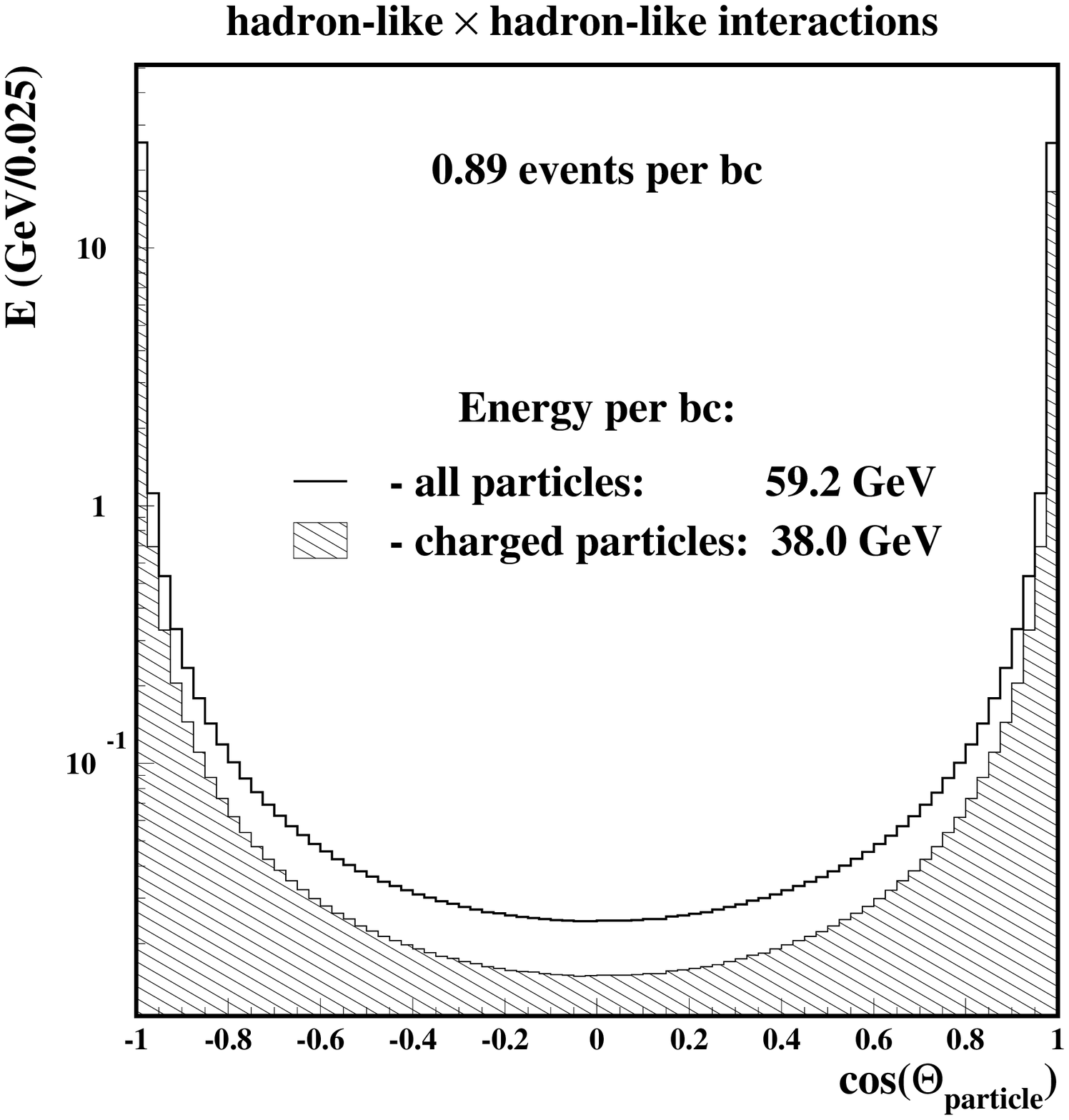}  \includegraphics{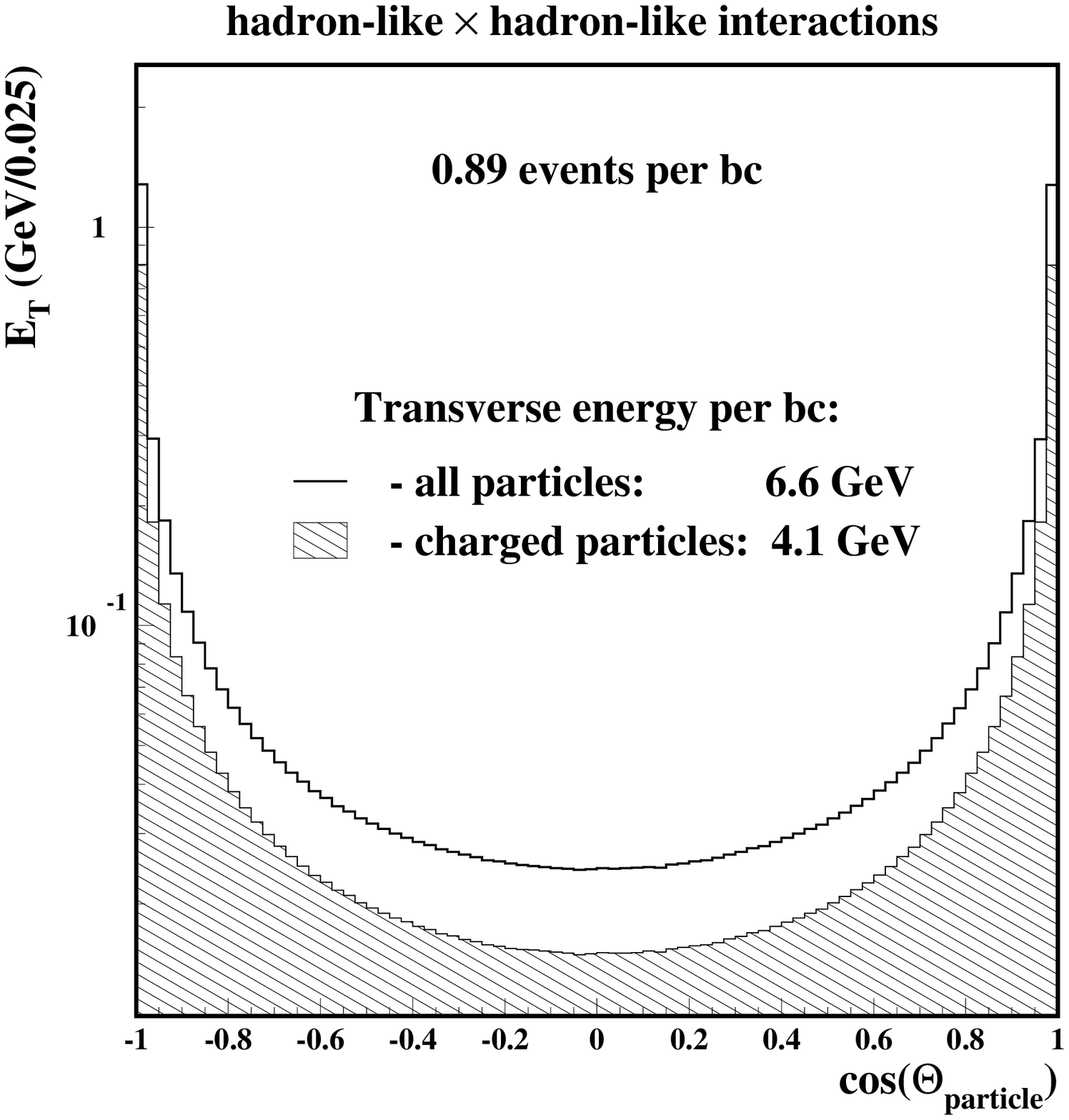} }{fig:gagahad_hadron_hadron}{
Distributions of particle, energy and transverse energy flow in $\cos(\theta_{\mathit{particle}})$ 
for \emph{hadron-like$\times$hadron-like} case for \sqrtseeeq 210~GeV.}

\pnfiggeneral{p}{5.5cm}{\includegraphics{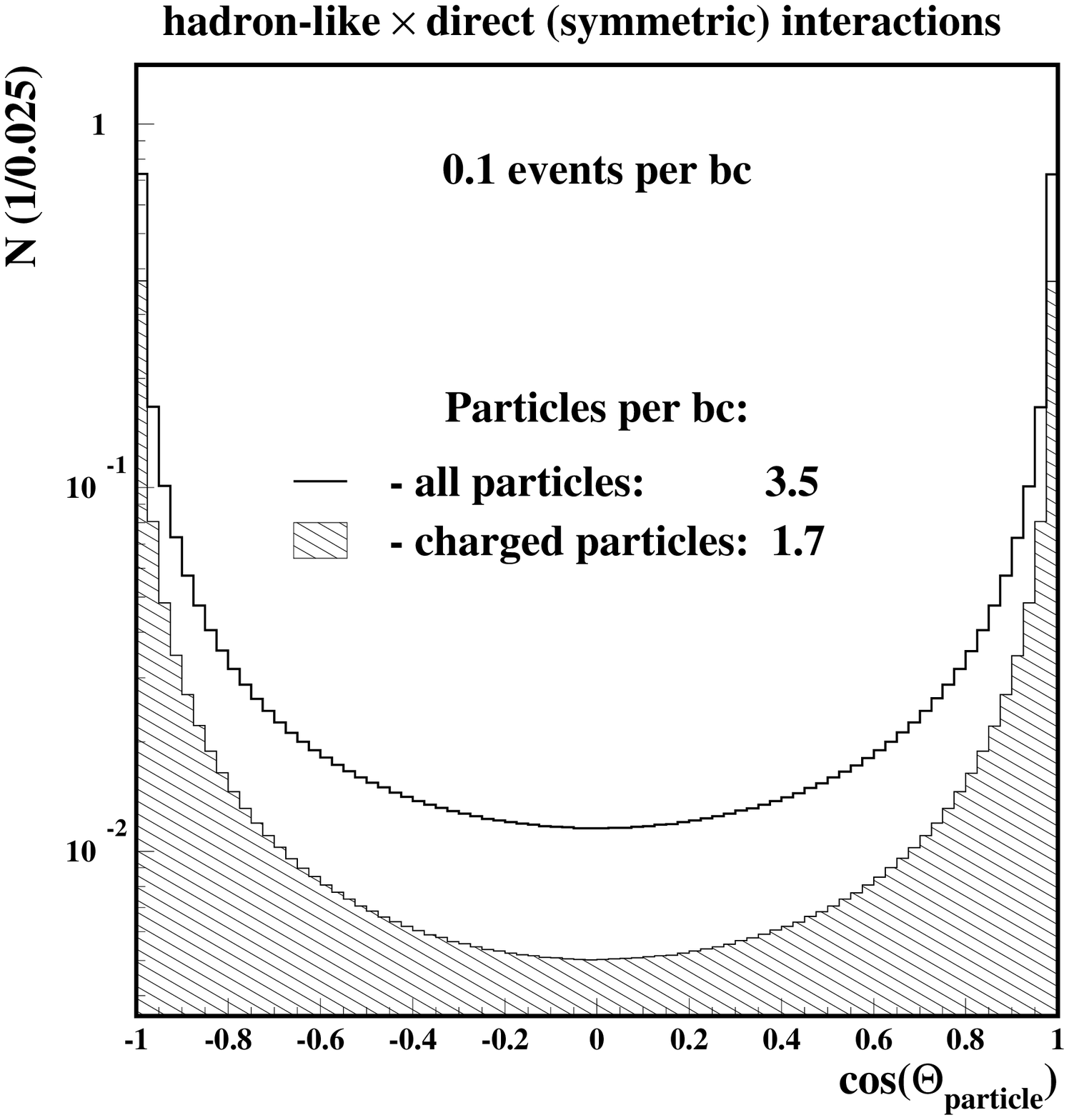} \includegraphics{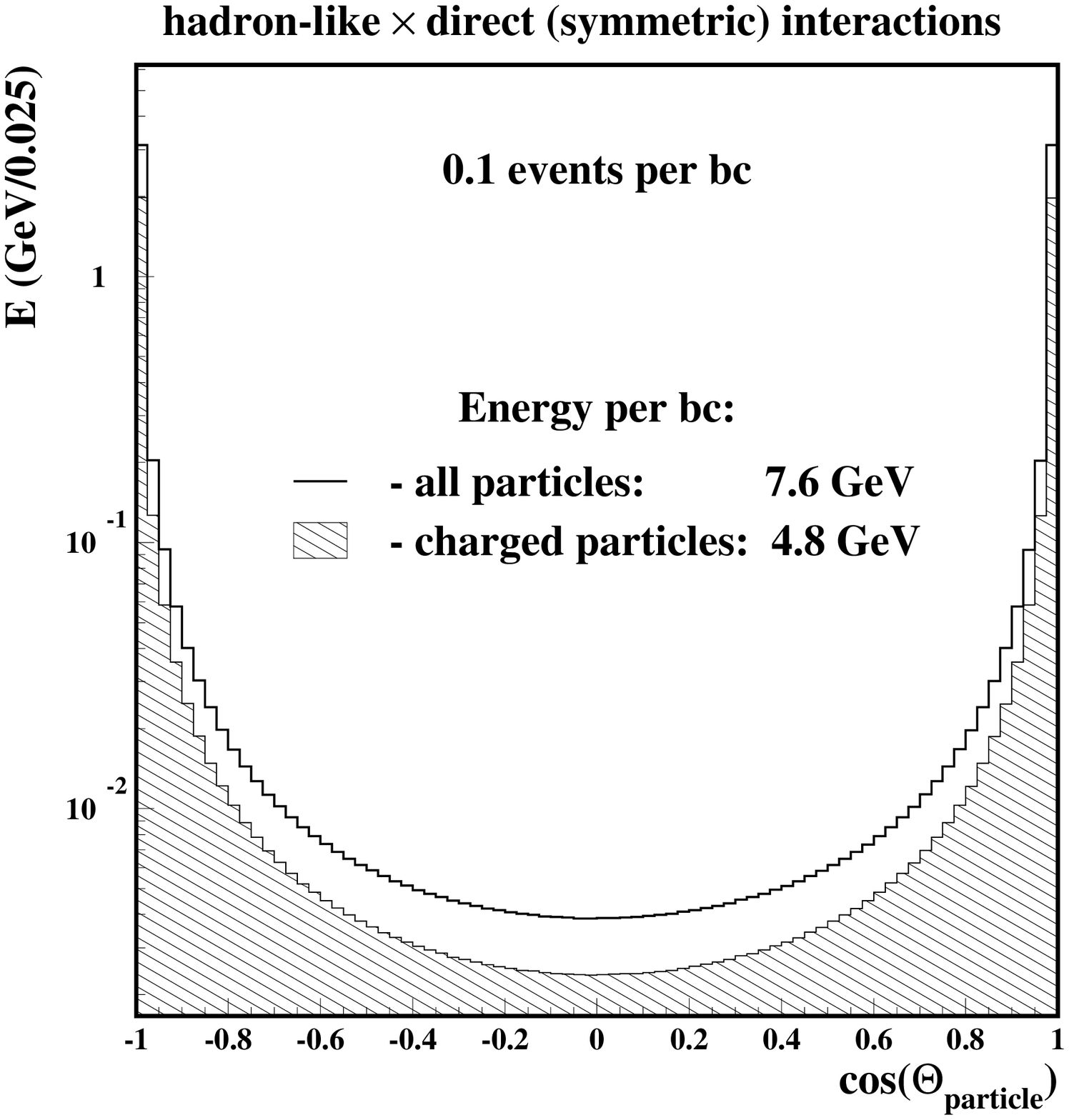}  \includegraphics{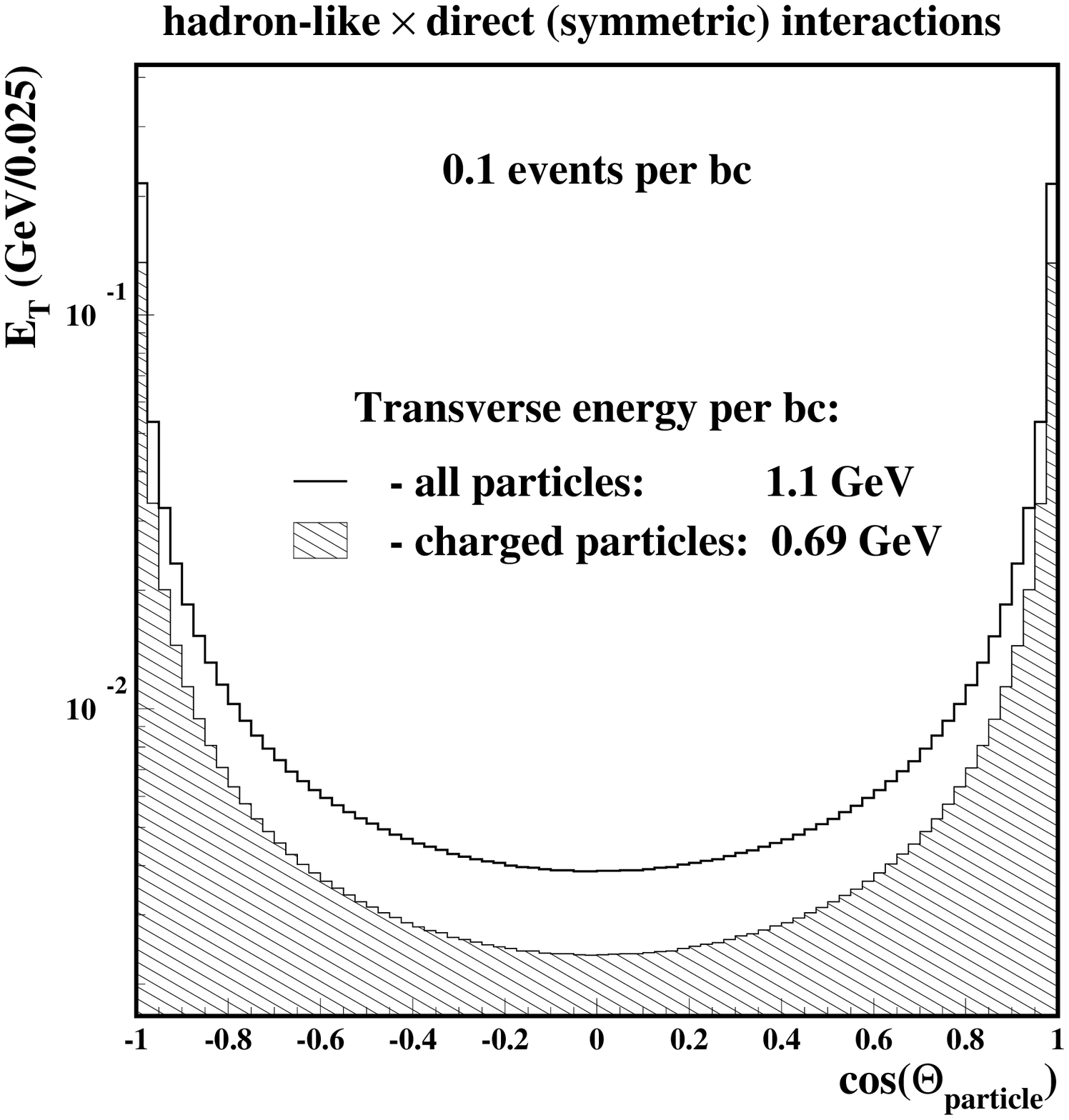} }{fig:gagahad_hadron_direct}{
Distributions of particle, energy and transverse energy flow in $\cos(\theta_{\mathit{particle}})$ for \emph{hadron-like$\times$direct} case for \sqrtseeeq 210~GeV.}

\pnfiggeneral{p}{5.5cm}{\includegraphics{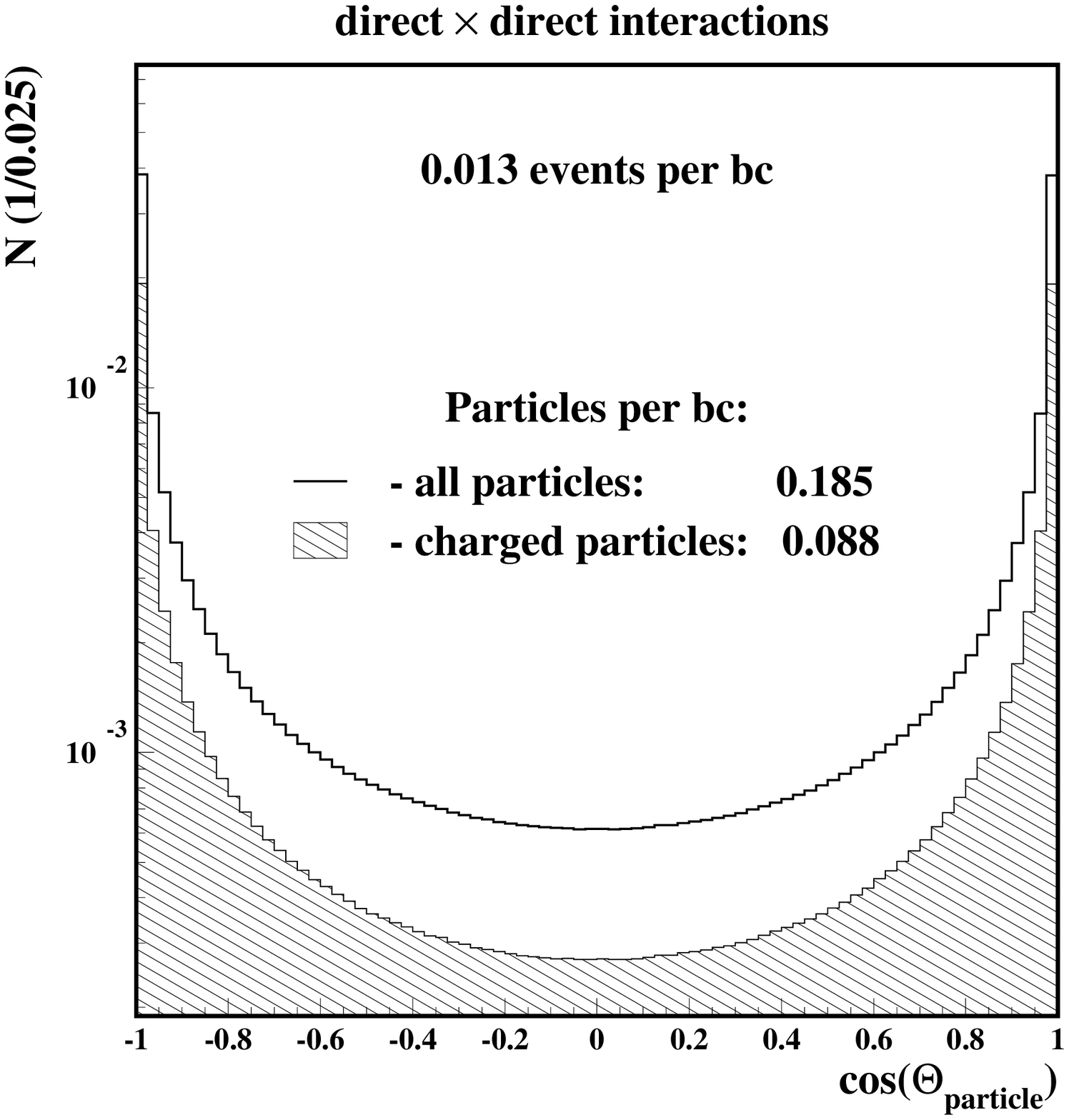} \includegraphics{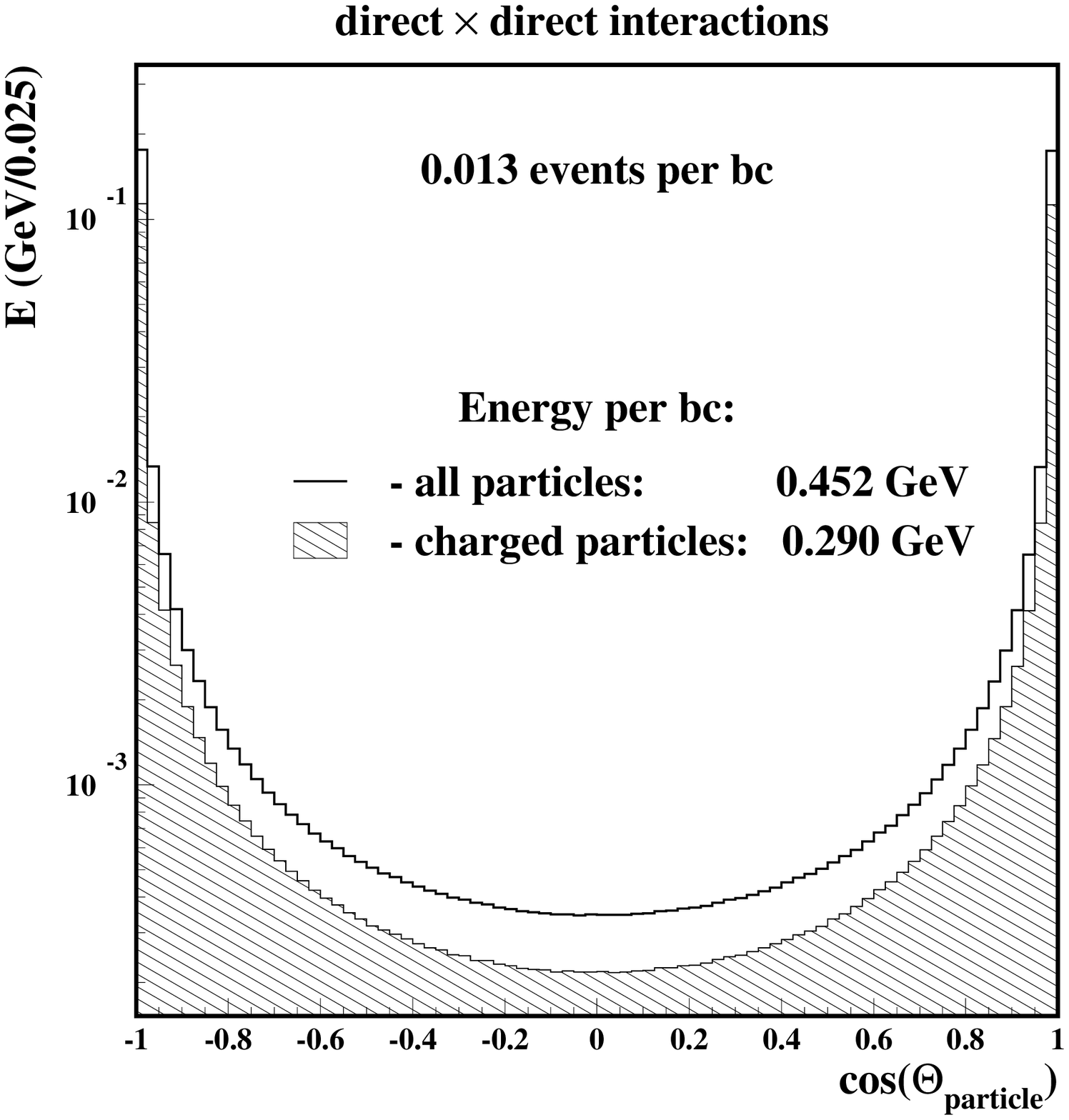}  \includegraphics{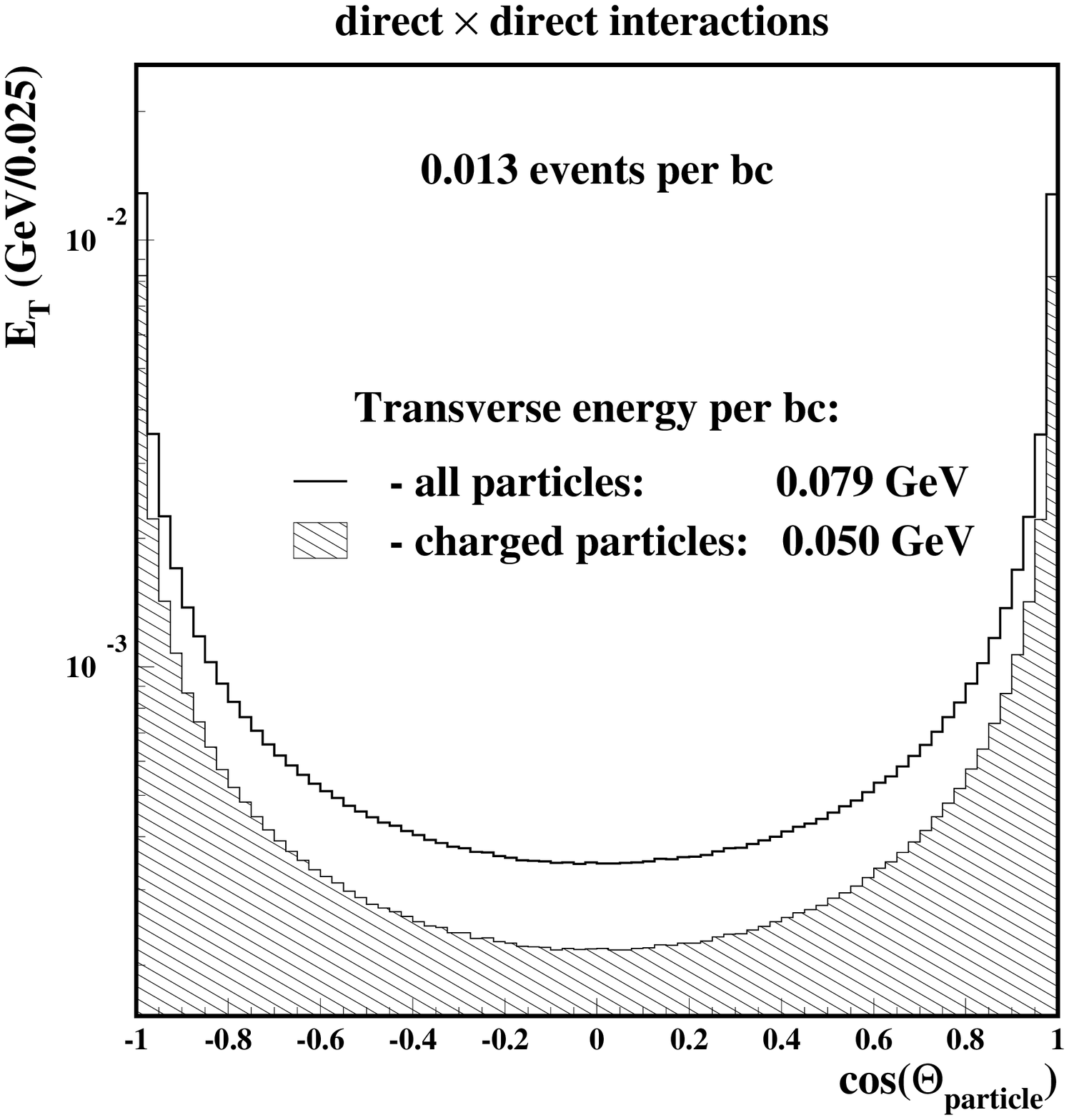} }{fig:gagahad_direct_direct}{
Distributions of particle, energy and transverse energy flow in $\cos(\theta_{\mathit{particle}})$  for \emph{direct$\times$direct} case for \sqrtseeeq 210~GeV.}


\pnfiggeneral{p}{\twofigheight}{\includegraphics{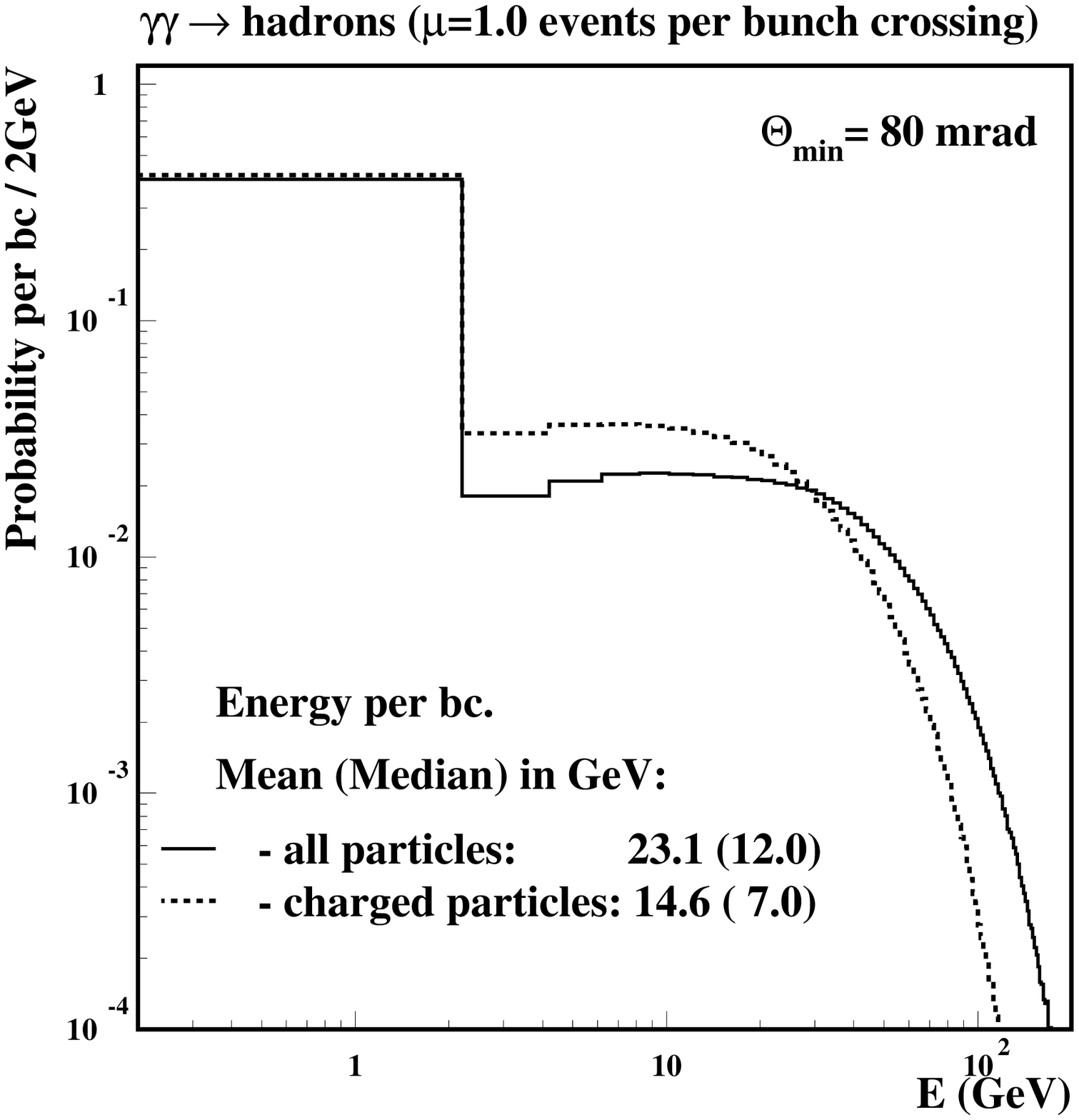} \includegraphics{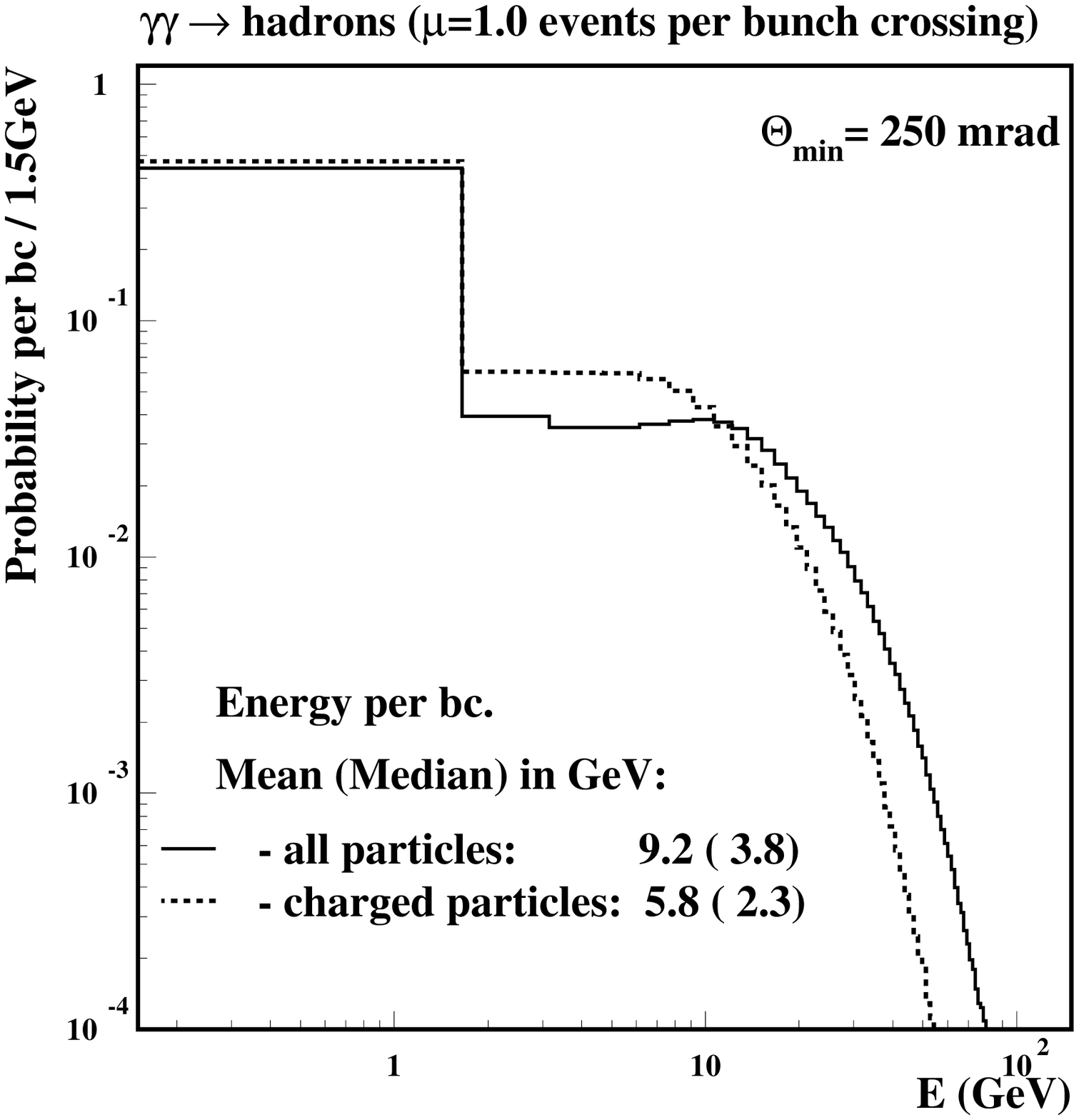}}{fig:gagahad_e_dist}{
Energy distributions for \gagahad{} event with $\theta_{min} = $ 80 and 250 mrad.}

\pnfiggeneral{p}{\twofigheight}{\includegraphics{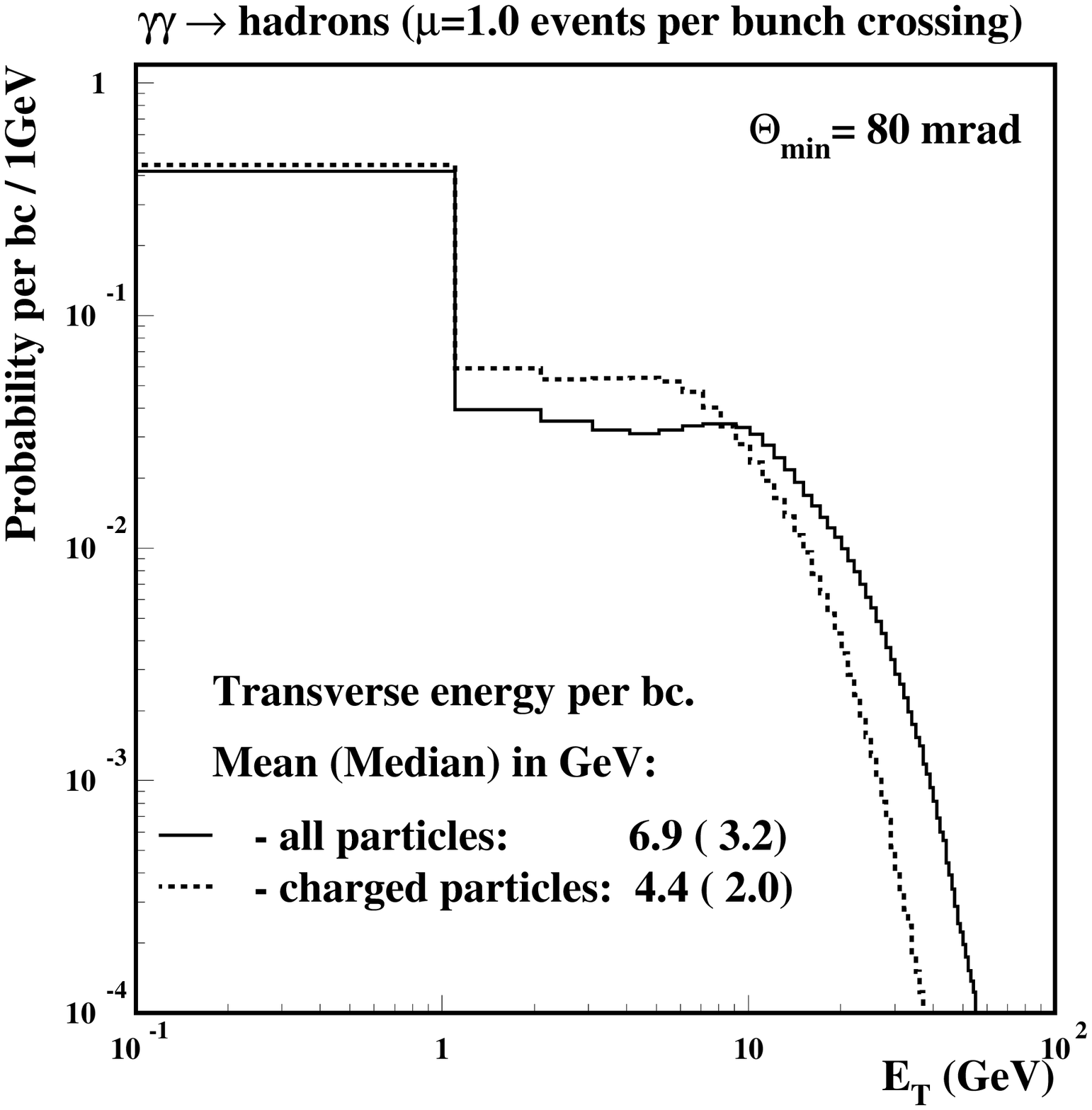} \includegraphics{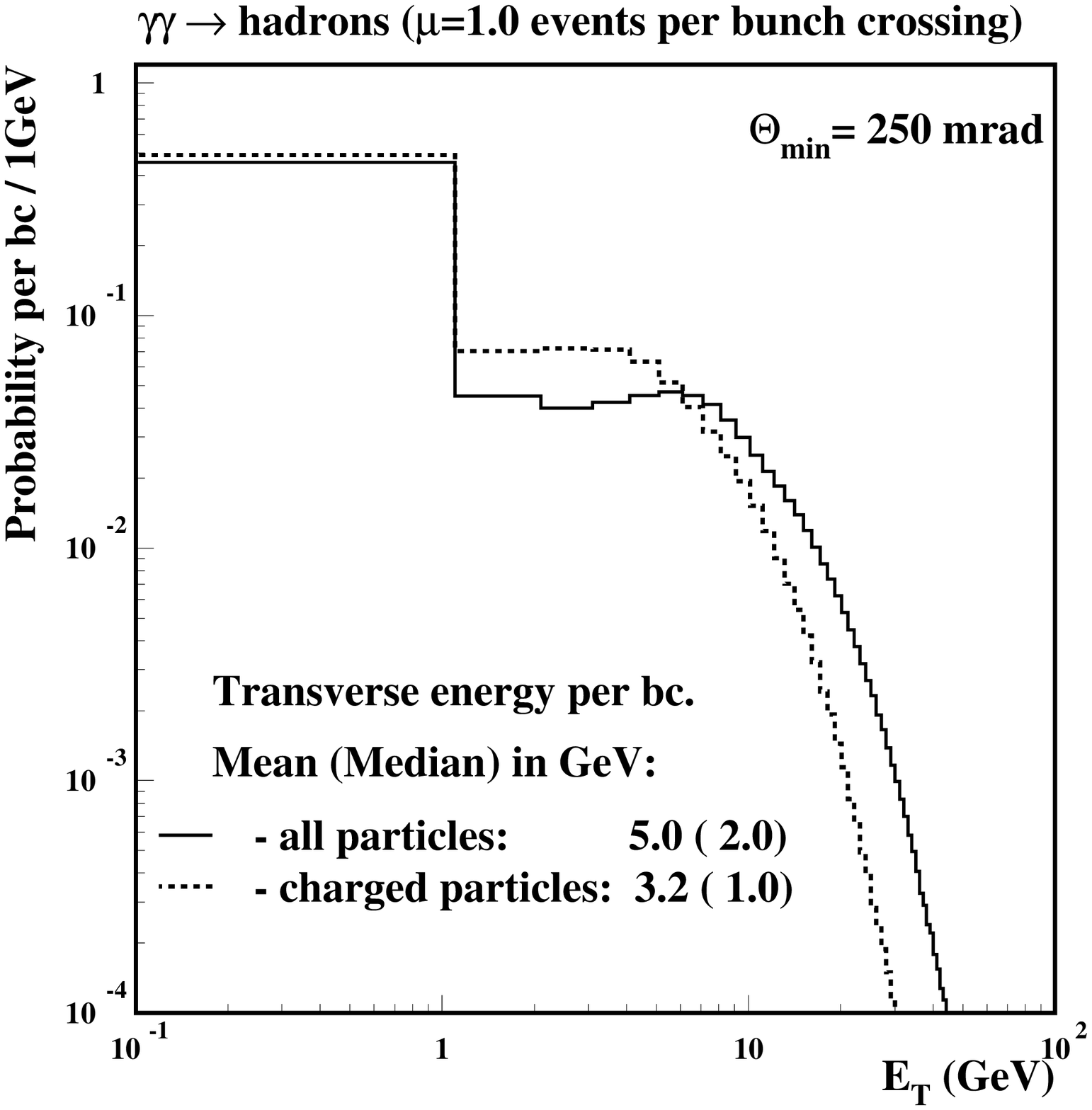}}{fig:gagahad_et_dist}{
Transverse energy distributions for \gagahad{} event with $\theta_{min} = $ 80 and 250 mrad.}


After these promising results on the generator level, we studied the detector
response to the overlaying events.
%
%
The influence of low-angle particles can be reduced 
by ignoring tracks and clusters with 
polar angle less than $\thetamindet$.
Following values of $\thetamindet$ were considered: 131, 315, 450,  555, 643 and 722 mrad,
corresponding to $\costhmindet$ values of (approximately) 0.99,  0.95, 0.90, 0.85, 0.80 and 0.75. 
The lowest value \thetamindeteq 131 mrad corresponds to the expected size of the mask, $\thetamask$.
%
 Probability distributions for reconstructed energy, $E_{rec}$, 
for various values of $\thetamindet$ are shown 
in Fig.\ \ref{fig:gagahadrons_120_2} and \ref{fig:gagahadrons_300_2} for \sqrtseeeq 210 and 419~GeV, respectively.
Probability distributions of all buch crossings (multiple overlaying events possible) 
are compared with probability distributions of  buch crossing with only one \gagahad{} event
(visible in the detector). 
The low energy parts of the distributions are presented in detail in Fig.\ \ref{fig:gagahadrons_120_1} 
and \ref{fig:gagahadrons_300_1},
showing a dip due to the limited detector sensitivity  around $E_{rec} \approx 0.5 $~GeV.
The most significant reduction of \gagahad{} contamination is
obtained when increasing  $\thetamindet$  cut from $\thetamask$
to \costhmindeteq 0.95 and 0.9.
However, it is important to notice that the probability 
of observing no energy in the detector ($E_{rec} < $ 0.2~GeV)
does not increase significantly with increase of $\thetamindet$.
Most events remain visible in the detector although with highly reduced energy.

\pnfiggeneral{p}{\twofigheight}{\includegraphics{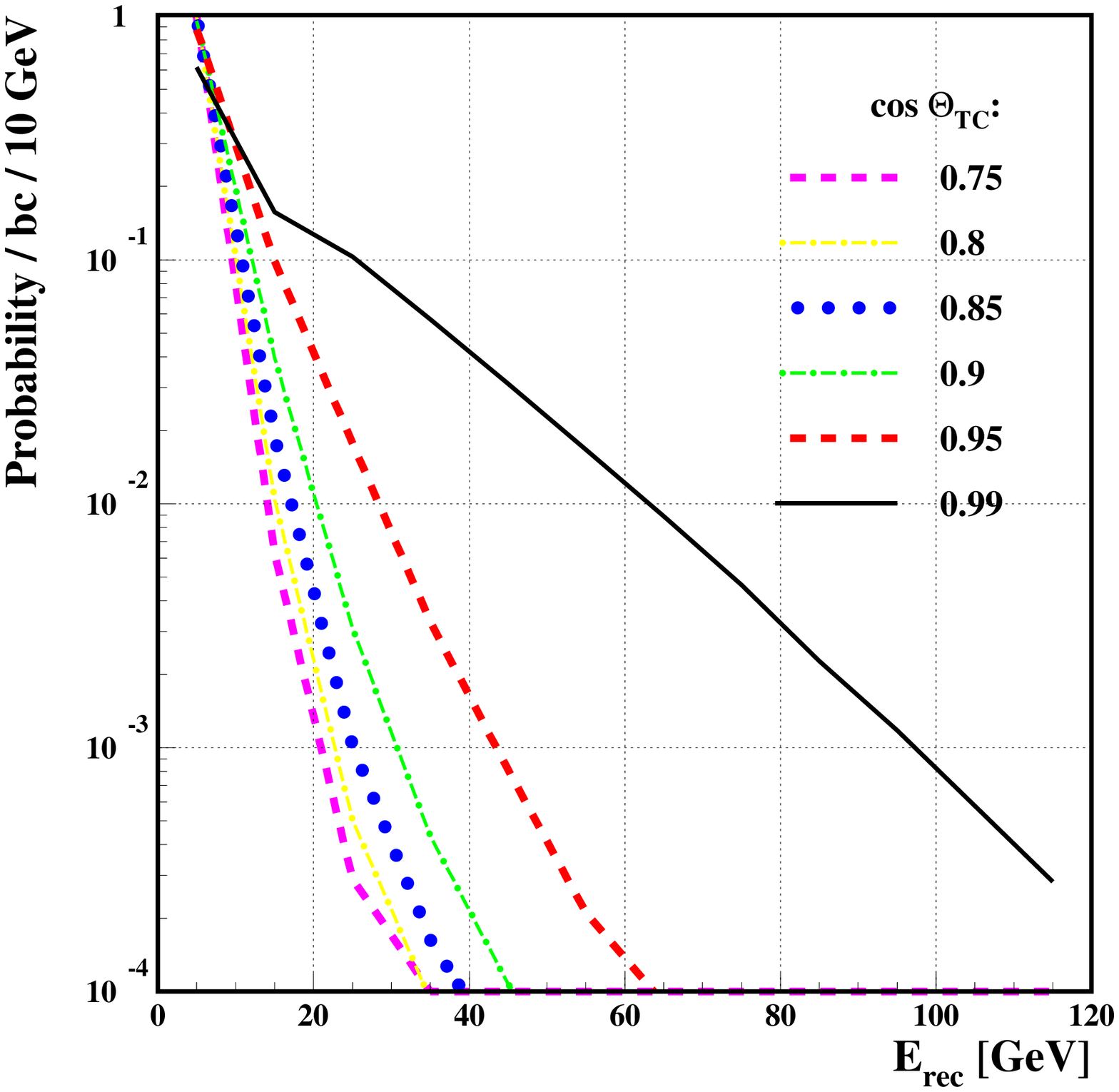} \includegraphics{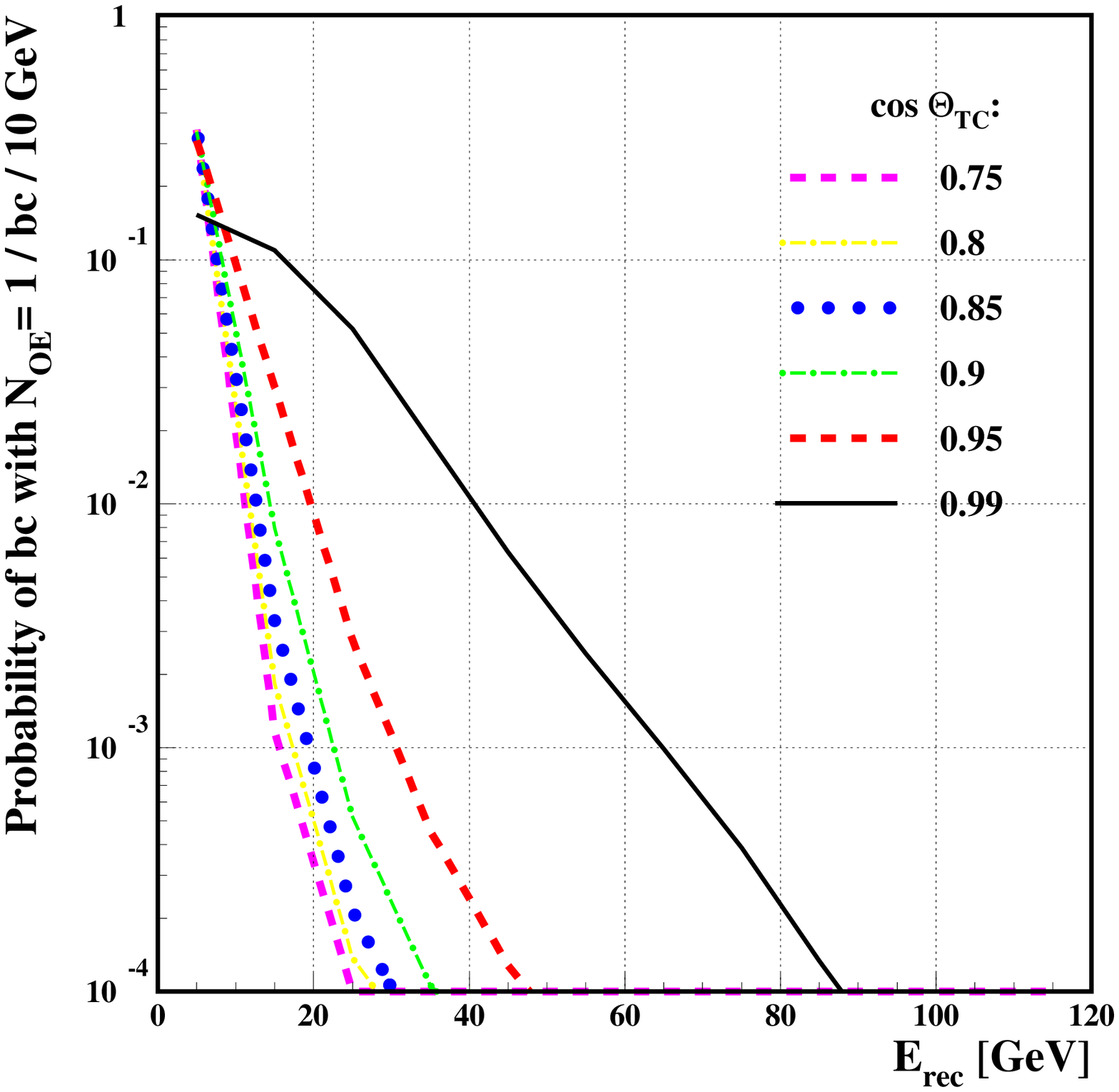}}{fig:gagahadrons_120_2}{
Probability distribution for reconstructed energy, $E_{rec}$, for various values of $\thetamindet$ and for \sqrtseeeq 210~GeV.
On the right plot the probability for a bunch crossing with only one \gagahad{} event
which on the generator level has some particles above the detector mask.
For clarity only centers of bins are connected by lines.
}

\pnfiggeneral{p}{\twofigheight}{\includegraphics{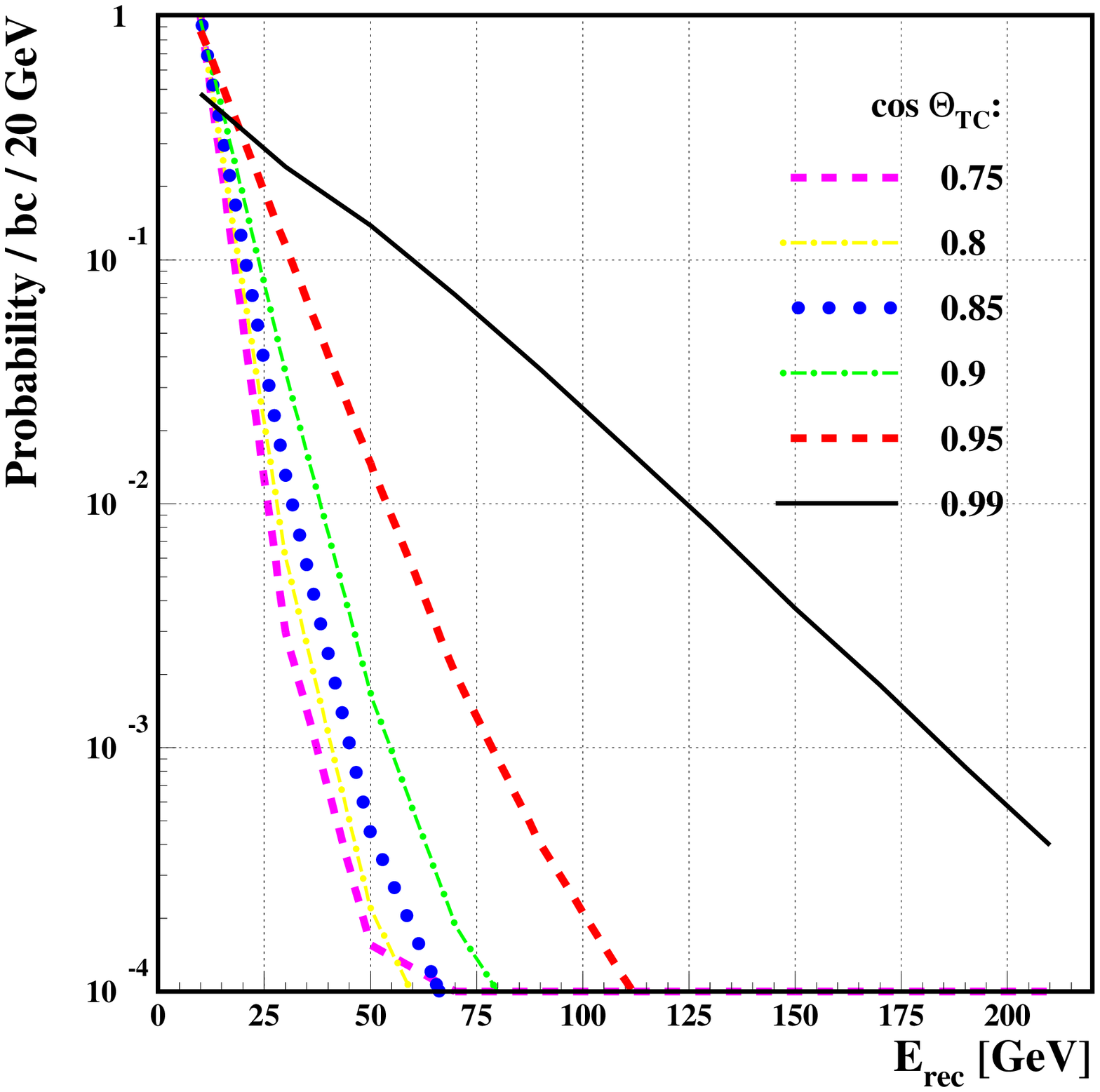} \includegraphics{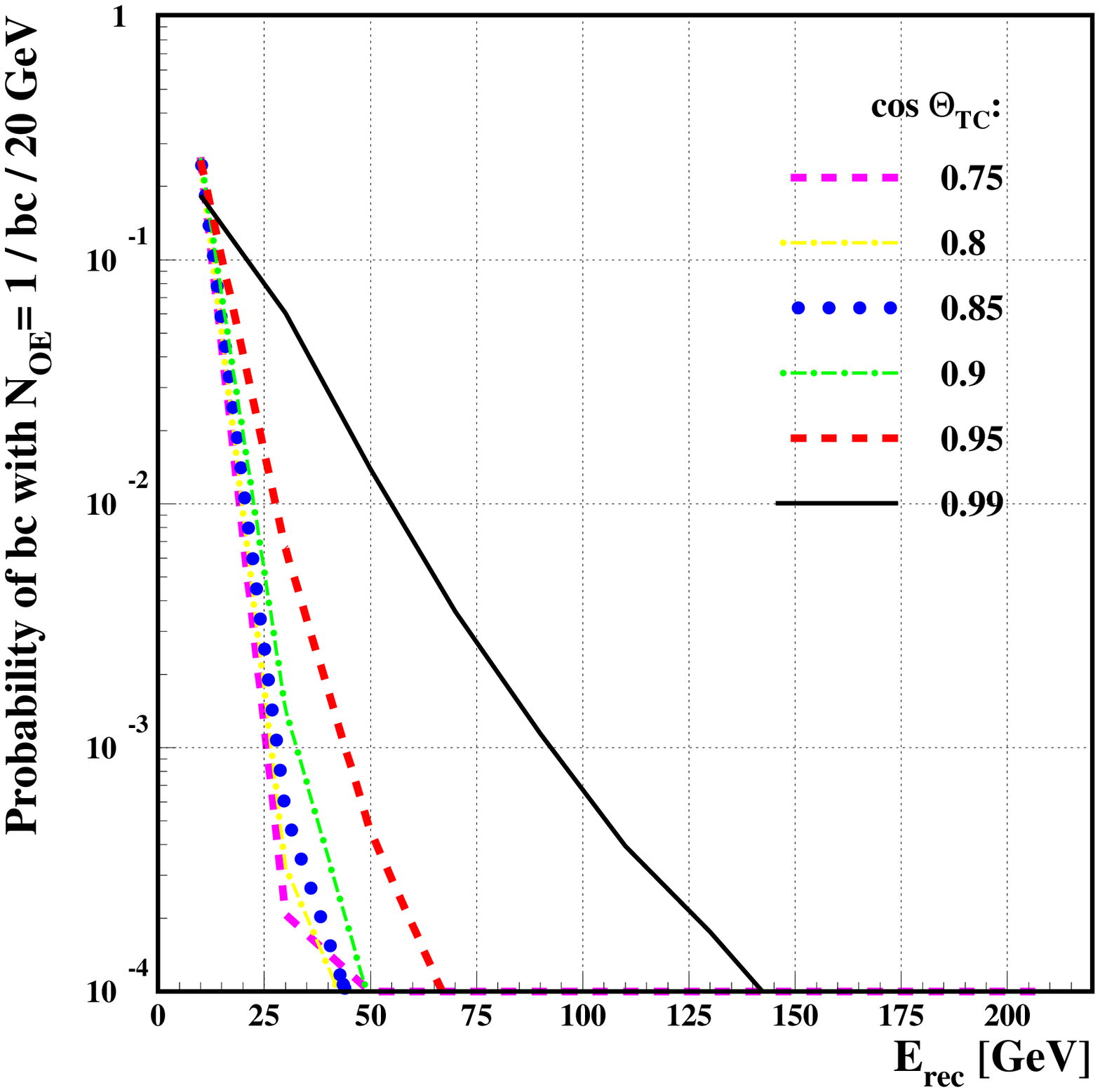}}{fig:gagahadrons_300_2}{
Probability distribution for reconstructed energy, $E_{rec}$, for various values of $\thetamindet$ and for \sqrtseeeq 419~GeV.
On the right plot the probability for a bunch crossing with only one \gagahad{} event
which on the generator level has some particles above the detector mask.
For clarity only centers of bins are connected by lines.
}

\pnfiggeneral{p}{\twofigheight}{\includegraphics{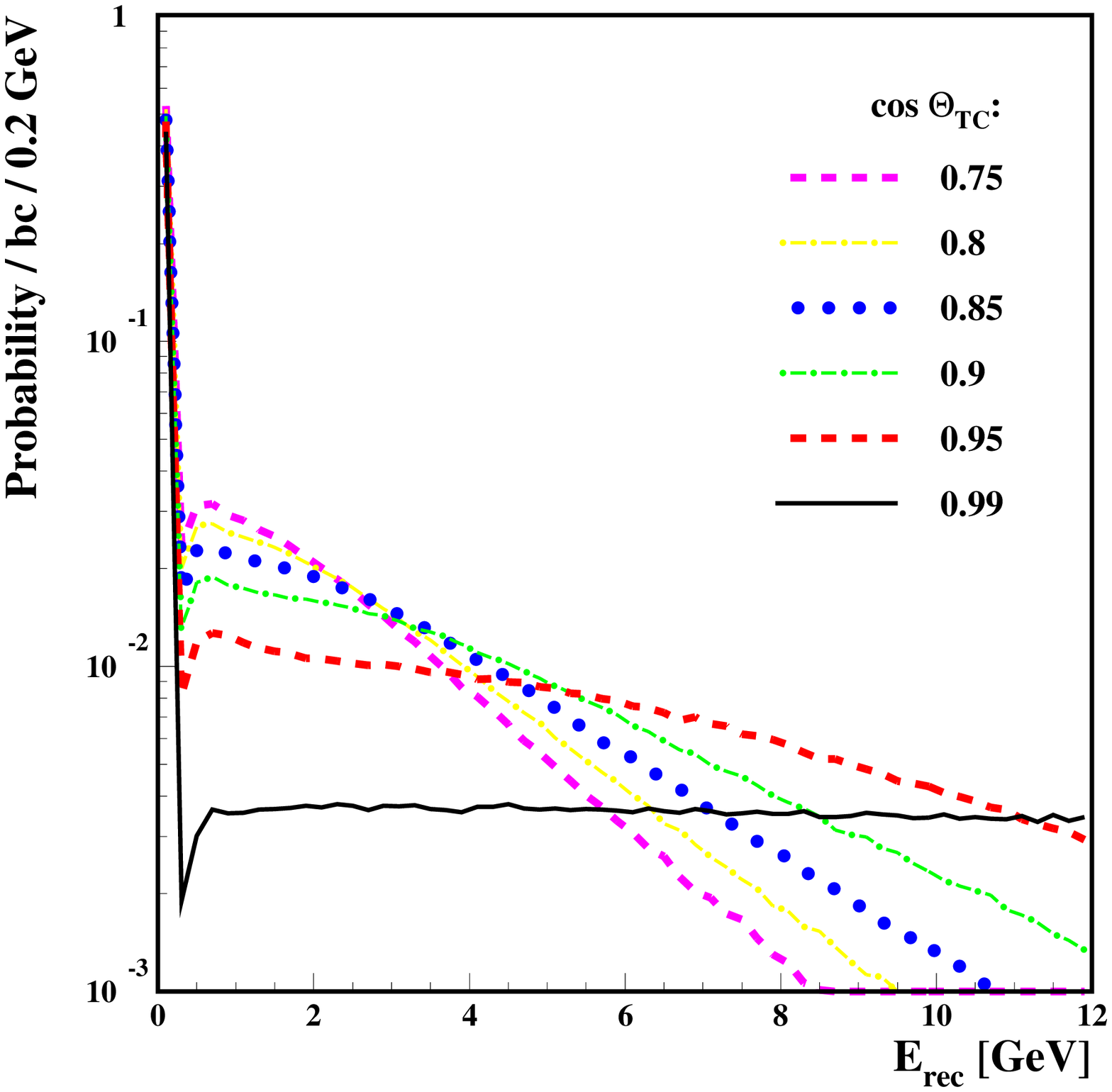} \includegraphics{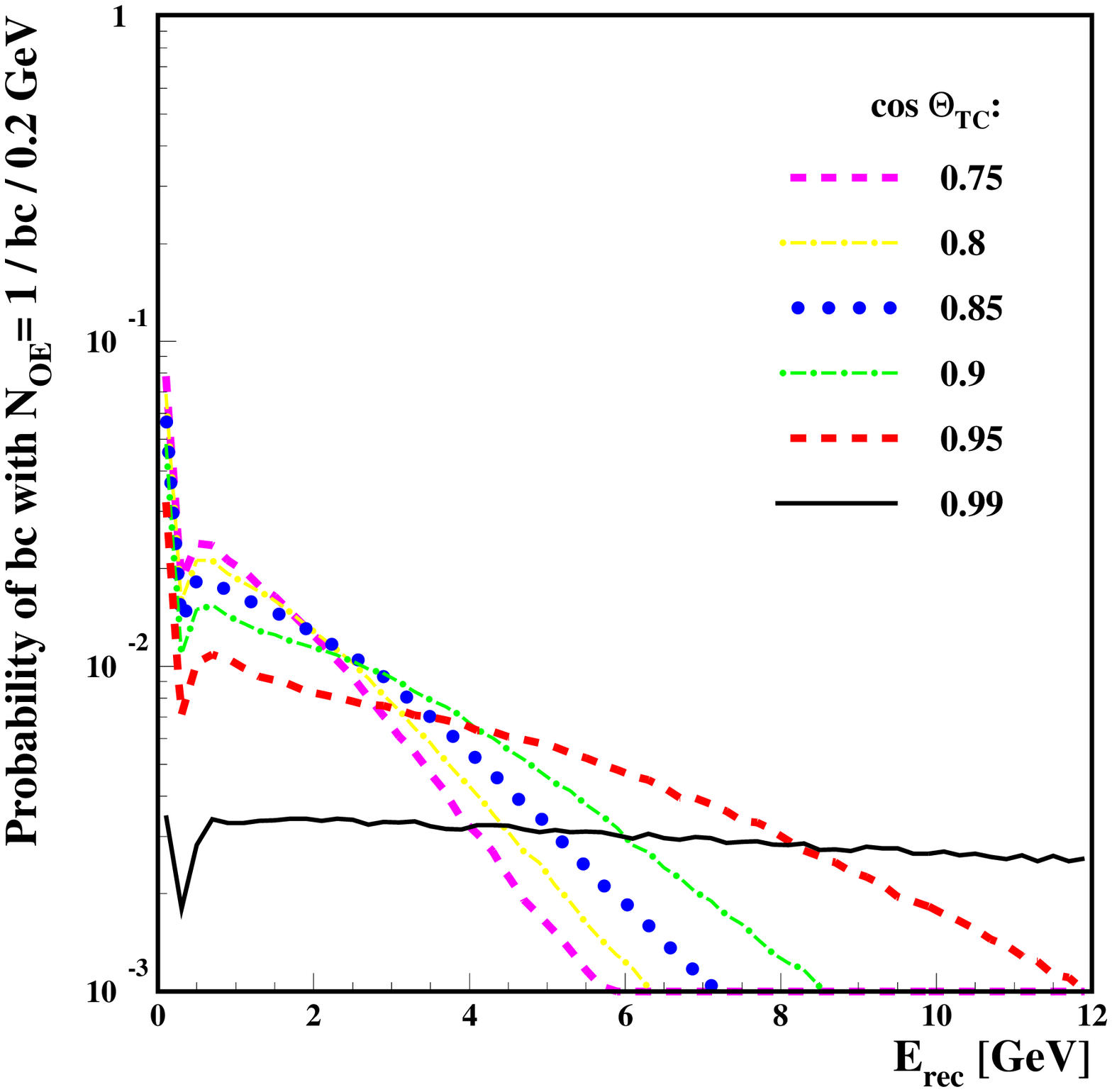}}{fig:gagahadrons_120_1}{
Probability distribution for reconstructed energy, $E_{rec}$, for various values of $\thetamindet$ and for \sqrtseeeq 210~GeV.
On the right plot the probability for a bunch crossing with only one \gagahad{} event
which on the generator level has some particles above the detector mask.
For clarity only centers of bins are connected by lines.
}

\pnfiggeneral{hbt}{\twofigheight}{\includegraphics{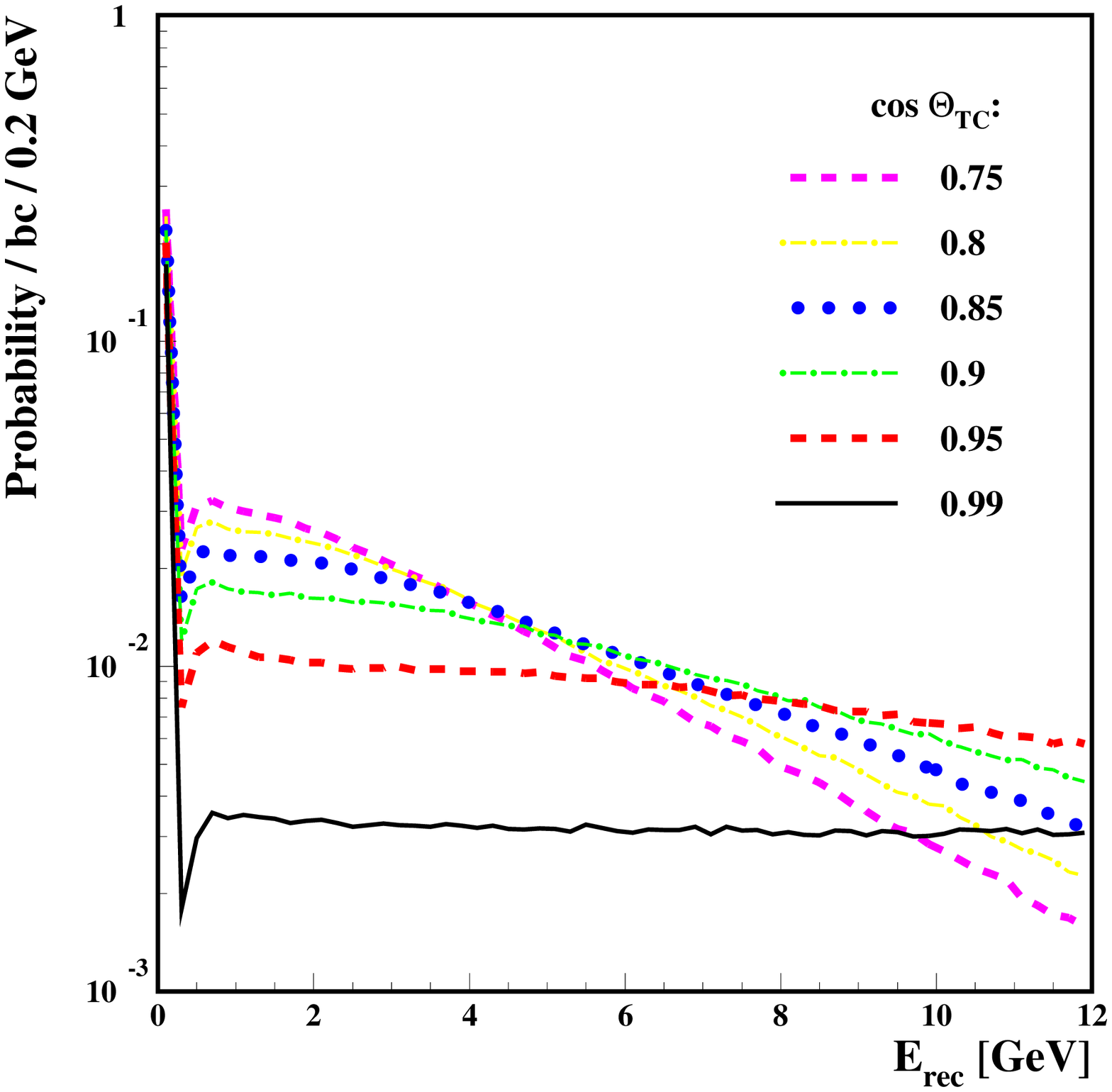} \includegraphics{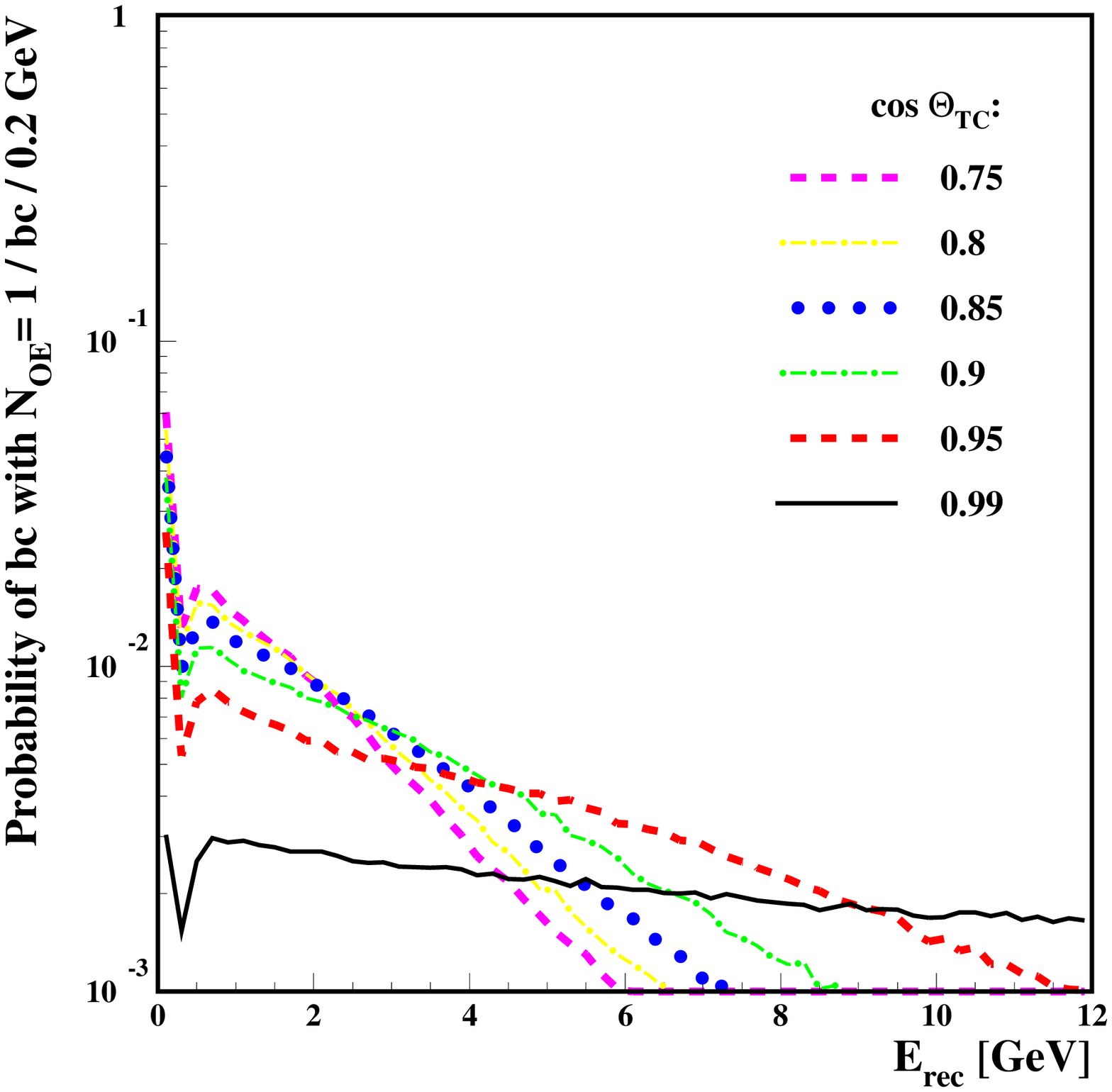}}{fig:gagahadrons_300_1}{
Probability distribution for reconstructed energy, $E_{rec}$, for various values of $\thetamindet$ and for \sqrtseeeq 419~GeV.
On the right plot the probability for a bunch crossing with only one \gagahad{} event
which on the generator level has some particles above the detector mask.
For clarity only centers of bins are connected by lines.
}

As the \Pythia{} generator does not produce  \gagahad{} events for $\Wgaga < 4 $~GeV,
we try to estimate the contribution of these events.
In Fig.\ \ref{fig:gagahadrons_3} average  energy reconstructed
per one \gagahad{} event, $<\!E_{rec}\!>$, 
is shown as  a function of $\Wgaga$,
for \sqrtseeeq 210~GeV and 419~GeV.
Again, the strong reduction of reconstructed energy is observed with increase of $\thetamindet$.
It is worth noticing that for \sqrtseeeq 210~GeV the average reconstructed energy is 
slightly higher at given $\Wgaga$ than for \sqrtseeeq 419~GeV. 
This is due to the larger boosts of the low energy $\gaga$ events at higher $\emem$-beam energies.
However, the reconstructed energy per bunch crossing  is higher for \sqrtseeeq 419~GeV
as the expected number of \gagahad{} events increases from about 1 to about 2 per bunch crossing.
From the average numbers of overlaying events per bunch crossing 
and the fact that the fraction of luminosity below $\Wgaga = 4 $~GeV is equal approximately to 30\%,
we estimated that 
the neglected low-$\Wgaga$ overlaying events would contribute 
to no more than 0.3~GeV (on average) to the energy measured in the detector 
if $\costhmindet \approx 0.85$ is chosen.
 Particles coming from overlaying events,  entering the detector 
 with the angle larger than $\thetamindet$, 
 are expected to form jets with small relative transverse momentum, $\ptjetET$,
 where $\ptjet$ is the transverse momentum of the jet 
 and $E_T$ is the transverse energy of all energy-flow objects above $\thetamindet$.
We studied reconstructed jet distributions for the $\gagahbb$ and $\gagaAbb$  events with overlaying events included, 
for various values of $\costhmindet$-cut.
For comparison the 'ideal' measurement, \ie without overlaying events (therefore using $\thetamindet = \thetamask$)
 was also considered.
For consistency with the final analysis we required that for the jet with the 
highest transverse momentum an angle cut was fulfilled: $|\cos\theta_{jet}|<0.7$,
where $\theta_{jet}$ is the polar angle of the jet.
The   $\ptjetET$  distributions for jets reconstructed in $\gaga \ar \hSM / \AO \ar \bbbar$  events are shown
for various values of $\costhmindet$ in Fig.\ \ref{fig:ptjet_mjet_120} 
( $\hSM$ production for \sqrtseeeq 210~GeV and \Mheq 120~GeV)
and in Fig.\ \ref{fig:ptjet_mjet_300} ($\AO$ production for \sqrtseeeq 419~GeV and \MAOeq 300~GeV).
For bunch crossing with $\thetamindet = \thetamask$ the contribution from overlaying events
is clearly seen for  $\ptjetET < 0.1$.
Already for the cut \costhmindeteq 0.95 this contamination is significantly  reduced 
and the distribution
approaches the 'ideal' one. 
With increasing the $\thetamindet$ cut
the influence of overlaying events on the Higgs-boson production events is decreasing.
Also the lower systematic uncertainty  due to \gagahad{} contributiuon 
coming from  relatively large uncertainties
of the cross sections for hadron-like interactions of photons 
can be expected.
Hovewer, some tracks and clusters coming from signal event can be ignored at the same time.
This will decrease mass resolution and worsen precision of the measurement.
In Fig.\ \ref{fig:ptjet_mjet_120} and \ref{fig:ptjet_mjet_300}
distributions of the reconstructed jet mass, $M_{jet}$,  are shown for different $\costhmindet$ cuts.
We see that for \costhmindeteq 0.75 and 0.8
some jets loose most of their content and
their mass is significantly underestimated ($M_{jet} \lesssim 3 $~GeV).
Distribution obtained for  \costhmindeteq 0.85
nearly recovers the  'ideal' jet-mass distribution
having almost the same mean value.

Finally, we checked the influence of the $\thetamindet$-cut 
on the Higgs-boson mass resolution.
For this comparison 
the mass resolution was defined as the dispersion, $\sigma$, of the Gaussian distribution
fitted to  $W_{rec}-\Wgaga$ or  $W_{corr}-\Wgaga$ distribution.
The fit was performed 
in the range $(\mu - \sigma,\; \mu + 1.5\sigma)$, 
where $\mu$ is the mean of the Gaussian distribution. 
As shown in Fig.\ \ref{fig:mres_120} and \ref{fig:mres_300},
mass resolutions do not change significantly
 for $\costhmindet \leq 0.9$.
Both, the obtained resolution values and  the shape of distributions show
that the cut values \costhmindeteq 0.99 and 0.95
are too weak to suppress the \gagahad{} contribution.
The full analysis (see Chapter \ref{ch_sm_analysis}) 
considering Higgs-boson production signal and
the $\gagaQQg$ background was performed for 
processes $\gagahbb$ (for \Mheq 120~GeV) and $\gagaAHbb$ (for \MAOeq 300~GeV), 
for all considered values of $\costhmindet$.
The best precision of the cross-section measurement was obtained for \costhmindeteq 0.85 and 0.9.
Because higher $\thetamindet$ cut does not deteriorate the precision,
but has the advantage of reducing the systematical uncertainty
due to \gagahad{} events,
the value \costhmindeteq 0.85 was chosen as the optimal one. 

\pnfiggeneral{!p}{\twofigheight}{\includegraphics{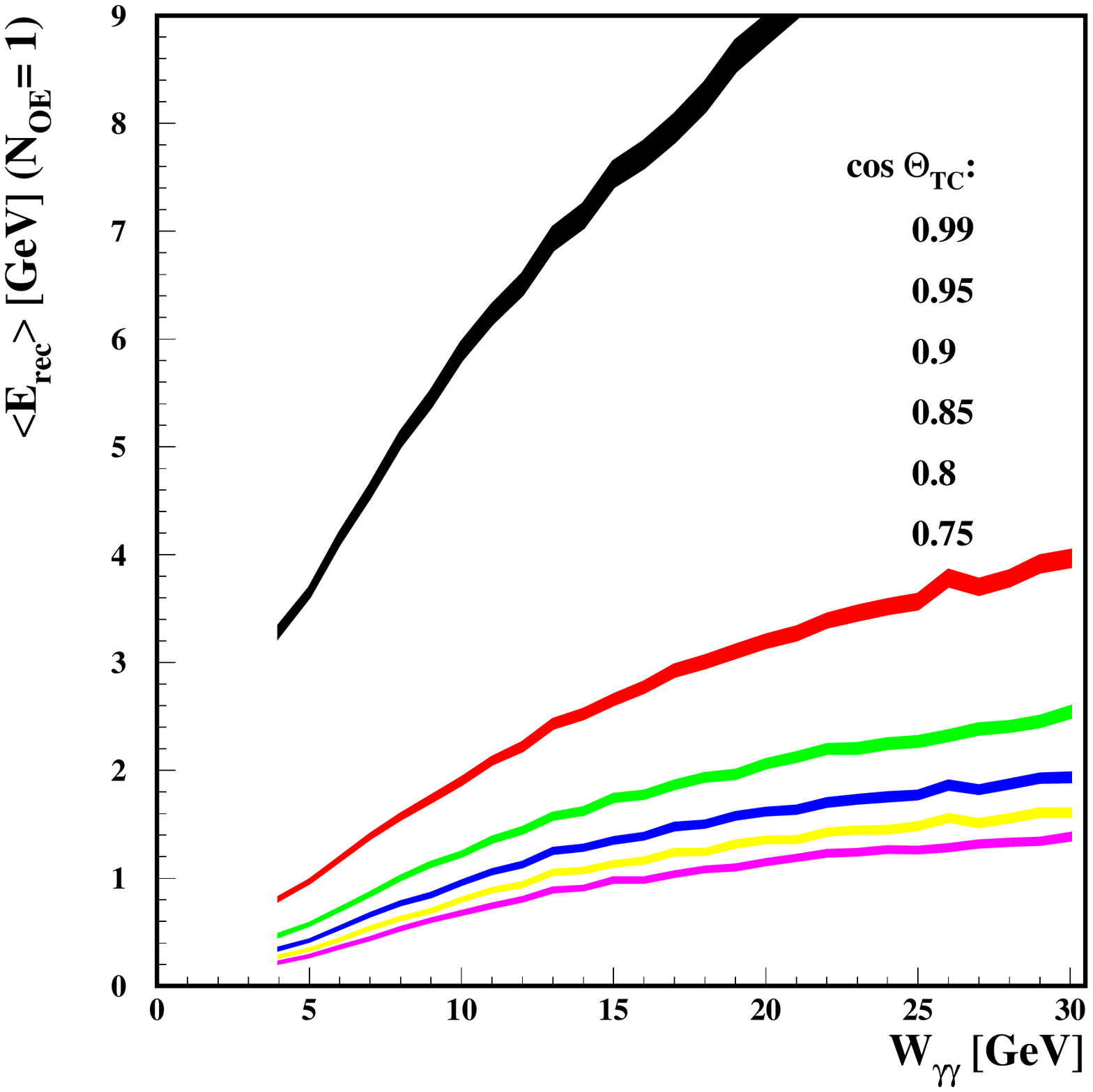} \includegraphics{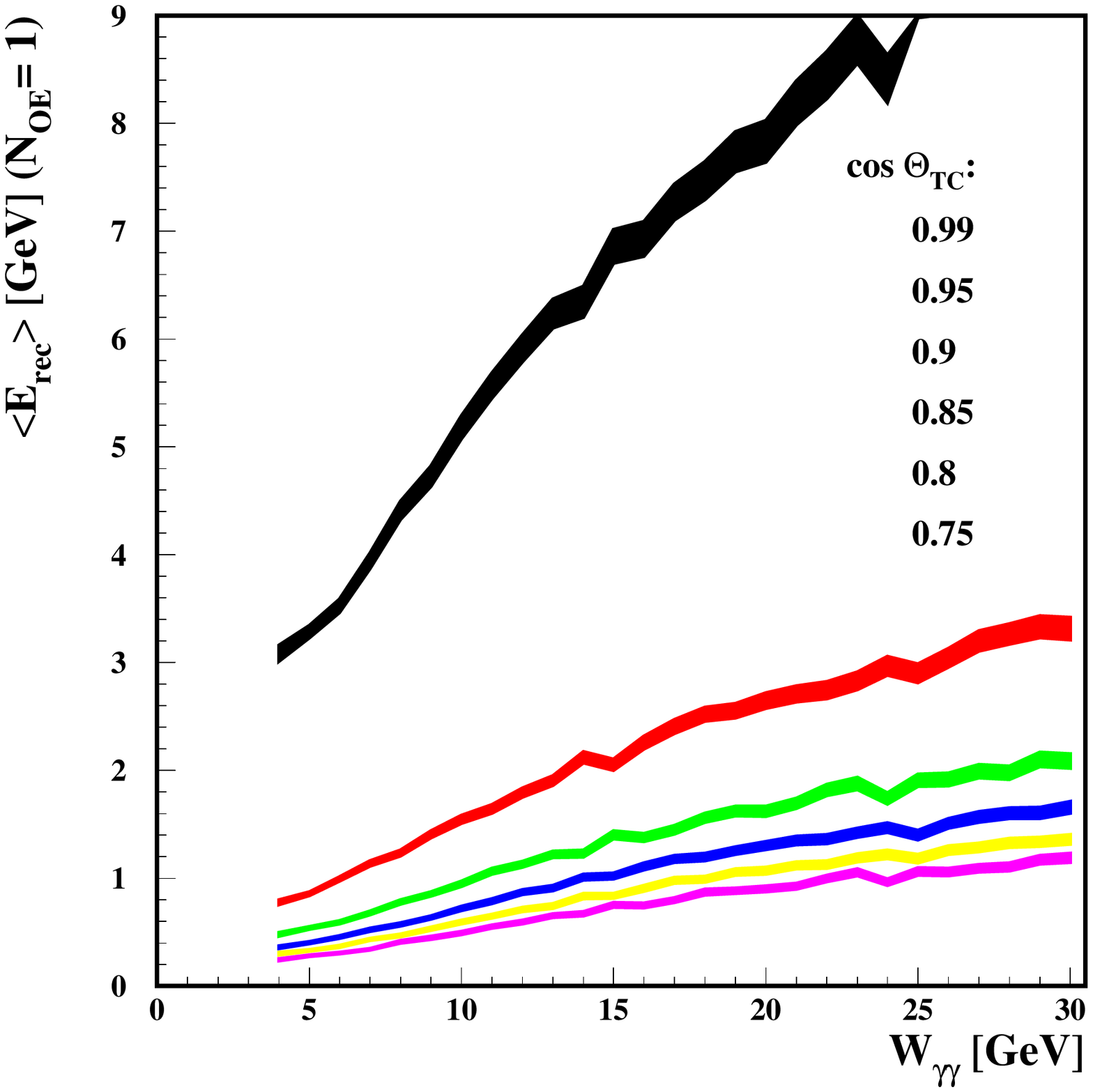}}{fig:gagahadrons_3}{
Average reconstructed energy, $<\!E_{rec}\!>$, dependence on $\gaga$ invariant mass, $\Wgaga$,
for bunch crossings with one \gagahad{} event.
Various values of $\thetamindet$  for \sqrtseeeq 210~GeV (left) and 419~GeV (right) are considered.
}

\clearpage

\pnfiggeneral{p}{\twofigheight}{\includegraphics{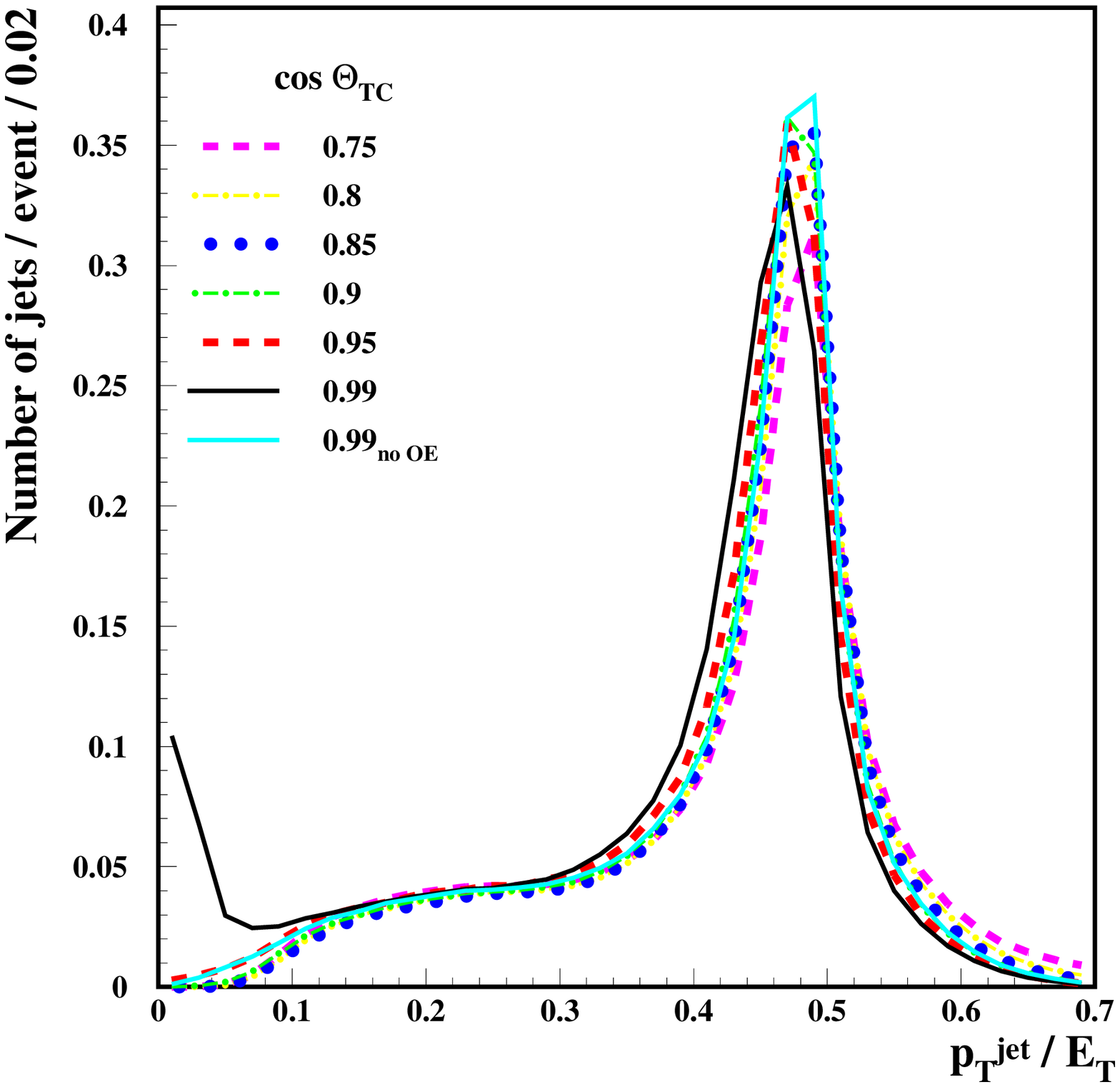} 
\includegraphics{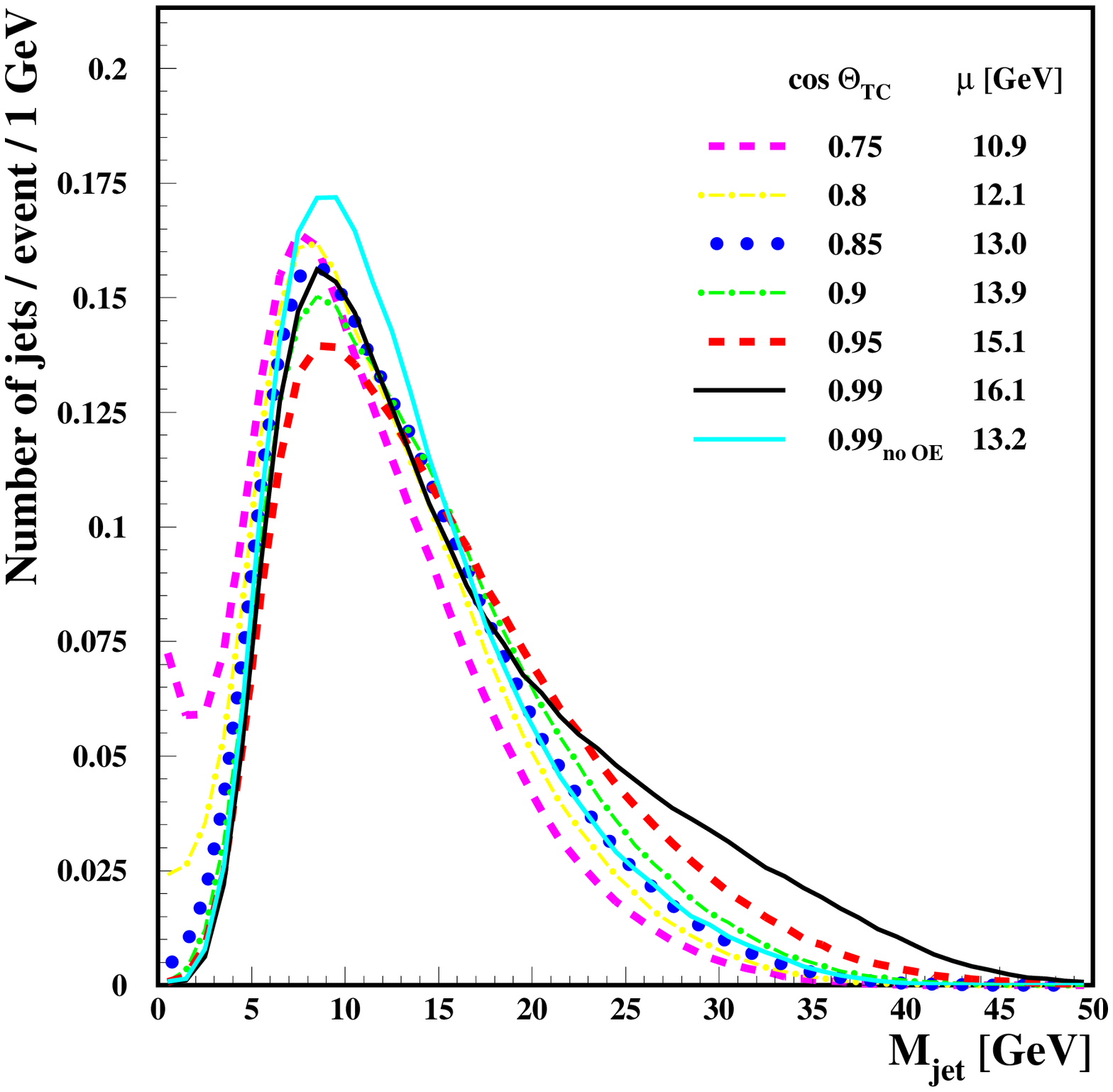}}{fig:ptjet_mjet_120}{
Distribution of  $\ptjetET$  (left) and $M_{jet}$ (right) for jets reconstructed in $\gagahbb$ event, 
for \sqrtseeeq 210~GeV and \Mheq 120~GeV.
Various values of $\thetamindet$  are considered. 
For comparison results for the ideal case without overlaying events are also shown (no OE).
The mean value of the  $M_{jet}$ distribution is presented for each case.
}
\pnfiggeneral{!}{\twofigheight}{\includegraphics{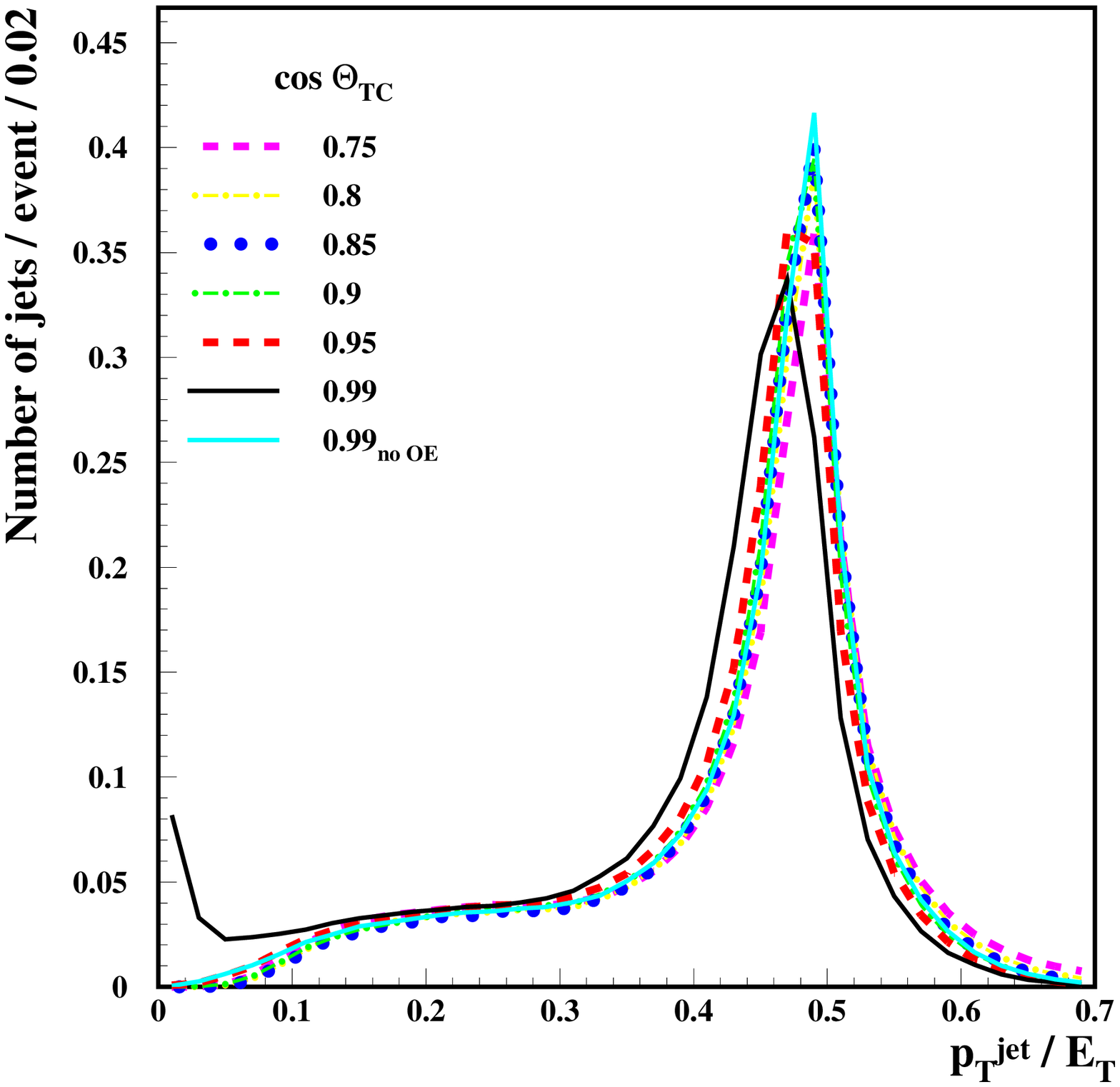} 
\includegraphics{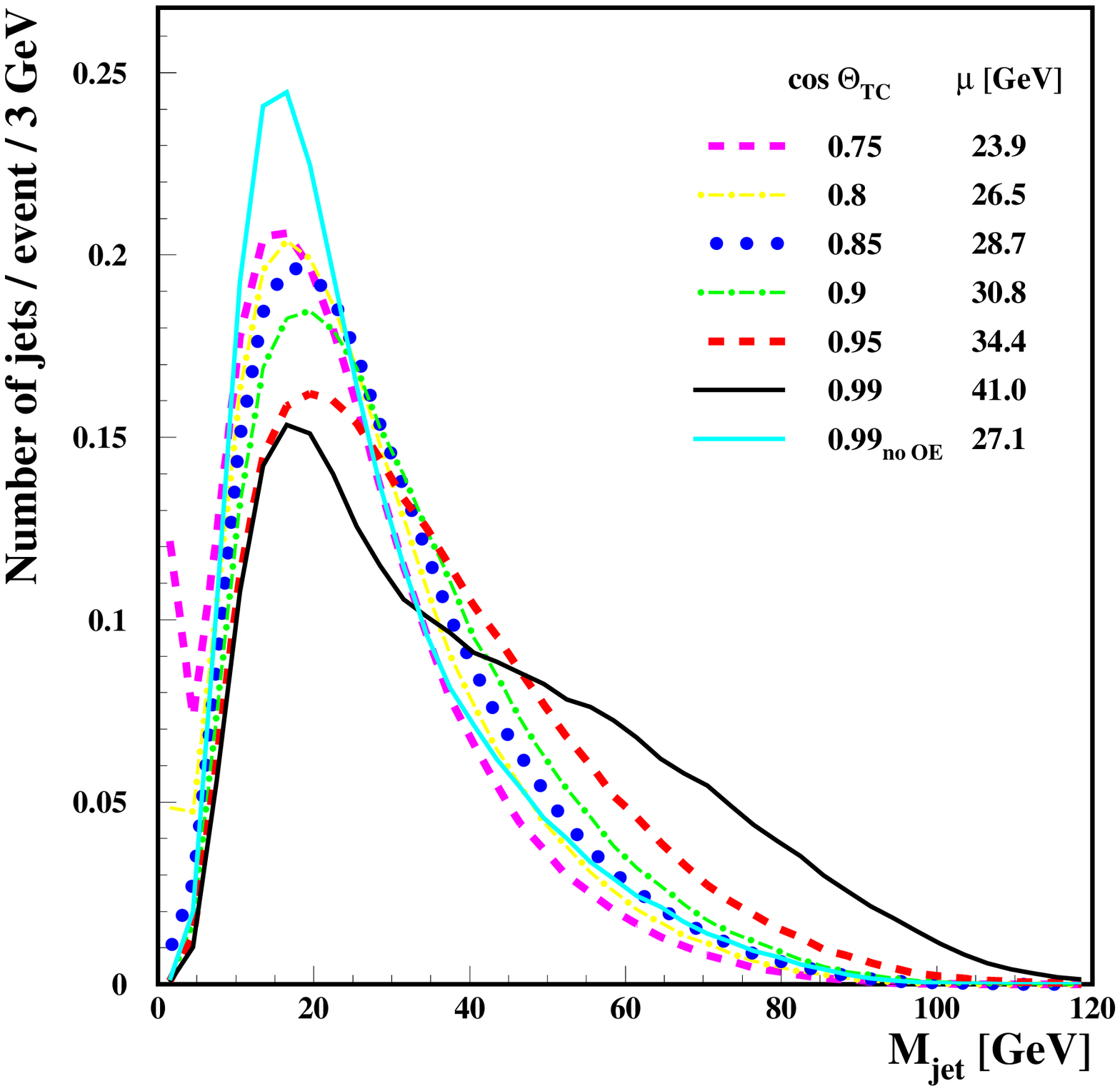}}{fig:ptjet_mjet_300}{
Distribution of   $\ptjetET$  (left) and $M_{jet}$ (right) for  jets reconstructed in $\gagaAbb$ event, 
for \sqrtseeeq 419~GeV and \MAOeq 300~GeV.
Various values of $\thetamindet$  are considered. 
Also results for the ideal case without overlaying events are shown (no OE).
The mean value of the  $M_{jet}$ distribution is presented for each case.
}

\pnfiggeneral{p}{\twofigheight}{\includegraphics{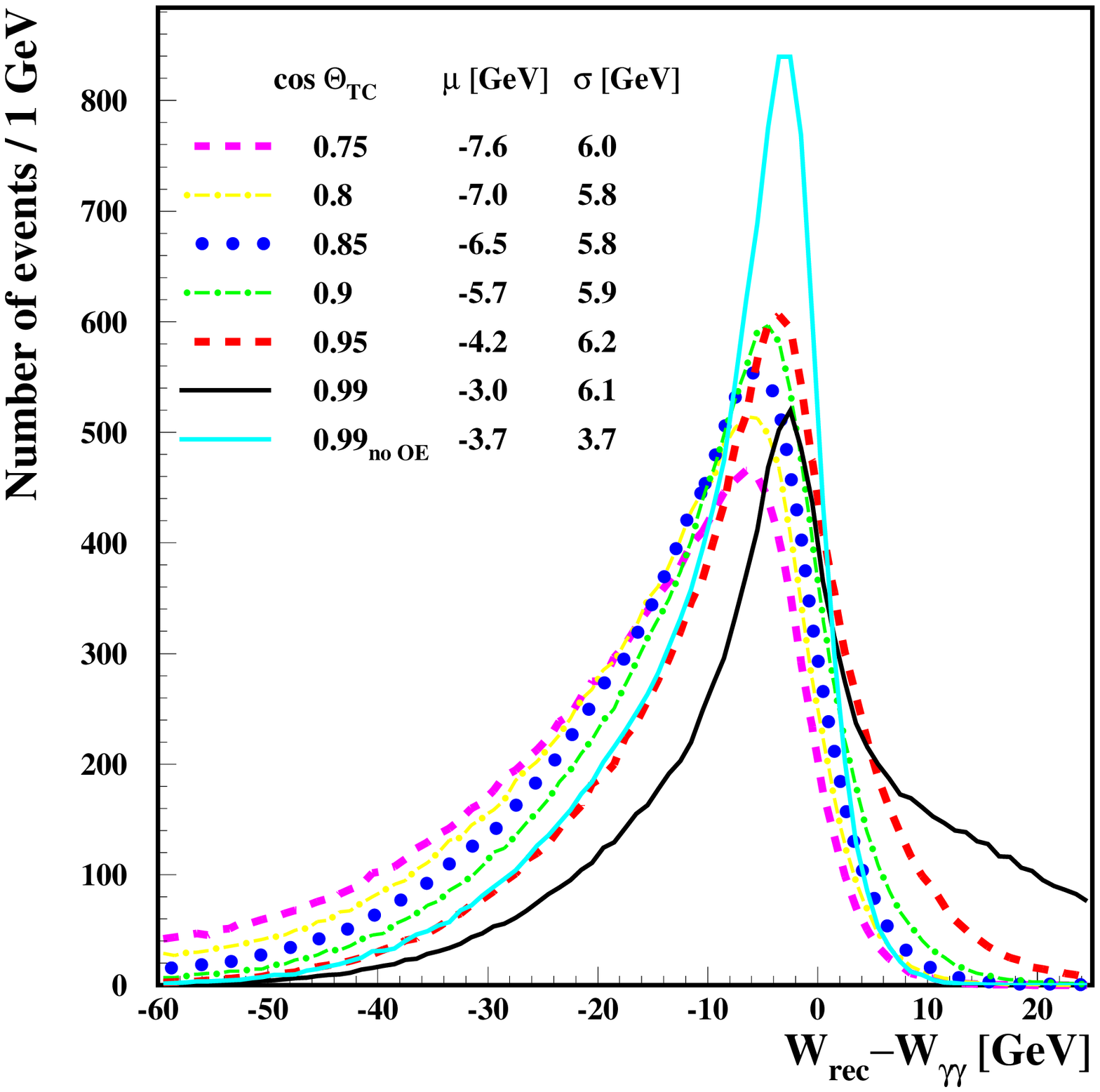} 
\includegraphics{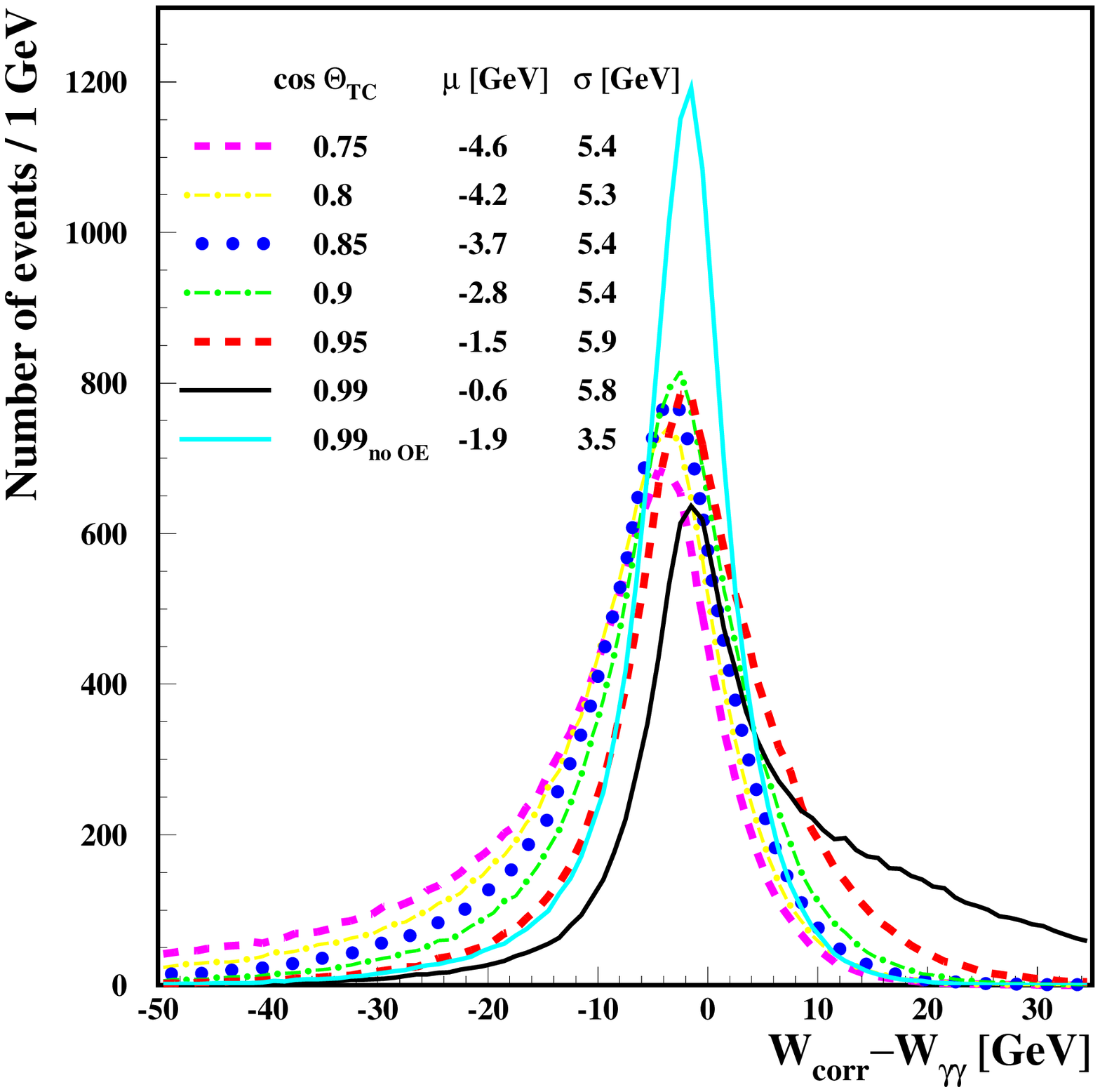}}{fig:mres_120}{
Distribution  of $W_{rec}-\Wgaga$  (left) and $W_{corr}-\Wgaga$ (right) 
for accepted $\gagahbb$ events, for \sqrtseeeq 210~GeV and \Mheq 120~GeV.
Various values of $\thetamindet$  are considered. 
Also results for the ideal case without overlaying events are shown (no OE).
The mean value, $\mu$, and the dispersion, $\sigma$,  
of the fitted Gaussian distribution is presented for each case.
The fit was performed in the range $(\mu - \sigma,\; \mu + 1.5\sigma)$.
}

\pnfiggeneral{p}{\twofigheight}{\includegraphics{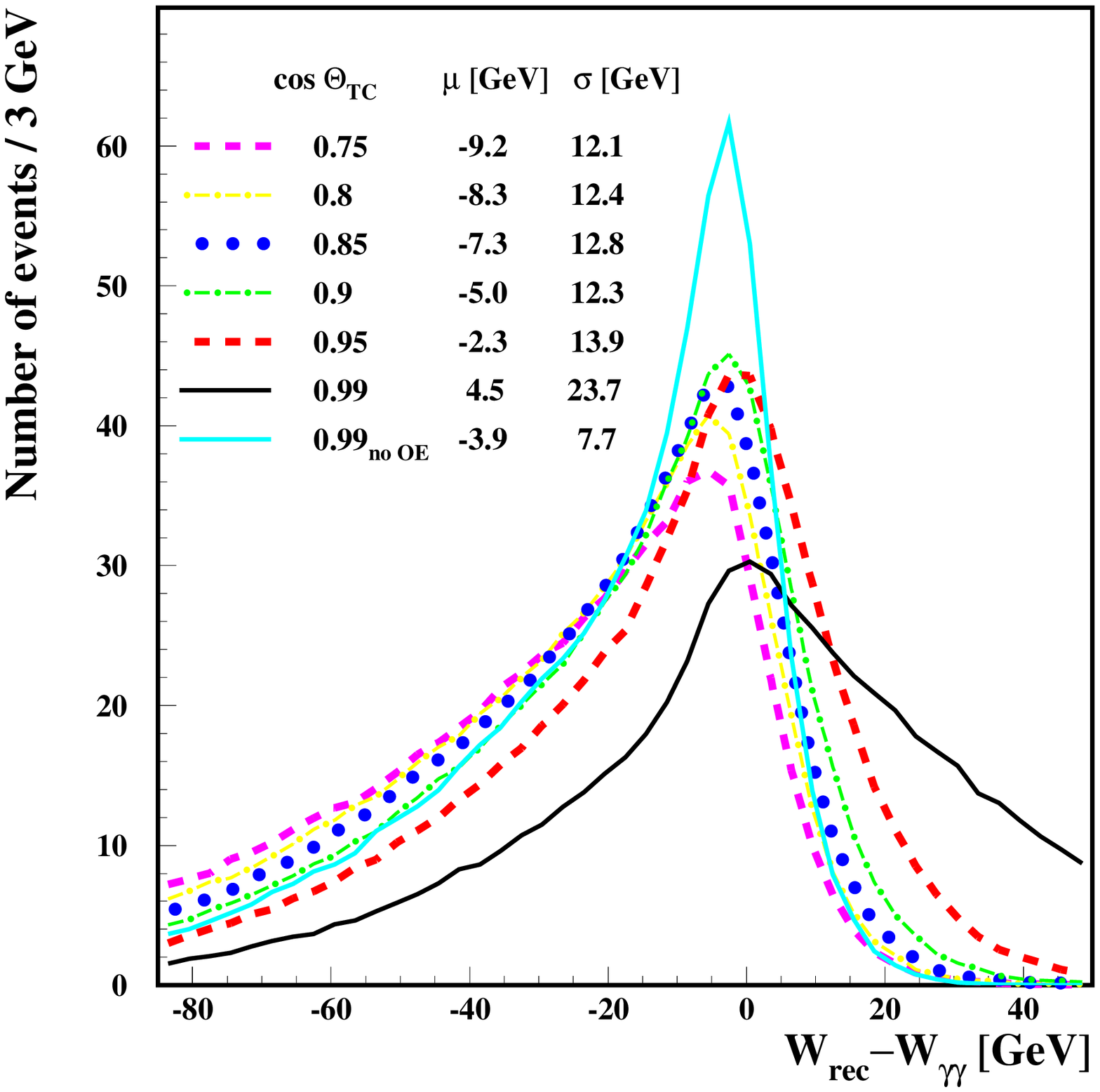} 
\includegraphics{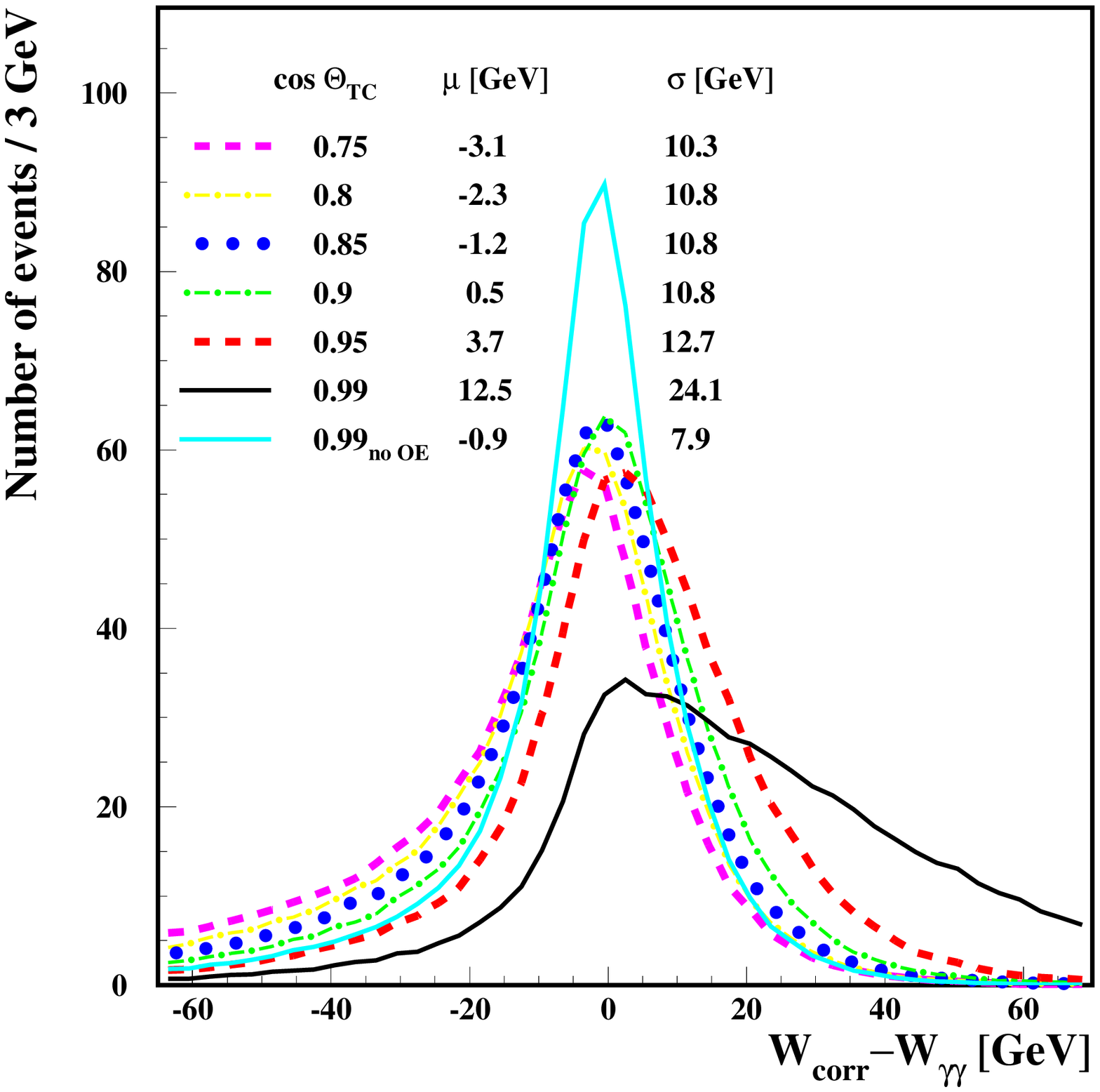}}{fig:mres_300}{
Distribution of $W_{rec}-\Wgaga$  (left) and $W_{corr}-\Wgaga$ (right)
for accepted $\gagaAbb$ events, for \sqrtseeeq 419~GeV and \MAOeq 300~GeV.
Various values of $\thetamindet$  are considered. 
Also results for the ideal case without overlaying events are shown (no OE).
The mean value, $\mu$, and the dispersion, $\sigma$,  
of the fitted Gaussian distribution is presented for each case.
The fit was performed in the range $(\mu - \sigma,\; \mu + 1.5\sigma)$.
}

\chapter{\Orlop{} \label{app_orlop}}

In this Appendix we describe
the package \Orlop{} 
(Ove{\textsc{\footnotesize RL}}aying events f{\textsc{\footnotesize O}}r {\textsc{\footnotesize P}}hoton collider) 
which was developed to take \gagahad{} overlaying events  into account
in our analysis.
The package includes subroutines which read in the luminosity spectra for $\gaga$ collisions, 
appropriately rescale energies of beam photons (if necessary),
generate \gagahad{} events  with 
variable beam energies, and add random number of the overlaying events
to the earlier generated events of the studied hard scattering process.
Luminosity spectra obtained from the full simulation of $\gaga$ collisions \cite{V.TelnovSpectra} 
are used to account properly for the low energy
part of the spectrum.
The results of the simulations are available only for three values of $\emem$ invariant mass,
\sqrtseeeq 200, 500 and 800~GeV.
The luminosity spectrum for user-defined value of $\sqrtspee$
is obtained by rescaling photon energies from the closest set 
by the factor $E'^{\max\!1}_{\ga}/E^{\max\!1}_{\ga}$,
where  $E'^{\max\!1}_{\ga}$ and $E^{\max\!1}_{\ga}$ are maximal energies
of photons coming from electrons scattering off one laser photon for $\sqrtspee$
and  $\sqrtsee$, respectively.
To obtain $E'^{\max\!1}_{\ga}$ and $E^{\max\!1}_{\ga}$  values the \CompAZ{} program is used \cite{CompAZ}. 
The \Orlop{} package can be easily extended to provide overlaying events also to $\epem$ collisions
if the beamstrahlung photon spectrum is implemented.
The main program  \verb'gen_orlop', running as a separate process, uses \Pythia{} to generate \gagahad{} events
with variable beam energies,
according to the realistic luminosity spectrum.
As \Pythia{} sometimes does not succeed to hadronize events with very low invariant mass,  
\verb'gen_orlop' uses only $\gaga$ collisions with $\Wgaga > \Wgaga^{\min} = 4$~GeV
(see Appendix \ref{app_thetatc} for detailed discussion).
The program sets following values for \Pythia{} steering variables: 
\verb'MSTP(14)=10' to properly include all kinds of real photon interactions,
and \verb'MSEL=2' to switch on elastic and diffractive processes.
The crossing angle and the interaction point distribution are taken into account.
At the beginning the program is run for 50000 events to calculate more precisely internal \Pythia{} weights which
influence contributions of various event classes.
The calculated total cross section is used to determine 
the average number of overlaying events per bunch crossing, $\mu$.
The subroutine  \verb'add_orlop' is called to include overlaying events 
during generation of the considered processes.
This subroutine adds   
particles which are provided by the program \verb'gen_orlop'
to the  \Pythia{} event record.
The call sequence of \verb'add_orlop' is shown in the following listing
where only parts relevant for use of \Orlop{} are present.

\singlespacing
\begin{verbatim}
*     Example of ORLOP use: 
*       Higgs-boson production 'gamma gamma -> h -> b bbar' 
*       with overlaying events 'gamma gamma -> hadrons'. 
      program produce_higgs_with_oe

*     Set Higgs-boson production 'gamma gamma -> h -> b bbar'

*     Initialization of ORLOP
      call add_orlop(1,ibeams,sqrts_ee,n_oe_per_bc)

*     Additional information (optional call)
      call add_orlop(2,n_collisions,luminosity,cross_sec)

*     Loop over bunch crossings
      do ibc=1,n_bc

*        Set photons energies for 'gamma gamma -> h -> b bbar'

*        Generate 'gamma gamma -> h -> b bbar' event
         call pyevnt
         
*        Add overlaying events 'gamma gamma -> hadrons'
         call add_orlop(3,n_oe,e_oe,pz_oe)
         
      enddo
*     End of loop over bunch crossings

*     Stop ORLOP
      call add_orlop(4,n_oe_tot,n_rewind,cross_sec)

      end
\end{verbatim}
\doublespacing

In calls of the subroutine \verb'add_orlop(iflag,ivar,var1,var2)' 
the action undertaken 
and also the meaning of variables \verb'ivar,var1,var2' 
depend on the value 
of the first parameter \verb'iflag' as explained below. 
The package \Orlop{}, which contains also a manual, 
shell scripts for convenient running of programs,
and two examples of generator-level analysis,
will be made publicly available  after publication of results obtained in this thesis. \\[1cm]
\verb'iflag=1'
\hspace*{3mm} \begin{minipage}[t]{0.85\textwidth}
Initialization. 
The main parameters for \gagahad{} generation are passed
to the \verb'gen_orlop' program. Variables \verb'ivar' and \verb'var1'
should be equal to the beam type and $\sqrtsee$, respectively.
These -- together with \verb'iflag' -- are the only input parameters.
Beam type is at the moment defined only for $\gaga$ collisions in the Photon Collider at TESLA.
Variable \verb'var2' is set to the average number of overlaying events per bunch crossing, $\mu$,
calculated by \verb'gen_orlop'. 
\end{minipage} \\ \\
\verb'iflag=2' 
\hspace*{3mm} \begin{minipage}[t]{0.85\textwidth}
Optional call allowing to obtain additional information from \Orlop{}.  
On return, variable \verb'ivar' is equal to the number of  beam collisions which are available in the simulated set.
Variables \verb'var1' and  \verb'var2' are equal to the luminosity of used spectrum [pb$^{-1}$s$^{-1}$]
and to the \gagahad{} cross section [pb]. 
\end{minipage} \\ \\
\verb'iflag=3' 
\hspace*{3mm} \begin{minipage}[t]{0.85\textwidth}
The call  which actually adds overlaying events to the event record.  
The returned value of variable \verb'ivar' is equal to the number of \gagahad{} events added.
This is a random number from Poisson distribution with mean   $\mu$ (see description of \verb'iflag=1').
Variables \verb'var1' and  \verb'var2' are set to the total energy and to the longitudinal
momentum of overlaying events. 
Before overlaying events are added all undetectable objects (except neutrinos) are removed from event record
with \verb'call pyedit(1)'.
\end{minipage} \\ \\
\verb'iflag=4' 
\hspace*{3mm} \begin{minipage}[t]{0.85\textwidth}
Terminate \gagahad{} generation by stopping \verb'gen_orlop'.  
Summary of overlaying events generation is returned.
Variable \verb'ivar' is set to the number of all added \gagahad{} events.
Variable \verb'var1' is equal to the number of times the available set of beam collisions was used.
Variable \verb'var2' is equal to the \gagahad{} final value of the cross section [pb] calculated by \Pythia{};
this can be compared to the value provided with \verb'iflag=2' to check stability of the result. 
\end{minipage} \\ \\



%






\
\newpage 
\

\end{document}